\pdfoutput=1
\documentclass[11pt,twoside,a4paper,cmspaper,final,collab]{cms-tdr}
\def\svnVersion{a6dcf6a}\def\svnDate{2025/04/07}\def\cmsCernNoTag{CERN-EP-2024-117}\def\cmsCernDate{\today}\def\cmsMessage{Published in Physics Reports as \href{http://dx.doi.org/10.1016/j.physrep.2024.11.005}{\doi{10.1016/j.physrep.2024.11.005}.}}
\begin{document}\cmsNoteHeader{SMP-23-004}

\newcommand{\invfb}{\fbinv}
\newcommand{\sqrts}{\ensuremath{\sqrt{s}}\xspace}

\newcommand{\mb}{\unit{mb}}
\newcommand{\pb}{\unit{pb}}
\newcommand{\fb}{\unit{fb}}

\newcommand{\scale}{\ensuremath{\,\text{(scale)}}\xspace}
\newcommand{\pdf}{\ensuremath{\,\text{(PDF)}}\xspace}
\newcommand{\tot}{\ensuremath{\,\text{(tot)}}\xspace}
\newcommand{\experr}{\ensuremath{\,\text{(exp)}}\xspace}
\newcommand{\fit}{\ensuremath{\,\text{(fit)}}\xspace}
\newcommand{\num}{\ensuremath{\,\text{(num)}}\xspace}
\newcommand{\param}{\ensuremath{\,\text{(param)}}\xspace}
\newcommand{\model}{\ensuremath{\,\text{(model)}}\xspace}

\newcommand{\Vzero}{\ensuremath{\text{V}^0}\xspace}

\newcommand{\PosWW}{\PWp\PWm}
\newcommand{\PssWW}{\PWpm\PWpm}

\newcommand{\HJMINLO} {{\textsc{HJ-MiNLO}}\xspace}
\newcommand{\JHUGEN}{\textsc{JHUGen}\xspace}
\newcommand{\NNLOPS}{\textsc{nnlops}\xspace}
\newcommand{\OPENLOOPS}{\textsc{OpenLoops}\xspace}
\newcommand{\MATRIX}{\textsc{matrix}\xspace}
\newcommand{\PHOJET}{\textsc{phojet}\xspace}
\newcommand{\QGSJET}{\textsc{QGSJET-II}\xspace}
\newcommand{\SIBYLL}{\textsc{sibyll}\xspace}
\newcommand{\EPOS}{\textsc{epos}\xspace}
\newcommand{\HELACOnia}{\textsc{HELAC-Onia}\xspace}
\newcommand{\JETPHOX}{\textsc{Jetphox}\xspace}
\newcommand{\COMIX}{\textsc{Comix}\xspace}
\newcommand{\CSSHOWER}{\textsc{CSShower}\xspace}
\newcommand{\DYTURBO}{\textsc{DYTurbo}\xspace}
\newcommand{\VBFNLO}{\textsc{vbfnl0}\xspace}
\newcommand{\LPAIR}{\textsc{lpair}\xspace}
\newcommand{\POWHEGBOX}{\textsc{powheg bpx}\xspace}
\newcommand{\NLOJETpp}{\textsc{NLOJet++}\xspace}
\newcommand{\FASTNLO}{\textsc{fastNLO}\xspace}
\newcommand{\NLLJET}{\textsc{NLLJet}\xspace}
\newcommand{\NNLOJET}{\textsc{NNLOJet}\xspace}
\newcommand{\CASCADE}{\textsc{cascade}\xspace}

\newcommand{\HERAFITTER}{\textsc{HERAFitter}\xspace}
\newcommand{\XFITTER}{\textsc{xFitter}\xspace}

\newcommand{\WW}{\ensuremath{\PW\PW}\xspace}
\newcommand{\WWSS}{\ensuremath{\PWpm\PWpm}\xspace}
\newcommand{\WWOS}{\ensuremath{\PWpm\PWmp}\xspace}
\newcommand{\WZ}{\ensuremath{\PW\PZ}\xspace}
\newcommand{\ZZ}{\ensuremath{\PZ\PZ}\xspace}

\newcommand{\ttH}{\ensuremath{\PQt\PQt\PH}\xspace}
\newcommand{\tH}{\ensuremath{\PQt\PH}\xspace}
\newcommand{\ptH}{\ensuremath{p_\text{T}^{\PH}}\xspace}
\newcommand{\PHj}{\ensuremath{\PH j}\xspace}
\newcommand{\PHjj}{\ensuremath{\PH jj}\xspace}

\newcommand{\absyw}{\ensuremath{\abs{y_{\PW}}}\xspace}
\newcommand{\afb}{\ensuremath{A_\mathrm{FB}}\xspace}
\newcommand{\ncube}{\ensuremath{\text{N}^3\text{LO}}\xspace}

\newcommand{\pp}{\ensuremath{\Pp\Pp}\xspace}
\newcommand{\fbinvns}{\mbox{\ensuremath{\text{fb}^{-1}}}\xspace}

\newlength\cmsTabSkip\setlength{\cmsTabSkip}{1ex}
\providecommand{\cmsTable}[1]{\resizebox{\textwidth}{!}{#1}}

\ifthenelse{\boolean{cms@external}}{
  \providecommand{\maybeCmsTable}[1]{{#1}}
}{
\providecommand{\maybeCmsTable}[1]{\resizebox{\textwidth}{!}{#1}}
}

\cmsNoteHeader{SMP-23-004}

\title{Stairway to discovery: a report on the CMS programme of cross section measurements from millibarns to femtobarns}

\date{\today}

\abstract{
The Large Hadron Collider at CERN, delivering proton-proton collisions at much higher energies and far higher luminosities than previous machines, has enabled a comprehensive programme of measurements of the standard model (SM) processes by the CMS experiment. These unprecedented capabilities facilitate precise measurements of the properties of a wide array of processes, the most fundamental being cross sections. The discovery of the Higgs boson and the measurement of its mass became the keystone of the SM. Knowledge of the mass of the Higgs boson allows precision comparisons of the predictions of the SM with the corresponding measurements. These measurements span the range from one of the most copious SM processes, the total inelastic cross section for proton-proton interactions, to the rarest ones, such as Higgs boson pair production. They cover the production of Higgs bosons, top quarks, single and multibosons, and hadronic jets. Associated parameters, such as coupling constants, are also measured. These cross section measurements can be pictured as a descending stairway, on which the lowest steps represent the rarest processes allowed by the SM, some never seen before.\\[2ex]
\textit{We dedicate this work to the memory of Prof.\ Peter Ware Higgs, whose transformative 
and groundbreaking ideas laid the foundation for the physics of the standard model and 
of the Higgs particle, which are the subjects of this Report.}
}

\hypersetup{
pdfauthor={CMS Collaboration},
pdftitle={Stairway to discovery: a report on the CMS programme of cross section measurements from millibarns to femtobarns},
pdfsubject={CMS},
pdfkeywords={top quark,Higgs boson,cross section,standard model,electroweak,QCD}}

\maketitle 

\tableofcontents

\section{Introduction}\label{sec:intro}
The Large Hadron Collider (LHC) at CERN, colliding protons at much higher energies and delivering far higher luminosities than previous machines, has enabled comprehensive measurements of the standard model (SM) of particle physics by the general-purpose experiments, CMS and ATLAS. The Higgs boson plays a special role in the SM, being the particle predicted by the Brout--Englert--Higgs (BEH) spontaneous electroweak (EW) symmetry-breaking mechanism. The discovery of the Higgs boson and the measurement of its mass became the keystone of the SM. This allowed significantly tightening the constraints on the theory and facilitated precision comparison of predictions with the corresponding measurements.

\begin{figure}[ht]
	\centering 
	\includegraphics[width=0.98\textwidth]{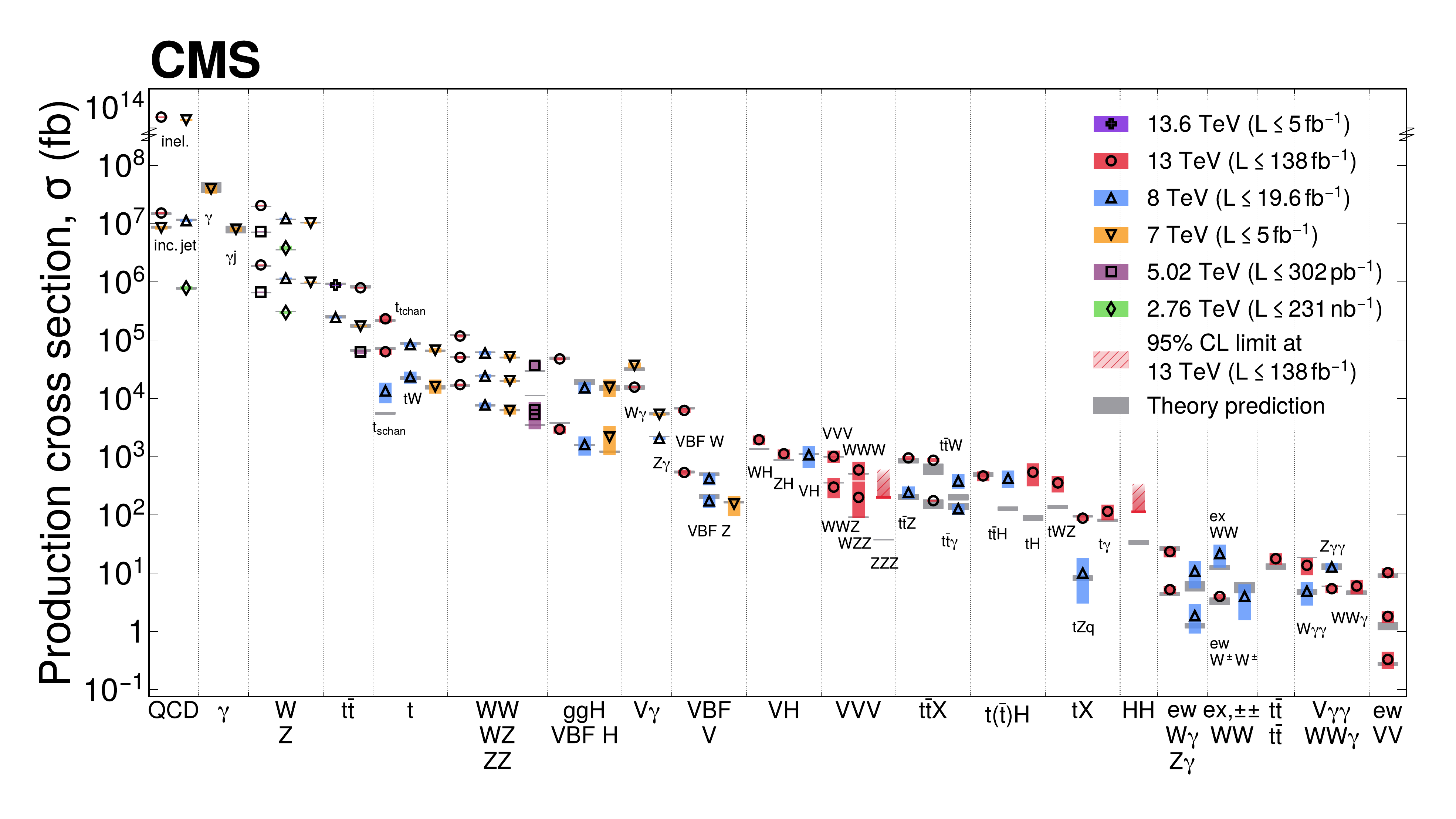}	
	\caption{Cross sections of selected high-energy processes measured by the CMS experiment.  Measurements performed at different LHC \pp collision energies are marked by unique symbols and the coloured bands indicate the combined statistical and systematic uncertainty of the measurement.   Grey bands indicate the uncertainty of the corresponding SM theory predictions.   Shaded hashed bars indicate the excluded cross section region for a production process with the measured 95\% \CL upper limit on the process indicated by the solid line of the same colour.}\label{fig:XsAll}
\end{figure}

The unprecedented capabilities of the LHC detectors have enabled precise measurements of the properties of a wide array of processes. The most fundamental of the properties is the cross section, which quantifies the probability of two particles interacting and producing a particular final state. Figure~\ref{fig:XsAll} shows the cross sections of
selected high-energy processes measured by the CMS experiment spanning some fourteen orders of magnitude, stepping from the total inelastic proton-proton (\pp)~cross section to the production of hadronic jets, single and multibosons, top quarks, Higgs bosons, down to the rarest processes, such as vector boson scattering of \PZ boson pairs and the production of four top quarks, the most massive of the SM particles. Since the start of operation, the LHC has operated at several increasing energies allowing the experiments to map the change of cross sections with energy. The agreement in Fig.~\ref{fig:XsAll} between the SM predictions and the measurements is remarkable.

In this Report, we exemplify the full spread of the CMS experimental programme in measuring cross sections involving high-energy quantum
chromodynamics (QCD) and EW processes, including those involving the top quark and those involving the Higgs boson. We point out the fundamental aspects of the SM elucidated by these cross section measurements, highlighting their importance. Accurate measurements of fundamental parameters, such as the Higgs boson mass, top quark mass, their production cross sections, along with the strong coupling constant and other SM parameters, play a pivotal role in refining the SM. They also contribute significantly to shaping a more accurate and comprehensive model of the origin of matter and of cosmology, \eg by understanding the features that affect the early universe and its eventual fate: the shape of the BEH vacuum potential and the EW vacuum stability, respectively.

The construction and operation of the LHC and the CMS and ATLAS detectors are a product of the accumulated experience of the high-energy physics community.
The instantaneous luminosity provided by the LHC exceeds that of the most recent previous hadron collider, the Fermilab Tevatron, by nearly two orders of magnitude.
The higher \pp collision energy significantly increases all production cross sections. This enables, for many processes, the collection of data sets, sometimes in only days, that match those of the entire experimental programme of previous experiments. For example, the precise measurement of the \PW and \PZ boson production cross sections can be performed in CMS with data collected in one day of LHC operation with a precision similar to that obtained during several years of operation of the UA1 and UA2 experiments that discovered the \PW and \PZ bosons.

The CMS detector at the LHC has performed both as a discovery instrument, observing a new particle---the Higgs boson---and new production processes, such as vector boson scattering and \ttbar\ttbar production, and as a cross section measuring device with the precision substantially exceeding that of previous experiments for a wide variety of final states. The CMS detector has a larger angular acceptance than the previous generation of hadron collider experiments. It measures physics objects, electrons, muons, tau leptons, photons, and jets, with higher efficiency, better precision, better purity, and fewer gaps in geometric coverage. These capabilities both expand the CMS potential and enable cross section measurements with high precision. The ability to measure new states in the SM allows CMS to study new aspects of the gauge structure of the theory, processes involving the top quark, explore the mechanism of EW symmetry breaking, and to search for beyond-the-SM (BSM) physics. The Higgs sector, currently only accessible at the LHC, is an ideal place to study the SM and to simultaneously look for signs of BSM physics signalled by deviations from the predictions of the SM.

For a given process, with a particular final state, the number of events produced, $n$, is given by the product of the instantaneous luminosity, \lumi, and the cross section, $\sigma$, integrated over the time during which the events are recorded, \ie $n = \int{\lumi \sigma \rd{t}}$. The instantaneous luminosity, which is expressed as an inverse cross section per
unit of time, $t$, depends on the number of protons in the colliding bunches, the frequency with which the bunches collide, and the lateral size and overlap of the bunches.
The unit of cross section used in particle physics is the barn, where the barn is defined as $10^{-24}\cm^2$. Cross sections of production processes involving heavy SM particles are typically of the order of nanobarns (nb), picobarns (pb), or femtobarns (fb). 

Not all events produced are observed due to limitations in the acceptance and efficiency of the detectors.
The acceptance, $A$, is the fraction of events in which the kinematics of the final state particles are such that they traverse, or impact, a detector with the capability to measure them. The efficiency, $\epsilon$, is the fraction of events within the acceptance that are detected.
Thus if $N$ signal events are observed $\sigma$ is given by:
\begin{center}
  $\sigma = N / \int (\lumi A \epsilon) \rd{t}$.
\end{center}

We frequently measure a ``fiducial'' cross section, that is the part of the cross section that corresponds to a defined set of kinematic requirements on the final-state particles for which the acceptance is high.
Measuring fiducial cross sections eliminates theoretical uncertainties related to the extrapolation from the fiducial phase space to the full phase space.

In the following sections, we first describe the LHC operation and the CMS detector; discuss the simulations and calculations used to predict cross sections; and then report cross sections, fiducial cross sections, and selected differential cross sections (cross sections as functions of kinematic variables) covering high-energy QCD and EW processes, including processes involving the top quark and the Higgs boson. Finally, we include projections for High-Luminosity LHC and conclude with a brief summary of the results.

The results shown in the primary summary plots of this report 
are tabulated in the HEPData record~\cite{hepdata}.

\section{The LHC and CMS}
\subsection{LHC operations, energies, and luminosities}\label{sec:lhcandcms}
The LHC has operated providing collisions to feed its physics programme over three runs, with long shutdowns in between for collider and detector maintenance, and upgrades. In Run~1 from 2010 to 2012, the LHC operated at 7\TeV (2010--2011) and 8\TeV (2012) providing 6.1\fbinv and 23.3\fbinv of \pp collision data, respectively, to the CMS experiment. In Run~2 from 2015 to 2018, the LHC increased the collision energy to 13\TeV and eventually more than doubled the peak luminosity providing 163.6\fbinv of \pp collision data to the CMS experiment. In Run~3, currently in progress (since 2022), the LHC has increased the collision energy to 13.6\TeV and also increased the peak luminosity. The Run~3 results presented in this Report use data collected during the first year of Run~3 operation.
Only a subset of Run~3 data has been analyzed and used in this Report.

The CMS experiment typically operates and records data for over 90\% of the LHC operational time, with the detector working at peak performance suitable for physics analysis 88\% of the LHC operational time. The LHC has additionally operated for short periods taking \pp collision data at collision energies of 2.76\TeV and 5.02\TeV as reference for heavy ion collision runs having those collision energies per nucleon pair.

\textit{CMS integrated luminosity:}
The integrated luminosities collected by the CMS experiment for each LHC running period are listed in Table~\ref{tab:lumi}. The integrated luminosity for 2016--2018 Run~2 period was reevaluated, achieving a lower uncertainty and an increase in the evaluated value from 137 to 138\fbinv. The total integrated luminosity of Run~2 is known with a better relative uncertainty than that of subperiods of data taking within Run~2. The integrated luminosities for the years 2015--2018 of LHC Run~2 data taking have individual uncertainties between 1.2 and 2.5\%~\cite{CMS-LUM-17-003,CMS-PAS-LUM-17-004,CMS-PAS-LUM-18-002}, and the overall uncertainty for the 2016--2018 period used in most of the analyses included in this Report is 1.6\%. The Run~1 absolute integrated luminosity of the \pp collisions at 7 and 8\TeV has been determined with a relative precision of 2.2\% and 2.6\%, respectively~\cite{CMS-PAS-SMP-12-008,CMS-PAS-LUM-13-001}.
The Run~3 integrated luminosity is measured using the techniques from the 2015--2016 Run~2 luminosity determination~\cite{CMS-LUM-17-003} and is estimated to be 2.1\%~\cite{CMS-PAS-LUM-22-001}.

Some measurements, for instance \PW and \PZ cross sections, were performed using short runs of \pp collision data with features such as low instantaneous luminosity in order to improve measurement uncertainties. These measurements use luminosity determinations specific to those runs with the uncertainties reported with the corresponding analysis.  Other measurements use partial data sets typically corresponding to specific calendar years. Finally, some measurements, for instance inelastic cross section measurements, use early data collected in a given run with luminosity determinations that are significantly less precise.

\begin{table}[htbp]
  \centering
  \caption{Integrated \pp collision luminosity \lumi, analyzed by the CMS experiment during LHC Runs~1, 2 and 3, as well as during \pp reference runs for the heavy ion physics programme at 2.76 and 5.02\TeV. Since the LHC Run 3 is in progress, the results presented in this Report use data only from the first year of data taking (2022).\label{tab:lumi}}
\renewcommand{\arraystretch}{1.2}
  \begin{tabular}{c c c c c }
    Run &  Energy~($\TeVns$) & $\lumi~({\fbinvns})$ & Uncertainty\\
    \hline
    1  & $ 7 $    & $5.0 $ & $ 2.2\%$    \\
    1  & $ 8 $    & $19.6 $ & $ 2.6\%$   \\
    2  & $ 13 $   & $138 $ & $ 1.6\%$   \\
    3  & $ 13.6 $ & $ 5.0 $ & $ 2.1\%$   \\
    1  & $ 2.76 $ & $ 2.31\times 10^{-4} $ & $ 3.7\%$    \\
    2  & $ 5.02 $ & $0.302 $ & $ 1.9\%$    \\
  \end{tabular}
\end{table}

\subsection{The CMS detector}\label{subsec:cmsdetector}
The central feature of the CMS apparatus is a superconducting solenoid of 6\unit{m} internal dia\-meter, providing a magnetic field of 3.8\unit{T}. The large size of the solenoid allows the inner tracker and almost all the calorimetry to be installed inside the solenoid.
Thus, within the magnetic volume are a silicon pixel and strip tracker, a lead tungstate crystal electromagnetic calorimeter (ECAL), and a brass and scintillator hadron calorimeter (HCAL), each composed of a barrel and two endcap sections. The geometric coverage of the ECAL and HCAL goes down to an angle of about 6$^\circ$ from the beamline, \ie at a pseudorapidity $\abs{\eta}$ of about 3.
The hadron forward (HF) calorimeter extends the $\eta$ coverage, using steel as an absorber with quartz fibres embedded in a matrix arrangement as the sensitive material.
Half of the fibers extend over the full depth of the detector (long fibers) while the other half does not cover the first 22\unit{cm} measured from the front face (short fibers). As the two sets of fibers are read out separately, electromagnetic showers can be distinguished from hadronic showers.
The two halves of the HF are located 11.2\unit{m} from the interaction region, one at each end, and together they provide coverage in the range $3.0 < \abs{\eta} < 5.2$.
They also serve as luminosity monitors. The very forward angles are covered at one end of CMS ($-6.6 < \eta < -5.2$) by the CASTOR calorimeter~\cite{Andreev:2010zzb}.
Muons are measured in gas-ionization detectors embedded in the steel flux-return yoke outside the solenoid. The precision proton spectrometer~\cite{Albrow:1753795} (PPS) is a system of near-beam tracking and timing detectors, located in Roman pots (RPs) at about 200\unit{m} from the CMS interaction point.  

\begin{figure}
	\centering 
	\includegraphics[width=0.98\textwidth]{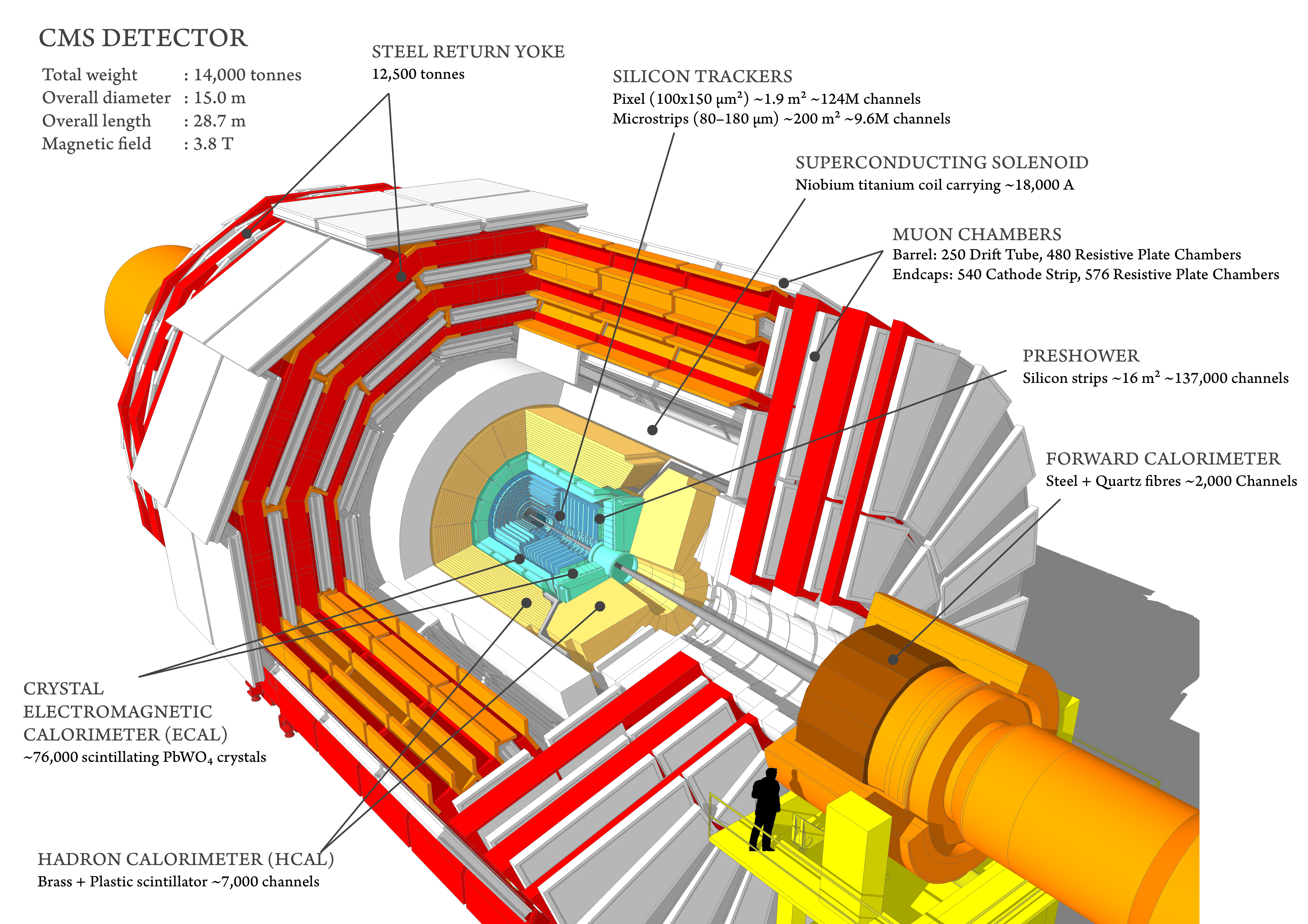}	
	\caption{The CMS detector for the data-taking period 2017--2018.}\label{fig:CMS}
\end{figure}

A detailed description of the CMS detector, together with a definition of the coordinate system used and the relevant kinematic variables, is given in Ref.~\cite{CMS:2008xjf}.  The upgraded configuration of the detector for the LHC Run~3 is given in Ref.~\cite{CMS:2023gfb}. The CMS detector as it was configured during 2017--2018 is shown in Fig.~\ref{fig:CMS}.

Calibration of the calorimeters and alignment of the tracking systems have played an important role in both maintaining and improving the performance of the detector as refined techniques are developed. The calorimeter calibration includes both relative calibration of the
detector elements, in particular following changes in
response (typically those resulting from radiation-induced effects on the scintillating materials), and also absolute calibration of the physics objects, electrons, photons, and jets, using, \eg the mass of the \PZ boson as a reference. Alignment of the tracker uses tracks of charged particles to improve upon the original information about the relative positions of the various detector modules and from the laser alignment system. 

As described in Section~\ref{sec:lhcandcms} there have been three periods of LHC operation: Runs 1-3. The Run~3 analyses covered in this Report typically rely on the methods developed for Run~2. In the description of the CMS event selection and reconstruction below, substantial differences in the CMS operation and methodology between these operational periods are noted.

\textit{Trigger:}
Events of interest are selected using a two-tiered trigger system. The first level, composed of custom hardware processors, uses information from the calorimeters and muon detectors to select events at a rate of around 100\unit{kHz} within a fixed latency of about 4\mus~\cite{CMS:2020cmk}. The second level, known as the high-level trigger, consists of a farm of processors running a version of the full event reconstruction software optimized for fast processing, and reduces the rate of selected events to around 1\unit{kHz} before data storage~\cite{CMS:2016ngn}.

\textit{Particle-flow:}
The global event reconstruction (also called particle-flow event reconstruction~\cite{CMS:2017yfk}) aims at reconstructing and identifying each individual particle in an event, with an optimized combination of all subdetector information. In this process, the identification of the particle type (photon, electron, muon, charged or neutral hadron) plays an important role in the determination of the particle direction and energy. Photons, both prompt, produced in parton-parton collisions, and nonprompt, \eg from \PGpz decays or electron bremsstrahlung, are identified as ECAL energy clusters not linked to the extrapolation of any charged-particle trajectory into the ECAL. Prompt electrons and nonprompt electrons, which come from photon conversions in the tracker material or \PQb hadron semileptonic decays, are identified as a primary charged-particle track with potentially more than one ECAL energy cluster, corresponding to the track, as extrapolated to the ECAL and possible bremsstrahlung photons emitted by the electron as it traverses the tracker material. Prompt muons and nonprompt muons, which come from \PQb hadron semileptonic decays, are identified as tracks in the central tracker consistent with either a track or several hits in the muon system, and associated with energy deposits in the calorimeter compatible with the muon hypo\-thesis. Charged hadrons are identified as charged-particle tracks neither identified as electrons, nor as muons. Finally, neutral hadrons are identified as HCAL energy clusters not linked to any charged-hadron trajectory, or as a combined ECAL and HCAL energy excess with respect to an expected charged-hadron energy deposit.

The energy of photons is obtained from the ECAL measurement. The energy of electrons is determined from a combination of the track momentum at the main interaction vertex, the corresponding ECAL cluster energy, and the energy sum of all bremsstrahlung photons assigned to the track. The energy of muons is obtained from the corresponding track curvature. The energy of charged hadrons is determined from a combination of the track momentum and the corresponding ECAL and HCAL energies, corrected for the response function of the calorimeters to hadronic showers. Finally, the energy of neutral hadrons is obtained from the corresponding corrected ECAL and HCAL energies. The reconstruction of each of these individual physics objects is described below.

\textit{Electrons:}
Electrons are identified and measured in the range $\abs{\eta} < 2.5$. The momentum resolution for electrons with transverse momentum $\pt \approx 45\GeV$ from $\PZ \to \Pe \Pe$ decays ranges 1.6--5.0\% in Run~2, and 1.7--4.5\% in Run~1. The resolution is better in the barrel region than in the endcaps, and also depends on the bremsstrahlung energy emitted by the electron as it traverses the material in front of the ECAL~\cite{CMS:2015xaf,CMS:2020uim,CMS-DP-2020-021}. 

The dielectron mass resolution for $\PZ \to \Pe \Pe$ decays is in the ranges 1.2--2.0\%  (1.9\% in Run~1) when both electrons are in the ECAL barrel, and 2.2--3.2 (2.9\% in Run~1) otherwise, the exact values depending on the bremsstrahlung energy emitted by the electrons and the data-taking year~\cite{CMS:2015xaf,CMS-DP-2020-021}.

Electrons in the HF are measured in the range $3 < \abs{\eta} < 5$ with an energy resolution of approximately 32\% at 50\GeV and a resolution of 0.05 in $\eta$ and $\phi$.

\textit{Photons:}
Photons are identified and measured in the range $\abs{\eta} < 2.5$. In the barrel section of the ECAL, an energy resolution of about 1\% is achieved for unconverted or late-converting photons in the tens of \GeV energy range. The energy resolution of the remaining barrel photons is about 1.3\% up to $\abs{\eta} = 1$, worsening to about 2.5\% by $\abs{\eta} = 1.4$. In the endcaps, the energy resolution is about 2.5\% for unconverted or late converting photons, and 3--4\% for the rest~\cite{CMS:2015myp}.

The diphoton mass resolution, as measured in $\PH \to \PGg\PGg$ decays, is typically in the 1--2\% range, depending on the topology of the photons~\cite{CMS:2020xrn}.

\textit{Muons:}
Muons are identified and measured in the range $\abs{\eta} < 2.4$, with detection planes made using three technologies: drift tubes, cathode strip chambers, and resistive-plate chambers. The single-muon trigger efficiency exceeds 90\% over the full $\eta$ range, and the efficiency to reconstruct and identify muons is greater than 96\%. Matching muons identified in the muon detection system to tracks measured in the silicon tracker results in a \pt resolution, for muons with \pt up to 100\GeV, of 1\% (1.3--2.0\% in Run~1) in the barrel and 3\% (6\% in Run~1) in the endcaps. For muons with \pt up to 1\TeV, the \pt resolution in the barrel is better than 7\% (10\% in Run~1)~\cite{Chatrchyan:2012xi,CMS:2018rym}.

\textit{Taus:}
Hadronic \PGt decays (\tauh) are reconstructed from jets, using the hadrons-plus-strips algorithm~\cite{CMS:2018jrd}, which combines one or three tracks with energy deposits in the calorimeters, to identify the tau lepton hadronic decay modes. Neutral pions are reconstructed as strips with a dynamic size in $\eta$--$\phi$ (where $\phi$ is the azimuthal angle about the beam axis, measured in radians) from reconstructed electrons and photons, where the strip size varies as a function of the $\pt$ of the electron or photon candidate.

To distinguish \tauh decays from jets originating from the hadronization of quarks or gluons, and from electrons or muons, the DeepTau algorithm is used~\cite{CMS:2022prd}. Information from all individual reconstructed particles near the (\tauh) axis is combined with properties of the ($\tauh$) candidate and the event.
The rate of a jet to be misidentified as \tauh by the DeepTau algorithm depends on the \pt and quark flavour of the jet. Based on simulated events from \PW boson production in association with jets, the misidentification rate has been estimated to be 0.43\% for an identification efficiency for genuine \tauh of 70\%.
The misidentification rate for electrons (muons) is 2.60 (0.03)\% for a genuine \tauh identification efficiency of 80 ($>$99)\%.

\textit{Isolation:}
Photons, electrons, muons and tau leptons directly produced in hard collisions or resulting from the decay of massive bosons are expected to be isolated.  Isolation variables are constructed based on the kinematic information of the particles in the geometric neighborhood of the candidate lepton or photon and used for selection.  For instance, in many analyses selection is performed by requiring a low relative magnitude of the sum of the \pt of all other reconstructed particles and unclustered calorimeter energy inside a narrow cone around the candidate particle as a ratio to the candidate particle \pt.

\textit{Primary vertex:}
In Run 2, the primary vertex (PV) is taken to be the vertex corresponding to the hardest scattering in the event, evaluated using tracking information alone, as described in Section 9.4.1 of Ref.~\cite{CMS-TDR-15-02}.
In Run~1, the reconstructed vertex with the largest value of summed charged-particle track $\pt^2$ was taken to be the PV.

\textit{Jets:} 
Using the particle-flow global event reconstruction, hadronic jets are clustered from the reconstructed particles, using the infrared- and collinear-safe anti-\kt algorithm~\cite{Cacciari:2008gp, Cacciari:2011ma}. Typically, a distance parameter that measures the angular separation between constituents in the jet is defined as $\Delta R = \sqrt{\smash[b]{(\Delta y)^2+(\Delta\phi)^2}}$ of 0.4 is used ($\Delta R = 0.5$ in Run~1), but also $\Delta R = 0.8$ is used to identify merged jets from hadronic decays of Lorentz-boosted particles, \eg the \PW boson. Jet momentum is determined as the vectorial sum of all particle momenta in the jet, and is found from simulation to be, on average, within 5--10\% of the true momentum over the entire \pt spectrum and detector acceptance.

Additional tracks and calorimetric energy depositions resulting from particles produced in additional \pp interactions within the same or nearby bunch crossings (pileup) can add to the jet momentum. To mitigate this effect, charged particles identified as originating from pileup vertices are discarded and an offset correction is applied to correct for remaining contributions. Jet energy corrections are derived from simulation to bring the measured response of jets on average to that of jets constructed directly from the simulated particles. In situ measurements of the momentum balance in dijet, $\PGg + \text{jet}$, $\PZ + \text{jet}$, and multijet events are used to correct any residual differences in the jet energy scale (JES) between data and simulation~\cite{CMS:2016lmd}.
Additional selection criteria~\cite{CMS-PAS-JME-16-003} are applied to each jet to remove jets that are potentially affected by anomalous contributions or reconstruction failures.
For central jets at higher \pt a jet energy scale calibration uncertainty of better than 1\% is achieved~\cite{CMS-DP-2020-019}.

In many cases, the pileup-per-particle identification (PUPPI) algorithm~\cite{CMS:2020ebo,Bertolini:2014bba} is used to mitigate the effect of pileup, utilizing local shape information, event pileup properties, and tracking information. A local shape variable distinguishes between collinear particles originating from the hard scatter and the (on average) softer diffuse particles originating from the additional \pp interactions. Charged particles identified as originating from pileup vertices are discarded. For each neutral particle, a local shape variable is computed using the surrounding charged particles compatible with the PV within the tracker acceptance ($\abs{\eta} < 2.5$), and using both charged and neutral particles in the region outside of the tracker coverage. The momenta of the neutral particles are then rescaled according to the probability that they originated from the PV deduced from the local shape variable, superseding the need for jet-based pileup corrections~\cite{CMS:2020ebo}.

In a few early Run~1 analyses, prior to the full deployment of the particle-flow global event reconstruction methodology, hadronic jets were reconstructed from the energy deposits in the calorimeter, clustered using the anti-\kt algorithm with a distance parameter of $\Delta R = 0.5$.

\textit{Missing transverse momentum:} 
The missing \pt vector \ptvecmiss is computed as the negative vector sum of the transverse momenta of all the particle-flow candidates in an event, and its magnitude is denoted as \ptmiss~\cite{CMS:2019ctu}.
The \ptvecmiss is modified to account for corrections to the energy scale of the reconstructed jets in the event.
In some cases, the PUPPI algorithm is applied to reduce the pileup dependence of the \ptvecmiss observable.
The \ptvecmiss is computed from the particle-flow candidates weighted by their probability to originate from the PV~\cite{CMS:2019ctu}.
Several early analyses used a \ptvecmiss calculated from the calorimeter information alone, using calorimeter towers.

\textit{Heavy-flavour identification:}
A variety of algorithms are used to identify jets that originate from heavy-flavour \PQb and \PQc quarks.
The algorithms may incorporate primary and secondary vertex information; track kinematics, impact parameter and quality information; decay product information that is indicative of a heavy-flavour hadron decay, such as the presence of charged leptons with high impact parameter; or partial or full reconstruction of heavy-flavour hadrons; and various combinations of these ingredients.

\phantomsection\label{hftaggers}
The heavy-flavour jet identification algorithms used in the analyses presented in this Report are listed below. 
Typically these algorithms are applied to the constituents of a particle-flow jet and produce an estimator for the probability of the jet to originate from a \PQb or \PQc quark.

\begin{itemize}
\item SSV, simple secondary vertex algorithm~\cite{CMS:2012feb}: SSV uses the significance of the displacement from the PV of a reconstructed secondary vertex (the ratio of the displacement to its estimated uncertainty) as the discriminating variable.  The SSV algorithm can be used in high-efficiency and high-purity modes where two and three tracks are required to be associated with the vertex, respectively. Some analyses select SSV vertices using displacement from the PV or requiring that one of the tracks associated with the secondary vertex be identified as a muon~\cite{CMS-PAS-BTV-09-001}.
\item IVF, inclusive vertex finder~\cite{CMS:2013xck,CMS:2011yuk}: IVF identifies vertices with high three-dimensional displacement significance independently of jet reconstruction, by examining vertices around seed tracks with high impact parameter significance $S_{\mathrm IP}$ (the ratio of the track impact parameter to its estimated uncertainty).
\item CSV, combined secondary vertex algorithm for 7\TeV~\cite{CMS:2012feb} and 8\TeV~\cite{CMS-PAS-BTV-13-001}: CSV uses secondary vertex information as in SSV, ``pseudo vertices'' formed from tracks with high $S_{\mathrm IP}$, in addition to directly using the track $S_{\mathrm IP}$ information to form a likelihood-based discriminator.
\item CSVv2, combined secondary vertex algorithm for 13\TeV~\cite{BTV-16-002}: CSVv2 is based on CSV and combines the information of displaced tracks with the information on secondary vertices associated with the jet using a multivariate technique.
\item \textsc{DeepCSV}~\cite{BTV-16-002}: A deep machine-learning-based secondary vertex algorithm using IVF vertices and tracks as input.  Probability outputs are provided for bottom-, charm- and light-flavoured or gluon jets and can be combined to form the bottom or charm jet discriminants.
\item \textsc{DeepJet}~\cite{BTV-16-002,CMS:2021scf}:  A deep neural network algorithm based on the properties of charged and neutral particle-flow jet constituents, as well as 12 properties of secondary vertices associated with the jet. 
\item D hadron tag: Identifies a fully reconstructed D hadron within a jet based on the secondary vertex and mass reconstruction of the decay products. 
\item $\mu$ tag: Identifies a muon found in the candidate jet with large $S_{\mathrm IP}$ and representing a significant portion of the total jet momentum.
\end{itemize}

\textit{Jet substructure:}
Finally, massive particles such as top quarks, Higgs bosons, and \PW and \PZ bosons that decay to jets can be identified in boosted topologies using algorithms that make use of jet substructure, based on jets reconstructed with a distance parameter of 0.8. These algorithms are described where the specific analyses that use them are discussed.

\textit{Intact scattered protons:}
The PPS makes it possible to measure the four-momentum of scattered protons, along with their time-of-flight from the interaction point (IP). The proton momenta are measured by the two tracking stations in each arm of the spectrometer.

\section{Event simulation and cross section calculation}\label{sec:simulation}
The measurement of cross sections and their comparison with the predictions of the SM requires precise calculation of cross sections and the production of simulated events using Monte Carlo (MC) techniques.
Monte Carlo simulation of signal and background events involves a sequence of distinct operations.
First, occurrences of the hard scattering process are generated modelling the full distribution of the possible kinematics of the partons (quarks and gluons) and other elementary particles (leptons, gauge bosons, and the Higgs boson) in the process of interest.
This can be achieved either by attaching a weight corresponding to the probability of the kinematic state generated or by producing the states according to their kinematic probability.
The calculations are performed by factorization of the problem into a perturbatively calculable parton scattering process, and generalized functions that are obtained semi-empirically with fits to data.
The most essential of these functions, used in every calculation, are the parton distribution functions (PDFs), which describe the momentum distribution of the partons within the colliding protons.
They represent the number densities of partons carrying a momentum fraction $x$ at a given energy scale (expressed as the squared momentum transfer $Q^2$),
and are derived from fits to a large number of cross section measurements, generally measurements made by many experiments, over a large range of $Q^2$ and $x$ values.
The hard scattering is modelled by sampling the probability distributions of the PDFs to take account of the kinematics of the incoming partons in the proton and the corresponding phase space of final-state particle kinematics as described by the matrix element of the hard scattering interaction.
The final-state partons produced by the hard scattering are evolved down to some energy scale limit in a ``parton shower'' (PS) process that simulates the radiation of additional quarks and gluons, using leading logarithmic approximations.
The resulting partons are then hadronized---assembled into hadrons---producing jets of final-state particles.   This full process is known as hadronization.
Short-lived particles are decayed.
An ``underlying event'' (UE), including, \eg multiparton interactions (MPI), is added simulating the production of particles from the partons in the colliding protons that were not directly involved in the hard scattering process (and properly accounting for the kinematics of the initial state partons of the process).
The UE parameters in event generators are tuned so that observed features of data particularly sensitive to the contribution of the underlying event, such as charged-particle multiplicity and transverse momentum densities, match those in simulated events, as described, \eg in Ref.~\cite{CMS:2019csb}.
Finally, the particles are tracked through the detector, modelling their interactions with the detector elements, followed by simulation of the generation of electrical signals and their digitization to form a recorded event.

Table~\ref{tab:simprogs} lists the MC simulation programs used for analyses included in this Report.
General-purpose MC event generators, such as \PYTHIA, which aim to describe all final state particles emerging from a \pp collision, usually rely on only the Born matrix element for the perturbative calculation of the hard scattering.
Increased precision may be achieved by using dedicated MC programs that aim to better model some subset of hard scattering processes, or some aspect of a process, usually by using an improved level of approximation in QCD perturbative expansion:
next-to-leading order (NLO), next-to-next-to-leading order (NNLO), or even \ncube (\ie adding another ``next-to'').
These generators modelling higher-order Feynman diagrams are thus usually called matrix element (ME) generators.
When dedicated generators are used, the hadronization, and provision of the UE must
be accomplished by a more general event-generator program, such as \PYTHIA or \HERWIG, that can model the hadronization, particle decay, final-state radiation, and UE, in addition to the hard-scattering process.
Simulation of the interactions of the particles with the detector is performed by \GEANTfour, using a detailed geometrical model of the CMS detector, whereas the simulation of signal generation and digitization is handled by the CMS software.

A list of the sets of PDFs used for analyses included in this Report is shown in Table~\ref{tab:PDFlist}, categorized by the collaboration that produced them.

\begin{table}[htbp]
  \centering
\caption{Monte Carlo programs used by analyses included in this Report.\label{tab:simprogs}}
\renewcommand{\arraystretch}{1.3}
\begin{tabular}{l l   }
       \multicolumn{2}{l}{\textit{Cross section calculation}} \\
   \hline
         \DYTURBO & {\cite{Camarda:2019zyx}} \\
         \FEWZ & {\cite{Melnikov:2006di,Melnikov:2006kv,Gavin:2010az}} \\
         \GAMJET & {\cite{Baer:1989xj,Baer:1990ra}} \\
         \HELACOnia & {\cite{Shao:2012iz,Shao:2015vga}} \\
         \MATRIX & {\cite{Grazzini:2017mhc}}   \\
         \NLLJET & {\cite{Liu:2018ktv}} \\
         \NLOJETpp (with \FASTNLO) & {\cite{Nagy:2001fj,Nagy:2003tz}}  (\cite{Kluge:2006xs,Britzger:2012bs}) \\
         \NNLOJET (with \FASTNLO) & {\cite{Currie:2016bfm,Currie:2018xkj,Gehrmann:2018szu}} (\cite{Kluge:2006xs,Britzger:2012bs}) \\
         \OPENLOOPS & {\cite{Schonherr:2016pvn}}   \\ [\cmsTabSkip]
      \multicolumn{2}{l}{\textit{Hard-scattering process generation}} \\
   \hline
         \BLACKHAT & {\cite{Berger:2008ag}} \\
         \COMPHEP & {\cite{Boos:1992ap}} \\
         \HJMINLO & {\cite{Luisoni:2013kna,Hamilton:2012rf,Becker:2669113}}  \\
         \JHUGEN & {\cite{Gao:2010qx, Bolognesi:2012mm, Anderson:2013afp, Gritsan:2016hjl,Gritsan:2020pib}}  \\
         \MCFM &  {\cite{Campbell:2010ff,Campbell:2011bn}} \\ 
         \MADGRAPH 5, \MGvATNLO & {\cite{Alwall:2011uj,Wiesemann:2014ioa,Alwall:2014hca}} \\
         \NNLOPS & {\cite{NNLOPS1,NNLOPS2,NNLOPS3}}  \\
         \OPENLOOPS  & {\cite{Buccioni:2019sur,Denner:2016kdg,Ossola:2007ax,vanHameren:2010cp}} \\
         \PHOTOS & {\cite{Golonka:2005pn}} \\
         \POWHEG, \POWHEGBOX & {\cite{Nason:2004rx,Frixione:2007vw,Alioli:2010xd, Alioli:2010xd}} \\
         \VBFNLO, \VBFNLO 2.7 & {\cite{Arnold:2008rz, Baglio:2014uba,Baglio:2011juf}} \\ [\cmsTabSkip]
       \multicolumn{2}{l}{\textit{Full particle event generation}} \\
    \hline
         \CASCADE 3 & {\cite{CASCADE:2021bxe}} \\ 
         \HERWIG 7, \HERWIGpp & {\cite{Bahr:2008pv,Bellm:2015jjp}}   \\
         \PHOJET & {\cite{Bopp:1998rc}} \\
         \PYTHIA 6, \PYTHIA  8 & {\cite{Sjostrand:2006za, Sjostrand:2007gs, Sjostrand:2014zea}} \\
         \SHERPA 1, \SHERPA 2 & {\cite{Gleisberg:2008ta,Gleisberg:2008fv,Schumann:2007mg,Hoeche:2009rj,Sherpa:2019gpd}}   \\ [\cmsTabSkip]
       \multicolumn{2}{l}{\textit{Particle transport and detector interaction}} \\
    \hline
         \GEANTfour & {\cite{GEANT4:2002zbu}} \\
  \end{tabular}
\end{table}

\begin{table}[htbp]
  \centering
\caption{Sets of PDFs used for analyses included in this Report.\label{tab:PDFlist}}
\renewcommand{\arraystretch}{1.2}
  \begin{tabular}{l l   }
      \multicolumn{2}{l}{\textit{ABKM/ABM/ABMP Collaboration}} \\
      \hline
       ABKM09 & {\cite{Alekhin:2009ni}} \\
       ABM11 & {\cite{Alekhin:2012ig}} \\
       ABMP16  & {\cite{Alekhin:2017kpj,Alekhin:2018pai}} \\
    [\cmsTabSkip]
     \multicolumn{2}{l}{\textit{CTEQ-Jefferson Lab Collaboration}} \\
     \hline
       CJ15 & {\cite{Accardi:2016qay}} \\ [\cmsTabSkip]
       \multicolumn{2}{l}{\textit{CTEQ-TEA Collaboration}} \\
       \hline
       CT10 & {\cite{Lai:2010vv,Gao:2013xoa}} \\
       CT14 & {\cite{Dulat:2015mca}} \\
       CT18 & {\cite{Hou:2019efy}} \\ [\cmsTabSkip]
      \multicolumn{2}{l}{\textit{HERAPDF Collaboration}} \\
      \hline
       HERAPDF1, 1.5 & {\cite{H1:2009pze}} \\
       HERAPDF2.0 & {\cite{H1:2015ubc}} \\ [\cmsTabSkip]
       \multicolumn{2}{l}{\textit{MSTW/MMHT/MSHT Collaboration}} \\
       \hline
       MSTW 2008 NLO, NNLO & {\cite{Martin:2009iq}} \\
       MMHT2014 & {\cite{Harland-Lang:2014zoa}} \\
       MSHT2020 NLO, NNLO & {\cite{Bailey:2020ooq}}  \\
       MSHT20an3lo & {\cite{McGowan:2022nag}} \\ [\cmsTabSkip]
       \multicolumn{2}{l}{\textit{NNPDF Collaboration}} \\
       \hline
       NNPDF 2.0 & {\cite{Ball:2010de}} \\
       NNPDF 2.1 & {\cite{Ball:2011mu}} \\
       NNPDF 2.3 & {\cite{Ball:2012cx}} \\
       NNPDF 3.0 & {\cite{NNPDF:2014otw}} \\
       NNPDF 3.1 & {\cite{NNPDF:2017mvq}} \\
       NNPDF 3.1luxQED & {\cite{Bertone:2017bme}} \\
       NNPDF 4.0 & {\cite{NNPDF:2021njg}} \\ [\cmsTabSkip]
       \multicolumn{2}{l}{\textit{Transverse momentum dependent PDFs}} \\
       \hline
      PB-TMD PDFs & {\cite{BermudezMartinez:2018fsv,Hautmann:2017xtx,Hautmann:2017fcj}} \\
   \end{tabular}
\end{table}\section{Measurements of quantum chromodynamics}
The strong interaction between quarks is mediated by the gluons and is described by QCD, which is a quantum gauge theory based on a non-Abelian $\mathcal{SU}(3)_\mathrm{C}$ symmetry group, operating with three colour charges. Quarks and gluons are the fundamental constituents of the proton, which makes QCD physics ubiquitous at a hadron collider.  
The non-Abelian nature of QCD, which leads to a self-coupling of the massless gluon, results in a renormalization scale dependence (running) of strong coupling, the leading of the two major properties of the strong interaction. On the one hand, the asymptotic freedom at large scales (or small distances) allows for a perturbative description of quasi-free quarks and gluons. On the other hand, at small scales (large distances), the coupling becomes too large for perturbative calculations to be applied. This large-$\alpS$ region of the confinement can be only described phenomenologically. In many cases of interest at the LHC, the interactions involve large momentum transfers, where the theory is perturbative. However, the nonperturbative aspects of QCD are still relevant for the understanding of large momentum transfer physics.

This section presents a selection of measurements essential for probing QCD in nonperturbative and perturbative regimes. The measurements include PDF constraints, determinations of the strong coupling constant $\alpS$, multiple-parton interaction (MPI) effective cross sections, and the total inelastic cross section. High-\pt measurements span total, inclusive differential, and exclusive differential measurements of jet production cross sections. In differential measurements regions of phase space can be chosen, typically involving high jet multiplicities, to test the predictions of recent higher-order QCD calculations. Also, the high-\pt jet data collected by the CMS experiment offer sensitivity to deviations from the SM predictions that may occur in a diverse set of BSM scenarios involving heavy new particles or new forces. Measurements of the QCD jet production in association with heavy objects, such as vector bosons (as discussed in Sections~\ref{subsubsec:vjets} and~\ref{subsubsec:hfjets}), top quarks (Section~\ref{subsec:topandjets}), and Higgs bosons (Section~\ref{subsec:Higgsdiff}) are detailed in the respective sections on those topics.

\subsection{Total inelastic cross sections}
The total \pp cross section includes elastic- and inelastic-scattering components. In elastic scattering, the protons scatter via QCD or quantum electrodynamics (QED) processes without the proton dissociating (breaking up) or producing any additional particles. Inelastic scattering includes diffractive and nondiffractive interactions. In the diffractive events, the protons may emerge intact, excited, or dissociate into low-mass states, and these interactions are mediated by the exchange of colour-singlet objects such as the Pomeron (for QCD-induced) or a photon (for QED-induced) processes~\cite{Khoze:2002dc}\cite{ParticleDataGroup:2022pth} (see section 20). Photon-induced diffraction, where a proton dissociates into low-mass states, is commonly called dissociative rather than diffractive.  Diffractive or dissociative physics processes may be soft, producing low-mass final states with small \pt, or hard, possibly producing massive colorless vector or scalar bosons at central rapidities~\cite{Harland-Lang:2012bcb}.   Measurements of photon-induced vector boson production via diffractive central exclusive production are described in the section on vector boson scattering~\ref{subsec:EW}.  In the nondiffractive case, the partons in the colliding protons interact with sufficient momentum transfer to break up the protons. Processes included in the inelastic component of the total \pp cross section are the primary subject of this Report. They encompass interactions with large momentum transfer ($Q$), and most cases where heavier SM particles and possibly BSM particles may be produced. The total cross section and its components are not analytically calculable and instead fit from lower-energy data, and extrapolated to the LHC energies. The components of the total \pp collision cross section can be described by nonperturbative phenomenological models based on unitarity and analyticity principles~\cite{Barone:2002cv}. These models have large uncertainties when extrapolating to \TeV-scale collision energies and the measurement of these cross sections at new energies is an essential input to improving the reliability of the predictions. The measurement of the inelastic \pp interactions is necessary to address many issues essential for measuring cross sections. For example, the inelastic cross section determines probability and properties of additional inelastic collisions in the same or adjacent bunch crossings, referred to as pileup, which is necessary for interpreting the performance of nearly all physics object reconstruction at hadron colliders. Similarly, it enhances our understanding of the hadronic recoil from hard interactions, which is essential in modelling the \pt distributions of massive SM particles. 

The CMS experiment has measured the inelastic component of the total \pp cross section, $\sigma_{\text{in}}$, in 7~\cite{CMS:2012gek} and 13\TeV~\cite{CMS:2018mlc} \pp collisions. The measurements were done for events with the dissociation system masses exceeding 15.7\GeV using the 7\TeV data. In the 13\TeV analysis, the thresholds were above 4.1\GeV and 13\GeV for dissociation masses at negative and positive pseudorapidities, respectively. The extension of the 13 TeV analysis phase space to include very low dissociated masses was enabled by utilizing the CMS CASTOR forward calorimeter. The measurements reported here are for a common phase space delineated by the requirement that the longitudinal momentum loss fraction from one proton, $\xi$, exceeds $5\times 10^{-6}$.    This corresponds to the mass of the larger disassociated proton system, $m_\PX$, being greater than 16\GeV, such that $\xi=m_{\PX}^2/s>5\times 10^{-6}$.   At 7\TeV, the CMS Collaboration measured $\sigma_{\text{in}} = 60.2 \pm 0.2\stat \pm 1.1\syst \pm 2.4 \lum \unit{mb}$ and at 13\TeV $\sigma_{\text{in}} = 67.5 \pm 0.8\syst \pm 1.6 \lum \unit{mb}$ with a negligible statistical uncertainty. These measurements are compared with predictions of general-purpose MC generators \PYTHIA 6.4~\cite{Sjostrand:2006za}, 8~\cite{Sjostrand:2007gs,Sjostrand:2014zea} for a variety of generator parameter tunes; generators specific to large rapidity gap physics \PHOJET~\cite{Bopp:1998rc}; and generators used in cosmic ray physics \QGSJET~\cite{Ostapchenko:2004ss,Ostapchenko:2010vb}, \SIBYLL~\cite{PhysRevD.50.5710}, and \EPOS~\cite{Werner:2005jf}.  
The agreement of the theory predictions with the data is good for almost all the generators at 7\TeV, whereas at 13\TeV most generators overestimate the cross section by about 10\%, which is attributed to the mismodelling of the low-mass diffractive processes. The results are consistent with those measured by the TOTEM Collaboration in the same fiducial phase space~\cite{Antchev:2011vs,TOTEM:2013vij,TOTEM:2012oyl,TOTEM:2017asr}. Fits to lower-energy cross section data performed before the start of the LHC operations~\cite{ParticleDataGroup:2010dbb} by the COMPETE Collaboration~\cite{COMPETE:2002jcr}, which predicted the total hadronic cross sections from \GeV energies to the 57\TeV energy measured by the Pierre Auger Collaboration~\cite{PierreAuger:2012egl}, are in agreement with these measurements. The CMS measurements of fiducial inelastic production cross sections are shown in Fig.~\ref{fig:XsAll} together with total or fiducial cross sections of all other processes covered in this Report.

\subsection{Jet production cross section measurements}
Jet production measurements at the LHC test QCD over a large range of energies. The statistical power of the data allows for comparison of QCD predictions to precise total, differential, and multidifferential measurements. State-of-the-art calculations in QCD jet physics extend to NNLO QCD and NLO EW accuracy in the perturbative expansion and may include additional final-state partons in the ME predictions at a given order. 

\subsubsection{Inclusive fiducial jet production cross section measurements}
Inclusive jet production cross sections have been measured as functions of basic kinematic distributions at 2.76~\cite{CMS:2015jdl,CMS:2016lna}, 5.02~\cite{CMS:2024iaa},  7~\cite{CMS:2011ab,CMS:2012ftr,CMS:2014nvq,CMS:2014qtp,CMS:2016lna}, 8~\cite{CMS:2016lna}, and 13~\cite{CMS:2016jip,CMS:2021yzl}\TeV. The measurements typically present the inclusive jet production cross section as a function of \pt in intervals of rapidity $y$. The measurement is inclusive in that each jet that meets the rapidity and \pt criteria contributes to the cross section of the corresponding bin. The events including those jets may contain any number of additional jets or other final-state particles. Multiple jets in a collision event may contribute to the cross section according to their transverse momenta and rapidity. These measurements have been used to test NLO and NNLO QCD predictions.  

The conceptually simplest possible observable in high-\pt QCD physics is a fiducial inclusive cross section for the total production of all jets above a given \pt threshold and within a given rapidity range. The jet cross sections at 2.76~\cite{CMS:2015jdl}, 7~\cite{CMS:2012ftr}, 8~\cite{CMS:2016lna}, and 13~\cite{CMS:2021yzl}\TeV for inclusive production of jets that satisfy $\pt > 133\GeV$ and $\abs{y} < 2.0$ are reported in Table~\ref{tab:jet}.  Jets are clustered from particle-flow objects using the anti-\kt algorithm with a distance parameter of $\Delta R = 0.7$. These cross sections are calculated by integrating the differential measurements presented in the original publications, taking into account the correlation of systematic uncertainties between the bins when calculating the total systematic uncertainty. These results are compared with NNLO QCD predictions calculated using the \NNLOJET program~\cite{Currie:2016bfm,Currie:2018xkj,Gehrmann:2018szu} with \FASTNLO~\cite{Kluge:2006xs,Britzger:2012bs} and the CT18~\cite{Hou:2019efy} PDF set, with nonperturbative (NP) corrections applied based on MC generators, such as \PYTHIA~6, \PYTHIA~8, or \HERWIGpp~\cite{Bahr:2008pv,Bellm:2015jjp} using the state-of-the-art UE generator parameter sets (so called ``tunes'') derived at the time of each publication. These generators simulate UE and hadronization effects. Several MC generators are used in each publication to derive NP corrections and associated uncertainties.  Finally, the QCD predictions are corrected for the EW effects~\cite{Dittmaier:2012kx}. These predictions in a single phase space region have improved statistical and systematic precision compared to what is achievable in more restricted phase space regions or differential measurements.

\begin{table}[htbp]
  \centering
\topcaption{The measured inclusive fiducial jet production cross sections for four \pp collision energies for inclusive production of anti-\kt $R = 0.7$ jets satisfying $\pt > 133\GeV$ and $\abs{y} < 2.0$. Results are compared with predictions at NNLO QCD and NLO EW precision. The statistical uncertainty in the theory predictions is negligible.\label{tab:jet}}
\renewcommand{\arraystretch}{1.3}
  \begin{tabular}{l l l}
    $\sqrt{s}$ ($\TeVns$) &  $\sigma(\text{jet})$ (pb) & $\sigma^\text{SM}(\text{jet})$ (pb) \\
    \hline
    2.76~\cite{CMS:2015jdl}  & $  787 \pm 7 \stat \pm 49 \syst $ & $  777\,^{+40}_{-33} \syst $\\
    7~\cite{CMS:2012ftr}     & $  8520  \pm 90 \stat \pm 610 \syst $ & $  8760\,^{+390}_{-440} \syst $\\
    8~\cite{CMS:2016lna}     & $  11\,220 \pm 40\stat\,^{+610}_{-600} \syst $ & $  11\,650\,^{+270}_{-330} \syst $\\
    13~\cite{CMS:2021yzl}    & $  15\,230 \pm 70 \stat\pm 700 \syst $ & $  14\,980\,^{+420}_{-570} \syst $\\
  \end{tabular}
\end{table}

\subsubsection{Inclusive differential jet production cross section measurements}\label{subsec:InclJet}
\begin{figure}
	\centering 
	\includegraphics[width=0.40\textwidth]{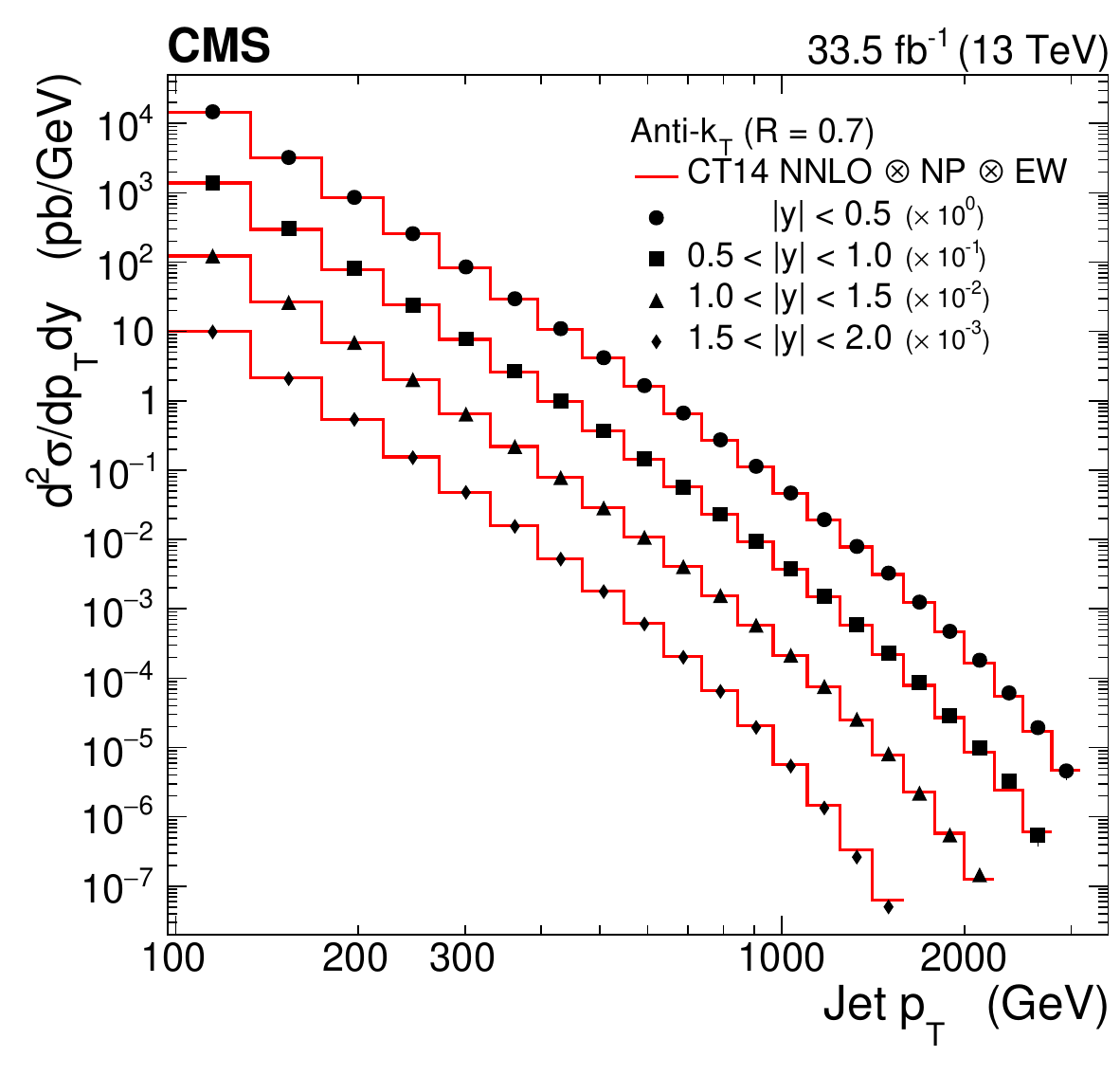}	
	\includegraphics[width=0.40\textwidth]{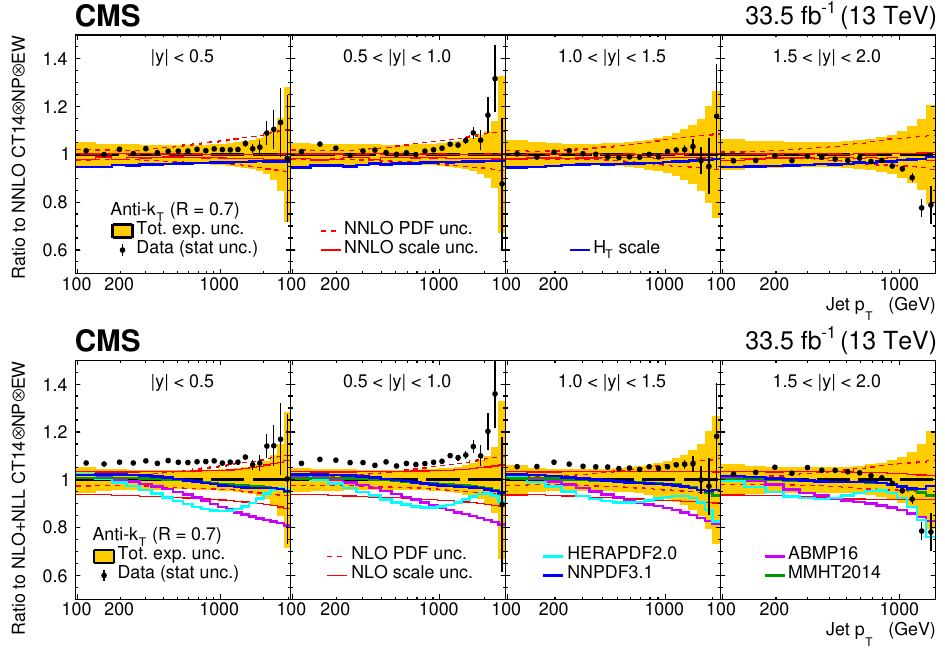}	
	\caption{The inclusive jet production cross sections as functions of the jet transverse momentum \pt measured in intervals of the absolute rapidity $\abs{y}$ (right). The cross section obtained for jets clustered using the anti-\kt algorithm with $\Delta R = 0.7$ is shown. The results in different $\abs{y}$ intervals are scaled by constant factors for presentation purposes. The data in different $\abs{y}$ intervals are shown by markers of different styles. The statistical uncertainties are too small to be visible; the systematic uncertainties are not shown. The measurements are compared with NNLO QCD predictions (solid line) using the CT14nnlo PDF set and corrected for EW and NP effects.
The double-differential cross section of inclusive jet production presented as ratios to the QCD predictions (left). The data points are shown by the filled circles, with statistical uncertainties shown by vertical error bars, while the total experimental uncertainty is centred at one and is represented by the orange band. In the upper panel, the data are divided by the NNLO prediction, corrected for NP and EW effects, using CT14nnlo PDF and with the renormalization and factorization scales jet \pt and, alternatively, \HT (blue solid line). In the lower panel, the data are shown as ratio to NLO+NLL prediction, calculated with CT14nlo PDF, and corrected for NP and EW effects. The scale (PDF) uncertainties are shown by the red solid (dashed) lines. NLO+NLL predictions obtained with alternative PDF sets are displayed in different colours as a ratio to the central prediction using CT14nlo. Figure and caption taken from Ref.~\cite{CMS:2021yzl}.}\label{fig:inclusivejets}
\end{figure}

The analysis of inclusive jet production at 13\TeV~\cite{CMS:2021yzl} includes comparisons to several perturbative QCD (pQCD) predictions.  
The NLO prediction using \NLOJETpp~\cite{Nagy:2001fj,Nagy:2003tz} and \FASTNLO~\cite{Kluge:2006xs,Britzger:2012bs} is further complemented by next-to-leading logarithmic (NLL) calculations using logarithmic resummation techniques.  Two classes of logarithmic terms are relevant to jet physics are resummed using the \NLLJET program~\cite{Liu:2018ktv}; those that depend on the jet radius and the so-called threshold logarithms. The latter involve logarithmic terms created when a jet just fails to pass the threshold to be considered as a jet. In addition, these cross section measurements are compared with the NNLO predictions obtained using the \NNLOJET program. This is the first analysis of jet production in \pp collisions that is compared to NNLO predictions. These QCD predictions at NLO+NLL and NNLO accuracy are computed by using different available PDF sets, \eg CT14~\cite{Dulat:2015mca}, NNPDF3.1~\cite{NNPDF:2017mvq}, MMHT2014~\cite{Harland-Lang:2014zoa}, ABMP16~\cite{Alekhin:2017kpj,Alekhin:2018pai}, and HERAPDF2.0~\cite{H1:2015ubc}, evaluated at NLO or NNLO, respectively. The pQCD predictions are augmented with the EW corrections~\cite{Dittmaier:2012kx}. Finally, the predictions are corrected for NP effects using a correction derived from the average of the \HERWIGpp (EE5C tune~\cite{CMS:2015wcf}) and \PYTHIA 8 (CP1 tune~\cite{CMS:2019csb}) simulations. The NP factors correct for the hadronization and UE effects that are not included in the pQCD predictions. The inclusive jet production cross section at 13\TeV, measured as a function of \pt in four bins of rapidity, is shown in Fig.~\ref{fig:inclusivejets}.  The agreement seen in the figure is excellent in all rapidity regions and spans nine orders of magnitude in cross section.   Also shown are the ratios of the measured jet cross sections to NLO and NNLO predictions.  The agreement observed with the NNLO perturbative QCD prediction with an NNLO PDF is better than the agreement with NLO predictions and NLO PDFs, where differences between the measured cross sections and the predictions are seen at in some rapidity regions and at high \pt and the different PDF predictions are inconsistent with each other.

\subsubsection{Exclusive differential measurements of jet production cross sections}
\begin{figure}
	\centering 
	\includegraphics[width=0.80\textwidth]{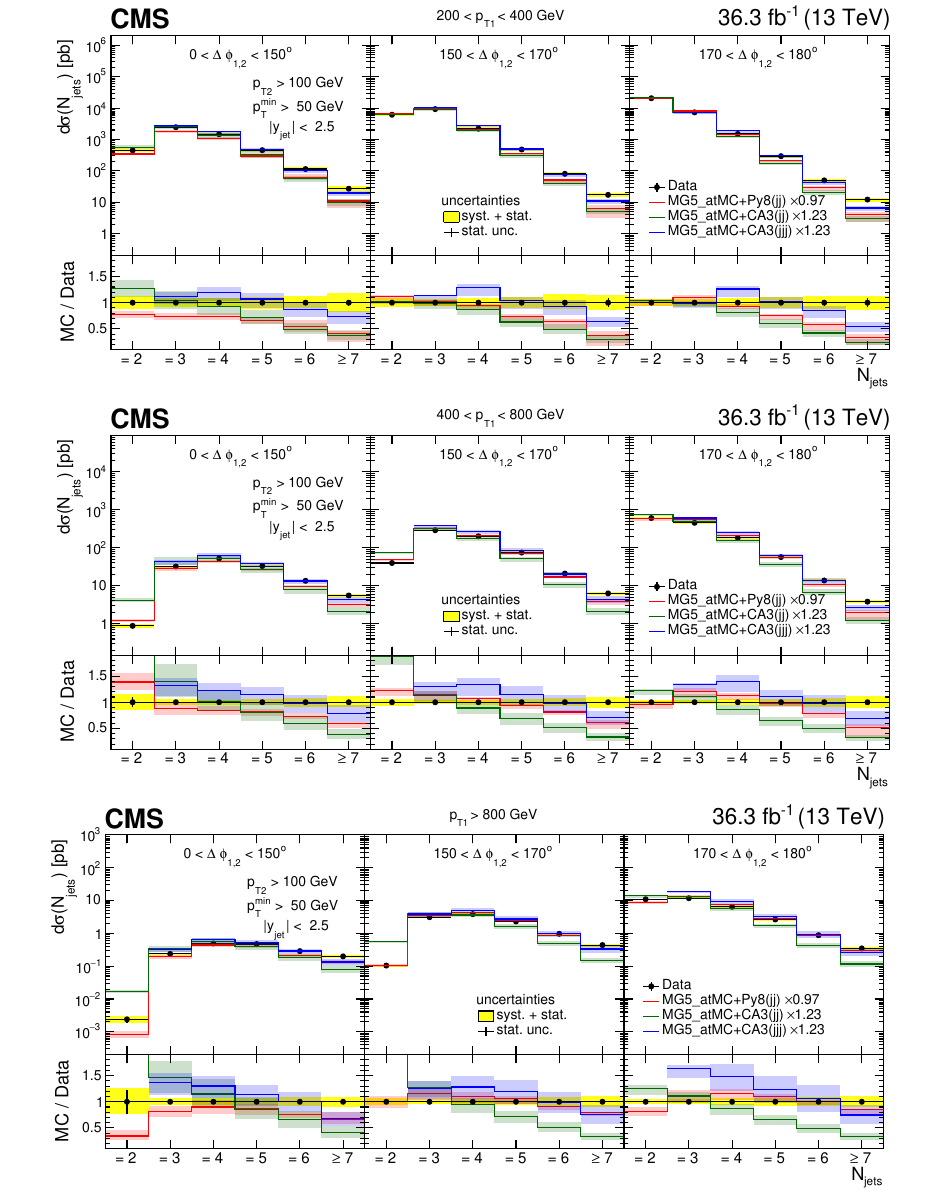}	
	\caption{Differential cross section of jet production as a function of the exclusive jet multiplicity (inclusive for 7 jets) in bins of \pt and $\Delta\phi_{12}$. The data are compared with the NLO dijet predictions from \MGvATNLO: MG5\_aMC+Py8 (jj) and MG5\_aMC+CA3 (jj), as well as the NLO three-jet prediction of MG5\_aMC+CA3 (jjj), where parton showering is performed by \PYTHIA 8 (Py8) and {{\textsc{Cascade}}\xspace}3~\cite{CASCADE:2021bxe} (CA3). The vertical error bars correspond to the statistical uncertainty, the yellow band shows the total experimental uncertainty. The shaded bands show the uncertainty from a variation of the renormalization and factorization scales. The predictions are normalized to the measured inclusive dijet cross section using the scaling factors shown in the legend. Figure taken from Ref.~\cite{CMS:2022drg}.}\label{fig:njet}
\end{figure}

The CMS experiment has performed a wide array of differential cross section measurements of jet production at all the collision energies at which the LHC operated. Of particular interest are measurements that isolate areas of phase space where current cross section calculations and MC simulations do not model the data well.  For instance, let's consider a case of high-\pt jets where the two highest \pt jets are not back-to-back because of multiple additional jet emissions. In this topology, no single MC prediction can model the jet multiplicity distribution for all ranges of azimuthal angle between the two highest \pt jets~\cite{CMS:2022drg} (as shown in Fig.~\ref{fig:njet}). The predictions shown in the figure use NLO MCs and matched PS generators at NLO including dijet predictions from \MGvATNLO: MG5\_aMC+Py8 (jj) and MG5\_aMC+CA3 (jj), as well as the NLO three-jet prediction of MG5\_aMC+CA3 (jjj). The NLO prediction  includes MEs with one additional real emission of a parton at LO accuracy, effectively generating events with up to three or four hard partons. Parton showering is performed with \PYTHIA 8 (Py8) and {{\textsc{cascade}}\xspace}3~\cite{CASCADE:2021bxe} (CA3). 
The CA3 prediction uses transverse momentum dependent (TMD) PDFs~\cite{BermudezMartinez:2018fsv} based on the parton-branching method (PB-TMD PDFs)\cite{Hautmann:2017xtx,Hautmann:2017fcj} in the PS model. In this analysis initial-state \pt is generated and PB-TMD PDF-dependent PS is performed using the \CASCADE 3 MC simulation~\cite{CASCADE:2021bxe} and compared with predictions using standard PS simulations. The TMD PDFs assess the \pt of hard-scattering system as it recoils against the UE physics involving the rest of the partons. These TMD PDFs implemented in the CA3 PS describe the data as well as do the standard PS methods, but without the need for tunable parameters. In general, the MC predictions fail to model the data for events with the jet multiplicity  greater than the number of hard partons generated in the ME predictions. Extending calculations and simulations to NNLO with matched NNLO PS generation and/or a larger number of partons simulated at the ME level would be expected to improve the agreement of the prediction with the data in high jet multiplicity topologies.   Improved agreement with the predictions would increase the sensitivity of BSM physics searches using final states with high jet multiplicities. However, improvements in methods of NNLO calculation for processes with high jet multiplicity are necessary to make them widely available for all \pp collision processes.

\subsubsection{Inclusive \texorpdfstring{\PQb}{b}-flavoured jet production}
\begin{figure}
	\centering 
      \includegraphics[width=0.48\textwidth]{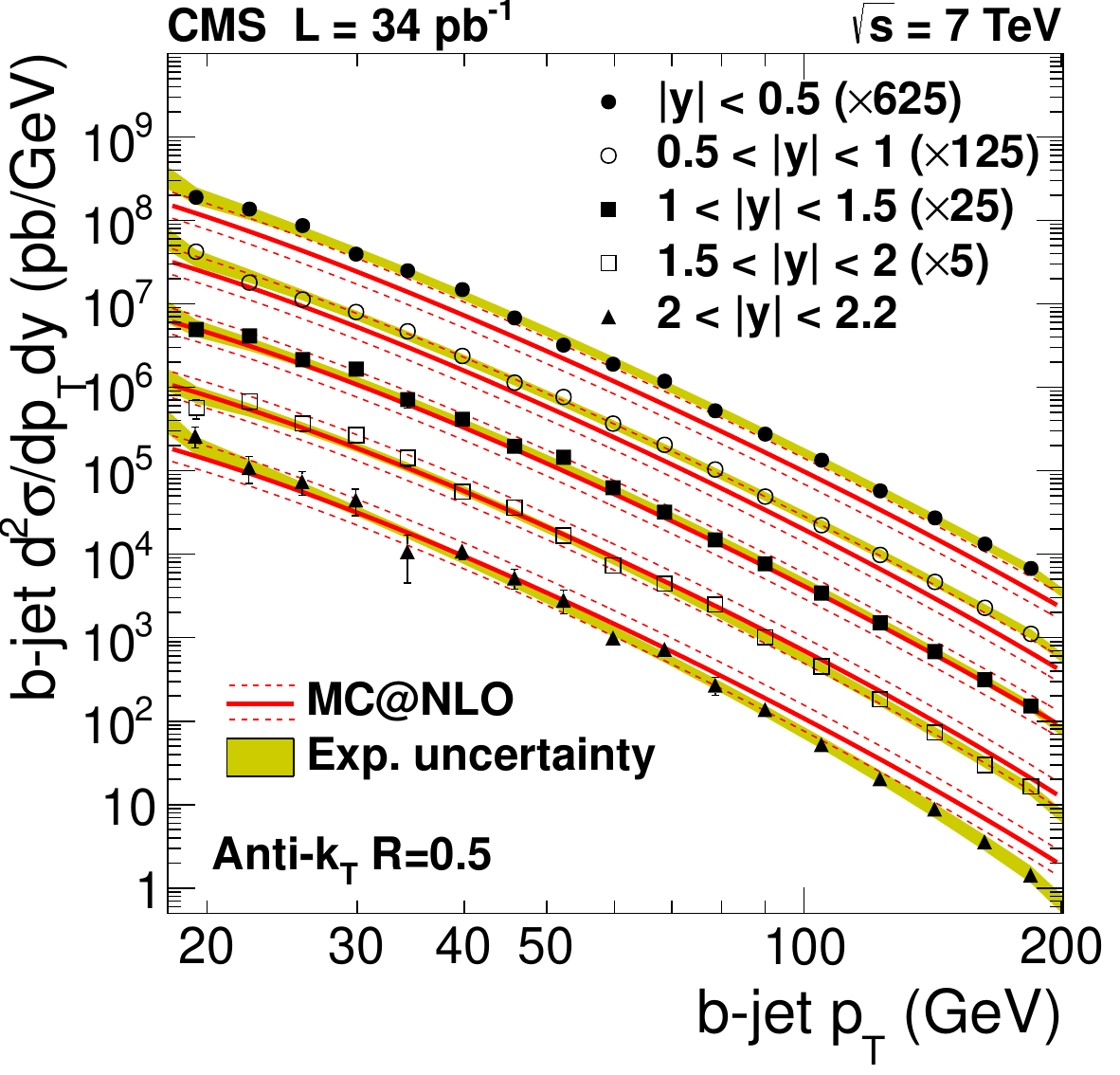}
      \includegraphics[width=0.48\textwidth]{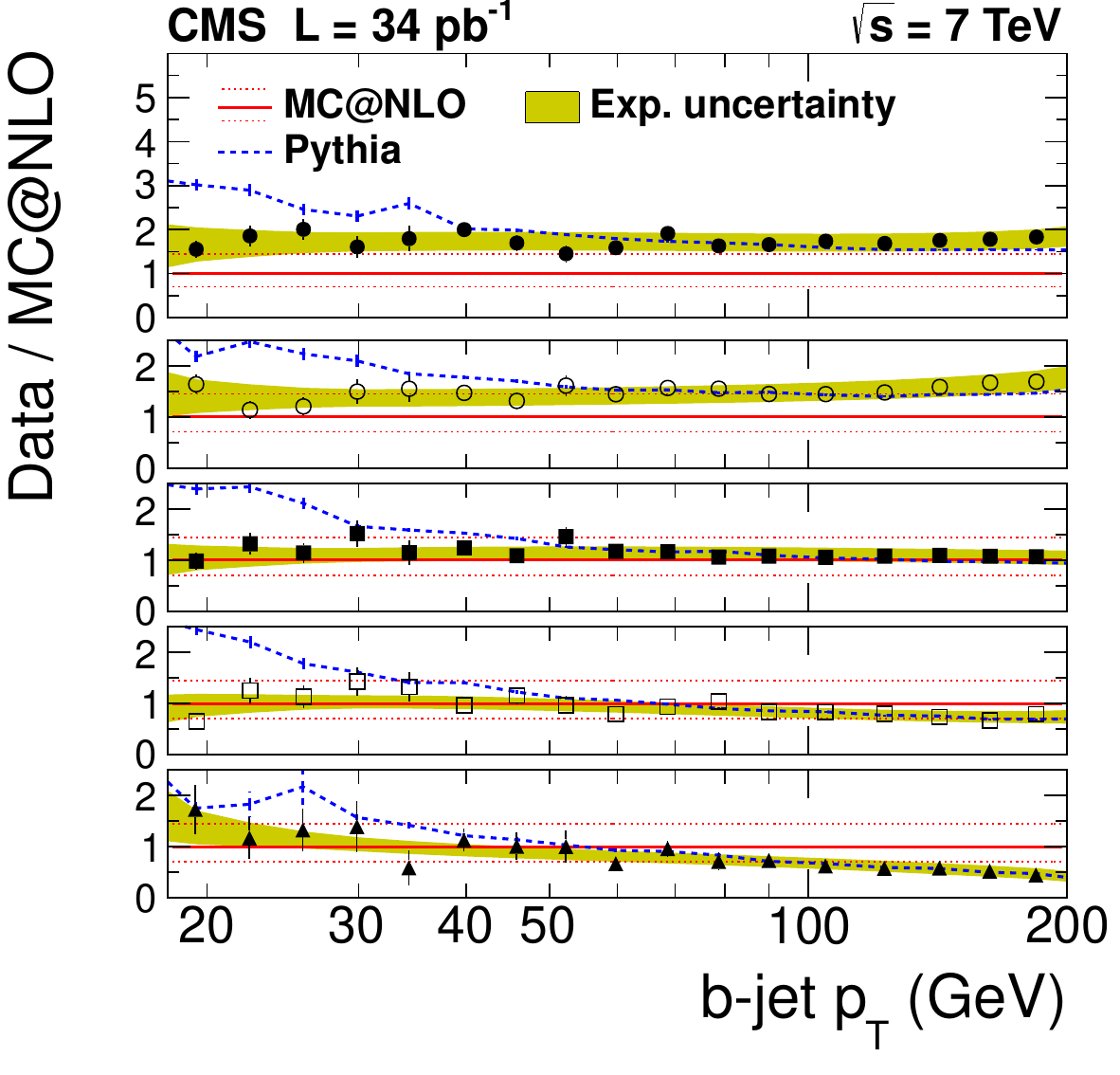}
    \caption{Measured $\PQb$ jet cross section from the jet analysis, multiplied by the arbitrary factors shown in the figure for easier viewing, compared to the \MCATNLO calculation (left) and as a ratio to the \MCATNLO calculation (right).  The experimental systematic uncertainties are shown as a shaded band and the statistical uncertainties as error bars. The \MCATNLO uncertainty is shown as dotted lines. The \PYTHIA prediction is also shown in the right panel.  Figure and caption taken from Ref.~\cite{CMS:2012pgw}.}\label{fig:bcrossvs}
\end{figure}

Inclusive b-flavoured jet production has been measured in 7\TeV \pp collisions as a function of \pt in five intervals of absolute rapidity $\abs{y}$~\cite{CMS:2012pgw}. In that publication jets are clustered from PF objects using the anti-\kt algorithm with a distance parameter of $\Delta R = 0.5$. The b-flavoured jets are selected using two methods. Both methods select reconstructed vertices displaced in three-dimensions from the PV using the SSV algorithm~\cite{CMS:2012feb}.  One method additionally requires one track associated with the secondary vertex be identified as a muon~\cite{CMS-PAS-BTV-09-001}. The two-dimensional cross section distributions are compared with MC predictions from \PYTHIA~6.4 and \MCATNLO in Fig.~\ref{fig:bcrossvs}. Also shown are comparisons of the cross section ratios to the MC predictions. Agreement is seen with the NLO prediction, except in the highest rapidity and \pt region where the simulation exceeds the measured cross section by up to two standard deviations for both the cross section and cross section ratio measurements.

\subsubsection{Additional differential measurements of jet production cross sections}
The full array of differential measurements performed by the CMS experiment is too extensive to report here. Only selected examples were discussed above. In addition, many measurements have been done that investigate lower-\pt QCD physics and flavour physics. Other differential measurements of high-\pt jet production cross sections performed by CMS not already discussed are listed below.  Each analysis includes a rich set of comparisons to state-of-the-art QCD predictions
\begin{itemize}
\item Differential dijet production \vs dijet invariant mass and jet rapidity at 7\TeV~\cite{CMS:2011nlq} 
\item Dijet azimuthal decorrelations at 7~\cite{CMS:2011hzb}, 8~\cite{CMS:2016adr}, and 13\TeV~\cite{CMS:2017cfb}.
\item Ratio of two- to three-jet cross sections as a function of the total jet transverse momentum at 7\TeV~\cite{CMS:2011maz}.
\item Shape, transverse size, and charged-hadron multiplicity of jets at 7\TeV~\cite{CMS:2012oyn}
\item Jet mass in dijet and $\PW/\PZ$+jet (7\TeV only) events, 7~\cite{CMS:2013kfv} and 13~\cite{CMS:2018vzn}\TeV.
\item Azimuthal separation between the second- and third-leading jets in nearly back-to-back topologies at 7\TeV~\cite{CMS:2013cvx}.
\item Study of hadronic event-shape variables, 7~\cite{CMS:2014tkl} and 13~\cite{CMS:2018svp}\TeV.
\item Topological observable in inclusive three- and four-jet events at 7\TeV~\cite{CMS:2015lzt}.
\item Jet charge at 8\TeV~\cite{CMS:2017yer}.
\item Azimuthal separation between the leading and second-leading jets in nearly back-to-back jet topologies in inclusive two- and three-jet events at  13\TeV~\cite{CMS:2019joc}.
\item Dependence of inclusive jet production on the anti-\kt distance parameter at 13\TeV~\cite{CMS:2020caw}.
\item Study of quark and gluon jet substructure in $\PZ$+jet and dijet events at 13\TeV~\cite{CMS:2021iwu}.
\end{itemize}

\subsection{Proton PDFs}\label{subsec:PDF}
Description of the proton structure, expressed in terms of PDFs, plays a central role in the interpretation of all the processes in \pp collisions at the LHC.  Protons are composite particles consisting of valence up- and down-flavoured quarks, gluons, and contributions from other quarks and antiquarks collectively known as the sea quarks. High-energy \pp collisions probe the structure of the proton at small distance scales. Proton-proton collisions at high energies are described by the QCD factorization theorem~\cite{Collins:1989gx}. At a certain factorization scale, the \pp cross section may be represented as a convolution of a (hard) partonic process, where individual, asymptotically-free partons from both colliding protons interact, with the parton distributions.
The parton (quark and gluon) distributions, are functions of the fraction $x$ of the proton momentum carried by the parton involved in the interaction, and the factorisation scale. The scale dependence is encoded in the Dokshitzer--Gribov--Lipatov--Altarelli--Parisi (DGLAP)~\cite{Gribov:1972ri,Lipatov:1974qm,Dokshitzer:1977sg,Altarelli:1977zs,Curci:1980uw,Furmanski:1980cm,Moch:2004pa,Vogt:2004mw} evolution equations, which are known up to NNLO and approximately at \ncube. The dependence of PDFs on $x$ needs to be extracted from the experimental data. Most of the information on the PDFs is provided by measurements in deep-inelastic scattering experiments data from either HERA-I~\cite{H1:2009pze} or the combined HERA-I and HERA-II data~\cite{H1:2015ubc}. Production of jets, top quarks, and weak bosons at the LHC provides additional sensitivity to the PDFs. Using corresponding cross section measurements, the PDFs and the strong coupling constant \alpS can be extracted with improved precision. 
PDFs have been extracted at LO, NLO, NNLO, and even at approximate \ncube, as well as in more complex systems, such as nuclei.

In practice, the PDFs are obtained in a course of a QCD analysis, assuming a certain x-dependence of the PDFs at a starting evolution scale. In such a QCD fit, the measurements are confronted with the corresponding pQCD predictions at highest available order and the parameters driving the $x$ behaviour of each PDF are obtained.  Besides a comprehensive QCD analysis where the PDFs are fitted, sometimes it is useful to investigate a possible impact of a new measurement on an uncertainty in already existing PDF without reevaluating the PDF. This is done by performing a so-called profiling analysis. 
In the CMS experiment, the open-source QCD analysis framework \XFITTER (former \HERAFITTER)~\cite{Alekhin:2014irh,xFitter:2022zjb} is used for PDF fits and profiling.  In a full PDF fit, together with the PDFs, further QCD or EW parameters  such as quark masses, strong coupling or EW mixing angle, can be obtained and the correlations of these parameters with the PDFs are mitigated. Furthermore, once contributions of new physics are included (\eg via methods of effective field theory) in addition to the SM cross section prediction, their couplings can be constrained together with the PDFs and SM parameters.  

\subsubsection{Overview of CMS constraints on PDFs}
The CMS Collaboration has explored the sensitivity of different processes to the PDFs and SM parameters. The CMS Drell--Yan measurements have improved constraints on the valence quark distributions, while production of $\ttbar$ and (multi)jets is particularly sensitive to the mass of the top quark, the gluon distribution, and the \alpS. The associated production of \PW boson with a charm quark ($\PW$+$\PQc$) is the only process at a hadron collider directly probing the strange content of the proton quark sea. The CMS experiment has pioneered the measurement of $\PW$+$\PQc$ production at a hadron collider and its interpretation in terms of the strangeness distribution. A list of CMS analyses used to constrain PDFs is given in Table~\ref{tab:PDFfits}. For each analysis the QCD order of the analysis and a PDF distribution of interest that is constrained by the inclusion of CMS data is listed. To date, the majority of these measurements are used by the global PDF fit collaborations. Finally, comparisons of cross section measurements with the predictions employing various PDFs are discussed in the relevant sections. 

\begin{table}[htbp]
  \centering
\topcaption{The CMS analyses where PDF fits were performed. The table lists the final state and distributions considered, the \pp collision energy, the HERA data set used or global PDF provided, the QCD perturbative order of the fit, and the most constrained PDFs. Whenever data from multiple analyses are used, the first analysis listed contains the PDF extraction.  In the 13\TeV analysis the inclusive jet data are used in an NNLO PDF fit, whereas the inclusive jet and \ttbar data are used in an NLO PDF fit.\label{tab:PDFfits}}
\renewcommand{\arraystretch}{1.2}
  \maybeCmsTable{
  \begin{tabular}{l l l l l}
    Analysis    & $\sqrt{s}$ & HERA Data & QCD   & Best PDF\\
       & ($\TeV$)   & or PDF     & order & constraint \\
          \hline
  \PW charge asym.~\cite{CMS:2013pzl}, $\PW$+$\PQc$~\cite{CMS:2013wql}  & 7 & HERA-I & NLO & \PQu, \PQd, \PQs \\
    Inclusive jet~\cite{CMS:2014qtp} & 7 & HERA-I & NLO & gluon \\
  \PW charge asym.~\cite{CMS:2016qqr}   & 8 & HERA-I\,+\,II  & NLO & \PQu and \PQd \\
   Inclusive jet~\cite{CMS:2016lna}  & 8 & HERA-I\,+\,II & NLO & gluon \\
   3D dijet~\cite{CMS:2017jfq}  & 8 & HERA-I\,+\,II & NLO & gluon \\
    Inclusive jet~\cite{CMS:2021yzl}, \ttbar~\cite{CMS:2019esx} & 13 & HERA-I\,+\,II, CT14nnlo & NNLO,NLO & gluon\\
    Dijet mass~\cite{CMS:2023fix}& 13 & HERA-I\,+\,II & NNLO & gluon\\
  \end{tabular}}
\end{table}

\subsubsection{The PDF constraints from jet production measurements}
\begin{figure}
	\centering 
	\includegraphics[width=0.80\textwidth]{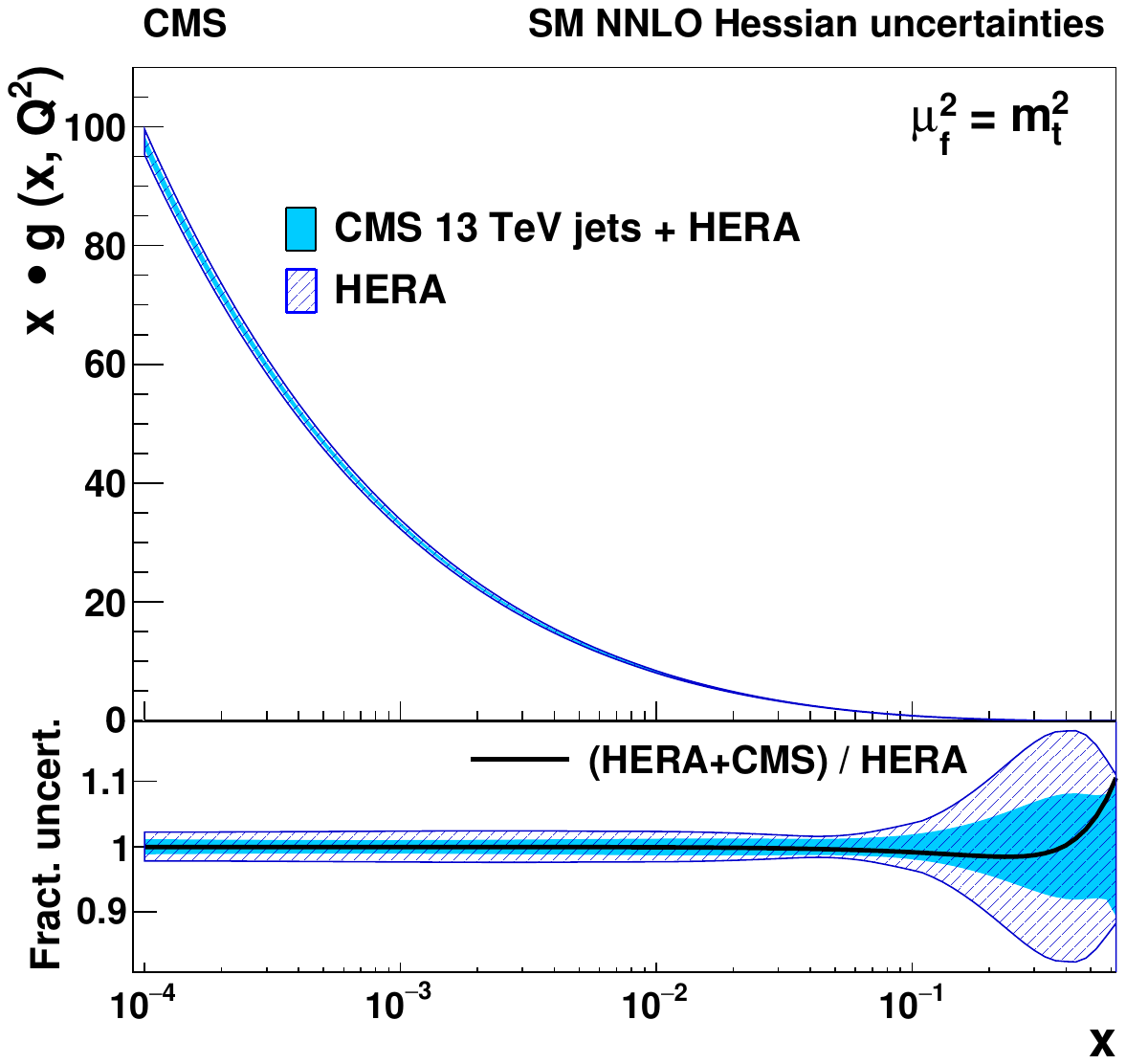}	
	\caption{The gluon distribution, shown as a function of $x$ for the factorization scale $\mu_{\mathrm f} = m_\PQt$. The filled (hatched) band represents the results of the NNLO fit using HERA DIS and the CMS inclusive jet cross section at $\sqrt{s} = 13\TeV$ (using the HERA DIS data only). The PDFs are shown with their total uncertainty. In the lower panel, the comparison of the relative PDF uncertainties is shown for each distribution. The solid line corresponds to the ratio of the central PDF values of the two variants of the fit. Figure and caption taken from Ref.~\cite{CMS:2021yzl}.}\label{fig:jetPDF}
\end{figure}

CMS measurements of multi-differential inclusive jet and dijet cross sections at different centre-of-mass energies were extensively used to constrain the PDFs and the value of \alpS (presented in Section~\ref{subsec:alpha_s}). They include double-differential inclusive jet analysis at 7~\cite{CMS:2014qtp}, 8~\cite{CMS:2016lna}, and 13\TeV~\cite{CMS:2021yzl}; triple-differential dijet analysis at 8\TeV~\cite{CMS:2017jfq}; and an analysis of dijet mass at 13\TeV~\cite{CMS:2023fix}.
These data were included in comprehensive QCD analyses together with the measurements of the DIS cross sections, available at the date of each analysis. Since the NNLO predictions in a form suitable for the PDF fit became available only recently, the fits to 7 and 8\TeV measurements were performed only at NLO QCD, while the QCD analysis of 13\TeV data were performed at NNLO. The CMS inclusive jet and dijet measurements provide a substantial additional constraint on the gluon PDF at all values of x, as illustrated in Fig.~\ref{fig:jetPDF} taken as an example from the results obtained with inclusive jet cross sections at 13\TeV. In the same analysis, the value of \alpS was extracted simultaneously with the PDFs.  That paper also presents a PDF analysis including 13\TeV~\ttbar data performed at NLO~\cite{CMS:2023fix}. 

\subsection{The strong coupling constant, \texorpdfstring{$\alpS$}{alpha s}, and its running}\label{subsec:alpha_s}
Important tests of QCD are the precise extraction of the value of \alpS at the scale of the \PZ boson mass, $\alpS(m_\PZ)$, and the illustration of the running \alpS as a function of the renormalization scale $Q$, usually taken as \pt of the jet in proton collision, or momentum transfer in DIS. The scale dependence is encoded in the renormalization group equation (RGE) of QCD and represents a basic demonstration of our understanding of the dynamics of the strong interaction~\cite{Deur:2016tte}. 

Jet production is an ideal instrument for determination of \alpS, since its cross section is proportional to \alpS already at LO QCD. The first CMS determination of \alpS was performed by investigating the ratio of jet cross sections in three- and 2-jet topologies $R_{32}$~\cite{CMS:2013vbb}, which is linearly proportional to the value of \alpS.  In high-\pt collisions involving the production of jets, $\alpS$ is typically of order 0.1--0.2, which, as calculated using pQCD, corresponds to a probability for additional jet emissions in any \pp hard-collision event of the same order. Two-jet and multijet events with three or more jets are common, allowing for statistically precise determinations of \alpS. The $R_{32}$ analysis used events with jets with \pt in the range 0.42 to 1.39\TeV and conducted the first determination of $\alpS$ at TeV scale energies. Simultaneous extraction of \alpS together with PDFs was performed using inclusive jet and di-jet measurements and exploring the jet substructure. The uncertainties in \alpS extracted using jet production at hadron colliders are dominated by missing higher-order pQCD calculations, usually estimated by varying the renormalization and factorization scales by a factor of 2. Most of the aforementioned measurements were performed at NLO and suffer from a large theory uncertainty. Simultaneously CMS has pioneered extraction of \alpS using \ttbar production cross section measurements, which resulted in higher precision than jet-based extractions, due to availability of NNLO calculations for \ttbar production cross section. In addition, other physics processes such as weak boson production have been used to make precise determinations of $\alpS$. Since the NNLO calculation for jet production in \pp collisions have become available, the theory uncertainty in \alpS extraction using jet production is significantly reduced. The most precise measurement of $\alpS(m_\PZ)$ to date of $\alpS(m_\PZ) = 0.1166 \pm 0.0014 \fit \pm  0.0007 \model \pm 0.0004 \scale \pm 0.0001 \param = 0.1166 \pm 0.017 \tot$ is obtained in a simultaneous fit of PDFs and \alpS at NNLO using double-differential inclusive jet production data at 13\TeV~\cite{CMS:2021yzl}. 
The most recent CMS determination of \alpS uses jet substructure\cite{CMS:2024mlf}, performed by comparing with NLO plus approximate next-to-next-to-leading-logarithmic (aNNLL)~\cite{Dixon:2019uzg,Lee:2022ige,Chen:2023zlx} predictions of two- and three-point energy correlators inside jets. The most precise value of $\alpS(m_\PZ)$ in substructure measurements is achieved and the running of \alpS is probed. 

\begin{figure}
	\centering 
	\includegraphics[width=0.90\textwidth]{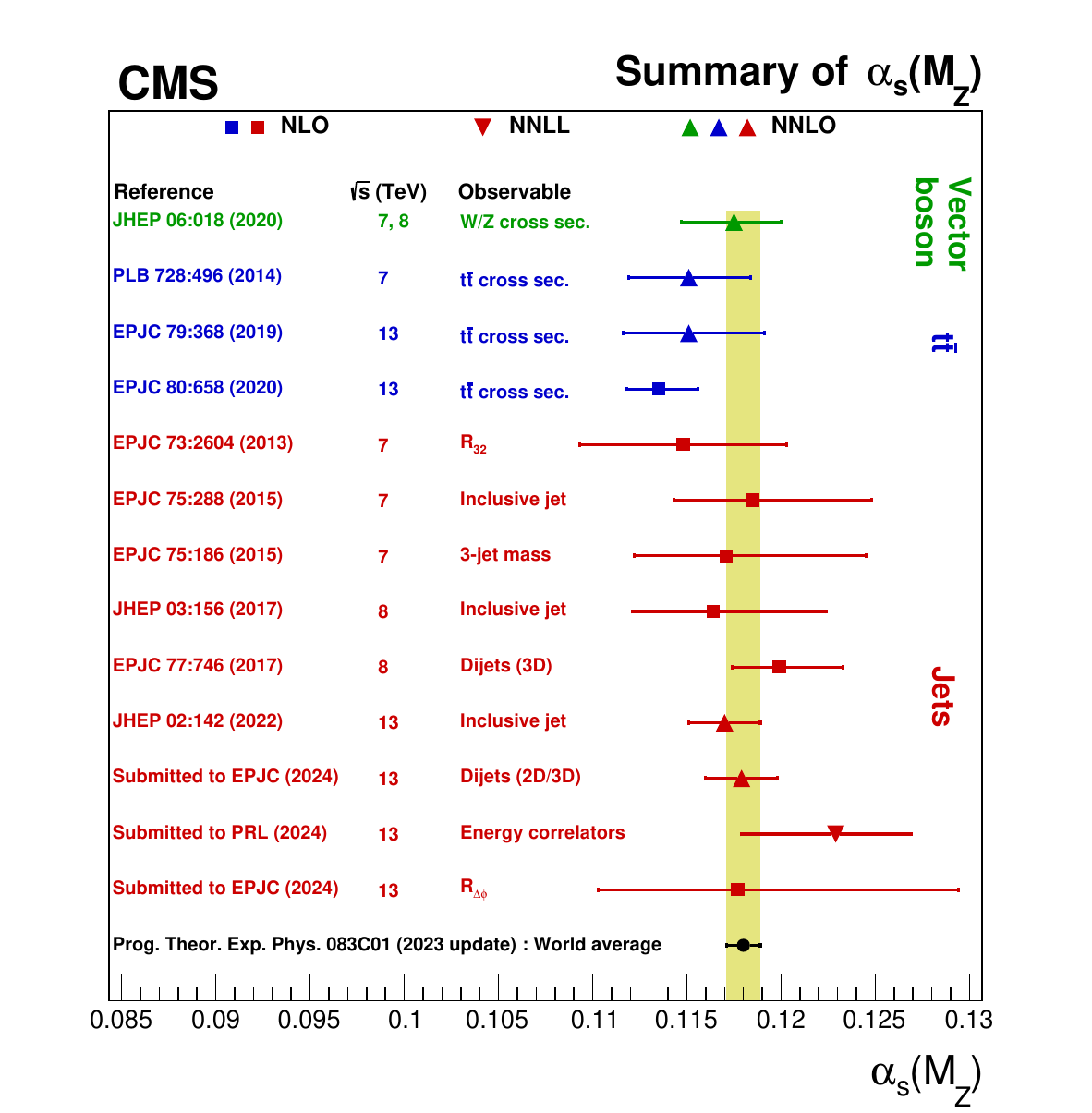}	
	\caption{A summary of $\alpS(m_\PZ)$ extractions from the CMS experiment compared with the 2023 PDG world-average. For each measurement, \pp collision energy and the QCD perturbative order of the $\alpS(m_\PZ)$ extraction are listed. Results are grouped by the type of the final state used: vector boson, \ttbar, and jets.}\label{fig:alphas}
\end{figure}

The CMS extractions of $\alpS$ are listed in Table~\ref{tab:alphas} and displayed in Fig.~\ref{fig:alphas}. For comparison, the results are presented by extrapolating $\alpS$ to the energy scale of the $\PZ$ boson mass, $\alpS(m_\PZ)$. Uncertainties are grouped together by type and further descriptions of the uncertainty types are reported in the glossary of terms in~\ref{sec:glossary}.

\begin{table}[p]
    \topcaption[Overview of \alpS results from CMS]{
        Overview of $\alpS(m_\PZ)$ from CMS analyses. Results where \alpS is determined by profiling a global PDF set, list the set used. The other results were obtained using a combined PDF and \alpS fit of the CMS and HERA data as described in the text. The 2D inclusive jet~\cite{CMS:2014qtp} analysis only uses the HERA-I data, whereas the other combined PDF and \alpS fits use the combined HERA-I and HERA-II data. The QCD perturbative order (pQCD order) of the determination is also given. For publications where more than one value is extracted, only one is reported. Whenever data from other analyses are used in the \alpS determination, the first analysis listed documents the \alpS extraction.  Uncertainties are grouped together by type and further descriptions of the uncertainty types are reported in the glossary of terms in~\ref{sec:glossary}.\label{tab:alphas}}
\renewcommand{\arraystretch}{1.2}
\cmsTable{
\begin{tabular}{l c c c c c c c c}
    Analysis &$\sqrt{s}$& $\alpS(m_\PZ)$    & fit unc.  & PDF unc. & scale unc. & other unc.    & PDF    & pQCD \\
            & ({\TeVns})    &          &           &          &  &                         &  & order \\
    \hline
    $R_{32}$~\cite{CMS:2013vbb}                    & 7    & $0.1148$ & $\pm0.0014$              & $\pm0.0018$               & \multicolumn{2}{c}{$\pm0.0050$ theo incl. scale}                                                                   & NNPDF2.1  &  NLO \\
    2D inclusive jet~\cite{CMS:2014qtp}~\cite{CMS:2012ftr}& 7    & $0.1185$ & $\pm0.0019$              & $\pm0.0028$               & ${}^{+0.0053}_{-0.0024}$  & $\pm0.0004$ NP                                   & \NA       &  NLO \\
    Inclusive 3-jet mass~\cite{CMS:2014mna}        & 7    & $0.1171$ & $\pm0.0013$              & $\pm0.0024$               & ${}^{+0.0069}_{-0.0040}$  & $\pm0.0008$ NP                                   & CT10      &  NLO \\
    \ttbar cross section~\cite{CMS:2014rml}                                & 7    & $0.1151$ & ${}^{+0.0017}_{-0.0018}$ & ${}^{+0.0013}_{-0.0011}$  & ${}^{+0.0009}_{-0.0008}$  & $\underbrace{\pm0.0013}_{m_\mathrm{t}} \underbrace{\pm0.0008}_{\sqrt{s}}$       & NNPDF2.3  & NNLO \\
    2D inclusive jet~\cite{CMS:2016lna}                      & 8    & $0.1185$ & ${}^{+0.0019}_{-0.0021}$ & $\underbrace{{}^{+0.0002}_{-0.0015}}_{\text{model}}\underbrace{{}^{+0.0000}_{-0.0004}}_{\text{param}}$
      & ${}^{+0.0022}_{-0.0018}$  &                                                                        & \NA       &  NLO \\
    3D dijet mass~\cite{CMS:2017jfq}                         & 8    & $0.1199$ & $\pm0.0015$              & $\underbrace{\pm0.0002}_{\text{model}}\underbrace{{}^{+0.0002}_{-0.0004}}_{\text{param}}$
   & ${}^{+0.0026}_{-0.0016}$  &                                                                        & \NA       &  NLO \\
    W, Z cross section~\cite{CMS:2019oeb}                                   & 7, 8 & $0.1163$ & $\underbrace{\pm0.0007}_{\text{stat}}\underbrace{\pm0.0010}_{\text{syst}}$              & ${}^{+0.0016}_{-0.0022}$  & $\pm0.0009$               & $\underbrace{\pm0.0013}_{\text{lumi}}\underbrace{\pm0.0006}_{\text{num}}$                                  & CT14      & NNLO \\
    \ttbar (dilepton)~\cite{CMS:2018fks}                     & 13   & $0.1151$ & \multicolumn{2}{c}{$\pm0.0035$ fit + PDF}                      & ${}^{+0.0020}_{-0.0002}$  &                                                                        & MMHT14    & NNLO \\
    Normalized \ttbar~\cite{CMS:2019esx}                & 13   & $0.1135$ & $\pm0.0016$              & $\underbrace{{}^{+0.0002}_{-0.0004}}_{\text{model}}\underbrace{{}^{+0.0008}_{-0.0001}}_{\text{param}}$
   & ${}^{+0.0011}_{-0.0005}$  &                                                                        & \NA       &  NLO \\
    2D inclusive jet~\cite{CMS:2021yzl}                      & 13   & $0.1166$ & $\pm0.0014$              & $\underbrace{\pm0.0007}_{\text{model}}\underbrace{\pm0.0001}_{\text{param}}$
   & $\pm0.0004$               &                                                                        & \NA      & NNLO \\
    2D \& 3D dijet mass~\cite{CMS:2023fix}                   & 13   & $0.1181$ & $\pm0.0013$              & $\underbrace{\pm0.0006}_{\text{model}}\underbrace{\pm0.0002}_{\text{param}}$
   & $\pm0.0009$               &                                                                        & \NA       & NNLO \\
    $R_{\Delta\phi}$~\cite{CMS:2024hwr}               & 13   & $0.1177$ & $\pm0.0013$              & $\underbrace{\pm0.0010}_{\text{NNPDF3.1}}\underbrace{\pm0.0020}_{\text{choice}}$
   & ${}^{+0.0114}_{-0.0068}$  & $\underbrace{\pm0.0011}_{\text{NP}}\underbrace{\pm0.0003}_{\text{EW}}$ & NNPDF3.1  &  NLO \\
    Energy correlators in jets~\cite{CMS:2024mlf}              & 13   & $0.1229$ & $\underbrace{{}^{+0.0014}_{-0.0012}}_{\text{stat}}\underbrace{{}^{+0.0023}_{-0.0036}}_{\text{syst}}$
   & \multicolumn{3}{c}{${}^{+0.0030}_{-0.0033}$}                 & \NA       &aNNLL \\
  \end{tabular}}
\end{table}

\begin{figure}
	\centering 
	\includegraphics[width=0.80\textwidth]{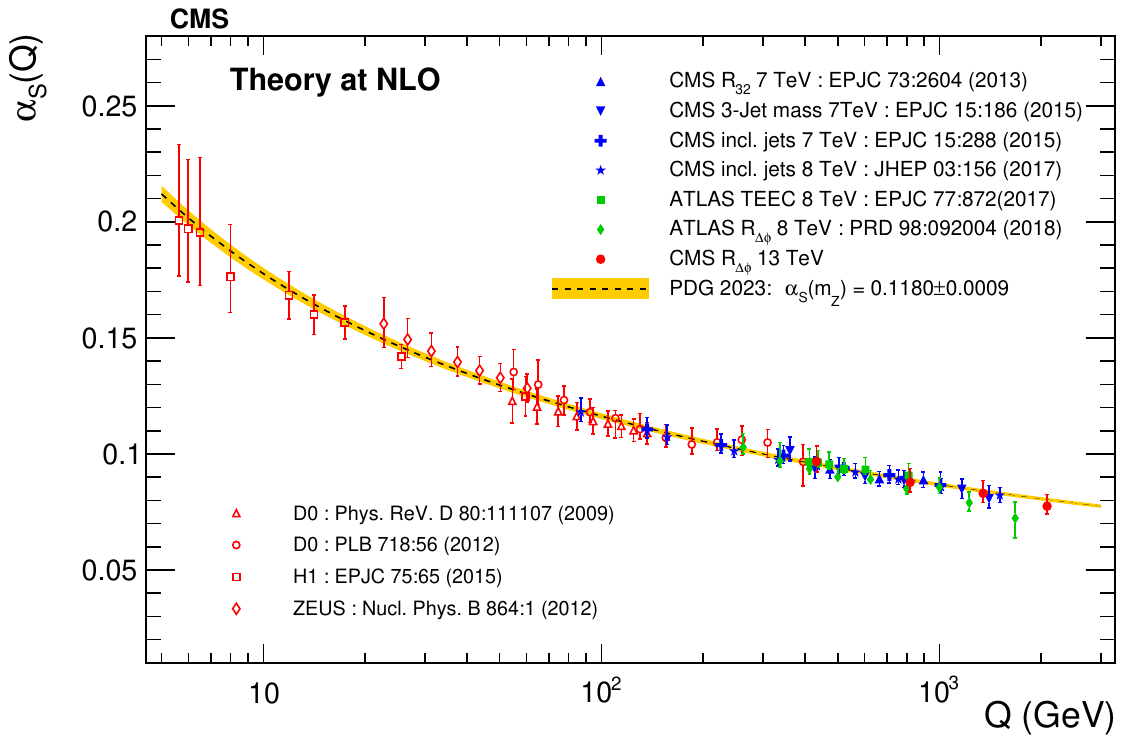}	
	\caption{Running of the strong coupling as a function of momentum transfer, $\alpS(\mathrm{Q})$ (dashed line), evolved using the 2023 world-average value, $\alpS(m_\PZ) = 0.1179 \pm 0.0009$, together with its associated total uncertainty (yellow band). The CMS extractions, which extend above 2\TeV, are compared with results from the H1, ZEUS, D0, and ATLAS experiments. The vertical error bars indicate the total uncertainty (experimental and theoretical). All the experimental results shown in this figure are based on predictions at NLO accuracy in perturbative QCD. Figure from Ref.~\cite{CMS:2024hwr}.}\label{fig:alphas_running}
\end{figure}

A summary figure of the running of \alpS, probed by several measurements shown in Fig.~\ref{fig:alphas_running} including CMS, ATLAS~\cite{ATLAS:2017qir,ATLAS:2018sjf}, and earlier determinations by the D0~\cite{D0:2009wsr,D0:2012xif}, H1~\cite{H1:2010mgp}, and ZEUS~\cite{ZEUS:2012pcn} Collaborations. For the CMS measurements $\alpS$ is determined in dijet \pt ($R_{32}$~\cite{CMS:2013vbb}), 3-jet mass~\cite{CMS:2014mna}, and jet \pt (inclusive jets 7\TeV~\cite{CMS:2014qtp}, inclusive jets 8\TeV~\cite{CMS:2016lna}, and $R_{\Delta\phi}$~\cite{CMS:2024hwr}) regions based on the average $Q$ of events in those regions.
The QCD RGEs, encoding the running of \alpS, are obtained using \textsc{NLOJet++} implemented in the \FASTNLO framework evolved from 2023 world-average value of $\alpS(m_\PZ) = 0.1179 \pm 0.0009$~\cite{ParticleDataGroup:2022pth}. The CMS determinations of \alpS agree well with the world-average and with the RGE at NLO predictions. 

\subsection{Double-parton scattering}
Double-parton scattering (DPS) is a process in which two parton-parton scattering interactions occur in a single hadron-hadron collision. The study of DPS is a test of our knowledge of the structure of the proton. For instance, DPS provides information on the energy evolution of the \pt profile of the partons in the proton, which is information that cannot be accessed in single-parton scattering (SPS) events~\cite{Diehl:2017wew}. Thus, where SPS interactions are widely used to measure the longitudinal PDFs of the partons in the proton, DPS events can measure the transverse PDFs. Also, since multiple  partons in each proton are colliding, DPS can be used to study the correlations between quantum numbers of the constituents of the proton. For instance, the spin of two partons in a single proton will be correlated and will have effects on the kinematics of a DPS collision.

\begin{figure}
	\centering 
	\includegraphics[width=0.80\textwidth]{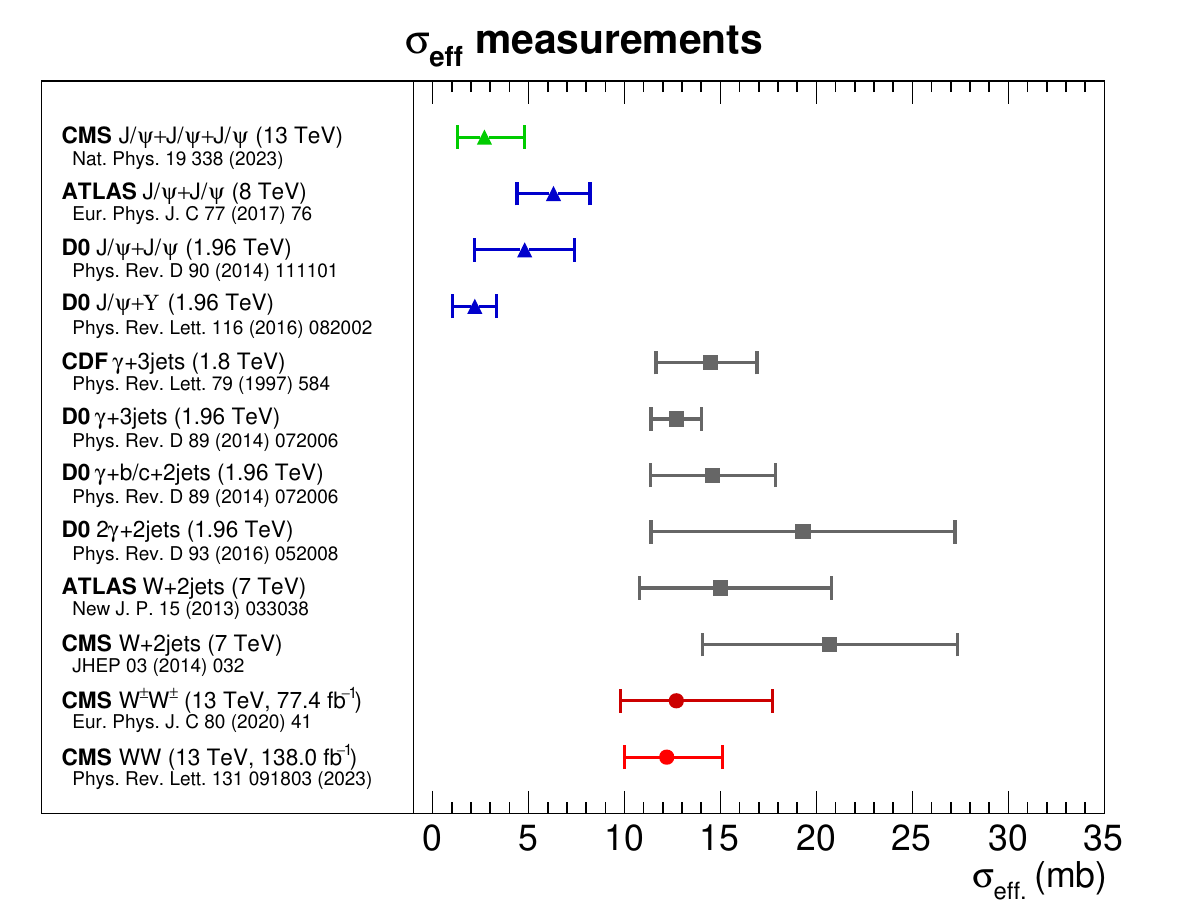}	
	\caption{Selected measurements of the effective DPS cross section in \pp collisions at the LHC by the CMS and ATLAS experiments, and in $\Pp\PAp$ collisions at the Tevatron by the CDF and D0 experiments. The horizontal bars indicate the combined statistical and systematic uncertainty for each measurement. Figure taken from Ref.~\protect\cite{CMS:2021qsn}.}\label{fig:DPS}
\end{figure}

The cross sections of DPS interactions are typically modelled as the product of the two independent SPS cross sections divided by an effective cross section, $\sigma_{\text{eff}}$, as shown in Eq.~(\ref{eq:DPSformula}). The ratio is multiplied by a combinatorial factor, $m$, that is equal to 2 when processes A and B are different and 1 when they are the identical. This effective cross section can be interpreted as the square of the average transverse distance between the interacting partons.
\begin{linenomath}
\begin{equation}
	\sigma^\text{DPS}_\text{A,B} = \frac{m}{2} \frac{\sigma_\text{A} \sigma_\text{B}}{\sigma_\text{eff}} \label{eq:DPSformula}
\end{equation}
\end{linenomath}
DPS has been extensively studied at the Tevatron by the CDF~\cite{CDF:1997lmq} and D0~\cite{D0:2014vql,D0:2015dyx,D0:2014owy,D0:2015rpo} experiments and at the LHC by CMS~\cite{CMS:2021qsn,CMS:2013huw,Sirunyan:2019zox,CMS:2022pio} and ATLAS~\cite{ATLAS:2016ydt,ATLAS:2013aph}. Figure~\ref{fig:DPS} shows the effective cross section values for DPS processes from the Tevatron and LHC experiments determined from measurements with quarkonium final states and from processes with jets, photons, and $\PW$ bosons. The expected relationships between the SPS, DPS and triple-parton scattering (TPS) cross sections from~\HELACOnia~\cite{Shao:2012iz,Shao:2015vga} are used to extract $\sigma_{\text{eff}}$ for DPS from the CMS measurement of triple-\JPsi production~\cite{CMS:2021qsn}. Distributions sensitive to DPS based on the \MGvATNLO and \PYTHIA 6 simulation of DPS physics are used to extract $\sigma_{\text{eff}}$ in \PW plus 2 jet events, whereas multivariate classifiers based on \PYTHIA 8 simulation with the CP5 and CUETP8M1 tunes of MPI parameters~\cite{CMS:2019csb} are used to extract $\sigma_{\text{eff}}$ in $\PW^\pm\PW^\mp$ and $\PWpm\PWpm$ events.  The effective cross sections obtained from quarkonium measurements favour values below 10\mb, as compared with effective cross sections derived from final states with harder scales, which favour values above 10\mb. Such apparent process-dependent $\sigma_{\text{eff}}$ values are suggestive of different parton transverse PDFs and/or correlations probed inside the proton at varying fractional momenta.

\subsection{Summary of QCD measurements}
The CMS Collaboration has conducted a broad array of QCD measurements across a large range of energies. The PDF measurements substantially constrain the gluon, valence quark, and sea quark (collectively and individually such as constraints on the \PQs quark) PDFs. The $\alpS(m_\PZ)$ extractions are competitive and agree with those of other experiments and measure the running of $\alpS(m_\PZ)$ up to \TeV energy scales. Together these measurements constrain important aspects of QCD that are essential for making predictions of high-\pt interactions at the LHC. Inclusive and multidifferential jet production measurements have been performed, testing the limits of the current generation of NNLO QCD and NLO EW perturbative predictions. In general, given the high probability of additional jet production in high-energy \pp collisions, the detailed QCD analyses produced by the LHC experiments and their comparisons with the most sophisticated theory predictions are essential for expanding our understanding of all aspects of high-\pt SM physics.

\section{Measurements in the electroweak sector of the standard model}
The EW sector involves the EW gauge bosons (the photon, and the $\PW$ and $\PZ$ bosons) and their interactions with other SM particles. The EW sector of the SM combines a $\mathcal{U}(1)_\text{Y}$ and a non-Abelian $\mathcal{SU}(2)_\text{L}$ gauge symmetries, with associated weak hypercharge and weak isospin charges, respectively. The electromagnetic force is based on a $\mathcal{U}(1)_\text{EM}$ symmetry, with electric charge, and the associated massless photon resulting from a linear combination of the $B$ and $W_3$ fields of the $\mathcal{U}(1)_\text{Y}$ and $\mathcal{SU}(2)_\text{L}$ gauge symmetries after the EW symmetry breaking. Similarly, the weak force, weak charges, and $\PW$ and $\PZ$ bosons result from linear combinations of the $W_1$ and $W_2$ fields of the $\mathcal{SU}(2)_\text{L}$ symmetry and a linear combination of the $B$ and $W_3$ fields, respectively. The combination of these gauge symmetries and the EW symmetry-breaking mechanism forms a unified EW theory. Electroweak physics measurements at the LHC test many aspects of the SM. These include the complex interactions between multiple EW gauge bosons predicted by the non-Abelian $\mathcal{SU}(2)_\text{L}$ portion of the EW gauge structure and the nature of EW symmetry breaking via the Brout--Englert--Higgs mechanism, which generates masses of the $\PW$ and $\PZ$ bosons.  The small values of the EW couplings imply that most EW processes at the LHC can be calculated perturbatively with good precision. The EW bosons are copiously produced at the LHC and can be measured with high precision by the LHC detectors.

For EW physics, the number of accessible final states at the LHC is without precedent.  They include states with single, double, or triple gauge bosons. Production of EW gauge bosons can occur via radiation from quarks, multi-gauge-boson interactions, such as vector boson scattering (VBS) and vector boson fusion (VBF), and from the decay of heavier particles, such as the Higgs boson and top quark.  Many processes have only been observed at the LHC, which is the first collider that allows access to processes such as VBS. In each subsection total and fiducial cross sections, cross sections including production of additional jets, and differential measurements are presented.   At the end of the section we briefly summarize the results.

Analysis of the EW physics at the CMS experiment is primarily conducted using physics objects, such as jets, photons, electrons, or muons. Neutrinos are inferred from the \ptvecmiss in the vector sum of objects reconstructed as originating from the PV. Jets are typically required to have $\pt > 30\GeV$.   Photons are required to satisfy $\pt > 25\GeV$ to remove lower-\pt photons originating from the decay of neutral pions. Electrons and muons are used to identify events with $\PW$ or $\PZ$ bosons. In EW analyses described in this Report, $\PW$ ($\PWp$ or $\PWm$) and $\PZ$ bosons are efficiently reconstructed via their leptonic decays, $\PWp \to \ell^+\nu_\ell$ (charge conjugate states are implied) and $\PZ \to \ell^+\ell^-$, where $\ell = \PGm$ or $\Pe$.  Backgrounds to $\PZ \to \ell^+\ell^-$ decays are very low. Muons and electrons with $\pt > 20\GeV$ are used in analysis with a single $\PW$ boson. Analyses with $\PZ$ bosons or multiple bosons often use thresholds as low as $\pt > 10\GeV$ for a second lepton and $\pt > 5\GeV$ for additional leptons. The $\PW$ bosons are also selected by identifying events with \ptvecmiss or selecting events with large transverse mass calculated using a lepton momentum and \ptvecmiss. The selection listed above is typical of CMS analyses, but higher thresholds are used in some cases to reject backgrounds, or lower thresholds to increase the acceptance. Generally, events using reconstructed $\PW$ and $\PZ$ candidates have low background caused by nonprompt leptons or other particles in jets misidentified as prompt leptons. The largest backgrounds (the so-called ``physics'' backgrounds) come from events with identical final-state particles. Flavour-tagging algorithms are used to identify bottom and charm jets. Reconstruction algorithms and identification criteria are described in Section~\ref{sec:lhcandcms}.

\subsection{Vector boson production}
Measurements of the production of single EW bosons are the simplest test of EW theory predictions. However, the prediction of the corresponding cross sections at a hadron collider is complicated by the necessity to understand the radiation of QCD jets and the PDFs of the proton, which describe the structure of the proton and predict the partonic luminosities of the colliding partons. Despite these complications, measurements of EW production cross sections can still be made with percent-level precision.  This makes physics involving single bosons both a precision test of EW theory and, in either inclusive production or production of vector bosons with jets, of perturbative QCD predictions. The low backgrounds when identifying vector bosons in the $\PWp \to \ell^+\nu_\ell$  and especially $\PZ \to \ell^+\ell^-$ decay modes and the size of the LHC data sets allows theoretical and experimental comparisons of total, differential, and often multidifferential distributions with good precision over wide ranges of energy, angle, and jet multiplicity. Together these processes provide a stringent test of SM predictions over a broad array of final states and kinematic configurations.

\begin{figure}
	\centering 
	\includegraphics[width=0.30\textwidth]{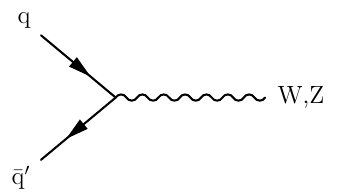}	
	\includegraphics[width=0.30\textwidth]{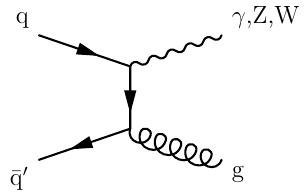}	
	\includegraphics[width=0.30\textwidth]{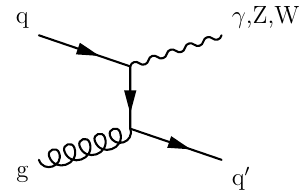}	
	\caption{The Feynman diagram for Drell--Yan production of $\PW$ and $\PZ$ bosons (left). The $\PZ$ boson production process involves annihilation of quark-antiquark pairs of  same flavour.  The $\PW$ boson production process requires different-flavour quarks, such as $\PQu\PAQd$ or $\PAQu\PQd$ pairs. The NLO diagrams with real emission of a jet for the production of single vector bosons and one jet with a final-state gluon jet (middle) or quark jet (right).}\label{fig:FD:DYVjet}
\end{figure}

Measurements of single-boson production constitute an essential test of our ability to predict SM parton-parton interaction cross sections using perturbative techniques.  Single photons are radiated off charged objects. Single weak boson production proceeds primarily through the Drell--Yan (DY) quark-antiquark annihilation process~\cite{PhysRevLett.25.316}, as shown in Fig.~\ref{fig:FD:DYVjet}.  The production of $\PZ$ bosons is sensitive to the sum of the \PQu and \PQd and the sum of the \PAQu and \PAQd PDFs and also the EW mixing angle $\theta_\PW$. The \PWp and \PWm boson production has sensitivity to the ratios of \PQu to \PQd and \PAQu to \PAQd contributions, especially when considering the charge asymmetry of the leptons from the $\PW$ boson decays as a function of their pseudorapidity. The DY process has been predicted at \ncube accuracy in perturbative QCD using matching \ncube PDF sets. The  PDF uncertainties, and higher-order QCD and EW radiative corrections limit the precision of current predictions.  Other sensitive comparisons are made using \ncube or NNLO predictions of ratios of production cross sections or in two-dimensional planes depicting pairs of the $\PZ$, \PWp, and \PWm boson cross sections. 

\subsubsection{Single photon production}
The photon is the longest known and most extensively studied vector boson. In high-energy \pp collisions the photon is observed as a promptly produced particle in a large number of SM processes and may also be produced in BSM topologies. Examples are Higgs boson decay to two photons~\cite{CMS:2012qbp} and monophoton searches for new physics, such as dark matter~\cite{CMS:2017qyo}. Photons are also produced in neutral pion decays and are radiated from final-state particles, leading to backgrounds in the study of prompt high-energy photons. The simplest measurement of photon production uses events with one or more prompt isolated photons above a given \pt threshold that are produced in the hard interaction. Unlike the situation with massive vector bosons, it is necessary to define a minimum momentum threshold, because singularities in the perturbative calculation of cross sections near zero momentum are not well defined.  Also, experimental constraints make it impossible to measure the lowest energy portion of photon production due to overwhelming backgrounds. A minimum threshold is required to reject both instrumental and physics backgrounds. In 7\TeV collision data the CMS experiment finds a production cross section of $39.6 \pm 0.7 \stat \pm 6.9 \syst$ \unit{nb} for photons with $\pt > 25\GeV$~\cite{CMS:2011nkw}. This cross section was calculated by integrating the differential cross section for photon production presented in that paper.    

\begin{figure}
	\centering 
	\includegraphics[width=0.80\textwidth]{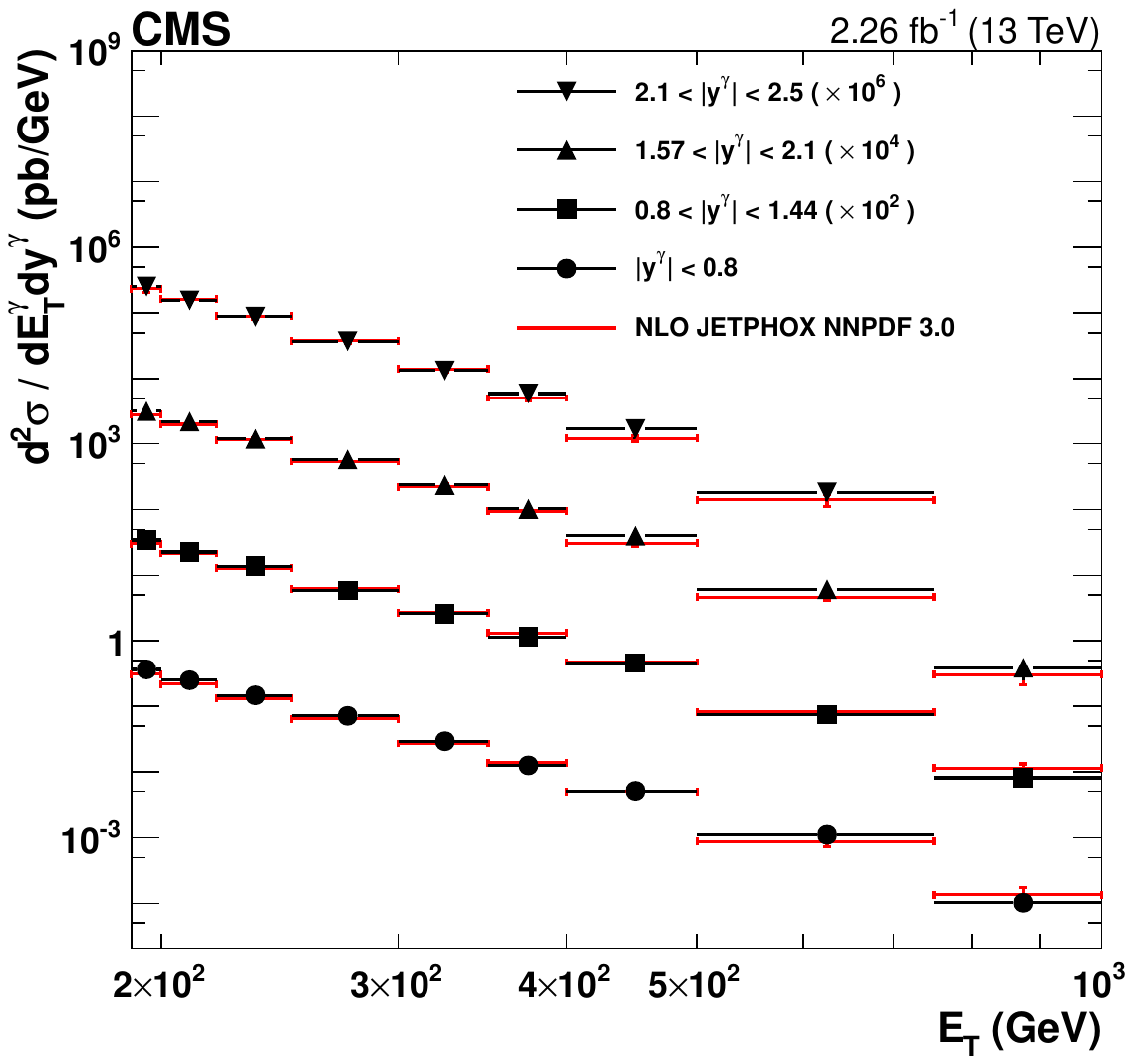}	
	\caption{Differential cross sections for isolated-photon production in four photon rapidity intervals. The points show the measured values and their total uncertainties; the lines represent the NLO \JETPHOX predictions with the NNPDF3.0 PDF set. Figure and caption taken from Ref.~\cite{CMS:2018qao}.}\label{fig:photonpteta}
\end{figure}

Inclusive photon production cross sections have been measured differentially  as functions of basic kinematic variables at 7~\cite{CMS:2010svd,CMS:2011nkw} and 13~\cite{CMS:2018qao}\TeV. As with jet production, the results are reported as functions of the photon \et in several intervals of rapidity. An example from the 13\TeV analysis of single-photon data is shown in Fig.~\ref{fig:photonpteta}. The measurements of differential and inclusive photon production cross sections are compared with the NLO calculations from \JETPHOX~\cite{Catani:2002ny} using the BFG~\cite{Bourhis:1997yu} fragmentation functions for quarks and gluons into photons, and found to be well modelled.  

\subsubsection{Single weak boson production}
The cross sections of single prompt massive vector bosons inclusively produced with any number of final-state quarks or gluons are among the most precisely measured at hadron colliders. The CMS experiment has measured single inclusive $\PW$ and $\PZ$ boson production in events where the boson decays to an electron or a muon and the corresponding antineutrinos, and $\Pe^+\Pe^-$ or $\PGm^+\PGm^-$ pairs, respectively. Inclusive cross section measurements have been made with 2\% precision primarily limited by the uncertainty in the integrated luminosity. This precision has been achieved because of several factors. The large data sets of \PW and \PZ bosons result in small to negligible statistical uncertainty in the measurements. Small systematic uncertainty is achieved due to large data sets for evaluating in granular detail the efficiency of lepton (electron and muon) detection; accurate MC simulations for estimating the acceptance for prompt leptons from \PW and \PZ boson decays, and predicting physics backgrounds involving prompt leptons from other sources; and low backgrounds and reliable methods to predict the rates of hadrons and leptons in jets being misidentified as prompt leptons based on control samples in data. The limiting integrated luminosity uncertainty has been extensively studied and minimized using techniques described in the references given in Section~\ref{sec:lhcandcms}.

\begin{figure}
        \centering 
        \includegraphics[width=0.80\textwidth]{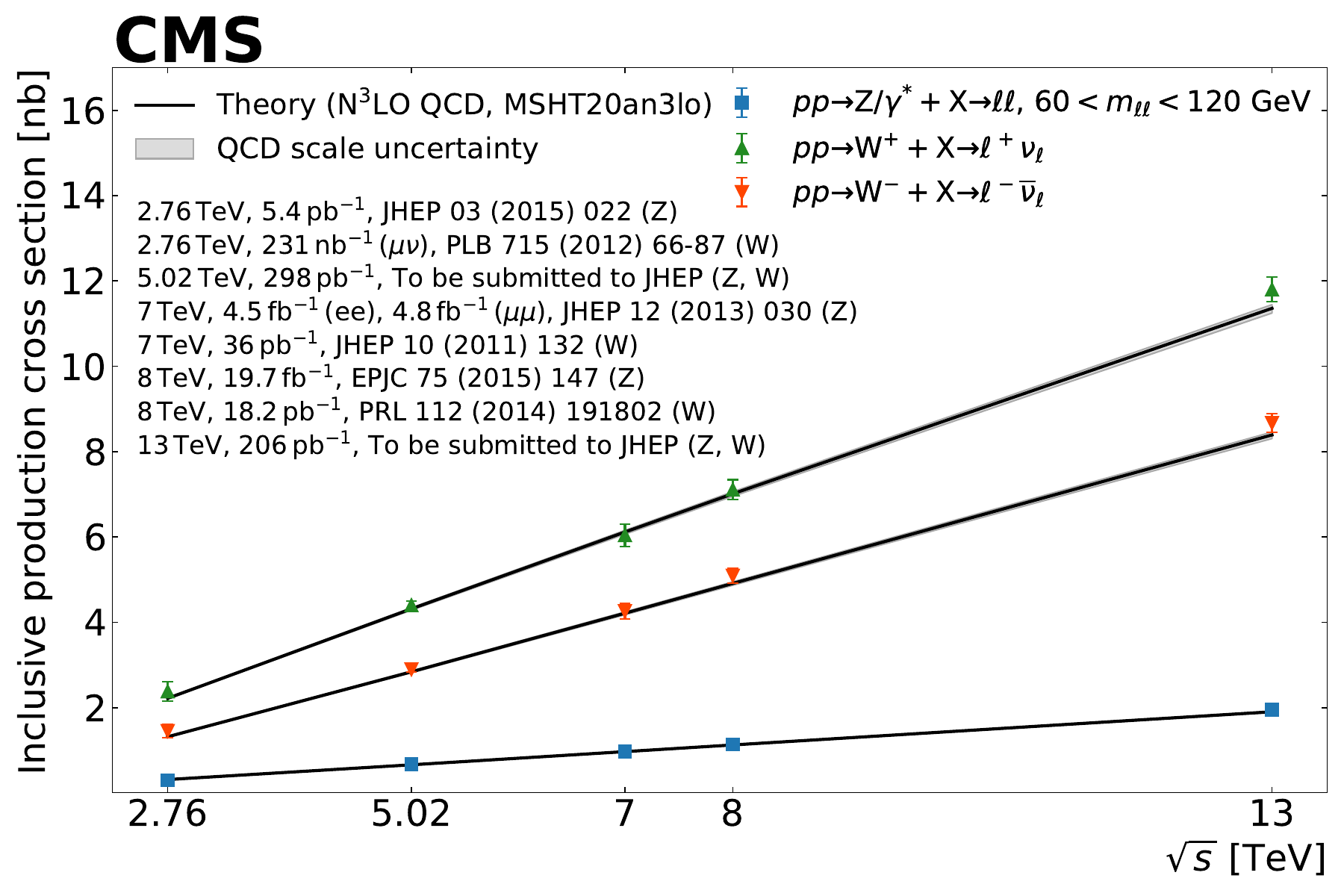} 
\caption{Summary of the production cross section of weak gauge bosons, measured by CMS, plotted against the $\Pp\Pp$ centre-of-mass energy ranging from 2.76 to 13\TeV. The error bars around the experimental data points represent the total uncertainty of the measurement. The measurements are compared with theoretical predictions (black lines) obtained at \ncube in QCD using the MSHT20a\ncube PDF set. The grey band shows the envelope from renormalization and factorization scale variations.}\label{fig:wzxs}
\end{figure}

\begin{figure}
	\centering 
	\includegraphics[width=0.80\textwidth]{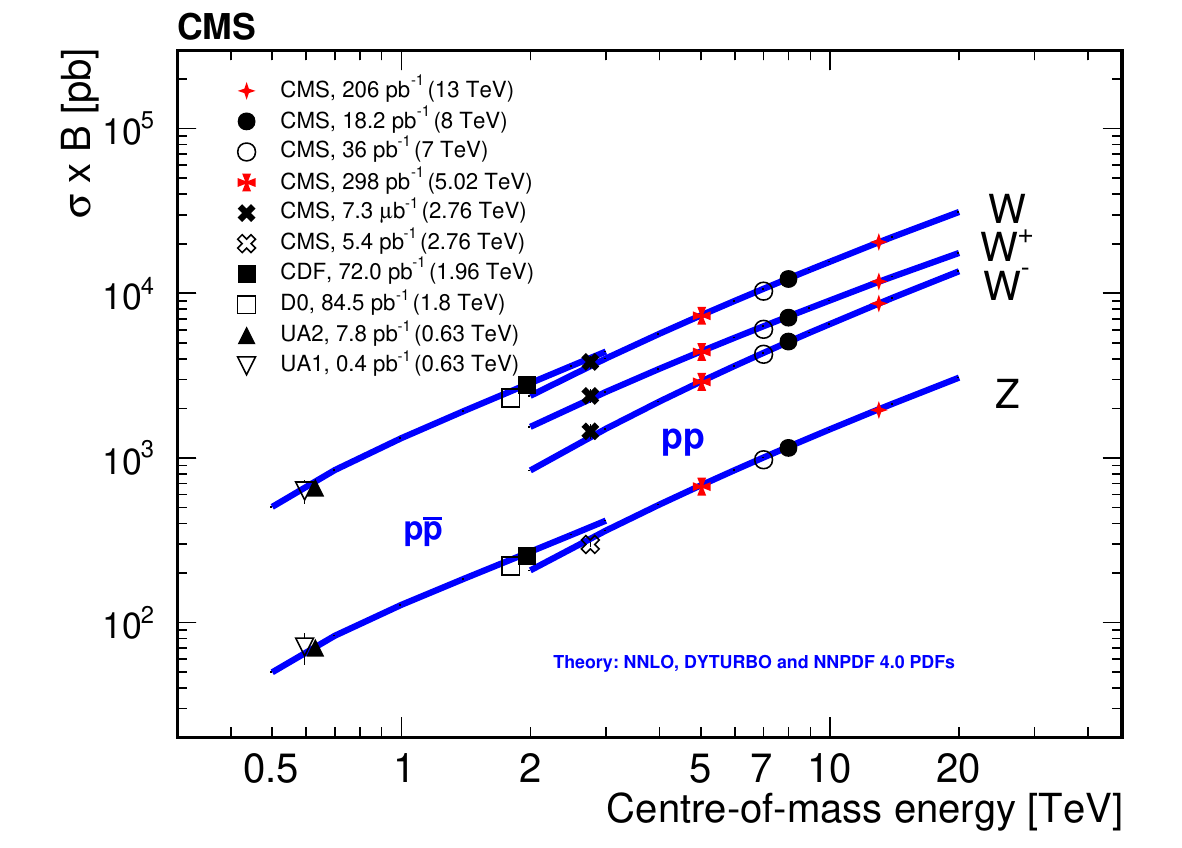}	
	\caption{Summary of the production cross section of weak gauge bosons in \pp collisions, measured by CMS, and in $\Pp\PAp$ collisions, by the UA1, UA2, CDF, and D0 experiments, plotted against the \pp or $\Pp\PAp$ centre-of-mass energy ranging from 0.63 to 13\TeV. The measurements are compared with theoretical predictions (blue lines) obtained at NNLO in QCD by using \textsc{DYTurbo} and the NNPDF4.0 PDF set. Figure taken from Ref.~\cite{CMS:2024myi}.}\label{fig:wzxswpp}
\end{figure}

These measurements have been made in fiducial phase spaces and extrapolated to the full production cross sections for both the $\PW$ and $\PZ$ bosons at each energy at which the LHC has operated.  Shown in Fig.~\ref{fig:wzxs} is a comparison of the CMS measurements of the full production cross section of $\PW$ and $\PZ$ bosons in leptonic decay channels at 2.76\TeV~\cite{CMS:2012fgk,CMS:2014dyj} ($\PW$ and $\PZ$ bosons, respectively), 5.02~\cite{CMS:2024myi},  7~\cite{CMS:2011aa,CMS:2013zfg}, 8~\cite{CMS:2014pkt,CMS:2014jea}, and 13~\cite{CMS:2024myi} compared with the predictions at \ncube~\cite{Baglio:2022wzu} in QCD using the MSHT20a\ncube~\cite{McGowan:2022nag} PDF set. The full production cross sections are presented as the cross sections times single leptonic branching fractions where the \PZ boson branching fraction is for a dilepton mass range of 60 to 120\GeV.  The measurement of the $\PZ$ boson cross section at~2.76\TeV uses the differential measurement versus rapidity presented in Ref.~\cite{CMS:2014dyj} integrating the results over the measured rapidity range and extrapolating to the full one using \textsc{DYTurbo}~\cite{Camarda:2019zyx} at \ncube. The measurements at 2.76 and 5.02\TeV are based on the \pp collision reference data for the heavy ion physics programme. The \ncube cross section predictions are the most accurate currently available and Fig.~\ref{fig:wzxs} illustrates the ability to make precise comparisons of cross sections between experimental measurements and theoretical prediction at a hadron collider. 
Figure~\ref{fig:wzxswpp} presents the CMS $\PW$ and $\PZ$ cross section measurements along with cross section measurements from previous $\Pp\PAp$ colliders including the UA1~\cite{ALBAJAR1987271} and UA2~\cite{ua21990measurement} experiments at the CERN $\mathrm{S\Pp\PAp S}$, where the $\PW$ and $\PZ$ bosons were first discovered, and the CDF~\cite{CDF:2004sns} and D0~\cite{D0:1999qdf} experiments at the Tevatron. The results are compared with the NNLO predictions computed using \textsc{DYTurbo} and the NNPDF4.0 PDF, which yields the smallest cross section uncertainties for weak boson production of the currently available global PDF sets. 
The CMS results are also presented in the full cross section summary Fig.~\ref{fig:XsAll}. The theoretical predictions for total, fiducial, and ratio measurements presented in the following tables are computed at NNLO using, for the 5 and 13\TeV predictions, \DYTURBO with the NNPDF3.1 PDF set; and, for 7\TeV, using \FEWZ with the NNPDF2.1 PDF set. The theoretical predictions for the 8\TeV ratio of cross sections are computed at NNLO using \FEWZ with the MSTW2008 PDF set. 

Table~\ref{tab:Zinclusive} presents the inclusive cross section for $\PZ$ production in \pp collisions at various energies. The largest source of uncertainty in the measurements is the integrated luminosity. The most precise cross section measurements have been made with low-pileup data sets collected in short time periods that allow a more precise determination of the luminosity.

\begin{table}[htbp]
  \centering
\topcaption{Measured inclusive cross sections for $\PZ$ boson production at \pp collision energies from 2.76 to 13\TeV. Total uncertainties in the experimental measurements are given in \pb and as a percentage. Separate components of the experimental statistical and systematic uncertainties other than the dominant integrated luminosity uncertainty were not published for the 2.76\TeV cross section measurement. The statistical uncertainties of the 7 and 8\TeV measurements are smaller than 1\pb and are not shown. The measurements are compared with theoretical predictions obtained at \ncube in QCD using the MSHT20a\ncube PDF set. The theoretical uncertainty is from renormalization and factorization scale variations.\label{tab:Zinclusive}}
\renewcommand{\arraystretch}{1.2}
  \begin{tabular}{l l l l}
    $\sqrt{s}$ ($\TeVns$)   &  $\sigma(\PZ)$ (pb) & Tot. exp. unc. & $\sigma^\text{SM}(\PZ)$ (pb)\\
      \hline
    2.76~\cite{CMS:2014dyj}  & $298  \pm 10 \stat \syst \pm  11 \lum $  & 5.0\%   & $313  \substack{+1 \\ -2}$ \\

    5.02~\cite{CMS:2024myi}  & $669  \pm 2 \stat \pm 6 \syst \pm 13 \lum $   & 2.2\%   & $674.7  \substack{+7.1 \\ -7.4}$ \\  
    7~\cite{CMS:2013zfg}     & $986  \pm 22\syst \pm 22 \lum $  & 3.1\%   & $968  \substack{+6 \\ -7}$ \\
 
    8~\cite{CMS:2014jea}     & $1138  \pm 26\syst \pm 30 \lum $  & 3.5\%   & $1124 \substack{+7 \\ -2}$ \\
 
    13~\cite{CMS:2024myi}    & $1952 \pm 4 \stat \pm 18 \syst \pm 45 \lum $  & 2.5\%   & $1940 \substack{+15 \\ -21}$ \\
  \end{tabular}
\end{table}

Measuring the cross section in a fiducial phase space reduces the total systematic uncertainty by removing or minimizing the additional uncertainty from the extrapolation of the cross section from the fiducial phase space region where it is measured to the full production phase space. Fiducial measurements of the $\PZ$ cross section are presented in Table~\ref{tab:Zfiducial}. The 8\TeV fiducial cross section measurement is from Ref.~\cite{CMS:2014pkt}.

\begin{table}[htbp]
  \centering
\topcaption{Measured fiducial cross sections for $\PZ$ boson production and decay to electrons and muons in \pp collisions at energies from 5.02 to 13\TeV. Total uncertainties in the experimental measurements are given in \pb and as a percentage.  The measurements are compared with theoretical predictions at NNLO in QCD described in the references above. In each case, the uncertainty in the CMS measurement of the fiducial $\PZ$ boson cross section is reduced compared with the inclusive measurement and the integrated luminosity uncertainty dominates the overall uncertainty of the measurements.\label{tab:Zfiducial}}
\renewcommand{\arraystretch}{1.2}
  \begin{tabular}{l l l l }
    $\sqrt{s}$ ($\TeVns$)   &  $\sigma_\text{fid.}(\PZ)$ (pb) & Tot. exp. unc. & $\sigma_\text{fid.}^\text{SM}(\PZ)$ (pb)\\
 \hline
    5.02~\cite{CMS:2024myi}  & $319.8 \pm 0.9 \stat \pm 1.2 \syst \pm 6.2 \lum  $ & 2.0\%   & $319.5 \pm 3.7$ \\
    7~\cite{CMS:2013zfg}     & $524.7  \pm 0.4 \stat  \pm 5.2\syst \pm 11.5 \lum $ & 2.4\%   & $525  \pm 6$ \\
    8~\cite{CMS:2014pkt}     & $410.0  \pm 10.0 \stat  \pm 10.0\syst \pm 10.0 \lum $ & 4.2\%   & $400  \pm 10 $\\
    13~\cite{CMS:2024myi}    & $754 \pm 2 \stat \pm 3 \syst \pm 17 \lum          $ & 2.3\%  & $743 \pm 18$ \\
  \end{tabular}
\end{table}

Table~\ref{tab:WZinclusiveratio} lists the measurements of ratios of the inclusive $\PW$ and $\PZ$ cross sections, and Table~\ref{tab:WZfiducialratio} lists the measurement of the ratios of fiducial cross sections. The measurements of the ratios of $\PW$ to $\PZ$ boson cross sections remove the dependence on the integrated luminosity determination and that of any other efficiencies or factors that apply to both measurements identically, substantially reducing the systematic uncertainty. For this reason, cross section ratios are among the most precise measurements performed by the CMS experiment.   

\begin{table}[htbp]
\centering
\topcaption{Measured ratios, $R_\text{exp}$, of inclusive cross sections for $\PW$ and $\PZ$ boson production times the branching fractions $\mathcal{B}(\PW \to \ell\nu)$ and $\mathcal{B}(\PZ \to \ell^+\ell^-)$ (with the dilepton mass between 60 and 120\GeV), respectively. Ratios $R_{\PWp/\PWm} = \sigma(\PWp) \mathcal{B}(\PWp \to \ell^+\nu)/ \sigma(\PWm) \mathcal{B}(\PWm \to \ell^-\PAGn)$ and $R_{\PW/\PZ} = \sigma(\PW) \mathcal{B}(\PW \to \ell\nu)/\sigma(\PZ) \mathcal{B}(\PZ \to \ell^+\ell^-)$ are shown for \pp collision energies from 5.02 to 13\TeV. The total uncertainty in the experimental measurement is shown in the standard and percentage forms. The measurements are compared with theoretical predictions, $\mathrm{R_{\text{SM}}}$,  obtained at NNLO in QCD. The theoretical uncertainties, expressed as percentages, are from renormalization and factorization scale variations, $\alpS$, and the PDF uncertainty.\label{tab:WZinclusiveratio}}
\renewcommand{\arraystretch}{1.2}
\maybeCmsTable{
  \begin{tabular}{lllll}
    $\sqrt{s}$ ($\TeVns$) &  Ratio & $R_\text{exp}$ & Tot. exp. unc. & $R_\text{SM}$ \\
      \hline
    5.02~\cite{CMS:2024myi}  & $R_{\PWp/\PWm}$ & $1.519\pm 0.002 \stat \pm 0.010 \syst  $ & $0.67\%$  & $1.5240  \substack{+0.33\% \\ -0.31\%}$ \\
    7~\cite{CMS:2011aa}     & $R_{\PWp/\PWm}$ & $1.421 \pm 0.006 \stat \pm 0.032\syst  $ & $1.8\%$  & $ 1.43 \pm 0.7\%$ \\
    8~\cite{CMS:2014pkt}     & $R_{\PWp/\PWm}$ & $1.39 \pm 0.01 \stat \pm 0.02 \syst    $ & $1.6\%$  & $ 1.41 \pm 0.7\% $ \\
    13~\cite{CMS:2024myi}    & $R_{\PWp/\PWm}$ & $1.3615\pm 0.0018\stat \pm 0.0094\syst $ & $0.70\%$  & $1.3536 \substack{+0.37\% \\ -0.33\%}$ \\
   5.02~\cite{CMS:2024myi}   & $R_{\PW/\PZ}$ & $10.905\pm 0.032 \stat \pm 0.054 \syst $ & $0.58\%$  & $ 10.777 \substack{+0.33\% \\ -0.34\%}$\\
    7~\cite{CMS:2011aa}     & $R_{\PW/\PZ}$   & $10.54 \pm 0.07 \stat \pm 0.18 \syst   $ &  $2.3\%$  & $ 10.74 \pm 0.4\%$\\
    8~\cite{CMS:2014pkt}     & $R_{\PW/\PZ}$   & $10.63\pm 0.11 \stat \pm 0.25 \syst    $ & $2.6\%$  & $ 10.74 \pm 0.4\%$\\
    13~\cite{CMS:2024myi}    & $R_{\PW/\PZ}$   & $10.491\pm 0.024\stat \pm 0.083\syst  $ & $0.82\%$  & $ 10.341 \substack{ +0.41\% \\ -0.38\% }$\\
  \end{tabular}}
\end{table}

\begin{table}[htbp]
  \centering
\topcaption{Measured ratios, $R_\text{exp}$, of fiducial cross sections for $\PW$ and $\PZ$ boson production times the branching fractions $\mathcal{B}(\PW \to \ell\nu)$ and $\mathcal{B}(\PZ \to \ell^+\ell^-)$, respectively. Ratios $R_{\PWp/\PWm} = \sigma(\PWp) \mathcal{B}(\PWp \to \ell^+\nu)/ \sigma(\PWm) \mathcal{B}(\PWm\to \ell^-\PAGn)$ and $R_{\PW/\PZ} = \sigma(\PW) \mathcal{B}(\PW \to \ell\nu)/ \sigma(\PZ) \mathcal{B}(\PZ\to \ell^+\ell^-)$ are shown for at \pp collision energies from 5.02 to 13\TeV. The total uncertainty in the experimental measurement is shown in the standard and percentage forms. The measurements are compared with theoretical predictions, $R_\text{SM}$, obtained at NNLO in QCD. The theoretical uncertainties, expressed as percentages, are from renormalization and factorization scale variations, $\alpS$, and PDF uncertainty.\label{tab:WZfiducialratio}}
\renewcommand{\arraystretch}{1.2}
  \begin{tabular}{lllll}
    $\sqrt{s}$ ($\TeV$) &  Ratio & $R_\text{exp}$ & Tot. exp. unc. & $R_\text{SM}$ \\
     \hline
    5.02~\cite{CMS:2024myi}  & $R_{\PWp/\PWm}$ & $1.6232\pm 0.0026 \stat \pm 0.0065 \syst $ & $0.43\%$  & $1.631  \pm 0.98\%$ \\
    8~\cite{CMS:2014pkt}     & $R_{\PWp/\PWm}$ & $1.40 \pm 0.01 \stat \pm 0.02 \syst $ & $1.6\%$  & $ 1.42 \pm 1.4\% $ \\
    13~\cite{CMS:2024myi}    & $R_{\PWp/\PWm}$ & 1.3159$\pm 0.0017\stat \pm 0.0053\syst $ & $0.43\%$  & $ 1.307 \pm 1.3\%$ \\
   5.02~\cite{CMS:2024myi}  & $R_{\PW/\PZ}$  & $12.505\pm 0.037 \stat \pm 0.032 \syst $ & $0.39\%$  & $ 12.51 \pm 0.96\%$\\
    8~\cite{CMS:2014pkt}     & $R_{\PW/\PZ}    $ & $13.26\pm 0.15 \stat \pm 0.21 \syst $ & $1.9\%$  & $ 13.49 \pm 2.1\%$\\
    13~\cite{CMS:2024myi}    & $R_{\PW/\PZ}    $ & 12.078$\pm 0.028\stat \pm 0.032\syst $ & $0.35\%$  & $ 12.02 \pm 2.3\%$\\
  \end{tabular}
\end{table}

The recent cross section results at 5.02\TeV are the most precise because they feature an improved integrated luminosity uncertainty of 1.9\%. Comparisons of theoretical predictions to the total, fiducial, and the ratios of the measured 5.02 and 13\TeV \PW to \PZ cross sections are reported in Ref.~\cite{CMS:2024myi}, computed at NNLO in QCD using \DYTURBO~\cite{Camarda:2019zyx,Camarda:2021ict,Camarda:2021jsw} and the NNPDF3.1 NNLO PDF set. These predictions were improved to next-to-next-to-leading-logarithmic (NNLL) accuracy using resummation~\cite{Balazs:1995nz,Catani:2015vma}, which better models the \pt distribution of the $\PZ$ bosons at low \pt values.  This reduces systematic uncertainties associated with the extrapolation from the measurement in the fiducial region to the total cross section.
For instance, in 5.02\TeV \pp collisions the $\PZ$ and $\PW$ boson cross sections with a subsequent decay to leptons were measured in a fiducial phase space as:
$\sigma(\PZ) = 319.8 \pm 0.9 \stat \pm 1.2 \syst \pm 6.2 \lum\pb$ (2.0\% total uncertainty),
and $\sigma(\PW) = 4000 \pm 3 \stat \pm 11 \syst \pm 76 \lum \pb$  (1.9\% total uncertainty), 
which are the most precise single cross section measurements performed by the CMS experiment. 
Ratios of cross sections can be measured with better than 0.5\% precision in fiducial phase space, since the dependence of the measurement on the integrated luminosity and the understanding of some reconstruction efficiencies is removed by forming a ratio of cross sections of similar production processes. For 13\TeV \pp collisions the same analysis measured $\sigma(\PWp)/\sigma(\PWm) = 1.3159 \pm 0.0017\stat \pm 0.0053\syst $ (0.43\% total uncertainty), and $\sigma(\PW)/\sigma(\PZ) = 12.078 \pm 0.028\stat \pm 0.032\syst$ (0.35\% total uncertainty). The effort by the LHC experiments to make precise measurements has been matched by progress in theory in producing higher QCD and EW perturbative order predictions, and improving our understanding of PDFs. As with the experimental measurements precise predictions can be made of ratios of production cross sections. For comparison theoretical prediction of $\sigma(\PWp)/\sigma(\PWm)$ at 13\TeV, computed at NNLO, has a precision 0.35\% for the ratio of total cross sections and 1.3\% for the ratio of fiducial cross sections (using one PDF set) due to larger renormalization and factorization scale uncertainties when computing the ratio in a restricted phase space. These ratios are sensitive to the quark content of the protons as described above and,  in general, vector boson production measurements are a strong input to determining the proton PDFs.   

\begin{figure}
	\centering 
	\includegraphics[width=0.80\textwidth]{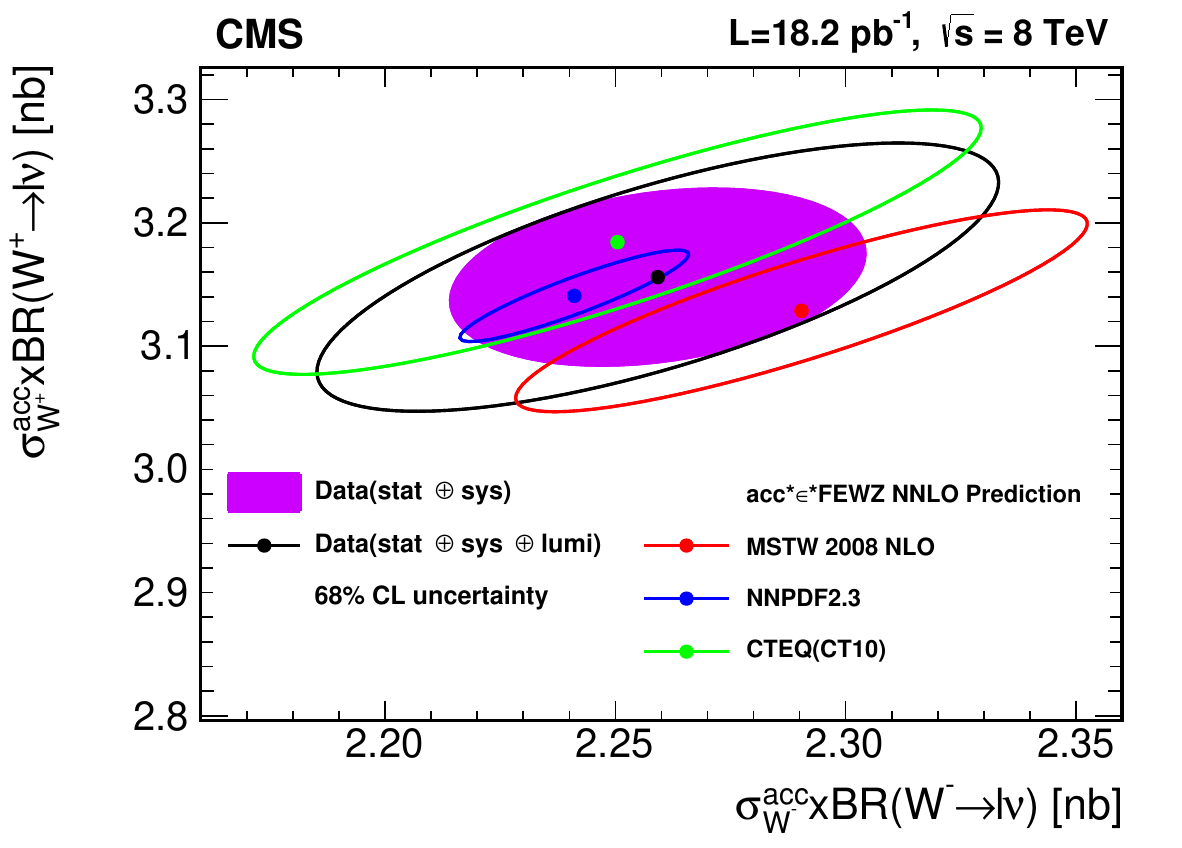}	
	\caption{Measured and predicted \PWp versus \PWm production fiducial cross sections times branching fractions. The ellipses illustrate the 68\% \CL coverage for total uncertainties (open) and excluding the integrated luminosity uncertainty (filled). The uncertainties in the theoretical predictions correspond to the PDF uncertainty components only and are evaluated for three PDF sets: NNPDF2.3, CTEQ CT10, and MSTW 2008 NLO. Figure taken from Ref.~\cite{CMS:2014pkt}.}\label{fig:wpwmratio}
\end{figure}

In Fig.~\ref{fig:wpwmratio} a 2D comparison of the \PWp and \PWm boson cross sections in 8\TeV \pp collisions is shown, illustrating the improved precision of ratios of both the experimental measurements~\cite{CMS:2014pkt} and theoretical predictions calculated at NNLO in QCD using \FEWZ~\cite{Melnikov:2006di,Melnikov:2006kv}. The large integrated luminosity uncertainty and its cancellation in the ratio are clearly seen in the shape of the uncertainty ellipse. 

\subsubsection{Differential measurements of vector boson production}
The CMS experiment has measured the differential cross sections of photons, and $\PW$ and $\PZ$ bosons \vs a variety of kinematic variables considered in up to three dimensions. Of particular interest are analyses that differentially measure the rapidity or other angular variables of the weak bosons or their leptonic decays. In $\PW$ boson decays, these measurements have direct sensitivity to the PDFs of the quarks in the proton of the same charge sign as the $\PW$ boson. The DY production of $\ell^+\ell^-$ pairs, when considering a wider range of masses around the $\PZ$ boson peak, has the sensitivity to the EW mixing angle $\theta_W$. The measurements are often reported as asymmetries comparing the positive and negative $\PW$ boson or lepton distributions as a function of rapidity in $\PW$ boson production or as a forward-backward asymmetry of the negative lepton direction in DY production of $\ell^+\ell^-$ pairs.

The $\PW$ production charge asymmetry can be measured as:
\begin{linenomath}
  \begin{equation}
    \mathcal{A}(\absyw) = \frac{\rd\sigma/\rd\absyw(\PWp\to\ell^+\PGn) - \rd\sigma/\rd\absyw(\PWm\to\ell^-\PAGn)}
 {\rd\sigma/\rd\absyw(\PWp\to\ell^+\PGn) + \rd\sigma/\rd\absyw(\PWm\to\ell^-\PAGn)},\label{eq:asymmetry}  
  \end{equation}
\end{linenomath}
where $\rd\sigma/\rd\absyw$ is the differential cross section for the absolute value of the $\PW$ boson production rapidity in the laboratory frame.

\begin{figure}
	\centering 
	\includegraphics[width=0.80\textwidth]{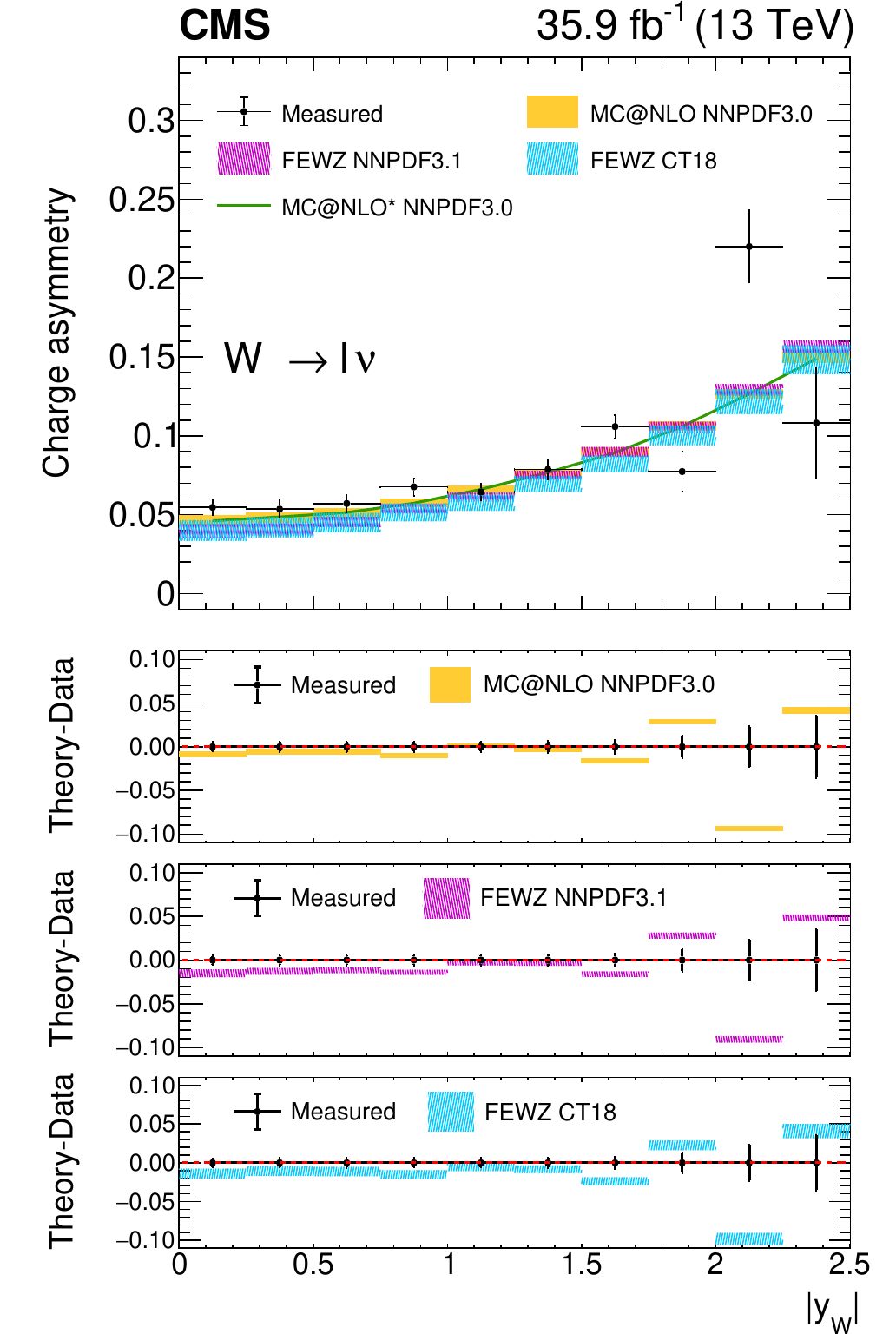}	
	\caption{Measured $\PW$ boson charge asymmetry as a function of $\absyw$ from the combination of the electron and muon channels (black dots), compared with different theoretical predictions. The vertical errors bars around the experimental data points show the total uncertainty of the measurements. The yellow band represents the default generator used in this analysis, MG5\_aMC with the NNPDF3.0 PDF set, the pink band represents the \FEWZ generator with the NNPDF3.1 PDF set, and the cyan band represents the \FEWZ generator with the CT18 PDF set. The uncertainty bands of the prediction include the PDF uncertainties only, which are dominant with respect to $\alpS$, or renormalization and factorization scale variations for this quantity. Figure taken from Ref.~\cite{CMS:2020cph}.}\label{fig:WAsym}
\end{figure}

The charge asymmetry in leptonic $\PW$ boson decays has been measured in \pp collisions at 7~\cite{CMS:2011bet,CMS:2013pzl,CMS:2012ivw}, 8~\cite{CMS:2016qqr}, and  13~\cite{CMS:2020cph}\TeV, where the charge asymmetry was also separately reported for the left- and right-handed $\PW$ boson helicity states. The $\PW$ boson charge asymmetry as a function of the absolute value of the $\PW$ boson rapidity is shown in Fig.~\ref{fig:WAsym}. Comparisons are made to \MGvATNLO NLO simulation (denoted \MCATNLO) interfaced with \PYTHIA for PS and QED lepton FSR and normalized to NNLO calculations using \FEWZ 2.0~\cite{Gavin:2010az} with two PDF sets. For the NLO comparison, the \pt distribution of the generated $\PW$ boson is reweighted based on comparisons between the \pt distribution of $\PZ$ boson data and \MGvATNLO simulation.   Also, the QED lepton FSR distribution is corrected to that of \PHOTOS~\cite{Golonka:2005pn}. All predictions agree well with the data, except at high rapidity where some fluctuations are visible in the measurements relative to all three predictions. The PDF fits performed using the 7 and 8\TeV data were reported in Section~\ref{subsec:PDF}.

For DY production of $\ell^+\ell^-$ pairs, the forward-backward asymmetry, ${A}_\mathrm{FB}$, is computed in several regions of lepton pair mass as:
\begin{linenomath}
\begin{equation}
	{A}_\mathrm{FB} = \frac{\sigma_\mathrm{F}-\sigma_\mathrm{B}}{\sigma_\mathrm{F}+\sigma_\mathrm{B}},
\end{equation}
\end{linenomath}
where $\sigma_\mathrm{F}$ ($\sigma_\mathrm{B}$) is the total cross section for the forward (backward) events, defined by $\cos\theta^{*} > 0$ ($\cos\theta^{*} < 0$), where $\cos\theta^{*}$ is the angle between the negatively charged lepton and the $\PZ$ boson momentum vector direction (in the laboratory frame) measured in the lepton pair centre-of-mass frame. The \afb depends on $m(\ell^+\ell^-)$, quark flavour, and the EW mixing angle $\theta_\PW$. Near the $\PZ$ boson mass peak, the \afb is close to zero because of the small value of the charged-lepton vector coupling to $\PZ$ bosons. Due to weak-electromagnetic interference, \afb is large and negative for  $m$ below the $\PZ$ boson peak ($m < 80$\GeV) and large and positive above the $\PZ$ boson peak ($m > 110$\GeV).

The DY ${A}_\mathrm{FB}$ measurements are reported for \pp collision data at 7~\cite{CMS:2011utm,CMS:2012zgh}, 8~\cite{CMS:2016bil,CMS:2018ktx}\TeV, and 13~\cite{CMS:2024ony}\TeV.  In Ref.~\cite{CMS:2024ony},  $\sin^2\theta^\text{eff}_{\text{lept}}$ was measured using the angular-weighted asymmetry ${A}^\mathrm{W}_\mathrm{FB}$ around the $\PZ$ boson peak as function of the dilepton mass and rapidity.   The measurement uses both dimuon and dielectron events including forward electrons in the range $3.14 < \abs{\eta} < 4.36$ which have greater sensitivity to determine $\sin^2\theta^\text{eff}_{\text{lept}}$.  The angular weighting method~\cite{Bodek:2010qg} employs simple event weights which are functions of the $y$ and $\cos\theta^{*}$ to yield the best estimate of the acceptance corrected ${A}^\mathrm{W}_\mathrm{FB}$.   As modelled for different $\sin^2\theta^\text{eff}_{\text{lept}}$ values using \POWHEG v2 incorporating the MiNNLO method~\cite{Monni:2019whf,Monni:2020nks}, and higher order EW corrections~\cite{Barze:2013fru,Chiesa:2019nqb,Chiesa:2024qzd}  with the CT18Z PDF set, the effective leptonic EW mixing angle was extracted as $\sin^2\theta^\text{eff}_{\text{lept}} = 0.23157\pm0.00010\stat\pm0.00015\syst\pm0.00009\thy\pm0.00027\pdf = 0.23157 \pm 0.00031$.

\subsubsection{Measurements of vector boson production in association with jets}\label{subsubsec:vjets}
Many vector boson analyses also consider associated jet production. As with pure QCD jet analysis, the production of vector bosons in association with jets is an excellent test of perturbative QCD predictions. Production of  $\PW$ and $\PZ$ in association with jets, followed by the $\PWp \to \ell^+\nu_\ell$ and $\PZ \to \ell^+\ell^-$ decays, respectively, allows for some of the most stringent perturbative QCD tests. Figure~\ref{fig:FD:DYVjet} shows Feynman diagrams for the radiation of a photon, $\PZ$ boson, or $\PW$ boson from a quark where the boson is produced in association with one jet. These NLO QCD diagrams for vector boson production can either involve a gluon in the initial state or the radiation of a gluon in the final state. The addition of new initial states, in this case involving a gluon, means that NLO production almost always increases the expected inclusive cross section and including NLO diagrams is always necessary to get reasonably accurate cross section predictions. Topologies with up to 8 jets have been analyzed and compared with MC generators at LO, NLO, and NNLO accuracy.  The MC generators achieve NLO or NNLO accuracy up to a limited number of additional jets, and further jets are simulated at lower perturbative accuracy.

\begin{figure}
	\centering 
	\includegraphics[width=0.80\textwidth]{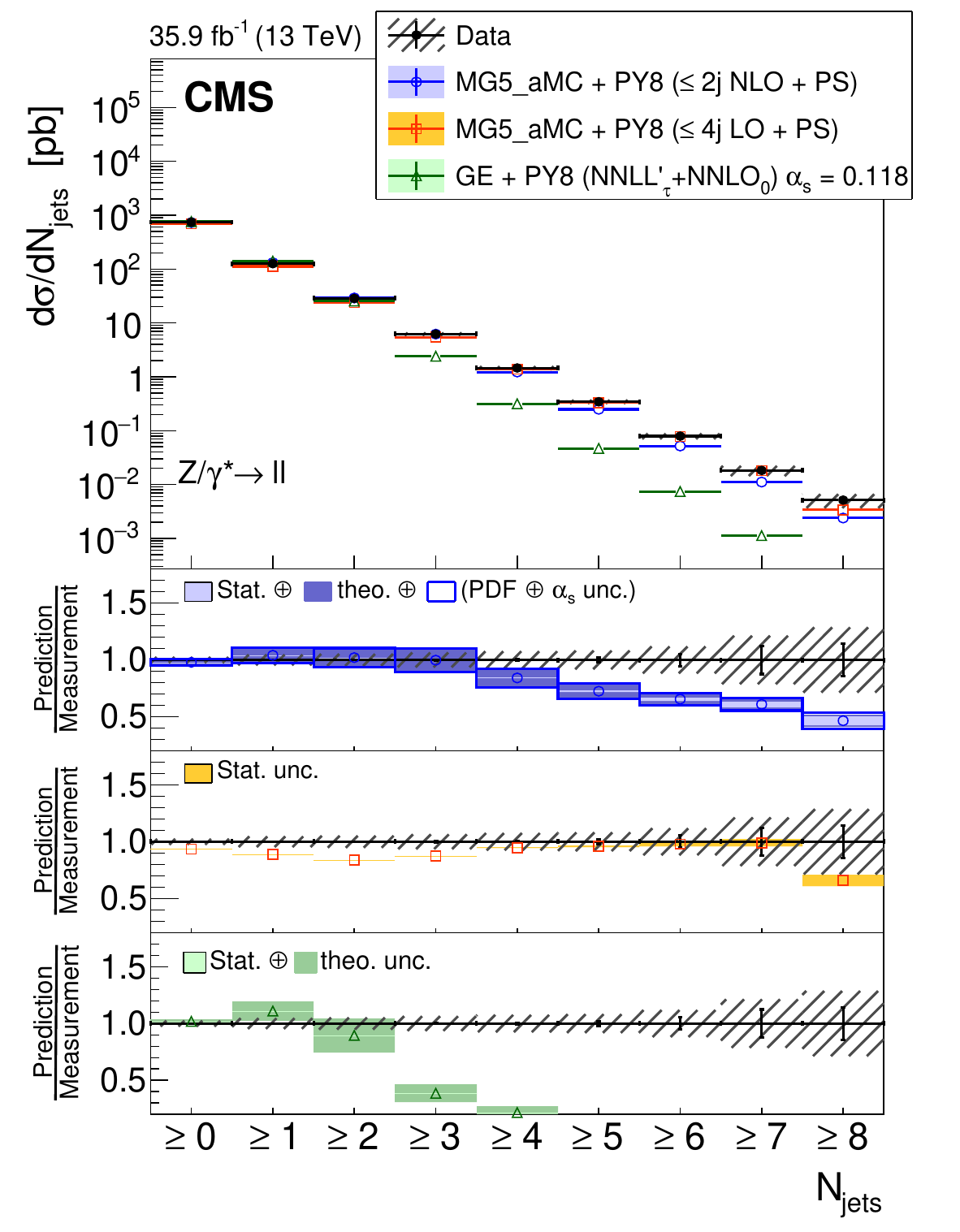}	
	\caption{The differential cross section of $\PZ \to \ell^+\ell^-$+jets production as a function of inclusive jet multiplicity, compared with the predictions calculated with \MGvATNLO (LO) + \PYTHIA~8, \MGvATNLO (NLO) + \PYTHIA~8, and \GENEVA. The lower panels show the ratios of the theoretical predictions to the measurements. The measurement statistical (systematic) uncertainties are presented with vertical error bars (hashed areas). The boxes around the \MGvATNLO (NLO) + \PYTHIA 8 to measurement ratio represent the uncertainty in the prediction as listed in the legend. Figure taken from Ref.~\cite{CMS:2022ilp}.}\label{fig:zjets}
\end{figure}

The most recent 13\TeV $\PZ$+jets measurement~\cite{CMS:2022ilp} is shown in Fig.~\ref{fig:zjets} with comparisons to three fixed-order MC generator predictions. Fixed-order predictions generate at a given level of perturbative accuracy all tree-level production diagrams for the selected process and all diagrams with additional partons up to a given number.   
In the analysis, jets are required to have $\pt > 30\GeV$ and $\abs{\mathrm y} < 2.4$.  The first comparison is to \MGvATNLO generated with $\le$4 partons at LO accuracy interfaced with \PYTHIA 8 for PS using the MLM~\cite{Alwall:2007fs,Alwall:2008qv} ME-PS jet merging scheme. The second comparison is to \MGvATNLO generated with $\le$2 partons at NLO accuracy interfaced with \PYTHIA 8 for PS using the FxFx~\cite{Frederix:2012ps} ME-PS jet merging scheme. As an NLO QCD prediction, one-loop diagrams are included, as well as diagrams with real emission of an additional parton (in this case a third parton) at LO accuracy. The samples are normalized to the inclusive NLO cross section prediction produced using \MCFM. The final comparison is to the \GENEVA~\cite{Alioli:2015toa,Alioli:2012fc} MC which combines an NNLO ME calculation with an NNLL accuracy resummation of the zero-jettiness $\tau$ variable, also known as the beam thrust~\cite{Stewart:2010tn}. The NNLO matrix elements include the real emission of two additional partons. Thus the \MGvATNLO prediction effectively includes three-jet topologies at LO accuracy, and the \GENEVA NNLO prediction effectively includes one-jet topology at NLO accuracy and two-jet topology at LO accuracy.  The results show that modelling additional jets using ME calculations produces the best agreement with predictions at higher jet multiplicities. In fact, the \MGvATNLO (NLO) and \GENEVA (NNLO) predictions exhibit disagreement for all jet multiplicities that exceed the number of jets included in the ME calculations. The \MGvATNLO L0 generator, with up to 4 partons in the ME calculations, models the entire distribution well. In this analysis, \PYTHIA 8 uses the CUETP8M1~\cite{CMS:2015wcf} tune of UE physics based on the MONASH~\cite{Skands:2014pea} tune, which was trained to improve modelling of a wide variety of data sets including DY production at lower LHC energies.  

A complete set of cross section measurements for $\PW$ and $\PZ$ production in association with jets is displayed in Fig.~\ref{fig:XsJetsEW}. The analyses, and the MC generators and configurations used to evaluate the theory comparisons shown in the plot are given in Table~\ref{tab:WZXJets}. The figure includes cross section measurements for topologies with vector bosons, multiple vector bosons, Higgs bosons, and top quark production in association with jets. The 8\TeV $\PZ$+jets results~\cite{CMS:2016php} are summed as necessary over the exclusive results per number of jets with uncertainties computed accounting for correlations of systematic sources.

\begin{table}[htbp]
  \centering
\topcaption{Measurements of $\PW$ and $\PZ$ boson production in association with jets and the MC generators used for comparison to the measured cross sections. All measurements are inclusive cross sections for the vector boson produced in association with the listed or higher number of jets. For each measurement, the \pp collision energy, ME generator, largest number of hard partons generated, largest number of hard partons generated at NLO accuracy, PS generator, and the ME-PS matching scheme are given. Events generated with greater than the number of NLO partons have LO accuracy. If no matching scheme is listed the comparison was done directly to the parton-level cross section predictions after applying a correction for NP effects. For the 7 and 8\TeV results the \SHERPA with \BLACKHAT (\SHERPA 1/2, BH) NLO comparison was done only for lower parton multiplicities. The \MADGRAPH 5 or \MGvATNLO (denoted MG5\_aMC) comparisons are shown for higher jet multiplicities.\label{tab:WZXJets}}
\renewcommand{\arraystretch}{1.2}
  \begin{tabular}{l l l l l l l}
    Boson  & $\sqrt{s}$  & Generator & partons & partons & PS & ME-PS \\
    \# Jets & ({\TeVns})  &           & total   & NLO    &    & scheme\\
     \hline
    \PW 1--5j~\cite{CMS:2014oon} & 7   & \SHERPA 1,BH  & 5 & 5 & \NA   & \NA   \\
 
    \PW 6j~\cite{CMS:2014oon}    & 7   & \MADGRAPH 5 & 4 & \NA  & Py6 & CKKW~\cite{Alwall:2007fs}\\
 
    \PW 1--4j~\cite{CMS:2016sun}  & 8   & \SHERPA 2,BH  & 4 & 4  & \NA   & \NA \\
 
    \PW 5,6j~\cite{CMS:2016sun}  & 8   & MG5\_aMC   & 3 & 2 & Py8 & FxFx\\
 
    \PW 1--6j~\cite{CMS:2017gbl}  & 13  & MG5\_aMC   & 4 & 2 & Py8 & FxFx\\
 
    \PZ 1--6j~\cite{CMS:2014bkk}  & 7   & \SHERPA 1,BH  & 4 & 1 & CS  & MEPS@NLO \\
 
    \PZ 1--7j~\cite{CMS:2016php}  & 8   & \SHERPA 2,BH  & 4 & 2 & CS  & MEPS@NLO \\
 
    \PZ 1--6j~\cite{CMS:2018mdf}  & 13  & MG5\_aMC   & 4 & 2 & Py8 & FxFx\\
  \end{tabular}
\end{table}

\begin{figure}
	\centering 
	\includegraphics[width=0.80\textwidth]{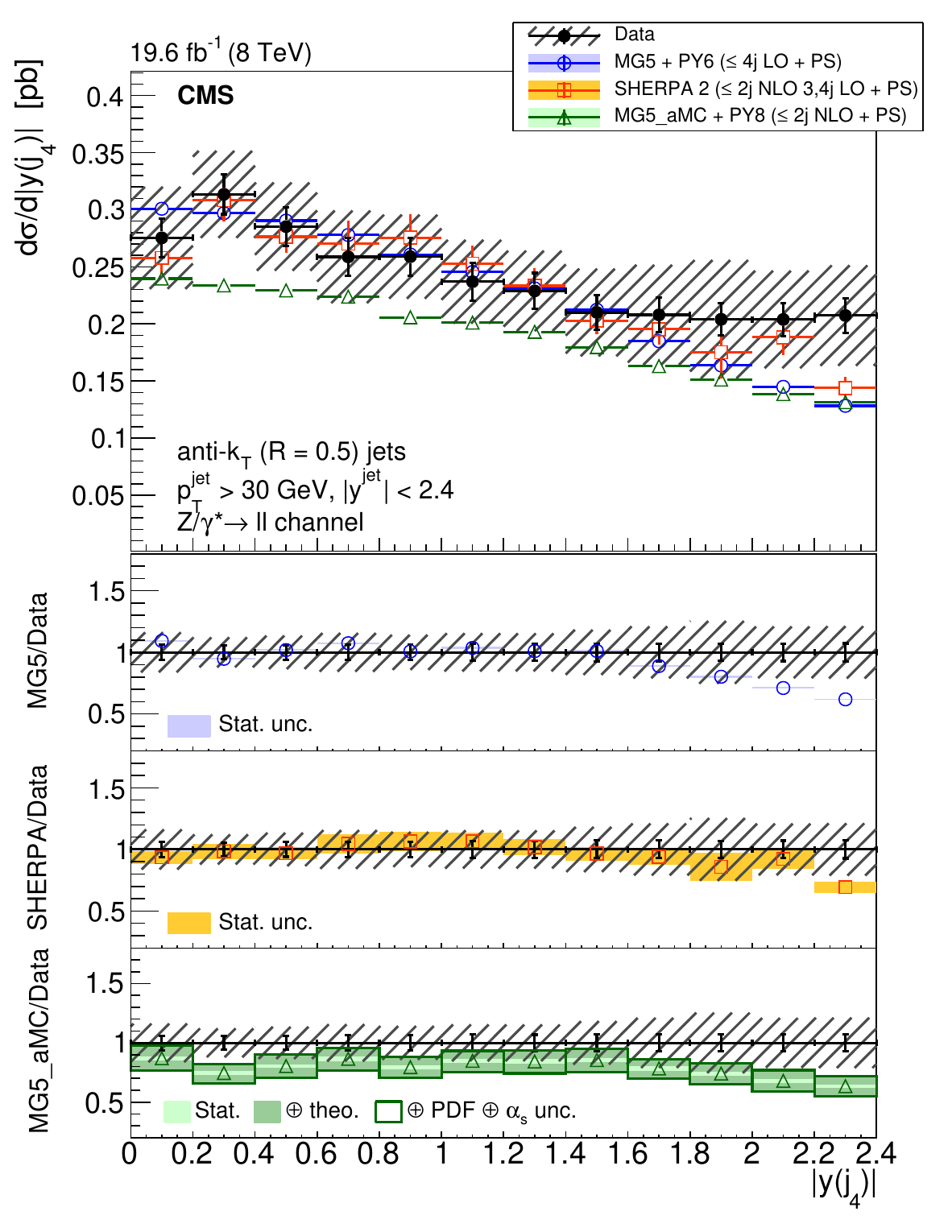}	
	\caption{The differential cross section for $\PZ \to \ell^+\ell^- +$ jets production as a function of the absolute value of the 4$\text{th}$ jet's rapidity compared with the predictions calculated with \MADGRAPH~5+\PYTHIA~6, \SHERPA~2, and MG5\_aMC +\PYTHIA~8. The lower panels show the ratios of the theoretical predictions to the measurements. Error bars around the experimental points show the statistical uncertainty and the cross-hatched bands indicate the statistical and systematic uncertainties added in quadrature. The boxes around the MG5\_aMC + \PYTHIA 8 to measurement ratio represent the uncertainty in the prediction, including statistical, theoretical (from scale variations), and PDF uncertainties. The dark green area represents the statistical and theoretical uncertainties only and the light green area represents the statistical uncertainty alone. Figure taken from Ref.~\cite{CMS:2016php}.}\label{fig:zjetsrap}
\end{figure}

Differential properties of vector boson production in association with jets are a complex and stringent test of our understanding of perturbative QCD physics. An illustrative example is shown in Fig.~\ref{fig:zjetsrap} of the jet rapidity of the $4^\text{th}$ jet from the 8\TeV analysis of $\PZ$+jets data~\cite{CMS:2016php}. This 8\TeV $\PZ$+jets measurement includes comparisons to three MC generators. The first comparison is to \MADGRAPH 5 generated with $\le$4 partons with LO accuracy interfaced to \PYTHIA 6 for PS (denoted MG5 + PY6). The parameters of \PYTHIA 6 are set to the Z2* tune~\cite{CMS:2013yli}, which are designed to reproduce lower collision energy LHC data, and are found to model DY data well~\cite{CMS:2015wcf}. The \MADGRAPH 5 prediction is normalized to the \FEWZ NNLO cross section. The second comparison is to \MGvATNLO (denoted MG5\_aMC) generated with $\le$3 partons, at NLO accuracy for events with $\le$2 partons and LO accuracy for 3 partons.  The \MGvATNLO generator is interfaced with \PYTHIA 8 for PS using the FxFx ME-PS merging scheme. The final comparison is to \SHERPA 2 with \BLACKHAT~\cite{Berger:2008ag,Berger:2010gf} generated with $\le$4 partons, with NLO accuracy for events with $\le$2 partons and LO accuracy for 3 and 4 partons, PS using \CSSHOWER PS~\cite{Schumann:2007mg} based on Catani--Seymour dipole factorization, interfaced with NLO accuracy using the MEPS@NLO~\cite{Hoeche:2012yf} ME-PS merging scheme (the combination of which is denoted Sherpa 2). The NLO predictions are not normalized. In this measurement, an analysis of the rapidity of each jet, where the jets are ordered in \pt, is performed. The selected plot corresponds to the fourth \pt-ordered jet, which is the highest jet multiplicity for which the statistical power is sufficient for a precise comparison of the rapidity distribution with the simulation. As shown above, LO predictions do well with more inclusive properties, such as the simple production of a given number of jets. However, they do not perfectly model many kinematic features of the production of jets. Higher-order generators can capture more of the details of the production kinematics. In this analysis, the LO predictions of the rapidity distribution of jets disagree for the lower-\pt jets in $\PZ$ boson + multijet events with high multiplicities of jets. The best agreement is seen with the \SHERPA 2 predictions, which include LO MEs for four-jet production and NLO generation for lower numbers of jets. Differential analyses of complex final states are essential in pushing our understanding of QCD and combined EW and QCD physics. These are the types of analyses that most directly reveal the shortcomings in our ability to model complex physics interactions and show the need for higher perturbative order predictions of parton-parton interactions.    

Associated production of a photon and a jet has been measured triple-differentially at 7~\cite{CMS:2013myp}, 8~\cite{CMS:2019jlq} and 13~\cite{CMS:2018qao}\TeV as a function of photon \et, photon rapidity, and jet rapidity. The results are compared with the NLO calculations from \JETPHOX~\cite{Catani:2002ny} (7 and 13\TeV) and NLL calculations using \GAMJET~\cite{Baer:1989xj,Baer:1990ra} and the CJ15 PDF set~\cite{Accardi:2016qay} (8\TeV). Both calculations use the BFG~\cite{Bourhis:1997yu} fragmentation functions for quarks and gluons. The measurements are in good agreement with the predictions. In the same analysis, the inclusive production cross section of events with at least one photon and one jet has been measured. With a requirement of $\pt > 40\GeV$ for both objects, a cross section of $8.01 \pm 0.11 \stat \pm 0.74 \syst$\unit{nb}~\cite{CMS:2013myp} is measured consistent with theory predictions. This result was obtained by integrating over the differential $\eta$ and \pt cross sections presented in Ref.~\cite{CMS:2013myp}, accounting for correlations between systematic uncertainty sources.

\begin{figure}
	\centering 
	\includegraphics[width=0.7\textwidth]{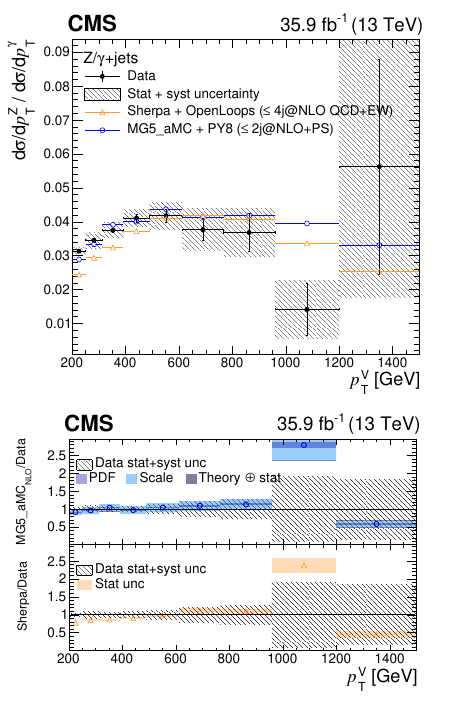}	
	\caption{Differential cross section ratio of $\PZ$+jets (with $\PZ \to \mu^+\mu^-$) to $\PGg$+jets as a function of the vector boson (V) transverse momentum compared with the theoretical prediction from \MGvATNLO and \SHERPA + \textsc{OpenLoops}. Only bosons produced centrally, with $\abs{y}< 1.4$, in association with one or more jets are considered. The panel shows the ratio of the theoretical prediction to the unfolded data. The vertical errors bars around the experimental data points show the statistical uncertainties of the measurements. The hatched band is the sum in quadrature of the statistical and systematic uncertainty components in the measurement. The dark (light) shaded band on the NLO prediction from \MGvATNLO represents the PDF (scale) uncertainties, which are treated as uncorrelated between $\PZ$+jets and $\PGg$+jets, whereas the statistical uncertainties are barely visible. The shaded band on the \SHERPA + \OPENLOOPS calculation is the statistical uncertainty. Figure taken from Ref.~\cite{CMS:2021fxy}.}\label{fig:Zphoton}
\end{figure}

Although the CMS experiment has not generally performed simple $\PGg$+jets counting analyses as in the $\PW$+jets and $\PZ$+jets cases, it has performed an array of differential analyses of $\PGg$+jets production. Among the most interesting of these analyses are comparisons between $\PGg$+jets and $\PZ$+jets production, where the $\PZ$ bosons decay to muons which is the lowest background decay mode. These allow us to study the similarities between these final states, which are leveraged in SM cross section analysis and BSM physics searches involving photons, by using our extensive understanding of low-background events with $\PZ$ bosons to better describe topologies involving a photon. The $\PGg$+jets and $\PZ$+jets comparisons have been performed at  7~\cite{CMS:2013jnr}, 8~\cite{CMS:2015onn}, and 13~\cite{CMS:2021fxy}\TeV.  Comparisons are made to MC simulations of the kinematic distributions of the bosons and the jets as functions of the number and type (light or \PQb-flavoured) jets. Cross section distributions are shown separately for events with $\PZ$ bosons and photons, and as ratios. Figure~\ref{fig:Zphoton} shows a comparison from the 13\TeV analysis~\cite{CMS:2021fxy} of the ratio of $\PZ$+jets and $\PGg$+jets production in events with at least one jet compared with NLO QCD with NLO EW theoretical predictions. Two fixed-order NLO MC generator comparisons are shown. The \MGvATNLO comparison (denoted MG5\_aMC) of Z production includes topologies with up to 3 hard partons and events with $\le$2 partons have NLO QCD accuracy, whereas events with 3 partons have LO accuracy. The \MGvATNLO $\PGg$+jets production is generated with up to one parton at NLO QCD accuracy. Matrix element to PS matching is performed using the FxFx prescription~\cite{Frederix:2012ps}. The cross section of the generated \PZ boson sample is normalized to the value of an NNLO prediction computed with \FEWZ. The \SHERPA + \OPENLOOPS~\cite{Sherpa:2019gpd,Schonherr:2016pvn} samples of $\PZ$ and $\PGg$ production are generated with $\le$4 partons, with NLO QCD accuracy for events with $\le$2 partons and LO accuracy for events with 3 and 4 partons. Approximate EW corrections are applied to these samples using the \COMIX~\cite{Gleisberg:2008fv} and \OPENLOOPS~\cite{Buccioni:2019sur,Denner:2016kdg,Ossola:2007ax,vanHameren:2010cp} ME generators. Parton showering is performed using \CSSHOWER~\cite{Schumann:2007mg} and ME-PS jets matching is performed using the \MCATNLO method~\cite{Hoeche:2011fd,Frixione:2002ik}. As the branching fraction of the $\PZ$ boson to muons is 3.4\%, Fig.~\ref{fig:Zphoton} is an illustration of EW unification at high energy, since the ratio of production cross sections and thus the coupling constants for the $\PZ$ bosons and photons is of order one and independent of energy above several times the $\PZ$  boson mass.

\subsubsection{Measurements of vector boson production in association with heavy-flavour jets}\label{subsubsec:hfjets}
\begin{figure}
	\centering 
	\includegraphics[width=0.31\textwidth]{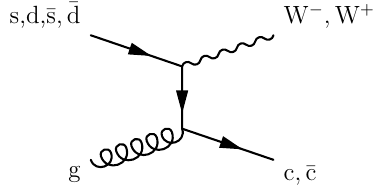}	
	\includegraphics[width=0.25\textwidth]{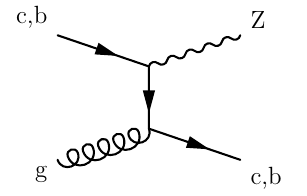}	
	\includegraphics[width=0.25\textwidth]{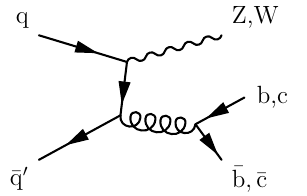}	
	\caption{Production of \PW or \PZ bosons with heavy-flavour quarks. Examples of lowest order Feynman diagrams include $\PW$ + charm (left), $\PZ$ + charm or bottom (middle), $\PW$ or $\PZ$ production with two heavy-flavour quarks (right).}\label{fig:FD:Vhf}
\end{figure}

The CMS experiment has performed many analyses of vector boson production in association with bottom- and charm-flavoured jets. Representative Feynman diagrams are shown in Fig.~\ref{fig:FD:Vhf}. Advancements in machine-learning techniques have resulted in the creation of highly efficient jet taggers for bottom and charm jets, demonstrating high accuracy and minimal backgrounds from light-flavour quark and gluon jets. Other effective techniques of identifying heavy-flavour jets include the reconstruction of exclusive final states for charm tagging. The measurement of  $\PW$ + charm jet events provides a direct probe of the strange quark content of the proton. The CMS PDF constraints from $\PW$ + charm measurements are competitive with those from the neutrino scattering and global PDF fits. The study of $\PW$ and $\PZ$ boson production with charm jets may eventually contribute to the endeavour to measure the second-generation quark Yukawa coupling to the Higgs boson using associated V\PH production with the Higgs boson decaying to charm quarks. The study of $\PZ$ + charm jets could contribute to studies of the intrinsic charm component of the proton PDF, where it would contribute to additional $\PZ$ + charm jet events at high \pt. Consequently, the CMS $\PZ$ + charm analyses measure the differential distribution of charm jet production \vs jet \pt. The V+\PQb or multiple \PQb jets production, where V is a $\PW$ or $\PZ$ boson, contains events sensitive to the \PQb quark content in the proton or gluon splitting to \PQb jets. The  CMS experiment has also studied $\WZ$ and $\ZZ$ production, with one $\PZ$ boson decaying to two \PQb jets~\cite{CMS:2014zmy}, yielding the V+2 \PQb jets signature, constitutes the dominant irreducible background to associated Higgs boson production ($\PW\PH$ and $\PZ\PH$), and provides important input to that study.

A complete set of cross section measurements for vector boson with heavy-flavour jet production is shown in Fig.~\ref{fig:XsJetsEW}. One of the most critical components of each analysis is the heavy-flavour jet tagging method. Table~\ref{tab:WZHFXJets} lists the production cross sections measured, the \pp collision energy, the heavy-flavour tagging technique, and the source of theory cross section calculation used for comparison of the vector boson with heavy-flavour jet production measurements. The heavy-flavour tagging techniques were explained in Section~\ref{hftaggers}. In addition, the table lists for each analysis other results produced, such as differential distributions and PDF constraints. As the measurement of the $\PZ$ + charm jet cross section at 8\TeV is performed in a fiducial region, the cross section is multiplied by the acceptance for leptonic $\PZ$ boson decays taken from the same Ref.~\cite{CMS:2017snu} to calculate the total cross section for comparison to the other results. The measurement and prediction of cross sections with jets have long been difficult at high-energy colliders with many discrepancies between data that were identified and later resolved with a better understanding of both detector calibration of quark and gluon jet momentum and the theoretical modelling of such processes. The good agreement between the experimental measurements and predictions for a high multiplicity of jets, including the production of heavy-flavour jets, is an important achievement of the LHC physics programme that found use in the discovery of the Higgs boson and in the searches for BSM physics.

\begin{table}[htbp]
  \centering
\topcaption{Table of measurements of $\PW$ and $\PZ$ boson production in association with heavy-flavour quarks. The table lists the measured production cross sections,  \pp collision energy,  heavy-flavour tagging technique, source of theory cross section calculation used for comparison, and other results of interest produced by the analysis.  In several cases, ratios of production cross sections are measured including 
$R_{\PWp\PAQc/\PWm\PQc} = \sigma(\PWp\PAQc)/\sigma(\PWm\PQc)$, $R_{\PW\PQc/\PZ\PQb} = \sigma(\PW\PQc)/\sigma(\PZ\PQb)$, $R_{\PZ\PQb/\PZ\PQq} = \sigma(\PZ\PQb)/\sigma(\PZ\PQq)$ and $R_{\PZ\ge 2\PQb/\PZ\ge 1\PQb} = \sigma(\PZ\ge 2\PQb)/\sigma(\PZ\ge 1\PQb)$.  Parton-level \MCFM NLO and NNLO predictions are corrected for NP effects. All predictions are computed at NLO QCD accuracy except for the $\PW$+$\PQc$ 13\TeV analysis, where the prediction is done at NNLO QCD and NLO EW accuracy~\cite{Czakon:2020coa,Czakon:2022khx}.\label{tab:WZHFXJets}}
\renewcommand{\arraystretch}{1.2}
  \begin{tabular}{l l l l l}
    Boson  & $\sqrt{s}$  & Heavy flavour & Theory      & Other \\ 
    \# Jets & ({\TeVns}) &  tagging     & calculation & results\\
     \hline
    $\PW$+1$\PQc$~\cite{CMS:2013wql} & 7    & D meson & \MCFM &  $R_{\PWp\PAQc/\PWm\PQc}$, $\pt(\mu)$\\
    $\PW$+1$\PQc$~\cite{CMS:2021oxn} & 8    & $\mu$, SSV, IVF & \MCFM & $R_{\PWp\PAQc/\PWm\PQc}$, $\pt(\mu)$, $\eta(\ell)$, s PDF\\
    $\PW$+1$\PQc$~\cite{CMS:2018dxg} & 13   & D meson & \MCFM&  $R_{\PWp\PAQc/\PWm\PQc}$, $\eta(\mu)$, s PDF\\
    $\PW$+1$\PQc$~\cite{CMS:2023aim} & 13   & SV tag: SSV IVF & NNLO &  $R_{\PWp\PAQc/\PWm\PQc}$, $\pt(\mu)$, $\eta(\mu)$\\
    $\PW$+2$\PQb$~\cite{CMS:2013xis} & 7    & CSV& \MCFM & \\
    $\PW$+2$\PQb$~\cite{CMS:2016eha} & 8    & CSV & \MCFM & \\
    $\PZ$+1$\PQc$~\cite{CMS:2017snu} & 8    & $\mu$+SV: SSV IVF, D & \MCFM & $R_{\PW\PQc/\PZ\PQb}$, $\pt(\PZ)$, $\pt(\PQc)$\\
    $\PZ$+1$\PQc$~\cite{CMS:2020cso} & 13   & \textsc{DeepCSV}+$m_{\text{SV}}$ & MG5\_aMC & $\pt(\PZ)$, $\pt(\PQc)$\\
    $\PZ$+1,2$\PQb$~\cite{CMS:2014jqj} & 7  & SSV & \MCFM & $R_{\PZ\PQb/\PZ\PQq}$ \\
    $\PZ$+1,2$\PQb$~\cite{CMS:2016gmz} & 8  & CSV & MG5\_aMC& $R_{\PZ\ge 2\PQb/\PZ\ge 1\PQb}$, $m_{\PQb\PQb}$, 20 dist.\\
    $\PZ$+1,2$\PQb$~\cite{CMS:2021pcj} & 13 & \textsc{DeepCSV} & MG5\_aMC & $R_{\PZ\ge 2\PQb/\PZ\ge 1\PQb}$, $m_{\PQb\PQb}$, 15 dist.\\
\end{tabular}
\end{table}

\subsection{Inclusive multiboson production and interactions}  
Multiboson production is typically categorized into inclusive production that is dominated by the radiation of vector bosons from initial-state quarks in the proton, and
EW  production in which the radiation of bosons is followed by pure EW interactions among the vector (and Higgs) bosons via scattering or fusion. These interactions are classified into the subsets of diboson production, triboson production, VBF, and VBS. Studying multiboson production provides a test of the gauge structure of the SM that uniquely predicts how the gauge bosons interact with each other by directly measuring triple gauge boson couplings (TGCs) and eventually quartic gauge boson couplings (QGCs). Studying VBS and the polarization of the bosons gives sensitivity to the features of EW symmetry breaking, which has been exclusively studied at the LHC and can provide a platform to search for BSM anomalous quartic gauge boson couplings (aQGCs). In addition, ratios of production rates have sensitivity to PDFs. Measurements are typically made either inclusively of a diboson
signature, including the EW processes, or of only the EW component, as described in Section~\ref{subsec:EW}. In principle, every multi-gauge-boson process in the SM with up to three gauge bosons can be observed at the LHC experiments. Several multiboson states can be observed in such pure samples for which cross section measurements are approaching the 3\% total uncertainty level, and they may eventually be measured with the accuracy approaching that of single vector boson production. Currently, only the rarest of the multivector boson processes, such as ZZ VBS production (which has been detected with 4$\sigma$ significance~\cite{CMS:2020fqz}) and most triboson production processes involving two or more weak bosons, have not been observed by the CMS experiment. Representative LO Feynman diagrams for $\PW\PZ$ production are shown in Fig.~\ref{fig:FD:WZ} including both radiative production, where the bosons are radiated off a quark, and TGC production, where $\PQq\PAQq$ annihilation results in an off-shell $\PW$ boson, which splits into the $\PW$ and $\PZ$ bosons. The interference of the amplitudes of these two processes dominates the production cross section for inclusive $\PW\PZ$ production.

\subsubsection{Diboson production}
\begin{figure}
	\centering 
	\includegraphics[width=0.40\textwidth]{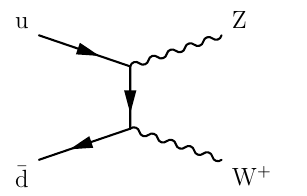}	
	\includegraphics[width=0.40\textwidth]{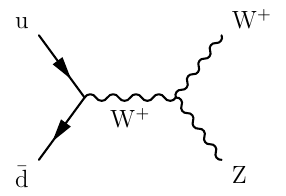}	
	\caption{Feynman diagrams for $\WZ$ diboson production. Shown are radiative production (left), where the vector bosons are radiated off a quark, and a TGC production (right), where a $\PW$ boson is created by $\PQq\PAQq$ annihilation and splits into $\PW$ and $\PZ$ bosons.  These diagrams are representative of all diboson production mechanisms that involve radiative or TGC processes. In the case of neutral final states TGCs are forbidden in the SM and only anomalous coupling due to new physics could lead to contributions from that type of diagram.}\label{fig:FD:WZ}
\end{figure}

The diboson production cross sections are among the most precisely measured by the CMS experiment. The combination of pure $\PWp \to \ell^+\nu_\ell$ and $\PZ \to \ell^+\ell^-$ samples and the large integrated luminosity delivered by the LHC and collected by the CMS experiment provide a precision rarely achieved previously by hadron collider experiments.   An understanding of diboson production is essential for the studies of the Higgs boson and searches for new physics where diboson production is often a significant SM background. Diboson production also has an indirect sensitivity to new physics that may occur in loop diagrams often characterised as anomalous additions to the SM TGC and QGC multiboson couplings.  The Feynman diagram shown in Fig.~\ref{fig:FD:WZ} (right) illustrates how $\WZ$ production has sensitivity to measure the SM $\PW\PW\PZ$ TGCs or anomalous TGCs (aTGCs) that could modify those couplings due to BSM physics contributions.  

\begin{figure}
	\centering 
	\includegraphics[width=0.40\textwidth]{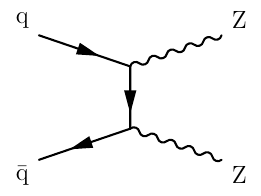}	
	\includegraphics[width=0.40\textwidth]{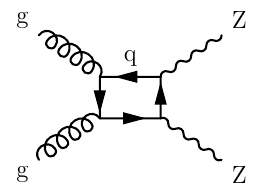}	
	\caption{Feynman diagrams for $\ZZ$ diboson production including radiative production (left) and NNLO production via a gluon-gluon initial state (right), which increases the total production cross section significantly.}\label{fig:FD:ZZ}
\end{figure}

\begin{figure}
	\centering 
	\includegraphics[width=0.95\textwidth]{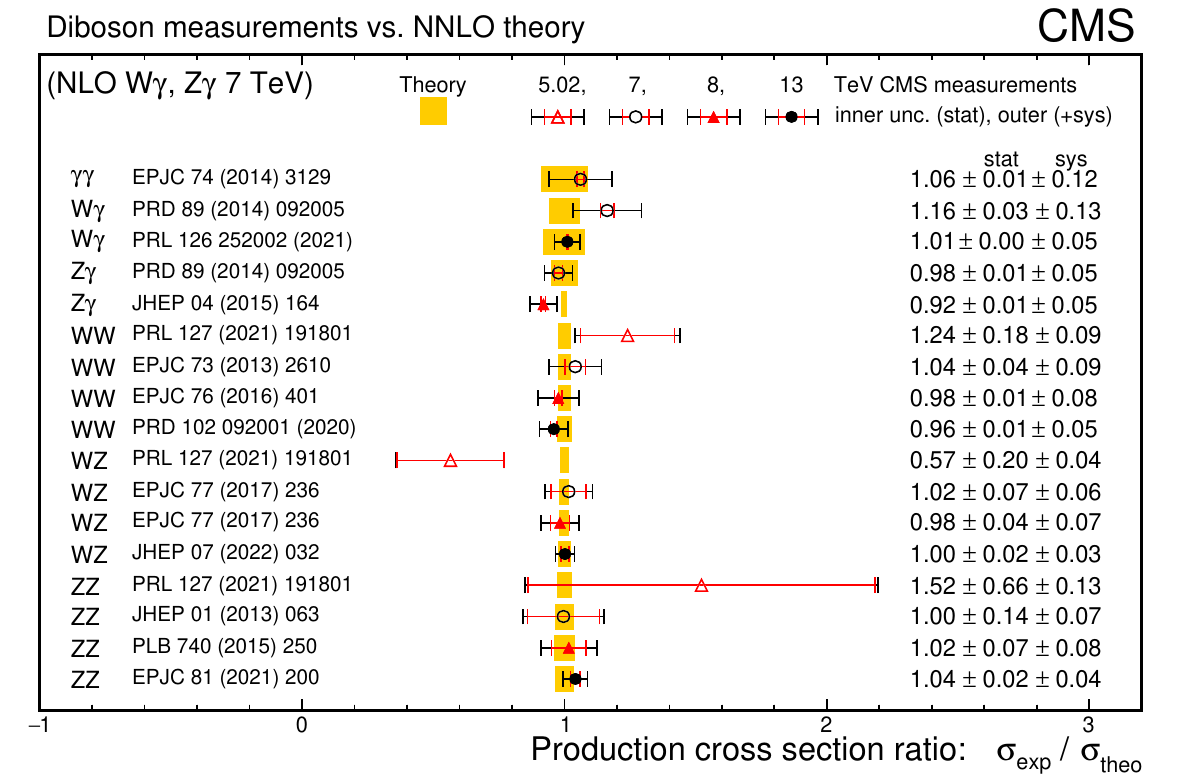}	
	\caption{Summary of cross section measurements for diboson production shown as a ratio over the NNLO or NLO QCD predictions. The yellow bands indicate the uncertainties in the theoretical predictions and the error bars on the points are the statistical uncertainties, whereas the outer bars are the combined statistical and systematic uncertainties.}\label{fig:diboson}
\end{figure}

In the first LHC 7\TeV run all the diboson states seen by previous experiments were observed, including $\PGg\PGg$~\cite{CMS:2014mvm}, $\PW\PGg$ and $\PZ\PGg$~\cite{CMS:2013ryd}, opposite-sign $\WWOS$~\cite{CMS:2013ant}, $\WZ$~\cite{CMS:2016qth}, and $\ZZ$~\cite{CMS:2012exm} signatures. The cross sections for diboson production have been measured at 5.02, 7, 8, and 13\TeV in Run 1 and Run 2 of the LHC.  
The diboson production processes measured at CMS are listed in Table~\ref{tab:diboson}.  Included is information on \pp collision energy, theory calculations used for comparison in Fig.~\ref{fig:diboson}, and other results of interest. For comparison NNLO QCD predictions are necessary to predict the cross sections and distributions of these processes with sufficient accuracy. This is both because NNLO production can introduce new initial states, such as the gluon-gluon initial state for $\ZZ$ (and $\WWOS$) production, shown along with the radiative production Feynman diagram in Fig.~\ref{fig:FD:ZZ}, and because the precision of the experimental diboson production measurement in many final states is at the several percent level, which requires NNLO QCD computations to achieve equivalent accuracy. These factors have pushed extensive developments in the theory to accurately predict these states and match the precision of the experimental measurements. The theoretical cross section for comparison to the measured $\PGg\PGg$ production rate is calculated using the 2$\PGg$\textsc{NNLO}~\cite{Catani:2011qz} program. Comparisons to theoretical cross section predictions for the 7\TeV $\PW\PGg$ and $\PZ\PGg$ production are calculated using parton-level \MCFM NLO predictions corrected for NP effects. The 8\TeV $\PZ\PGg$ result is compared with the NNLO prediction from Ref.~\cite{Grazzini:2013bna}. The \MATRIX predictions have NNLO QCD and NLO EW precision for $\PQq\PAQq$ processes, and NLO QCD accuracy for the $\Pg\Pg$ initial state processes that contribute to \WWOS and \ZZ production.  
Same-sign (SS) $\WWSS$ production has been measured as well and is discussed in Section~\ref{subsec:EW}.   

\begin{table}[htbp]
  \centering
\topcaption{Table of diboson production cross section measurements. Listed in the table are the final states studied, \pp collision energy, theory cross section calculation used for comparison, and selected additional results of interest from each paper.\label{tab:diboson}}
\renewcommand{\arraystretch}{1.2}
  \begin{tabular}{l l l l}
    Process  & $\sqrt{s}$  &  Theory      & Other results \\
             & ({\TeVns})  &  calculation & \\
     \hline
    $\PGg\PGg$~\cite{CMS:2014mvm}  & 7    & 2$\PGg$\textsc{NNLO} & $m_{\PGg\PGg}$, 4 dist. \\
    $\PW\PGg$~\cite{CMS:2013ryd}       & 7    & \MCFM NLO & aTGC, $\pt(\PGg)$ \\
    $\PW\PGg$~\cite{CMS:2021foa}       & 13    & MG5\_aMC 1p NLO & aTGC \\
    $\PZ\PGg$~\cite{CMS:2013ryd}       & 7    & \MCFM NLO & aTGC, $\pt(\PGg)$ \\
    $\PZ\PGg$~\cite{CMS:2015wtk}       & 8    & NNLO & aTGC, $\pt(\PGg)$ \\
    \WWOS~\cite{CMS:2021pqj}  & 5.02 & \MATRIX & \\
    \WWOS~\cite{CMS:2013ant}  & 7    & \MATRIX & aTGC\\
    \WWOS~\cite{CMS:2015tmu}  & 8    & \MATRIX &  aTGC, $\sigma$: with jet veto, 4 dist.\\
    \WWOS~\cite{CMS:2020mxy}  & 13   & \MATRIX & aTGC, $\sigma$: with jet veto\\
    $\PW\PZ$~\cite{CMS:2021pqj}              & 5.02 & \MATRIX & \\
    $\PW\PZ$~\cite{CMS:2016qth}              & 7    & \MATRIX & \\
    $\PW\PZ$~\cite{CMS:2016qth}              & 8    & \MATRIX & aTGC, $\pt(\PZ)$, $\pt(\mathrm{jet})$\\
    $\PW\PZ$~\cite{CMS:2021icx}              & 13   & \MATRIX & aTGC, boson polarization, 9 dist.\\
    $\PZ\PZ$~\cite{CMS:2021pqj}              & 5.02 & \MATRIX & \\
    $\PZ\PZ$~\cite{CMS:2012exm}              & 7    & \MATRIX & aTGC\\
    $\PZ\PZ$~\cite{CMS:2014xja}              & 8    & \MATRIX & aTGC, $m_{4\ell}$, 7 dist.\\
    $\PZ\PZ$~\cite{CMS:2020gtj}              & 13   & \MATRIX & aTGC, 6 dist.\\
  \end{tabular}
\end{table}

These measurements are summarized in Fig.~\ref{fig:diboson}. The figure shows that both experimental measurements and theory, typically at the level of NNLO QCD, agree over all of the diboson production states with percent-level precision. In papers with total and fiducial measurements (13\TeV \WWOS, \WZ and \ZZ), the fiducial cross section measurements have better precision and are used in the figure.

\begin{figure}
	\centering 
	\includegraphics[width=0.80\textwidth]{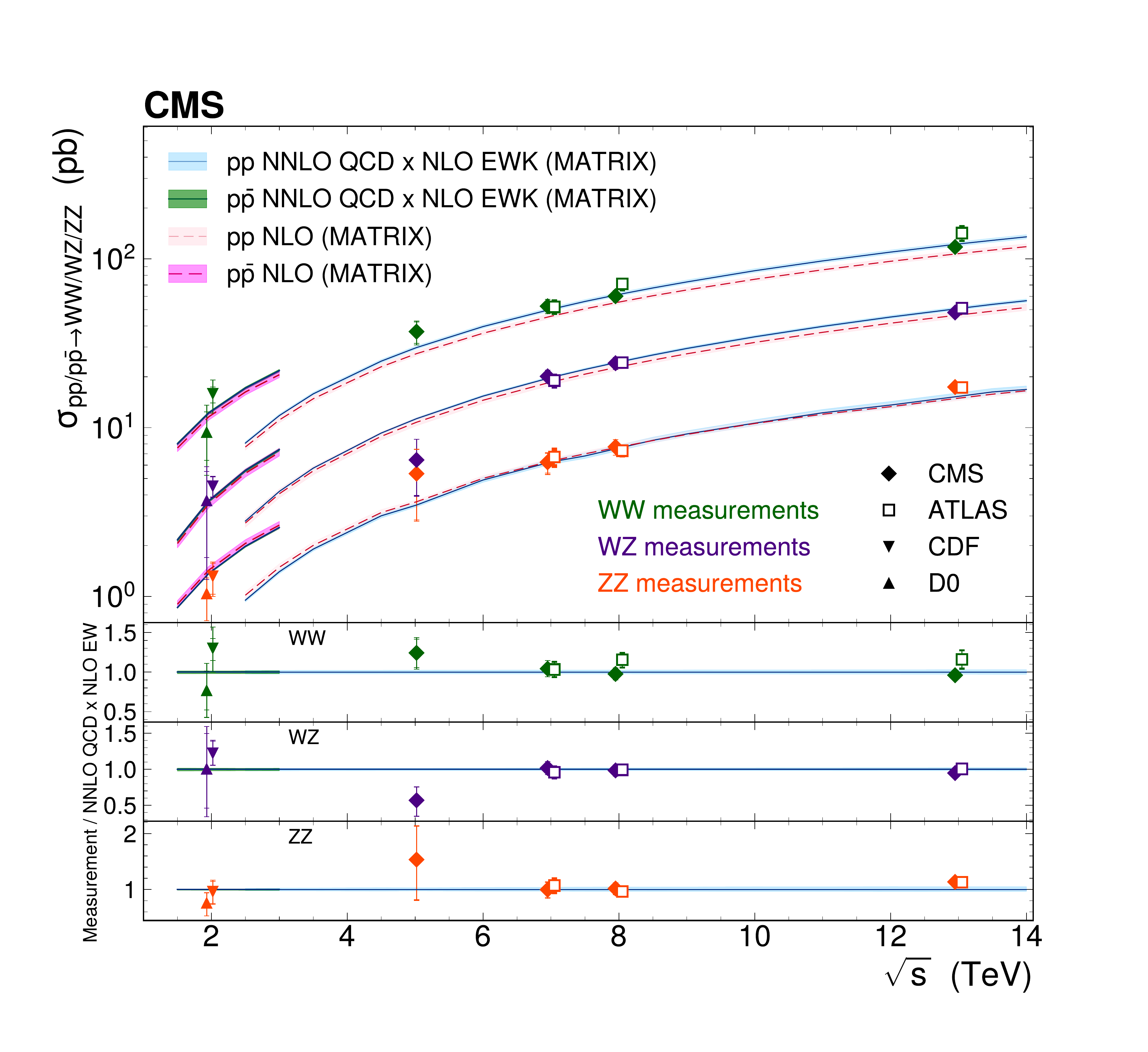}	
	\caption{The total $\WWOS$, $\WZ$ and $\ZZ$ cross sections as functions of the \pp centre-of-mass energy. Results from the CMS and ATLAS experiments for \pp collisions are compared with the predictions from \textsc{matrix} at NNLO in QCD and NLO in EW, and at NLO in QCD. Also shown are results from $\mathrm{p\overline{p}}$ collisions at the CDF and D0 experiments compared with \textsc{matrix} predictions as above. The inner vertical errors bars around the experimental data points show the statistical uncertainties of the measurements, whereas the outer bars show the total uncertainties. Measurements at the same centre-of-mass energy are shifted slightly along the horizontal axis for clarity. Figure taken from Ref.~\cite{CMS:2021pqj}.}\label{fig:VVxs}
\end{figure}

A plot focused on VV production, where V = $\PW$ or $\PZ$, is shown in Fig.~\ref{fig:VVxs} for four energies measured by the CMS experiment.  
The measured total cross sections of pairs of weak bosons agree with the theoretical predictions~\cite{CMS:2021pqj}. Also shown are results from the ATLAS experiment~\cite{ATLAS:2012mec,ATLAS:2016zwm,ATLAS:2017bbg,ATLAS:2012aid,ATLAS:2016bkj,ATLAS:2019bsc,ATLAS:2012bra,ATLAS:2016bxw,ATLAS:2017bcd}, and from the Tevatron CDF~\cite{CDF:2016zte,CDF:2014nef} and D0~\cite{D0:2011dez,D0:2012tku,D0:2013rca} experiments where the production of pairs of weak bosons in hadron collisions was first observed. The figure presents the inclusive total cross sections for weak boson pair production and, where necessary, results reported as production cross section times branching fraction to lepton final states have been scaled by the inverse of the appropriate branching fraction. Extrapolation from the fiducial measurement regions for the states involving $\PZ \to \ell^+\ell^-$ to total cross sections was done in mass ranges of 66--116\GeV and 60--120\GeV for ATLAS and CMS, respectively, leading to a 1.6\% (0.8\%) difference in the total cross sections calculated by ATLAS \vs CMS and the \MATRIX predictions for $\ZZ$ ($\WZ$) production. This effect is not corrected for in the plot and is not visible given the logarithmic scale. Diboson production cross sections are also summarized with other cross sections measured by CMS in Fig.~\ref{fig:XsAll} where, as above, the diboson results are presented as total cross sections.

The most precisely measured diboson cross sections at the CMS experiment are $\WZ$ and $\ZZ$ production. In the $\WZ$ case the high precision is possible because of the low background for $\PZ$ decays to electrons or muons and the higher branching fraction for leptonic $\PW$ decay. The $\WZ$ cross section in 13\TeV \pp collisions~\cite{CMS:2021icx} is measured as $\sigma_{\text{tot}}(\Pp\Pp \to \PW\PZ) = 50.6 \pm 0.8 \stat \pm 1.5 \syst \pm 1.1 \lum \pm 0.5 \thy \pb = 50.6 \pm 1.9$\pb.   The overall 3.7\% accuracy is dominated by the systematic and integrated luminosity uncertainties.  The cross section is also measured in a fiducial phase space, which reduces the extrapolation uncertainty to the full phase space, where a 3.4\% precision is achieved. At the time, the precision exceeded that of the single boson cross section measurements from the CMS. Despite having the lowest statistical precision of any diboson production process, the cross section for $\ZZ$ production is the next most accurately measured. The precision of the measurement is driven by the very low background to two fully reconstructed $\PZ$ boson decays to electrons and muons. The $\ZZ$ cross section for 13\TeV \pp collisions~\cite{CMS:2020gtj} is measured as $\sigma_{\text{tot}}(\mathrm{pp} \to \ZZ) = 17.2 \pm 0.3 \stat \pm 0.5 \syst \pm 0.4 \thy \pm 0.3 \lum$\pb. The combined overall uncertainty is 4.3\%. The cross section measured in a fiducial phase space has 3.7\% precision.   

\begin{figure}
	\centering 
	\includegraphics[width=0.80\textwidth]{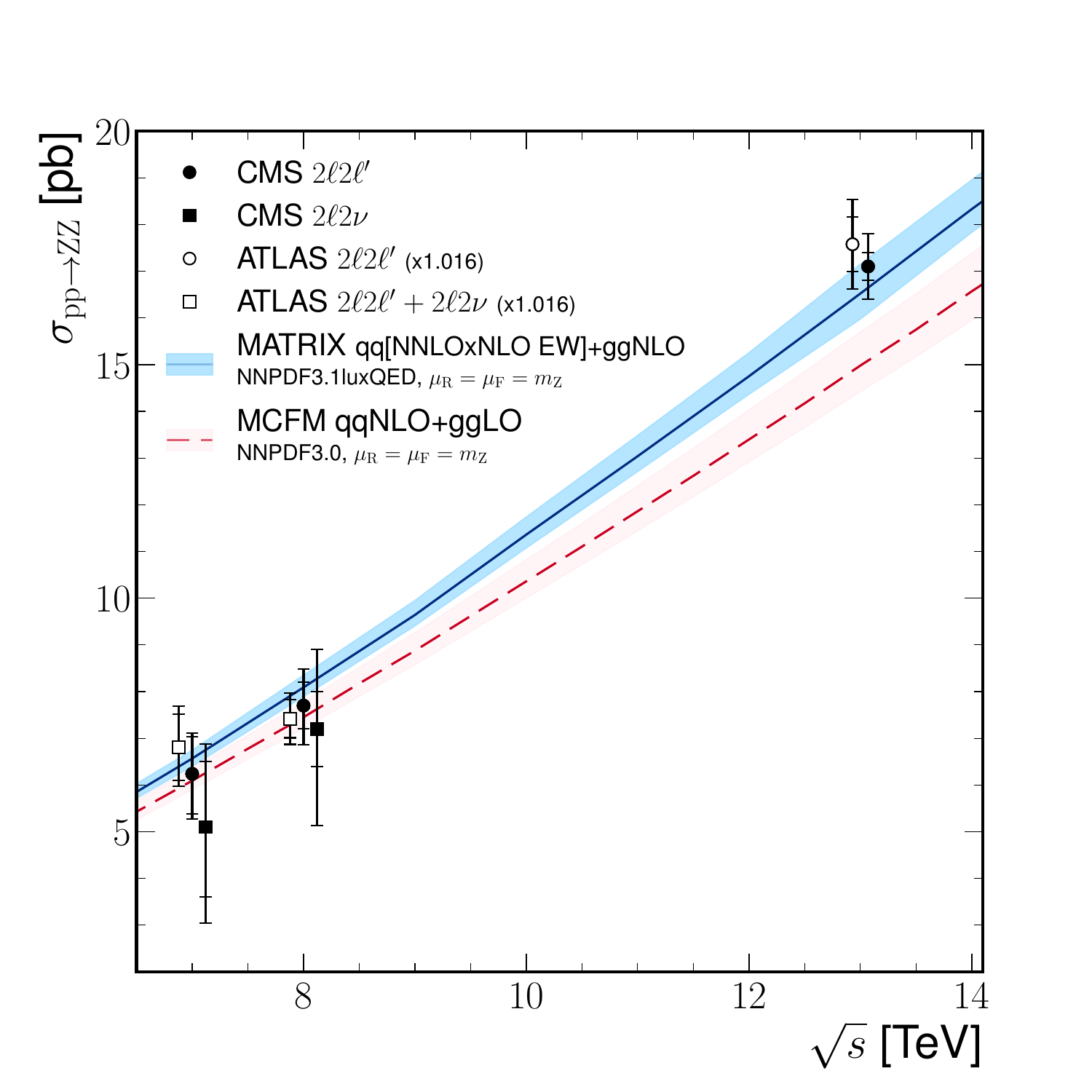}	
	\caption{The total $\ZZ$ cross section as a function of the \pp centre-of-mass energy. Results from the CMS and ATLAS~\cite{ATLAS:2012bra,ATLAS:2016bxw,ATLAS:2017bcd} experiments are compared with the predictions from \textsc{matrix} and \MCFM, as described in the text. The ATLAS measurements were performed with a $\PZ$ boson mass window of 66--116\GeV, instead of 60--120\GeV used by CMS, and are corrected for the resulting 1.6\% difference in acceptance.  The inner vertical errors bars around the experimental data points show the statistical uncertainties of the measurements, whereas the outer bars show the total uncertainties. Measurements at the same centre-of-mass energy are shifted slightly along the horizontal axis for clarity. Figure taken from Ref.~\cite{CMS:2020gtj}.}\label{fig:ZZ}
\end{figure}

The importance of NNLO QCD calculations is shown in Fig.~\ref{fig:ZZ} taken from Ref.~\cite{CMS:2020gtj}, where the measured $\ZZ$ cross sections are shown compared with two calculations. The first calculation is performed with \MCFM~\cite{Campbell:2010ff} at NLO in QCD 
for $\PQq\PAQq$ processes and LO QCD accuracy for $\Pg\Pg$ initial-state processes (denoted \MCFM qqNLO+ggLO). The second calculation is performed using \textsc{matrix}~\cite{Grazzini:2017mhc}, which includes both NNLO QCD and NLO EW contributions for $\PQq\PAQq$ processes and NLO QCD accuracy for $\Pg\Pg$ initial-state processes~\cite{Grazzini:2018owa} (denoted MATRIX qq[NNLOxNLOEW]+ggNLO). The predictions use NNPDF31\_nnlo\_as\_0118\_luxqed and NNPDF3.0 PDF sets, respectively, and fixed factorization and renormalization scales $\mu_\text{F}=\mu_\text{R} = m_\PZ$.
The CMS and ATLAS~\cite{ATLAS:2012bra,ATLAS:2016bxw,ATLAS:2017bcd} measurements are compared with the theoretical predictions. The ATLAS measurements were performed with a $\PZ$ boson mass window of 66--116\GeV, instead of 60--120\GeV used by CMS, and are corrected for the resulting 1.6\% difference in acceptance. Contributions from NLO and NNLO QCD diagrams substantially enhance the cross section of diboson production and are necessary to show agreement with the experimental data with measured total cross sections.  

\begin{figure}
	\centering 
	\includegraphics[width=0.80\textwidth]{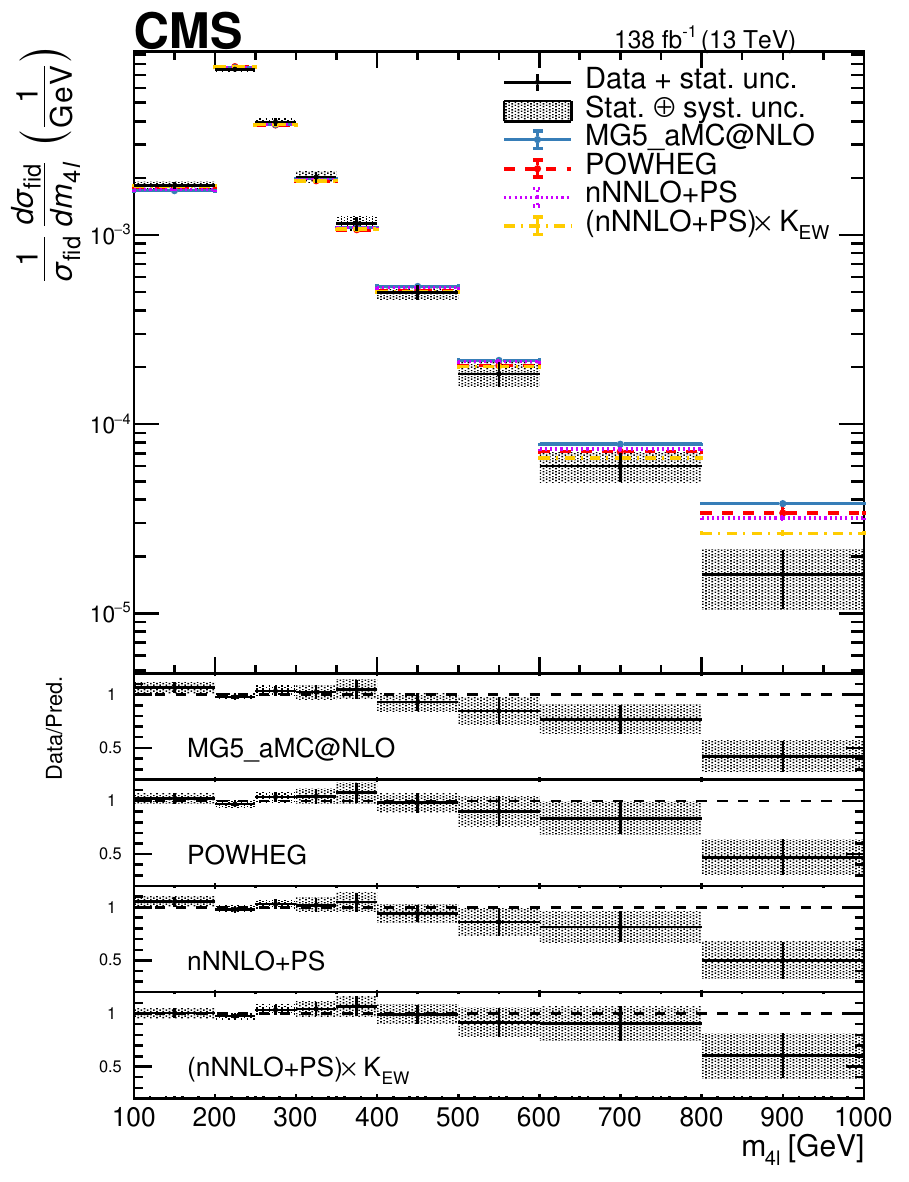}	
	\caption{Differential cross section normalized to the fiducial cross section as a function of $m_{4\ell}$. The on-shell $\PZ$ requirement $60 < m_{\PZ} < 120\GeV$ is applied for both \PZ boson candidates. Points represent the unfolded data, the solid lines the (\MGvATNLO $\qqbar \to \ZZ$) + (\MCFM $\Pg\Pg \to \ZZ$) + (\POWHEG $\PH \to \ZZ$) predictions, and red dashed lines the (\POWHEG $\qqbar \to \ZZ$) + (\MCFM $\Pg\Pg \to \ZZ$) + (POWHEG $\PH \to \ZZ$) predictions. The \MGvATNLO EW $\ZZ$ predictions are included. The purple dashed lines represent the nNNLO+PS predictions, and the yellow dashed lines represent the nNNLO+PS prediction with EW corrections applied. Vertical bars on the MC predictions represent the statistical uncertainties. The lower panels show the ratio of the measured to the predicted cross sections. The shaded areas represent the full uncertainties calculated as the sum in quadrature of the statistical and systematic uncertainties and the vertical bars around the data points represent the statistical uncertainties only. The overflow events are included in the last bin of the distributions. Figure and caption taken from Ref.~\cite{CMS:2024ilq}.}\label{fig:mZZ}
\end{figure}

Differential measurements have been made for all the diboson final states. A variety of distributions have been measured focusing on: basic kinematics, such as the \pt of leptons in leptonic vector boson decays and the \pt of the bosons; measurements of jets, including the number and \pt of associated jets; and quantities with sensitivity to possible BSM physics, such as the invariant mass of the diboson system or other quantities that assess the energy of the vector boson system. In differential measurements, areas of phase space can be identified that are particularly sensitive to higher-order QCD and EW perturbative predictions. For instance, variables that assess the energy of the diboson system, such as the diboson invariant mass, show large enhancements due to NLO and NNLO QCD effects at high mass. The NLO EW contributions tend to reduce the cross sections in the high-energy part of the distributions. As an illustration, Fig.~\ref{fig:mZZ} shows the $m_{4\ell}$ distribution from Ref.~\cite{CMS:2024ilq}. Comparisons are made to four MC generator predictions. The first prediction is from \MGvATNLO for $\qqbar \to \ZZ$ at NLO QCD,  \POWHEG $\PH \to \ZZ$ at NLO QCD, and \MCFM $\Pg\Pg \to \ZZ$ at LO QCD (denoted MG5\_aMC@NLO). The second prediction is from \POWHEG at NLO in QCD. The final two comparisons are calculated using nNNLO simulation, which performs NNLO QCD calculations matched to PS using the MiNNLO method~\cite{Buonocore:2021fnj} (denoted  nNNLO+PS). This simulation includes EW corrections that were applied as a multiplicative $K$-factor as a function of $m_{4\ell}$. The best agreement with data is seen with the nNNLO+PS with EW corrections applied, which are necessary to achieve better agreement at high $m_{4\ell}$.

\begin{figure}
	\centering 
	\includegraphics[width=0.80\textwidth]{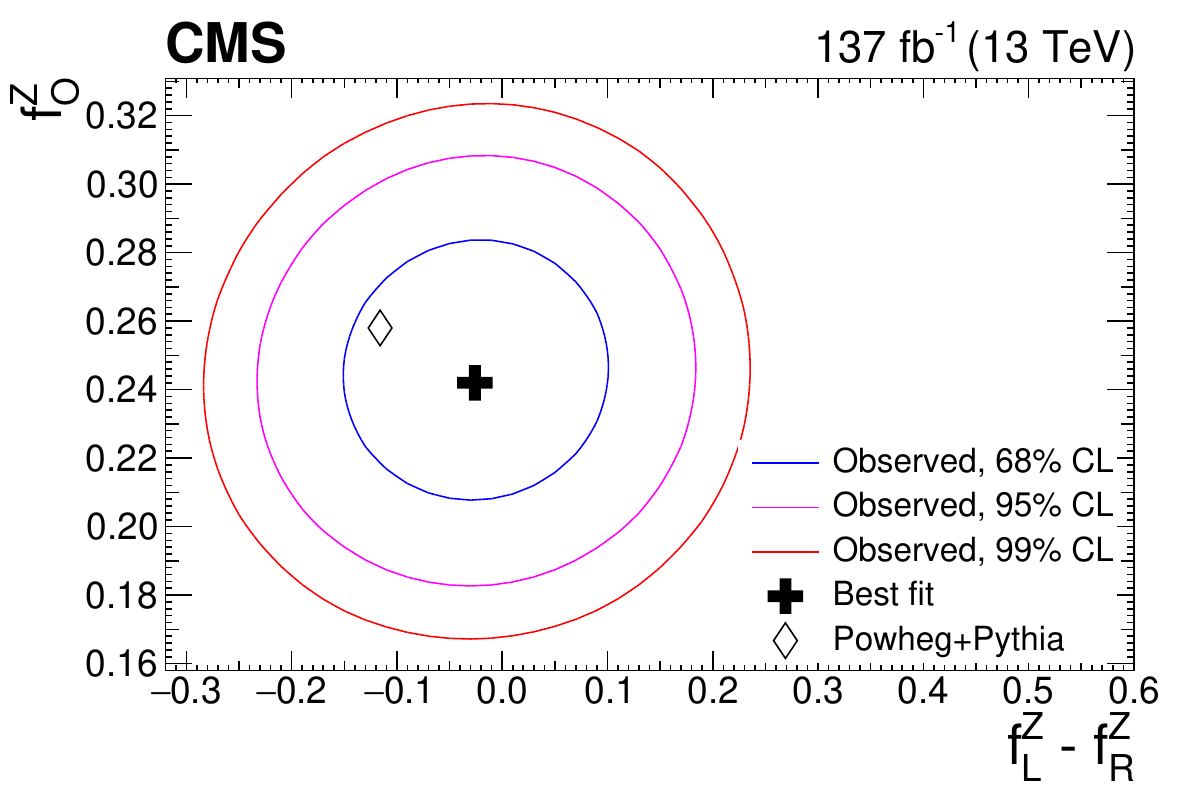}	
	\caption{Confidence regions in the $ f_{\mathrm{O}}^{\PZ} \vs f_{\mathrm{L}}^{\PZ} - f_{R}^{\PZ} $ parameter plane for the $\PZ$ boson polarization. The results are obtained with no additional requirement for the charge of the $\PW$ boson. The blue, magenta, and red contours present the 68, 95, and 99\% confidence levels, respectively. Figure from Ref.~\cite{CMS:2021icx}. The cross indicates the best fit to the observed data and the diamond shows the result of the \POWHEG+\PYTHIA simulation.}\label{fig:WZpol}
\end{figure}

An essential test of the EW interactions and the nature of the $\PW$ and $\PZ$ bosons is a measurement of their polarization. Through the EW symmetry-breaking Brout--Englert--Higgs mechanism, the $\PW$ and $\PZ$ bosons acquire longitudinal polarization and hence mass. The SM fractions of bosons produced in specific polarization states in \pp collisions in both single and multiboson production are predicted by the EW theory.  These fractions can be extracted from the angular distributions of the decay products of $\PW$ and $\PZ$ bosons. In cases with decays to charged leptons, the CMS experiment makes very accurate measurements of the angular distributions of the emitted leptons. The lepton emission angles in the boson rest frame relative to the boson momentum direction in the laboratory frame, which are approximately expected to have simple trigonometric probability distributions based on first- and second-order sine and cosine functions for each polarization state, can be precisely reconstructed and the polarization fractions extracted by fitting the expected distributions for the fraction of each polarization. In events with neutrinos, partial reconstruction of the full angular information can be used. The CMS experiment has measured boson polarization in the $\WWSS$ (discussed in Section~\ref{subsec:EW}) and $\WZ$ production~\cite{CMS:2021icx}. In the latter case, polarized production was observed. The fitted longitudinal polarization fraction versus the difference of left and right polarization fractions for $\PZ$ bosons in $\WZ$ production is shown in Fig.~\ref{fig:WZpol} demonstrating the ability of the measurement to distinguish the polarization states.

\begin{figure}
	\centering 
	\includegraphics[width=0.80\textwidth]{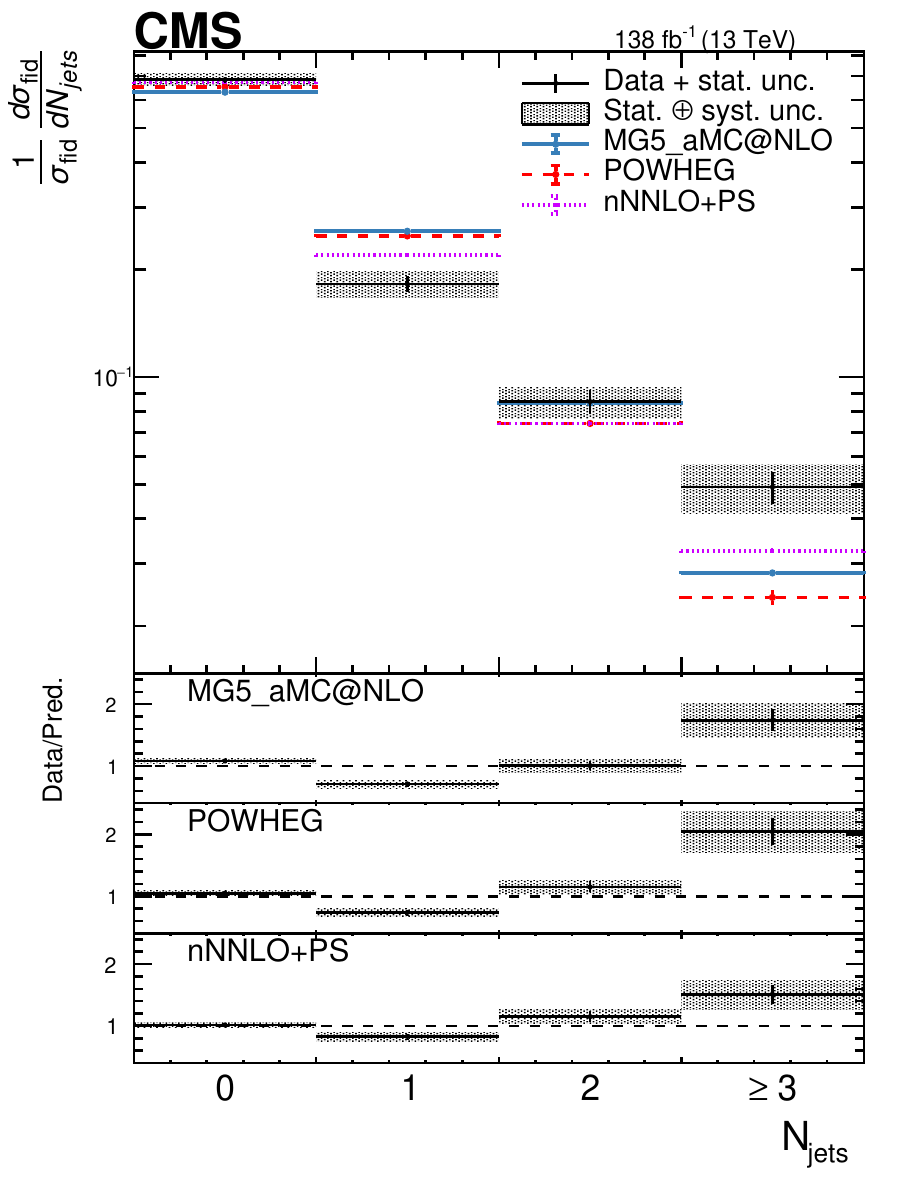}	
	\caption{The differential cross section normalized to the fiducial cross section as a function of the number of jets. The on-shell $\PZ$ requirement $60 < m_{\PZ} < 120\GeV$ is applied for both \PZ boson candidates. Points represent the unfolded data, the solid lines the (\MGvATNLO $\qqbar \to \ZZ$) + (\MCFM $\Pg\Pg \to \ZZ$) + (\POWHEG $\PH \to \ZZ$) predictions, and red dashed lines the (\POWHEG $\qqbar \to \ZZ$) + (\MCFM $\Pg\Pg \to \ZZ$) + (\POWHEG $\PH \to \ZZ$) predictions. The \MGvATNLO EW $\ZZ$ predictions are included. The purple dashed lines represent the nNNLO+PS predictions. Vertical bars on the MC predictions represent the statistical uncertainties. The lower panels show the ratio of the measured to the predicted cross sections. The shaded areas represent the full uncertainties calculated as the sum in quadrature of the statistical and systematic uncertainties and the vertical bars around the data points represent the statistical uncertainties only. The overflow events are included in the last bin of the distributions. Figure and caption taken from Ref.~\cite{CMS:2024ilq}.}\label{fig:ZZnjets}
\end{figure}

A test of perturbative QCD in a more complex signature involving EW vector bosons is the measurement of differential cross sections of diboson production versus the number of observed jets. Accurate predictions of these types of final states are essential for performing studies of diboson production through VBS, which is observed in the diboson + 2 jets final state; in Higgs physics where many signatures involve multiple vector bosons; and in searches for BSM physics involving multiple vector bosons. Previously, this type of analysis had only been performed by the CDF experiment, which observed $\WWOS$+jets production and measured the cross section for final states up to 2 jets~\cite{CDF:2015wtz}. The CMS experiment has measured $\PW\PGg$~\cite{CMS:2022yrl} and $\PZ\PGg$~\cite{CMS:2021gme} with two jets production in 13\TeV collisions; $\WWOS$+jets up to two jets at 13\TeV~\cite{CMS:2020mxy}; $\WZ$+jets up to three jets in 8\TeV collisions~\cite{CMS:2016qth}; and $\ZZ$+jets up to three jets at 8 and 13\TeV~\cite{CMS:2018ccg}.   
Details of the cross sections measured and generators used for comparison are given in Table~\ref{tab:VVJets}.
In the last case, a subsequent reanalysis of the 13\TeV $\ZZ$+jets data~\cite{CMS:2024ilq} with larger data samples showed that a more advanced nNNLO+PS simulation achieves better agreement at high jet multiplicities (as shown in Fig.~\ref{fig:ZZnjets}). The full description of the predictions in Fig.~\ref{fig:ZZnjets} is presented above in the discussion of the $m_{4\ell}$ distribution from the same analysis. The improved modelling of the data seen with the new nNNLO+PS simulation demonstrates the importance of continued development of advanced NNLO computations.

\begin{table}[htbp]
  \centering
\topcaption{Summary of measurements of diboson production in association with jets.   Listed are the diboson state, number of jets measured, generator(s) used with perturbative QCD order and $K$-factors used to scale the result to a higher order, total number of additional partons generated, number of partons generated at NLO, parton shower MC, and ME-PS jet merging scheme. The total number of partons includes additional real-emission partons generated by NLO or NNLO QCD matrix element calculations. The highest bin in the jet multiplicity includes events with a higher number of jets as well.\label{tab:VVJets}}
\renewcommand{\arraystretch}{1.2}
\cmsTable{
  \begin{tabular}{l l l l l l l l}
    Diboson  & $N_{\text{jets}}$& $\sqrt{s}$  & Generator &  Partons & Partons & PS & ME-PS \\ 
    State &  & ({\TeVns})  &          & total   & NLO    &    & scheme\\
     \hline
    $\PW\PGg$~\cite{CMS:2022yrl} & 2 & 13 & MG5\_aMC (NLO)  & 2 & 1 & Py8   & FxFx   \\
    $\PZ\PGg$~\cite{CMS:2021gme} & 2 & 13 & MG5\_aMC (NL0) & 2 & 1 & Py8   & FxFx   \\
    $\PosWW$~\cite{CMS:2020mxy} & 0--2 & 13 & (\POWHEG (NLO) + \MCFM (LO)) * $K_{\text{ NNLO}}$~\cite{Gehrmann:2014fva}& 1 & 0 & Py8   & \NA   \\
    $\PW\PZ$~\cite{CMS:2016qth} & 0--2 & 8 & (\MADGRAPH 5 (LO) + \MCFM (LO)) * $K_{\text{NLO}}$ \MCFM & 0 & \NA & Py6   & \NA   \\
    $\PZ\PZ$~\cite{CMS:2018ccg} & 0--3 & 8 & (MG5\_aMC (NLO)+ \MCFM (LO)) * $K_{\text{NLO}}$ \MCFM & 2 & 1 & Py8   & CKKW   \\
    $\PZ\PZ$~\cite{CMS:2023xxx} & 0--3 & 13 & nNNLO + \MCFM (NLO) & 2 & 1 & MiNNLO$_{\mathrm PS}$   & \NA   \\
  \end{tabular}}
\end{table}

The results for diboson production in association with jets are summarized in Fig.~\ref{fig:XsJetsEW} where they are presented as fiducial cross sections for leptonic final states. In the case of the $\WZ$+jets at 8\TeV~\cite{CMS:2016qth}, the result was multiplied by the leptonic branching fractions for easier comparison.

\subsubsection{Triboson production}
\begin{figure}
	\centering 
	\includegraphics[width=0.30\textwidth]{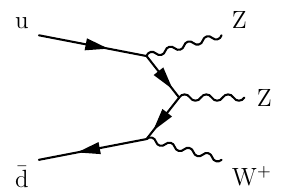}	
	\includegraphics[width=0.30\textwidth]{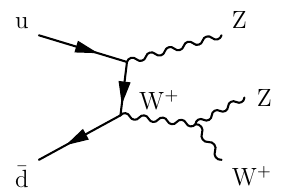}	
	\includegraphics[width=0.30\textwidth]{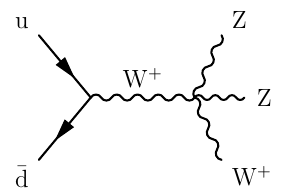}	
	\caption{Triboson $\PW\ZZ$ production via diagrams involving radiative production (left), TGCs (centre), and QGCs (right). This set of triboson Feynman diagrams is representative of most triboson signatures, with the caveat that neutral TGCs and some QGC combinations are not allowed in the SM.}\label{fig:FD:WZZ}
\end{figure}

\begin{figure}
	\centering 
	\includegraphics[width=0.80\textwidth]{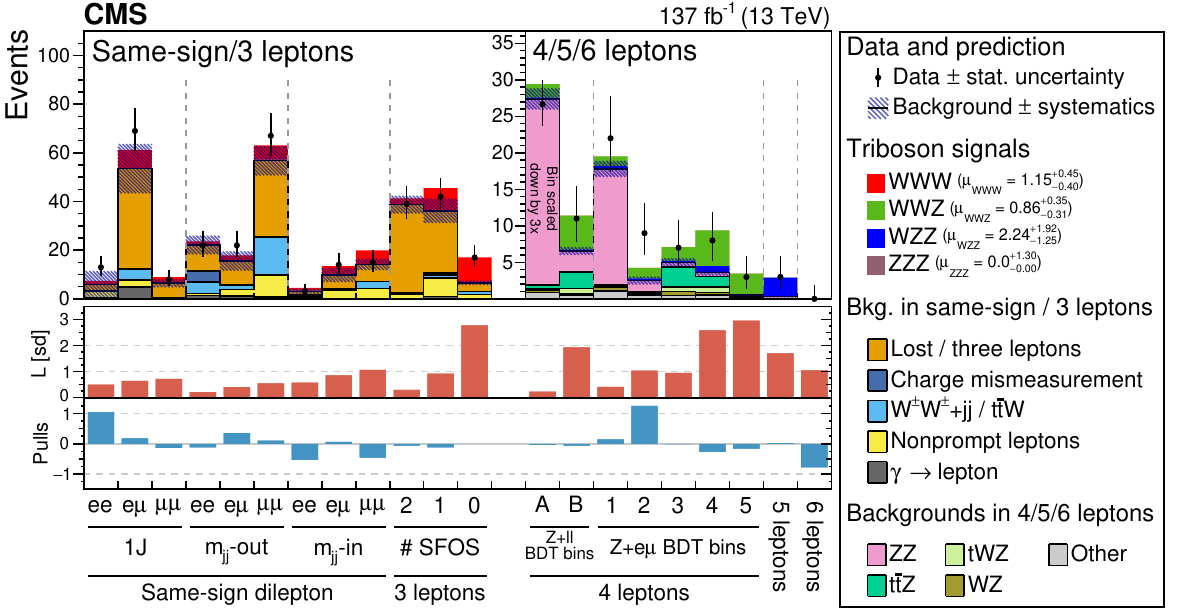}	
	\caption{Comparison of the observed numbers of events to the predicted yields. For the $\PW\WW$ and $\PW\WZ$ channels, the results from boosted decision tree (BDT) based selections are used. For the other results different categorizations based on the number of jets, whether dijet masses are inside or outside a selection window used to identify the boson, and specific lepton combinations or the number of same-flavour, opposite-sign (SFOS) leptons are shown. The VVV signal is shown stacked on top of the total background. The points represent the data and the error bars show the statistical uncertainties. The expected significance $L$ in the middle panel represents the number of standard deviations (sd) with which the null hypothesis (no signal) is rejected. The lower panel shows the pulls for the fit result. Figure taken from Ref.~\cite{CMS:2020hjs}.}\label{fig:VVV}
\end{figure}

The high centre-of-mass collision energy and the large integrated luminosity produced by the LHC  have made it possible to observe triboson production for the first time.   The most challenging measurements are those of the production of three massive vector bosons. The Feynman diagrams for $\PW\ZZ$ production are shown in Fig.~\ref{fig:FD:WZZ} including radiative production of three vector bosons and diagrams involving TGCs and QGCs. The sensitivity of triple gauge boson production to measure TGCs is weaker than that of diboson production because of the small production cross section, but the quartic coupling diagram gives this type of process direct sensitivity to QGCs. In a comprehensive analysis, CMS measured all possible massive triboson states simultaneously, categorizing them into all the possible final states involving electrons and muons, according to type and charge, and pairs of jets from hadronic boson decay. This analysis achieved collective observation of $\WW\PW$, $\WW\PZ$, $\WZ\PZ$, and $\ZZ\PZ$, and individual evidence for $\WW\PW$ and $\WW\PZ$ production at 3.3 and 3.4 standard deviations, respectively~\cite{CMS:2020hjs}.  
Figure~\ref{fig:VVV} depicts all of the analysis categories clearly showing the observed signal for all of the final states. The triboson production processes measured at CMS are listed in Table~\ref{tab:triboson}. Included in the table is information on \pp collision energy, theory calculations used for comparison in Fig.~\ref{fig:diboson}, and other results of interest in the paper. 

\begin{table}[htbp]
  \centering
\topcaption{Table of triboson production cross section measurements. Listed in the table are signatures studied, \pp collision energy, theory cross section calculation used for comparison, and selected additional results of interest from each measurement.\label{tab:triboson}}
\renewcommand{\arraystretch}{1.2}
  \begin{tabular}{l l l l}
    Process  & Energy ({\TeVns})  &  Theory calculation     & Other results \\
    \hline
    $\PW\PGg\PGg$~\cite{CMS:2017tzy}      & 8   & MG5\_aMC Py6 NLO  & aQGC \\
    $\PW\PGg\PGg$~\cite{CMS:2021jji}      & 13   & MG5\_aMC Py8 NLO  & aQGC \\
    $\PZ\PGg\PGg$~\cite{CMS:2017tzy}      & 8   & MG5\_aMC Py6 NLO  & aQGC \\
    $\PZ\PGg\PGg$~\cite{CMS:2021jji}      & 13   & MG5\_aMC Py8 NLO  & aQGC  \\
    $\PW\PV\PGg$~\cite{CMS:2014cdf}      & 8   & MG5\_aMC Py8 NLO  & aQGC \\
    $\PW\PW\PGg$~\cite{CMS:2023rcv}      & 13   & MG5\_aMC Py8 NLO  & aQGC, $\PH\PGg$ search\\
    $\PV\PV\PV$~\cite{CMS:2020hjs}             & 13   & NLO~\cite{Lazopoulos:2007ix,Binoth:2008kt,Hankele:2007sb} & $\PV\PH$ production \\
    $\PW\PW\PW$~\cite{CMS:2020hjs}             & 13   & NLO~\cite{Lazopoulos:2007ix,Binoth:2008kt,Hankele:2007sb} & $\PV\PH$ production\\
    $\PW\PW\PZ$~\cite{CMS:2020hjs}             & 13   & NLO~\cite{Lazopoulos:2007ix,Binoth:2008kt,Hankele:2007sb} & $\PV\PH$ production\\
    $\PW\PZ\PZ$~\cite{CMS:2020hjs}             & 13   & NLO~\cite{Lazopoulos:2007ix,Binoth:2008kt,Hankele:2007sb} & $\PV\PH$ production\\
    $\PZ\PZ\PZ$~\cite{CMS:2020hjs}             & 13   & NLO~\cite{Lazopoulos:2007ix,Binoth:2008kt,Hankele:2007sb} & $\PV\PH$ production\\
  \end{tabular}
\end{table}

\subsection{Electroweak single-boson and multiboson production}\label{subsec:EW}
\begin{figure}
	\centering 
	\includegraphics[width=0.30\textwidth]{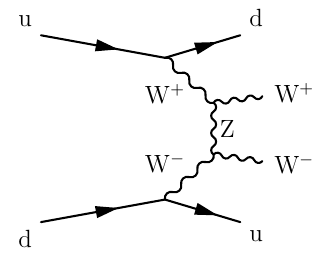}	
	\includegraphics[width=0.30\textwidth]{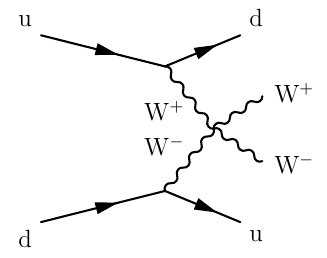}	
	\includegraphics[width=0.30\textwidth]{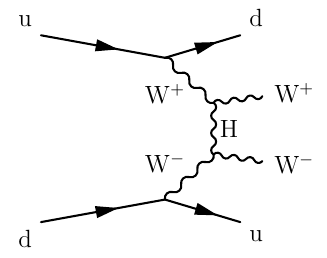}	
	\caption{Production of oppositely charged $\PW$ bosons via vector boson scattering.   Example Feynman diagrams include: scattering via $\PZ$ boson and two TGC vertices (left), a QGC vertex (middle), and scattering via a Higgs boson in $t$-channel (right). }\label{fig:FD:VBSWW}
\end{figure}

Pure EW production of single and multiple vector bosons with jets in collision events where bosons are radiated off incoming quarks and either fuse to a single boson (VBF) or scatter to pairs of bosons (VBS) is an essential test of the EW sector of the SM.  Vector boson fusion directly measures the TGCs of the SM. Vector boson scattering events can occur via the combination of double TGC interactions, in $t$- or $s$-channel; quartic coupling of bosons; or scattering via a Higgs boson, in $t$- or $s$-channel. The theoretical investigation of the Higgs boson scattering process was an important early component in understanding the essential role of the Higgs boson in the SM. The calculation of longitudinalVBS without the Higgs boson would predict an infinite cross section at high energy. Shown in Fig.~\ref{fig:FD:VBSWW} are representative VBS Feynman diagrams for \WWOS scattering. The features of these types of interactions are two scattered jets with large rapidity separation and one or two bosons produced centrally. The expected kinematic distributions from the different amplitudes contributing to VBS and their interference can be used to study the scattering kinematics and assess the polarization of the scattered bosons

\begin{figure}
	\centering 
	\includegraphics[width=0.30\textwidth]{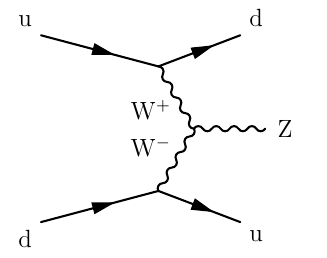}	
	\includegraphics[width=0.30\textwidth]{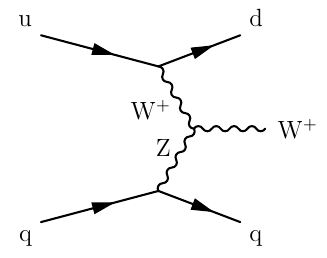}	
	\includegraphics[width=0.30\textwidth]{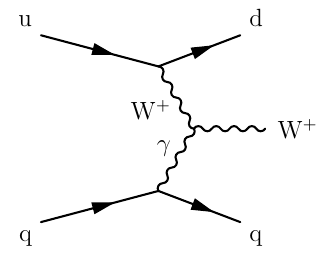}	
	\caption{Feynman diagrams for vector boson fusion production of $\PZ$ (left) and $\PW$ bosons (middle) via the $\WW\PZ$ TGC vertex and $\PW$ via the $\PW\PW\PGg$ TGC vertex (right).}\label{fig:FD:VBSZ}
\end{figure}

\begin{figure}
	\centering 
	\includegraphics[width=0.80\textwidth]{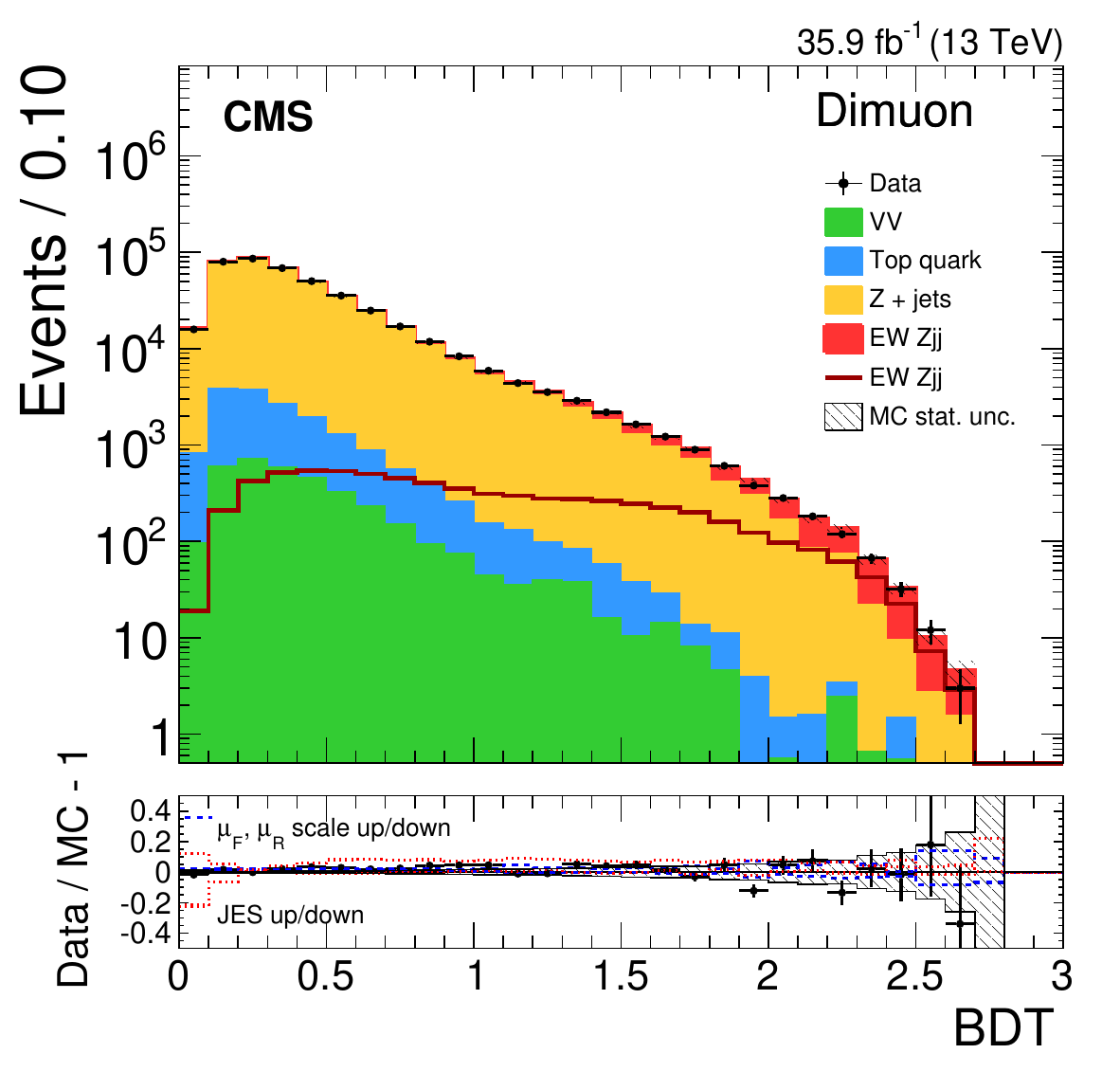}	
	\caption{Distribution for a BDT discriminant used to select VBF $\PZ$ events in dimuon events. The contributions from the different background sources and the signal are shown stacked, with data points superimposed. The vertical errors bars around the experimental data points show the total uncertainties. The expected signal-only contribution is also shown as an open histogram. The lower panel shows the relative difference between the data and expectations, as well as the uncertainty envelopes for the jet energy scale, and renormalization and factorization scale uncertainties. Figure taken from Ref.~\cite{CMS:2017dmo}.}\label{fig:VBFZBDT}
\end{figure}

The CMS experiment has measured VBF of single $\PW$ or $\PZ$ bosons in 7 ($\PZ$ only)~\cite{CMS:2013fqx}, 8~\cite{CMS:2016myt,CMS:2014sip}, and 13~\cite{CMS:2019nep,CMS:2017dmo}\TeV \pp interactions. The Feynman diagram for VBF production of a $\PZ$ boson is depicted in Fig.~\ref{fig:FD:VBSZ} showing direct sensitivity to the $\WW\PZ$ TGC. The extraction of the signal from a very large background of standard single boson + jets production requires the use of a multivariate discriminant. An example BDT distribution from the measurement of EW $\PZ$ production at 13\TeV is shown in Fig.~\ref{fig:VBFZBDT} demonstrating the performance of machine- learning techniques to separate the signal over an overwhelming $\PZ$+jets background with the same final state but slightly different kinematics~\cite{CMS:2017dmo}. These analyses have been used to set stringent limits on deviations from the expected SM TGC values.

\begin{figure}
	\centering 
	\includegraphics[width=0.95\textwidth]{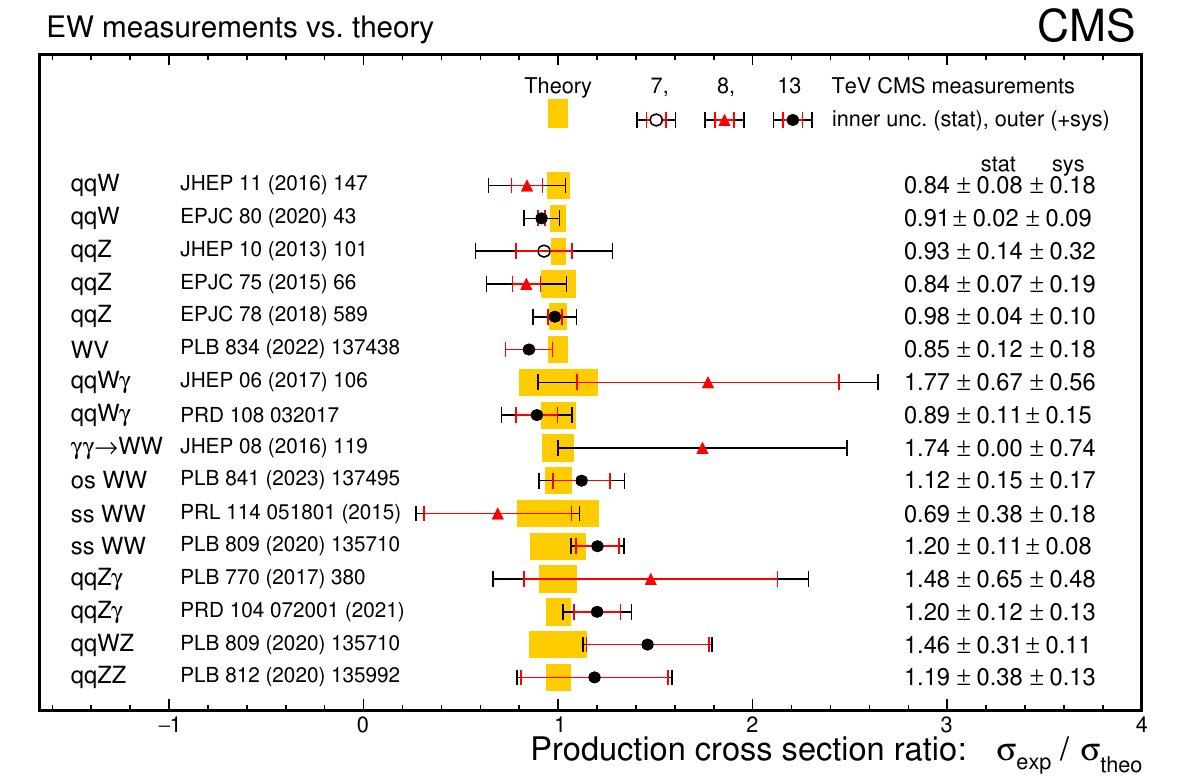}	
	\caption{Summary of cross section measurements of EW single or diboson production processes including vector boson fusion, vector boson scattering, and scattering via exclusive processes. Production of pairs of $\PW$ bosons can occur in same-sign (ss) $\WWSS$, opposite-sign (os), $\WWOS$, or exclusive production where photons are radiated from the incoming protons and form $\WWOS$ pairs via EW scattering.   Results are displayed as a ratio of the experimental measurement over the SM prediction. The yellow bands indicate the uncertainties in the theoretical predictions and the error bars on the points are the experimental uncertainties, with the outer bar being the combined statistical and systematic uncertainty.}\label{fig:EWxs}
\end{figure}

The EW production processes measured at CMS are listed in Table~\ref{tab:ewprod}.  Included is information on \pp collision energy, theory calculations used for comparison in Fig.~\ref{fig:EWxs}, and other results of interest. Good agreement with theoretical calculations is observed for all of these purely EW production processes.

\begin{table}[htbp]
  \centering
\topcaption{Purely EW production cross section measurements. Listed in the table are signatures studied, \pp collision energy, theory cross section calculation used for comparison, and selected additional results of interest from each paper.\label{tab:ewprod}}
\renewcommand{\arraystretch}{1.2}
\cmsTable{
  \begin{tabular}{l l l l}
    Process  & Energy   &  Theory          & Other results \\ [-0.5ex]
             & ({\TeVns})   &  calculation     & \\         
              \hline
       VBF $\PW$~\cite{CMS:2016myt}      & 8   & MG5\_aMC Py6 LO  & \NA \\
       VBF $\PW$~\cite{CMS:2019nep}      & 13  & MG5\_aMC Py8 LO  & aQGC \\ 
       VBF $\PZ$~\cite{CMS:2013fqx}      & 7   & \VBFNLO NLO      & central hadronic activity \\ 
       VBF $\PZ$~\cite{CMS:2014sip}      & 8   & MG5\_aMC Py6 LO  & jet activity\\
       VBF $\PZ$~\cite{CMS:2017dmo}      & 13  & MG5\_aMC Py8 LO  & aQGC, jet, central hadronic activity\\ 
       EW \WWOS,WZ~\cite{CMS:2021qzz}      & 13   & MG5\_aMC Py8 LO  & aQGC \\ 
       $\PGg\PGg \to \mathrm{W}^{\pm}\mathrm{W}^{\pm}$~\cite{CMS:2016rtz} & 13  & \MADGRAPH 5 LO rescaled  & aQGC \\
     EW $\PW\PGg$~\cite{CMS:2016gct}      & 8   & \MADGRAPH 5 Py6 \VBFNLO NLO         & aQGC \\     
     EW $\PW\PGg$~\cite{CMS:2022yrl}      & 13  & MG5\_aMC Py8 LO                     & aQGC, $m_{\mathrm{jj}}$, 6 dist. \\     
     EW $\PZ\PGg$~\cite{CMS:2017rin}      & 8   & \MADGRAPH 5 Py6 LO                  & aQGC \\     
     EW $\PZ\PGg$~\cite{CMS:2021gme}  & 13  & MG5\_aMC Py8 LO                     & aQGC, $m_{\mathrm jj}x\Delta\eta({\mathrm{jj}})$ + 3 1D dist. \\     
     EW \PssWW~\cite{CMS:2014mra}     & 8   & \MADGRAPH 5 Py6 \VBFNLO 2.7 NLO     & aQGC\\ 
     EW \PssWW~\cite{CMS:2017fhs}  & 13  & MG5\_aMC Py8 corr NLO QCD and EW~\cite{Biedermann:2016yds,Biedermann:2017bss} & aQGC, $m_{\mathrm{jj}}$, 3 dist. \\  
     EW \PosWW~\cite{CMS:2022woe}     & 13  & MG5\_aMC Py8 LO                     & \NA \\ 
     EW $\PW\PZ$~\cite{CMS:2020gfh}        & 13  & MG5\_aMC Py8 corr NLO QCD and EW~\cite{Denner:2019tmn} & aQGC, $m_{\mathrm{jj}}$ \\
     EW $\PZ\PZ$~\cite{CMS:2020fqz}        & 13  & \POWHEGBOX NLO~\cite{Jager:2013iza} & aQGC\\    
  \end{tabular}}
\end{table}

The first observed VBS process was $\WWSS$. 
The distinctive same-sign signature and significant \ptmiss in leptonic decays of the $\PW$ bosons, as well as the smaller cross section for the QCD-induced $\WWSS$ process, where the $\PW$ bosons are radiated off incoming quarks that scatter via a gluon, made it possible to observe this process in the initial year of LHC Run 2 at 13\TeV. Similarly, these characteristics made this mode the first place where polarized vector boson production in VBS could be studied~\cite{CMS:2020etf}.   The observation of the scattering of longitudinal vector bosons would be a clear sign of the presence of the Higgs boson scattering interaction as a component of VBS and is considered one of the essential tests of the EW symmetry-breaking mechanism. A first measurement has been made of longitudinal VBS in this mode using 13\TeV collision data where a 2.3 standard deviation signal consistent with respect to the SM expectation was measured. A summary of all the measured EW production cross sections presented as a ratio to the SM prediction is shown in Fig.~\ref{fig:EWxs} showing the ability of the CMS experiment to see clear, well-measured signals in never before observed VBS production modes.    

Among the listed results is the purely EW process of exclusive scattering to $\PW$ boson pairs, $\PGg\PGg \to \WWOS$, for which evidence is reported using 8\TeV collision data~\cite{CMS:2016rtz}. The calculation of the expected theory cross section for exclusive $\PGg\PGg \to \mathrm{W}^{\pm}\mathrm{W}^{\pm}$ is performed using \MADGRAPH 5 using the equivalent photon approximation~\cite{Budnev:1974de} and rescaled to account for proton dissociation, as studied in the same analysis using a comparison of $\PGg\PGg \to \mu^+\mu^-$ to a MC sample generated using \LPAIR~\cite{Vermaseren:1982cz,Baranov:1991yq}. The CMS experiment has also searched for the high-mass exclusive scattering of $\PGg\PGg \to \mathrm{W}^{\pm}\mathrm{W}^{\pm}$ and $\PGg\PGg \to \PZ\PZ$ using intact forward proton reconstruction in the precision proton spectrometer and set limits on these processes~\cite{CMS:2022dmc}.  

A combination of production mechanisms is necessary to unitarize the cross section of the overall VBS processes. Contributions from new scalar or vector particles could cause large deviations in the cross section, especially at the highest energies where the unitarization of the divergent contributions to the cross section would be modified. In CMS, analyses of most VBS modes have used that sensitivity to search for anomalous couplings and differential measurements have been made of related kinematic distributions.

\subsection{Summary of EW measurements}
The CMS Collaboration has carried out a broad array of QCD EW measurements. The precision of some measurements has reached the percent level and \ncube perturbative QCD theory computations are necessary to test the measurements at a similar level of precision. Differential measurements are also testing our ability to model SM processes and NNLO QCD, NLO EW, and integrated PDF and parton shower computations at the same perturbative order are necessary to model the data. In general, SM predictions model the data well. At the level of both inclusive and fiducial cross sections, all the measurements are well modelled, within statistical expectations, across a large number of signatures involving single or multiple vector bosons and up to two jets, as would be expected with correct modelling of the physics using computations of at least NLO accuracy. Also, the modelling of differential distributions is generally good with discrepancies observed only in complex final states involving larger numbers of additional jets. The theory community is actively engaged in confronting the LHC data, and in many cases, new computations have improved the modelling of the data where previously there was disagreement. Measurements with percent-level accuracy and studies of complex final states along with improved theoretical modelling are constantly extending our ability to further investigate the complexities of the SM and search for BSM physics indirectly and in complex final states. A visual summary of the results of the standard model QCD, EW, top quark, and Higgs boson measurements of individual cross sections and cross sections of processes including jets is presented in Figs.~\ref{fig:XsJetsEW}, and~\ref{fig:XsAll}, respectively.\clearpage

\section{Top quark measurements}\label{sec:topsection}

The large mass of the top quark, $m_\PQt=172.5\GeV$~\cite{CMS:2023wnd}, and, as a consequence, its short lifetime of about $0.5\times 10^{-24}\unit{s}$, drive the phenomenology associated with this particle.
Its properties make the top quark stand out amongst all the elementary fermions. The top quark lifetime is so short that it decays before hadronizing~\cite{Bigi:1986jk}, making it the only quark whose physical properties can be studied as if it were ``bare'', which, in turn, makes it a unique probe for constraining several extensions of the SM. 
Its mass attracts particular attention  also from a BSM physics perspective, for two main reasons: because it is the largest known for an elementary particle, by orders of magnitude with respect to any other elementary fermion; and because its Yukawa coupling to the Higgs boson ($y_\PQt$) is remarkably close to unity. 
These two facts have inspired a very rich theoretical literature, in which the top quark is surmised to hold the key to the spontaneous EW symmetry breaking of the SM~\cite{Hill:1991at,Weinberg:1975gm,Hill:1994hp}, and, in general, to be a promising window on BSM physics, contributing to the EW oblique parameters~\cite{Peskin:1991sw} and, potentially, coupling to new physics with a rich phenomenology, as discussed in a recent review~\cite{Franceschini:2023nlp}.
The top quark is also a privileged probe of the proton PDFs, since, due to its large mass, its production is very sensitive to the gluon density at high values of $x$.
Moreover, the relatively abundant production rates, the variety of final states, and the large kinetic energy of its decay products, make top quark processes a significant background for several other studies at particle colliders. 
The measurements of the production cross sections, its decay parameters, 
and the properties of the top quark are key areas of study at the LHC
and have been explored by the CMS Collaboration since the beginning of Run~1.

At the LHC, the top quark is predominantly produced in top quark-antiquark pairs (\ttbar) through the strong interaction, with a relatively large cross section that translates to a rate of about 8\unit{Hz} at an instantaneous luminosity of $10^{34}\unit{cm}^{-2}\unit{s}^{-1}$ at 13\TeV.
Other production modes include mixed EW and QCD, or pure EW vertices, which yield either single top quarks, or top quarks produced in association with other particles, such as vector bosons, Higgs bosons, or additional quarks.

The top quark decays through an EW process, and hence its natural width is primarily determined by $m_\PQt$, $m_\PW$, and the Fermi constant ($G_\text{F}$), receiving relatively small higher-order corrections from $\alpS$~\cite{ParticleDataGroup:2022pth}.
The $\PQt\to\PW\PQb$ decay channel dominates, since the value of the $V_{\PQt\PQb}$ element of the Cabibbo--Kobayashi--Maskawa (CKM) matrix is very close to unity, and thus $\abs{V_{\PQt\PQb}}\gg \abs{V_{\PQt\PQd}}, \abs{V_{\PQt\PQs}}$.
As a result, top quark events are characterized by final states with \PQb jets and the decay products of the \PW~bosons, \ie charged leptons and neutrinos, or light-quark jets. Additional jets, stemming from gluon radiation, may also be present in the events, and add to the complexity of the event signature.

Experimentally, the kinematics of the parent top quark are reconstructed using dedicated algorithms. Challenges arise from the presence of neutrinos originating in the decays of the \PW~bosons, as well as from combinatorial ambiguities in associating hadronic jets and charged leptons to form top quark or antiquark candidates; both difficulties are typically addressed by exploiting mass constraints. 
The CMS Collaboration has explored different techniques in fully hadronic~\cite{CMS:2015auz,CMS:2013lqq}, single leptonic~\cite{CMS:2012sas,CMS:2015rld}, and dileptonic~\cite{CMS:2012tdr,CMS:2019esx,CMS:2020djy} final states, and in boosted topologies~\cite{CMS:2020tvq,CMS:2022kqg}, or in associated production with bosons~\cite{CMS:2015uvn}. 
Top quark cross section measurements at the LHC are often presented as differential cross sections, obtained using an unfolding procedure~\cite{Conway:2011in,RooUnfold,Schmitt:2012kp} in which corrections for detector resolutions and efficiencies, as well as PS and hadronization effects are applied,  to obtain a measurement at the level of stable particles or at parton level.
At the particle level, so-called pseudotops~\cite{LHCTOPWGPUB} have been defined, which are reconstructed from generator-level final-state particles with a lifetime greater than $0.3 \times 10^{-10}\unit{s}$. The particle level simplifies the definition of detector-independent cross section acceptances and minimizes the impact of theory assumptions.
Parton-level measurements of top quark cross sections and properties, although affected by uncertainties stemming from nonperturbative models and PS uncertainties, are crucial inputs for comparison of the data with fixed-order calculations and the extraction of fundamental theoretical parameters, such as \alpS or $m_\PQt^\text{pole}$, the top quark pole mass~\cite{CMS:2024irj}.
CMS has often made measurements at both particle and parton level. Conceptual definitions and technical details for both these approaches are described in Refs.~\cite{LHCTOPWGPUB,Collaboration:2267573}.

The following subsections focus on cross section measurements performed by CMS using \pp collisions at centre-of-mass energies ranging from 5.02 to 13.6\TeV. The first cross section measurements with proton-lead ($\Pp$Pb) and lead nuclei (PbPb) collisions are also described. A detailed report of top quark mass measurements in CMS has recently been published in Ref.~\cite{CMS:2024irj}.

An overview of the measurements of inclusive single top quark and \ttbar production is presented in Sections~\ref{subsec:singletop} and~\ref{subsec:ttpairs}. In Section~\ref{subsec:ttpairsdiff}, a few examples of differential \ttbar cross sections are presented. The first measurements of top quarks in heavy ion collisions are described in Section~\ref{subsec:ttheavyion}. The processes of top quark production in association with vector bosons or with additional jets are reviewed in Sections~\ref{subsec:tvecbos} and~\ref{subsec:topandjets}, and the four top quark production process is presented in Section~\ref{subsec:fourtops}. Finally, the extraction of fundamental SM parameters from inclusive top quark cross sections is briefly discussed in Section~\ref{subsec:topandsmconstants}. A summary of the quark cross section measurements spanning several orders of magnitude (10\fb to 1\,nb) is presented in Section~\ref{subsec:topsummary}.

\subsection{Electroweak top quark production}\label{subsec:singletop}

\begin{figure*}[!htp]
\centering
\includegraphics[width=.24\textwidth]{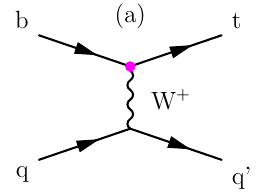}
\includegraphics[width=.24\textwidth]{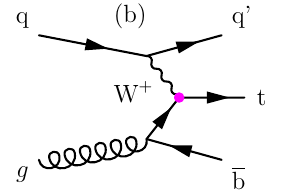}
\includegraphics[width=.24\textwidth]{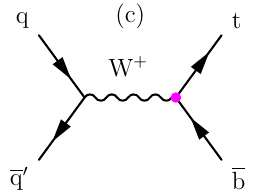}
\includegraphics[width=.24\textwidth]{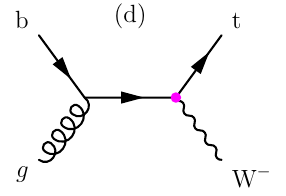}
\caption{
Feynman diagrams illustrating the pure EW contributions to single top quark production at the LHC at Born level. Charge conjugate states are implied. From left to right: the $t$-channel production, (a) with and (b) without a \PQb{} quark in the initial state; (c) the $s$-channel; and (d) the $\PQt\PW$-production.
In all diagrams the $\PQt\PW\PQq$ vertex is marked with a purple dot.
}\label{fig:Figure_000_top}
\end{figure*}

The production and decay of single top quark events occur through the EW $\PQt\PW\PQq$ vertex.
Figure~\ref{fig:Figure_000_top} represents the dominant Feynman diagrams for single top quark production in the SM.
In single top quark measurements, the properties of the $\PQt\PW\PQq$ vertex, marked in Fig.~\ref{fig:Figure_000_top} as a purple dot, are probed, including its magnitude, the CKM matrix elements ($V_{\PQt\PQq}$), and the polarization of the top quark. As a result of the V--A coupling structure of the EW interaction, the top quarks are expected to be almost 100\% polarized. 
Additional contributions from flavour-changing neutral currents~\cite{PhysRevLett.107.092002} and other BSM-induced effects~\cite{Aguilar-Saavedra:2008quj} are other aspects that are uniquely probed by these processes.

\begin{figure*}[!htp]
\centering
\includegraphics[width=.8\textwidth]{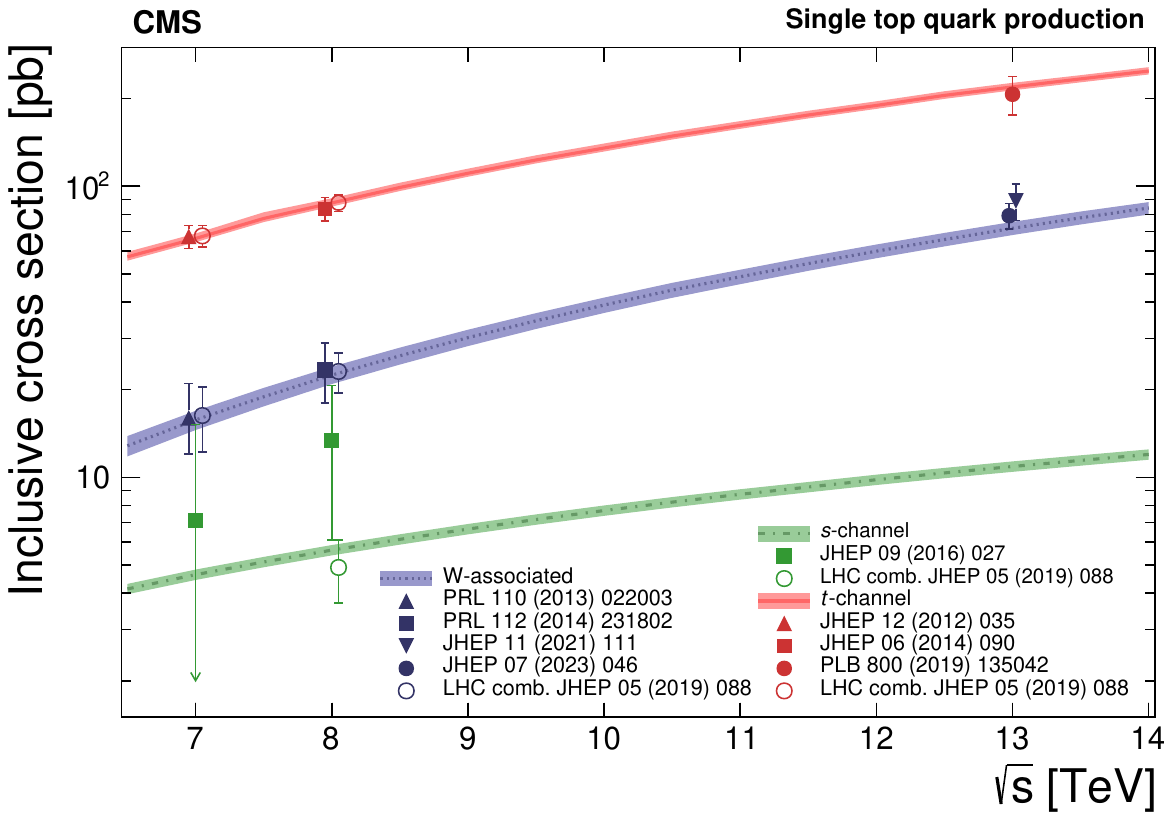}
\caption{
Single top quark cross section summary of CMS measurements as a function of the \pp centre-of-mass energy.
Where available the results from the full LHC combination are also overlaid for comparison.  
The theoretical calculations for $t$-channel, $s$-channel, and \PW-associated production are from Refs.~\cite{Aliev:2010zk,Kant:2014oha,Kidonakis:2010ux,Kidonakis:2013zqa}.
}\label{fig:Figure_001_top}
\end{figure*}

Figure~\ref{fig:Figure_001_top} summarizes the measurements of EW top quark production performed by CMS at different centre-of-mass energies.
At the LHC, the $t$-channel, represented in Figs.~\ref{fig:Figure_000_top} (a) and (b), has the highest cross section of the EW top quark production processes. The cross section at 13\TeV, calculated at NNLO in QCD, is expected to be
$\sigma_{\PQt}=214.2\,^{+2.4}_{-1.7}\,(\text{scale})\,^{+3.3}_{-2.0}\,(\text{PDF}+\alpS)\unit{pb}$,
where ``scale'' refers to the contributions from the uncertainties in the QCD factorization and renormalization scales~\cite{Campbell:2020fhf}.
The $t$-channel signature is characterized by the production of a top quark with a recoil jet that is typically produced at large rapidity. The large rapidity gap between the top quark and the forward jet is depleted in additional QCD emissions.
In cross section measurements, this signature is exploited to separate the $t$-channel signal from the background, which is dominated by top quark pair production.
Depending on whether the \PQb quarks are considered part of the proton or not, measurements in the $t$-channel can be compared with  predictions in the 5-flavour ($\PQqu\PQd\PQs\PQqc\PQb$) scheme (5FS), or in the 4-flavour ($\PQqu\PQd\PQs\PQqc$) scheme (4FS)~\cite{Frederix:2012dh}. 

CMS has measured the $t$-channel cross section at 7\TeV~\cite{CMS:2012xhh}, 8\TeV~\cite{CMS:2014mgj}, and 13\TeV~\cite{CMS:2018lgn}, as depicted in the upper curve of Fig.~\ref{fig:Figure_001_top}. In general, the measurements indicate that the 5FS predicts the rate more accurately, as expected from the resummation of initial-state large logs in the \PQb quark PDF, improving the stability of the calculations~\cite{Maltoni:2012pa}. On the other hand, the 4FS yields a more precise description of the kinematic distributions. These conclusions are supported by additional measurements of the differential $t$-channel cross sections~\cite{CMS:2019jjp}.

The selections and background estimations used in the measurement of the $t$-channel reflect the evolution of the data-taking conditions and event reconstruction techniques in CMS and of the theoretical (MC) predictions. Analyses make use of the single-lepton final states. To discriminate the signal from the main backgrounds (\ttbar, $\PW+$\,jets, and multijets), the events are categorized according to the jet and \PQb jet multiplicity. The region of two jets and one \PQb jet is expected to be enriched in signal events. Backgrounds arise from multijet events, typically estimated from data, $\PW+$\,jets events, and top quark pair production.
Two different approaches have been explored for the signal-to-background separation: a simple robust variable (the pseudorapidity of the forward jet, $\eta_{j'}$), or a multivariate-analysis (MVA) approach. Already the experience with the 7\TeV data recorded in 2011 showed that both approaches lead to accurate measurements of the $t$-channel cross section. The MVA approach improves the statistical precision by up to 40\% with respect to $\eta_{j'}$ but suffers slightly more from signal-modelling uncertainties. 

The relative uncertainty achieved in the measurements varies from 15\% to 9\%, after fitting the variable of interest in different categories. In the latest measurements the dominant uncertainties are related to the signal modelling, most notably the variation of the PS and the matching PS-ME matching algorithm. 
The most precise measurement of this process is attained in combination with results from the ATLAS Collaboration, yielding a 6.6\% relative uncertainty~\cite{ATLAS:2019hhu}, where the dominant contribution is still related to modelling uncertainties.
Additional mitigation of this uncertainty is expected from using higher-order accuracy predictions, employing better reconstruction algorithms, and, in general, using larger data sets.
Fiducial and ratio measurements, are also expected to have reduced extrapolation uncertainties~\cite{Campbell:2020fhf}.

\begin{figure*}[!htp]
\centering
\includegraphics[width=.8\textwidth]{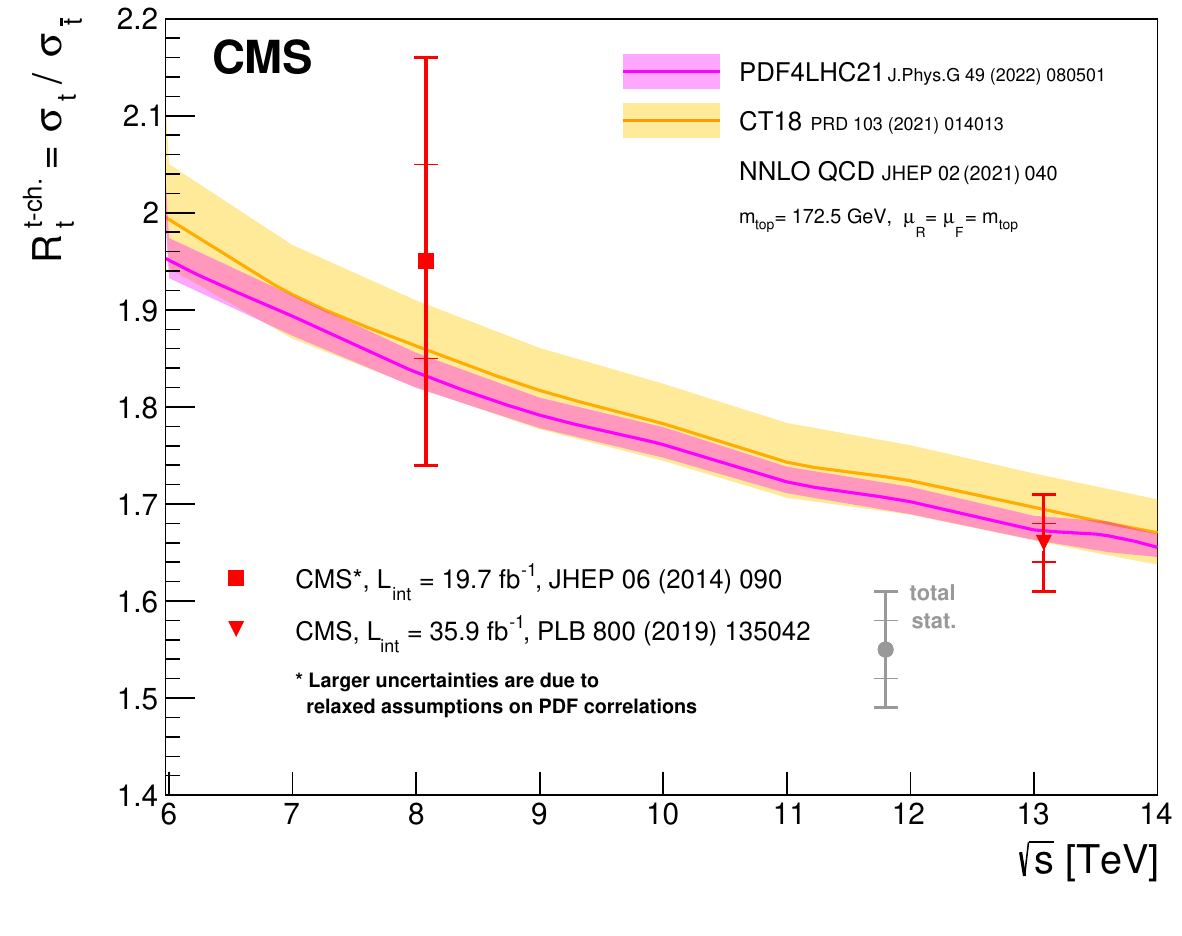}
\caption{
Summary of the CMS measurements of $R_\PQt=\sigma_\PQt/\sigma_\cPaqt$,
the cross section ratio between $t$-channel top quark and $t$-channel top antiquark production. The measurements are compared with NNLO QCD calculations using the PDF sets \textsc{CT18} and \textsc{PDF4LHC21}. The coloured bands represent the uncertainties in the theoretical predictions (scale and PDF uncertainties). The PDF uncertainties are estimated using the \textsc{PDF4LHC21} prescription~\cite{PDF4LHCWorkingGroup:2022cjn}. 
}\label{fig:Figure_002_top}
\end{figure*}

The flavour of the initial light quark defines the charge of the produced top quark: $\PQqu(\PQd)$ quarks in the initial state result in $\PQt(\cPaqt)$ quarks in the final state. Given this simple property,
the cross section inherits a charge asymmetry from the proton PDF of the quarks involved in the production. This asymmetry is typically quantified by the ratio of cross sections $R_\PQt = \sigma_\PQt / \sigma_\cPaqt$, which is predicted to be about 1.7 at 13\TeV~\cite{Campbell:2020fhf,Kidonakis:2021vob}.
In the measurement of the ratio, most systematic uncertainties cancel or are significantly reduced, resulting in a significantly more precise test of the PDF than the absolute cross section measurement. Figure~\ref{fig:Figure_002_top} summarizes the different $R_\PQt$ measurements compared with the predictions. Overall a good agreement is found for various PDFs.

From the experimental point of view, the $s$-channel production, shown in Fig.~\ref{fig:Figure_000_top}(c), is the most challenging of the purely EW processes at the LHC. This is due to the large backgrounds from \ttbar, $t$-channel, and \PW~boson production in association with heavy-flavour quarks, with respect to the small expected $s$-channel signal cross section of $10.32\,^{+0.29}_{-0.24}\,(\text{scale})\,^{+0.27}_{-0.26}\,(\text{PDF}+\alpS)\unit{pb}$, as calculated at NLO in QCD for 13\TeV~\cite{Kant:2014oha,Aliev:2010zk}. The CMS Collaboration has searched for the $s$-channel top quark production at both 7 and 8\TeV~\cite{CMS:2016xoq}, and the result is included in Fig.~\ref{fig:Figure_001_top}.

The analysis relies on MVAs for discriminating the signal process from the backgrounds. A combined fit to the MVA output distributions in the categories of different jet and \PQb jet multiplicities yields a measurement with an uncertainty of about 45\% in the signal strength, corresponding to an observed significance of 2.5~s.d.\,with 1.1~s.d.\,expected.
Although the measurement has a significant statistical uncertainty (11\%), its total uncertainty is dominated by the choice of the factorization and normalization scales, the matching scale in the modelling of the backgrounds (33\%), as well as by the jet energy scale and \PQb tagging uncertainties (25\%). An experimental observation of this channel is expected with improvements in the higher-order predictions, state-of-the art \PQb tagging, jet-energy scale uncertainties, as well as machine-learning based algorithms.

Finally we discuss the associated $\PQt\PW$ production, shown in Fig.~\ref{fig:Figure_000_top}(d), which can be interpreted as a more global set of double, single, and nonresonant $\PWp\PWm\PQb\cPaqb$ diagrams including both the $\PQt\PW$ and the \ttbar processes described in the next Section~\ref{subsec:ttpairs}. 
Establishing the single-resonant $\PQt\PW$ process is interesting in itself, as it is well defined at Born level and sensitive to CKM matrix elements and possible BSM effects. Most measurements in Run~1 and Run~2 have focused on isolating this process from the double-resonant (\ttbar) production by using distinctive features, such as lower jet multiplicity and the balance in the transverse plane between the top quark and the \PW~boson decay products. The predicted cross section of $\PQt\PW$ production in \pp collisions at 13\TeV is $\sigma(\PQt\PW)=79.3\,^{+1.9}_{-1.8}\,(\text{scale})\pm 2.2\,(\text{PDF}+\alpS)\unit{pb}$ at NLO+NNLL in QCD~\cite{Kidonakis:2021vob}, and thus about 10\% of the cross section for \ttbar.

Evidence for $\PQt\PW$ production was attained at 7\TeV~\cite{CMS:2012pxd} and observation at 8\TeV~\cite{CMS:2014fut}. Measurements with improved precision were made at 13\TeV~\cite{CMS:2018amb,CMS:2021vqm}.
With the exception of Ref.~\cite{CMS:2021vqm}, the measurements have focused on dilepton final states with one \PQb jet. 
A fit to the output of the MVA discriminator (or ancillary variables such as the subleading jet \pt in the two-jet-two-\PQb-tag bin) in the different categories resulted in improved precision from 31\% (7\TeV) to 11\% (13\TeV).  
Run~1 measurements were combined with those performed by the ATLAS Collaboration, and the final result is in agreement with the SM prediction with a total uncertainty of 16.5\%~\cite{ATLAS:2019hhu}.
The improvements obtained in Run~2 were due to the increased sample size and accuracy in the predictions, improved identification algorithms, and a better calibration of the CMS detector~\cite{CMS:2017yfk,CMS:2020uim,CMS-DP-2020-021,CMS:2018rym,CMS:2019ctu,CMS:2020ebo,BTV-16-002}. 

CMS has also measured the $\PQt\PW$ process in the single-lepton channel at 13\TeV~\cite{CMS:2021vqm}.
Although this channel offers the advantages of larger branching fractions and the possibility to fully reconstruct the top-quark system, it suffers from more numerous and larger backgrounds. The result, shown in the middle curve of Fig.~\ref{fig:Figure_001_top}, is in agreement with that obtained in the dilepton channel.

\subsection{Top quark pair production}\label{subsec:ttpairs}

\begin{figure*}[!htp]
\centering
\includegraphics[width=.24\textwidth]{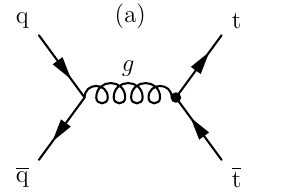}
\includegraphics[width=.24\textwidth]{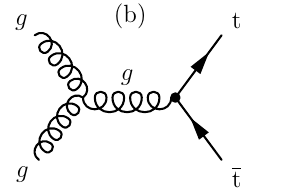}
\includegraphics[width=.24\textwidth]{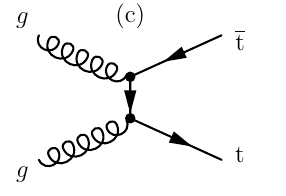}
\includegraphics[width=.24\textwidth]{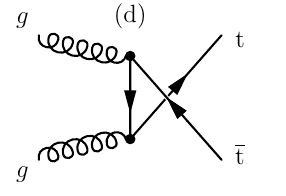}
\caption{
Leading order Feynman diagrams for \ttbar production.
}\label{fig:Figure_003_top}
\end{figure*}

The LO Feynman diagrams, depicted in Fig.~\ref{fig:Figure_003_top}, illustrate the main \ttbar production modes at the LHC, where the gluon fusion (diagrams b, c, and d) are dominant contributions to the cross section (about 85\% at 13\TeV).
At the lowest order in perturbation theory, the partonic cross section is proportional to $(\alpS/m_{\PQt})^2$ and it is dominated by the region where the rapidity difference of the pair is relatively small.
Parton distribution functions are sensitive to the determination of $\sigma_{\ttbar}$: the formation of a \ttbar pair requires high energy transfer ($Q > 2 m_\PQt$) and thus a relatively high momentum fraction of the incoming partons $x > 0.03\,(0.07)$ at 13\,(7)\TeV; the rapidity of the \ttbar system $y(\ttbar)$ is related to the momentum fraction via $y(\ttbar)\sim 1/2\log(x'/x)$, where $x$ and $x'$ are the fractional momenta of the initial-state partons. 
Precise cross section measurements of $\sigma_{\ttbar}$ have the potential to improve the knowledge of the gluon PDF, of \alpS, and of the top quark pole mass $m_{\PQt}^\text{pole}$~\cite{CMS:2024irj}, which are crucial ingredients to predictions for LHC physics such as the Higgs boson production cross section, and hence the Higgs boson couplings.
In addition, \ttbar is a background for many BSM searches and in some cases a final state.

Within the top quark sector the prediction for $\sigma_{\ttbar}$ is currently amongst the most precise; it is calculated at NNLO and includes the resummation of soft gluon terms at NNLL. The expected cross section at 13\TeV is
$\sigma_{\ttbar}=833.9\,^{+20.5}_{-30.0}\,(\text{scale})\pm 21\,(\text{PDF}+\alpS)$\unit{pb}
computed with \textsc{Top++2.0}~\cite{Beneke:2011mq,Cacciari:2011hy,Barnreuther:2012wtj,Czakon:2012zr,Czakon:2012pz,Czakon:2013goa,Czakon:2011xx}.

The CMS Collaboration made early measurements of $\sigma_{\ttbar}$, in \pp collisions at each centre-of-mass energy, and in $\Pp$Pb and PbPb collisions. These were milestones in the extensive programme of precision measurements and searches for new physics. Examples are: 
the very first measurement which inaugurated the top quark physics programme at the LHC using as few as 11 events collected in 3\pbinv of 7\TeV data~\cite{CMS:2010uxk};
the first measurements at the various $\sqrts$~\cite{CMS:2015yky,CMS:2017xrt,CMS:2023qyl}; 
and the first measurements of top quark pair cross sections in $\Pp$Pb~\cite{CMS:2017hnw} and PbPb~\cite{CMS:2020aem} collisions.
High precision measurements, employing larger data samples and more accurate calibrations of the detector, have been performed, such as Refs.~\cite{CMS:2016yys,CMS:2018fks,CMS:2021vhb}, or in combination with the ATLAS Collaboration~\cite{ATLAS:2022aof}, reaching uncertainties as small as 2--3\%.

In the CMS detector, top quark events can be identified with high purity and their rich final state comprising \PQb jets and leptons also makes them standard candles for calibration purposes.
The measurements have made use of all the various \ttbar final states, which are generically classified according to the number of leptonically decaying \PW~bosons. 
Among the dileptonic final states that have been exploited to measure $\sigma_{\ttbar}$, the channel with one electron and one muon in the final state is particularly clean, whereas the channels containing $\PGt$ leptons are particularly challenging, as they require dedicated trigger and reconstruction algorithms. 

The top quark programme has benefited from the increasingly large data samples and it heavily draws on experimental techniques such as \PQb tagging~\cite{CMS:2011acs}, missing transverse energy~\cite{CMS:2011dzu}, reconstruction of boosted topologies~\cite{CMS:2016poo}, kinematics-based selections (from likelihoods to MVA-based approaches)~\cite{CMS:2011woj,CMS:2013nie}, fitting techniques using several control regions and variables~\cite{CMS:2011dzu}, profiling of systematic uncertainties~\cite{CMS:2011woj}, and, not least, the combination of results~\cite{CMS:2016yys}.

\begin{figure*}[!htp]
\centering
\includegraphics[width=.9\textwidth]{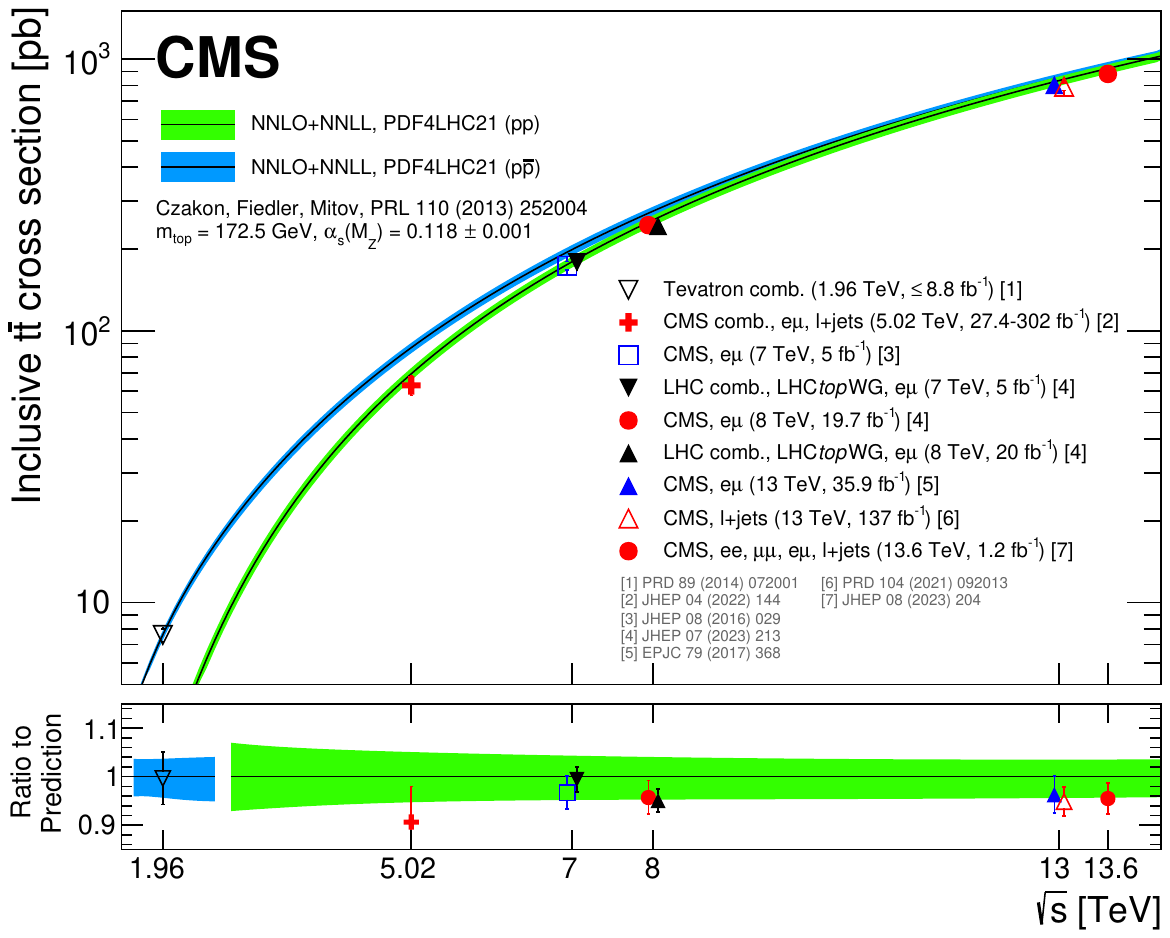}
\caption{
Summary of top quark-antiquark pair cross section measurements by the CMS Collaboration in comparison with the theory calculation at NNLO+NNLL accuracy. The Tevatron measurements are also shown. The lower panel displays the ratio between the different measurements and the theory prediction.
The coloured bands represent the theory uncertainty, while the error bars represent the uncertainty on the measurements.
}\label{fig:Figure_004_top}
\end{figure*}

\begin{figure*}[!htp]
\centering
\includegraphics[width=.8\textwidth]{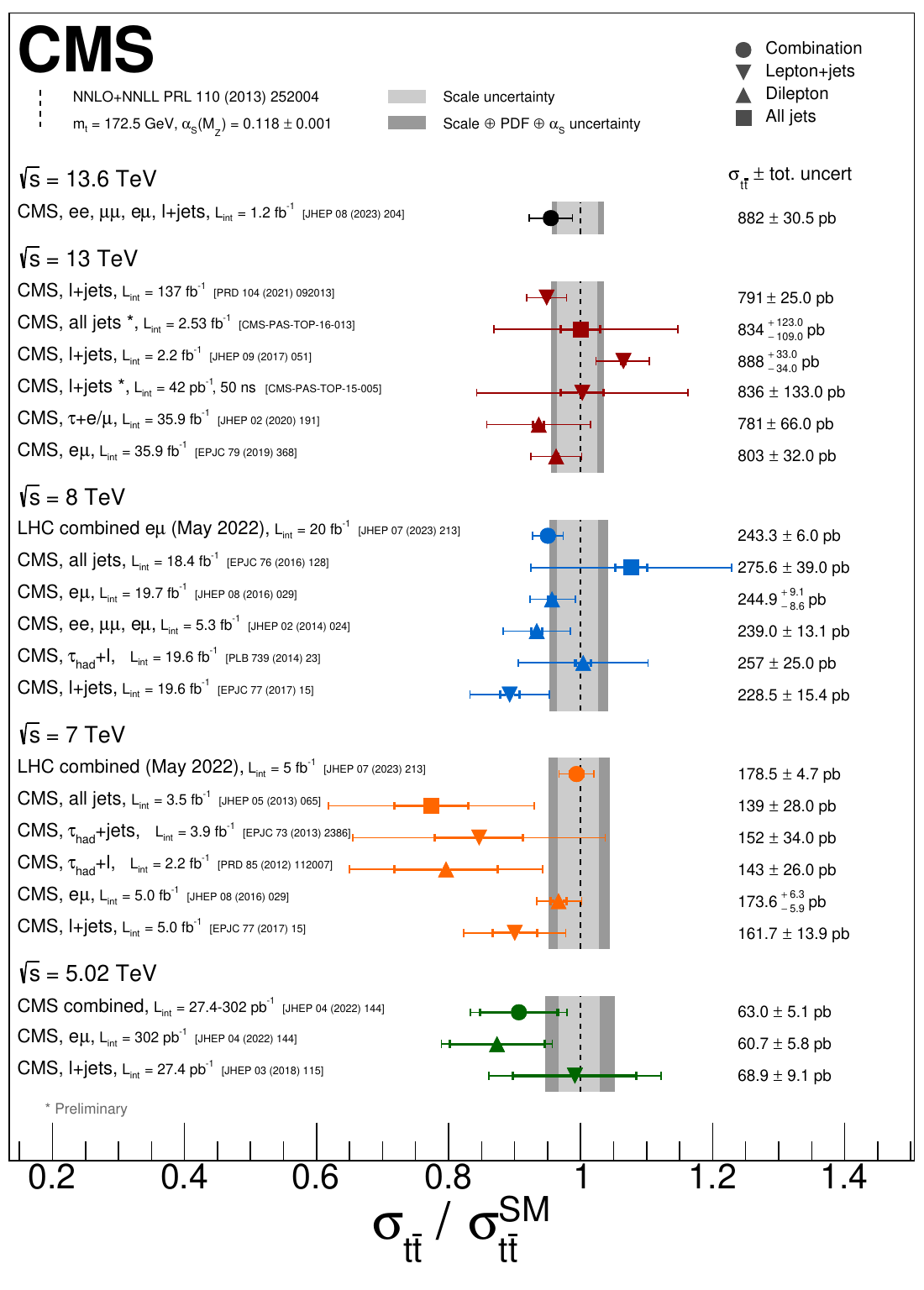}
\caption{
Summary of CMS top quark-antiquark pair cross section measurements at different $\sqrt{s}$, normalized to the theory calculation at NNLO+NNLL accuracy.
The different final states and $\sqrt{s}$ are respectively represented by various markers and colours. The total (statistical) uncertainty associated with the measurements is represented by the outer (inner) error bars. 
}\label{fig:Figure_005_top}
\end{figure*}

A summary of the $\sigma_{\ttbar}$ measurements performed by CMS is shown in Fig.~\ref{fig:Figure_004_top}. In this figure, the most precise results at each centre-of-mass energy are shown. Overall, all the results are compatible with each other and with the predictions. While consistent within the uncertainties, the data tend to be somewhat lower than most NNLO+NNLL predictions obtained for $m_{\PQt}=172.5\GeV$ and $\alpS=0.118$. Summaries of all the individual \ttbar measurements are shown in 
Fig.~\ref{fig:Figure_005_top}.

The precision of most top quark cross section measurements is limited by systematic uncertainties.
While the initial measurements at 7\TeV were limited by the trigger and selection uncertainties ($\approx$4\%), jet energy scale and \PQb tagging uncertainties (ranging from 7\% to 20\%), and the signal modelling, namely the choice of factorization and renormalization scales in the LO MC used, the most precise CMS measurements to date achieve a total relative uncertainty of 3.7\% (Run~1)~\cite{CMS:2016yys} and 3.9\% (Run~2)~\cite{CMS:2018fks}.
The latter measurements are performed in the $\Pe\Pgm$ final state in which a pure selection of events can be achieved with relatively loose lepton selection requirements. The analysis requires up to two \PQb jets (from the $\PQt\cPaqt$ decays) and counts the additional jets in the events. Categories are thus defined from the multiplicity of selected $\PQb$ and extra jets.

The categorization by \PQb-tagged jet multiplicity facilitates a fit procedure in which the \ttbar cross section and the \PQb-tagging efficiency are measured simultaneously, exploiting the binomial dependency of the \PQb-tagged jet multiplicity distribution on the \PQb-tagging efficiency. With this approach, the dominant uncertainties remain in the trigger and lepton selections, as well as the integrated luminosity ($\approx$2.2\%).

In the 13\TeV measurement, the signal was modelled using the NLO \POWHEG v2 MC generator~\cite{Alioli:2010xd,Frixione:2007vw,Nason:2004rx}. Although this change had reduced uncertainties from the theoretical point of view, it had no significant impact on the total uncertainty of the measurement since the experimental method effectively decreases uncertainties related to the ME-PS matching.

Variants of this approach have also been used with the $\ell$+jets final states at $\sqrt{s}=5.02\TeV$~\cite{CMS:2017zpm} and 13\TeV~\cite{CMS:2017xrt}, and more recently at 13.6\TeV, by combining both the dilepton and $\ell$+jets final states~\cite{CMS:2023qyl}. The relative uncertainties attained in these measurements are 12\%, 3.8\%, and 4\%, respectively. In the 5.02\TeV analysis, the uncertainty is larger because of the low integrated luminosity of that data set. 
These analyses have successfully applied the extra-jets categorization technique simply counting events in the different categories, or using variables such as $\Delta R(j,j')$, the distance between the two jets from the decays of \PW~bosons ($\PW\to j j'$), and $m(\ell\PQb)$, the invariant mass of the lepton-\PQb jet system.

In Ref.~\cite{CMS:2021vhb}, a total of 22 different measurements of $\sigma_{\ttbar}$ are performed, each based on the integration of a differential cross section measurement described below. The results are in general agreement with the SM and attain a total uncertainty of 3.2\%. The integrated luminosity is the dominant uncertainty (1.8\%) followed by lepton-selection uncertainties (1\%), \PQb tagging (0.9\%) and jet energy scale (1.4\%). 

Further improvements in the measurement of $\sigma_\ttbar$ require reduced uncertainties in the integrated luminosity, in the trigger, and in the lepton identification efficiencies. Luminosity measurements with an uncertainty of 1.2\% have been achieved for the CMS data recorded in 2015 and 2016~\cite{CMS-LUM-17-003}. Improved uncertainties are expected for the later data sets. In addition, the use of new luminosity detectors and novel techniques, such as \PZ~boson rates, can further improve the luminosity calibrations and their extrapolation uncertainties at high beam intensities~\cite{Dainese:2019rgk,CMS:2023pad}. 
Better measurements of the trigger and lepton identification efficiencies are expected from novel approaches. With larger sample sizes, efficiencies can be measured in finer categories, in turn leading to reduced uncertainties.

\subsection{Differential top quark cross sections}\label{subsec:ttpairsdiff}

\begin{figure*}[!htp]
\centering    
\includegraphics[width=.45\textwidth]{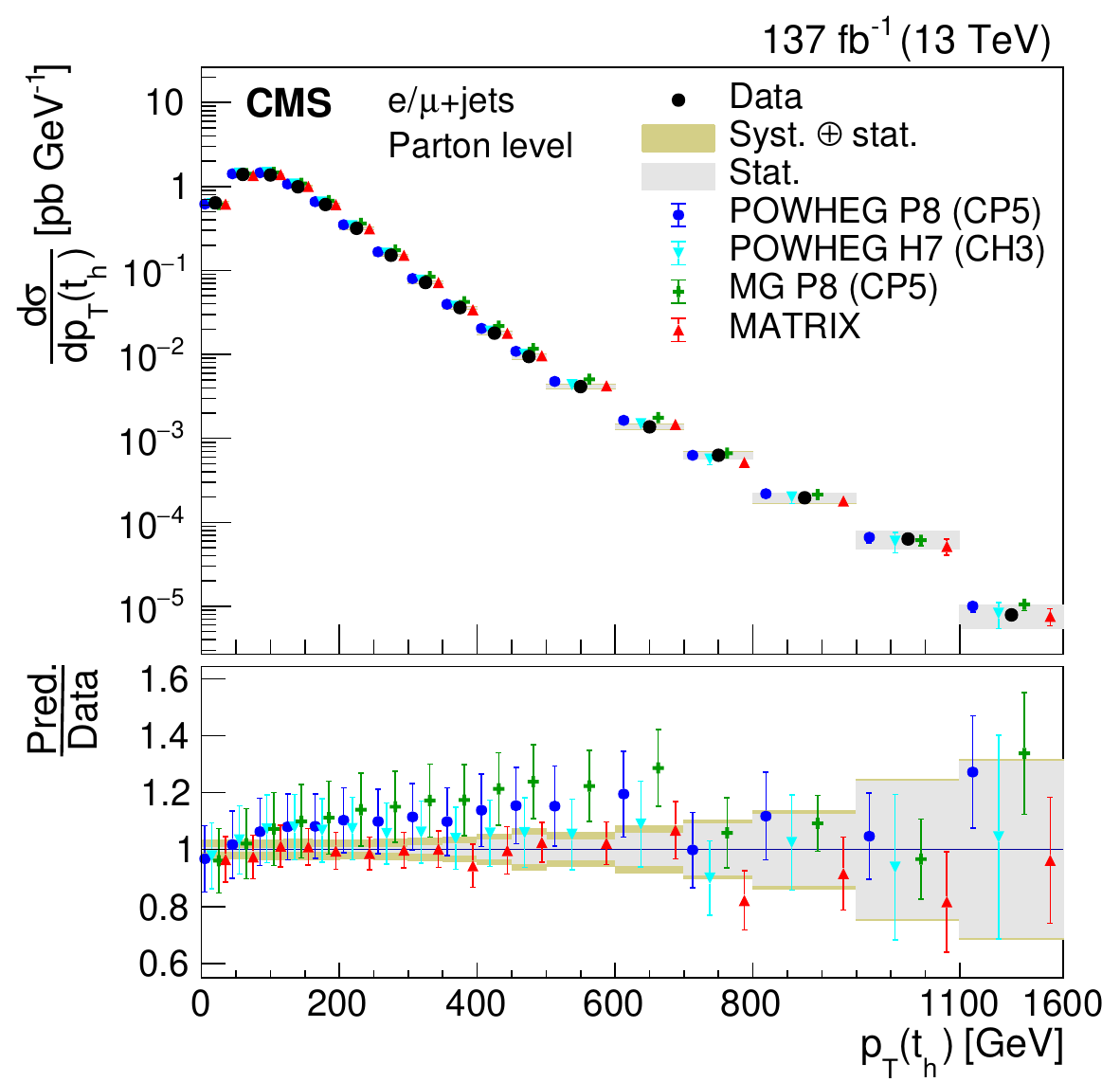}
\includegraphics[width=.45\textwidth]{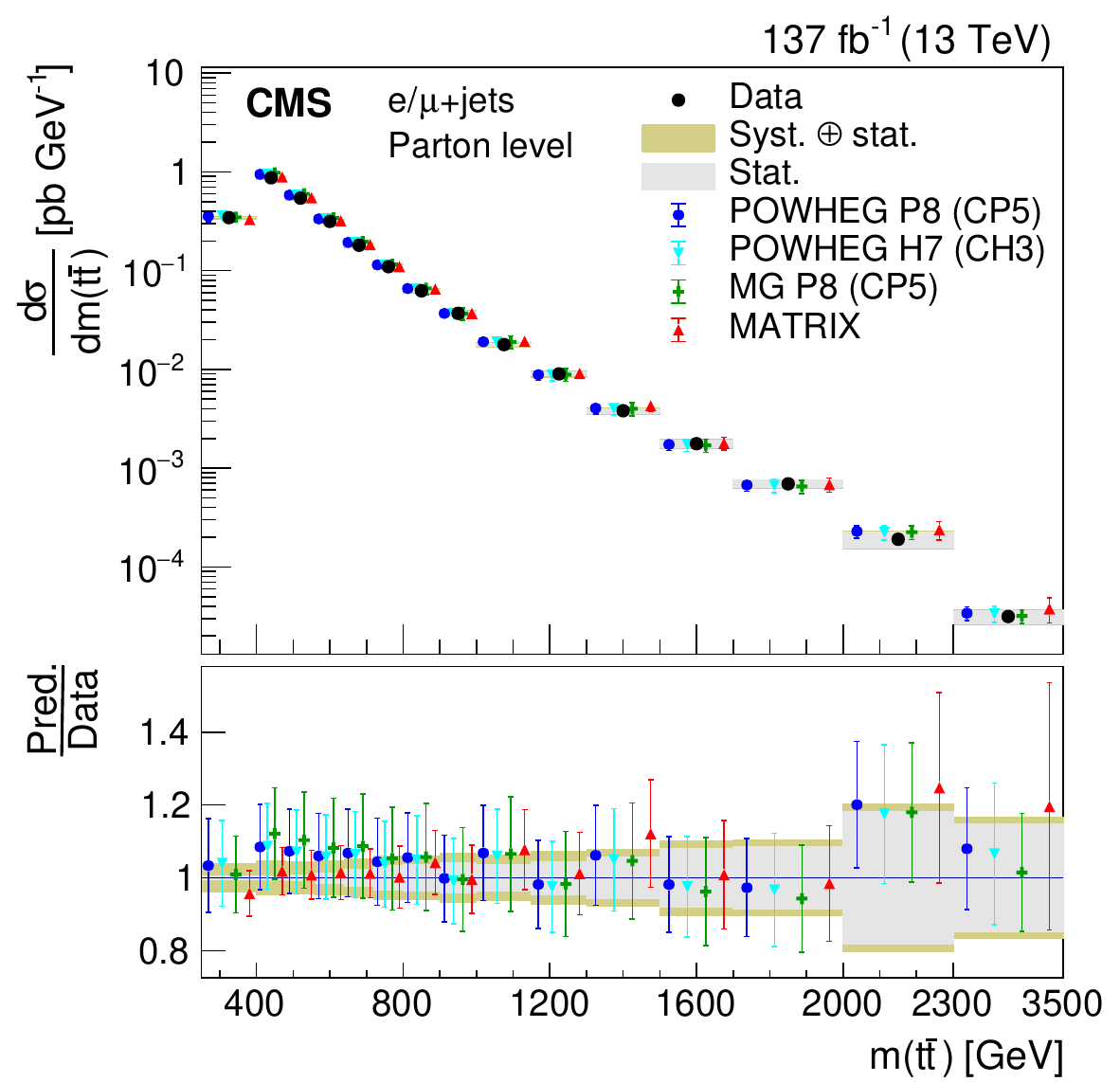}
\caption{
Differential cross sections at the parton level as a function of the hadronically decaying top quark \pt (left) and of the \ttbar invariant mass (right). The analysis was performed using \ttbar events in the $\ell$+jets final state. The data are shown as points with grey (yellow) bands indicating the statistical (statistical and systematic) uncertainties. The cross sections are compared with the predictions of \POWHEG combined with \PYTHIA (P8) or \HERWIG (H7), the multiparton simulation \MGvATNLO (MG)+\PYTHIA \textsc{FxFx}, and the NNLO QCD calculations obtained with \textsc{Matrix}.  The error bars represent the theory uncertainty in the predictions. The ratios of the various predictions to the measured cross sections are shown in the lower panels. Figure from Ref.~\cite{CMS:2021vhb}.}\label{fig:Figure_006_top}
\end{figure*}

Precise measurements of differential cross sections provide important information about the production process; the results have been used for detailed comparisons with theory predictions and to measure various SM and modelling parameters. In Fig.~\ref{fig:Figure_006_top}, a recent differential measurement of the \ttbar cross section is shown as a function of the top quark transverse momentum $\pt(\PQt)$ and the \ttbar invariant mass $m_\ttbar$~\cite{CMS:2021vhb}. These are only two of 22 differential distributions, which were also used to determine the inclusive cross section, as described in Section~\ref{subsec:ttpairs} above.

The \pt distribution of the top quark, shown in Fig.~\ref{fig:Figure_006_top} (left), shows a clear trend of most theory predictions to be somewhat harder than the data. Already early measurements of the top quark \pt in Run~1 identified this trend, as reported in Refs~\cite{CMS:2012hkm,CMS:2017iqf,CMS:2012hkm,CMS:2015rld,CMS:2016poo,CMS:2015auz}. Although it was found that the discrepancy is reduced by higher-order QCD and EW corrections~\cite{Catani:2019hip,Czakon:2017wor}, it still has a significant impact on precision measurements, most notably those where an extrapolation to the full phase space is needed to measure top quark properties. The uncertainty in the top quark \pt modelling is also relevant to searches in which the top quark is a background. 

An underlying challenge of differential measurements is the wide range of energy transfer at the LHC; although the \ttbar system is most often produced at rest, it is possible that it will also be produced at a large mass scale $Q\gg 2m_{\PQt}$, yielding boosted topologies in which the final state objects, jets and leptons, are merged. Experimentally, special techniques are used to retain high efficiency for boosted top quark jets~\cite{CMS:2016poo,CMS:2020poo}. On the theory side, additional modelling uncertainties arise. The most recent calculations achieve NNLO accuracy in perturbative QCD~\cite{Catani:2019iny,Catani:2019hip}, and include NNLL corrections~\cite{Beneke:2011mq,Cacciari:2011hy,Barnreuther:2012wtj,Czakon:2012zr,Czakon:2012pz,Czakon:2013goa,Czakon:2011xx}, and NLO EW corrections~\cite{Czakon:2017wor,Czakon:2017lgo}. 

\begin{figure*}[!htp]
\centering    
\includegraphics[width=.95\textwidth]{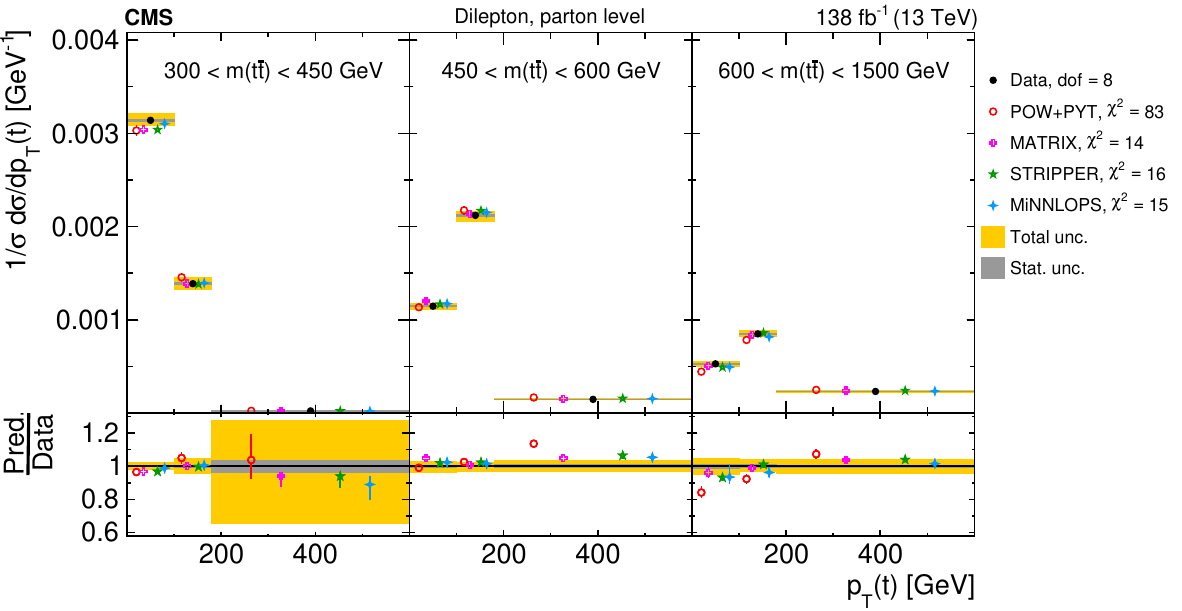}
\includegraphics[width=.95\textwidth]{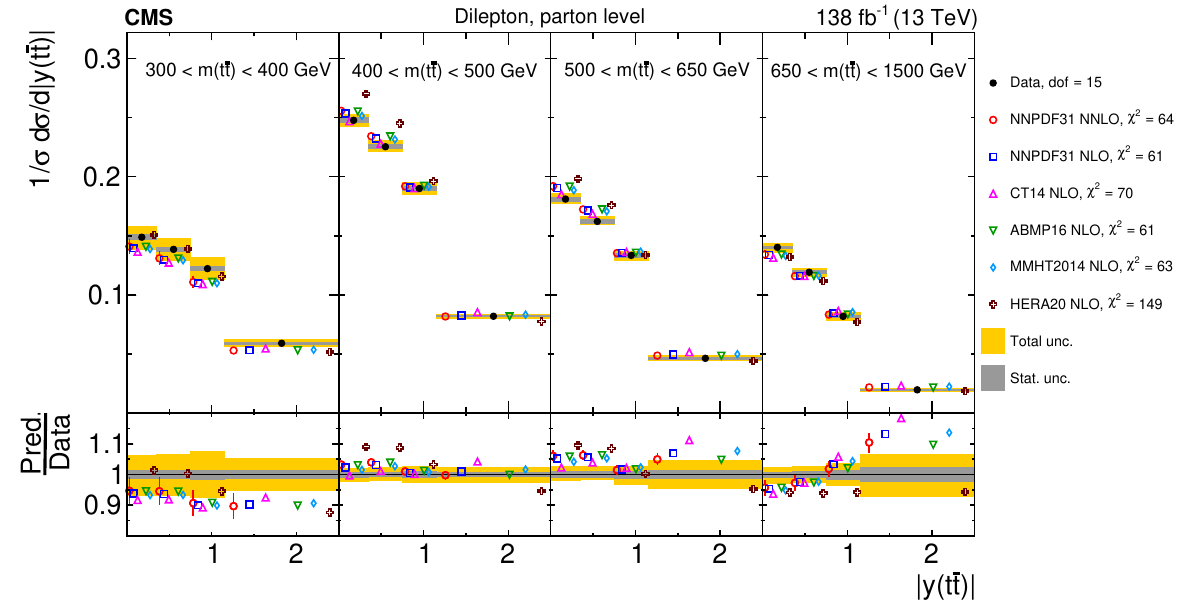}
\caption{Normalized differential cross sections as a function of $\pt(\PQt)$ in bins of $m(\ttbar)$ (upper), and as a function of $y(\ttbar)$ in bins of $m(\ttbar)$ (lower). The data, shown as bullets with grey and yellow bands indicating the statistical and total uncertainties, are compared with the prediction from \POWHEG+\PYTHIA 8 and various theoretical predictions (see text). The error bars represent the theory uncertainty in some of the predictions. The lower panel in each figure shows the ratios of the predictions to the data. Figure from Ref.~\cite{CMS:2024ybg}.}\label{fig:Figure_TOP20006}
\end{figure*}

CMS has also published a wealth of multidifferential distributions, such as those shown in Fig.~\ref{fig:Figure_TOP20006} for the dilepton channel~\cite{CMS:2024ybg}. Detailed comparisons are performed between the data and predictions up to approximate N$^3$LO. 
In Fig.~\ref{fig:Figure_TOP20006} (upper), especially in the bin of large $m(\ttbar)$, a clear improvement can be seen in the description of the data by the NNLO calculations MATRIX~\cite{Grazzini:2017mhc}, STRIPPER~\cite{Czakon:2017wor} and MiNNLOPS~\cite{Mazzitelli:2021mmm}.
In Fig.~\ref{fig:Figure_TOP20006} (lower), the data are compared with predictions from \POWHEG{}+\PYTHIA (P8) for various PDF sets. The differences between the PDF illustrate the sensitivity of the data to the parton distribution functions. In the region $300 < m(\ttbar) < 400\GeV$, the data are consistently higher than the NLO predictions for all PDFs.

\begin{figure*}[!htp]
\centering
\includegraphics[width=.6\textwidth]{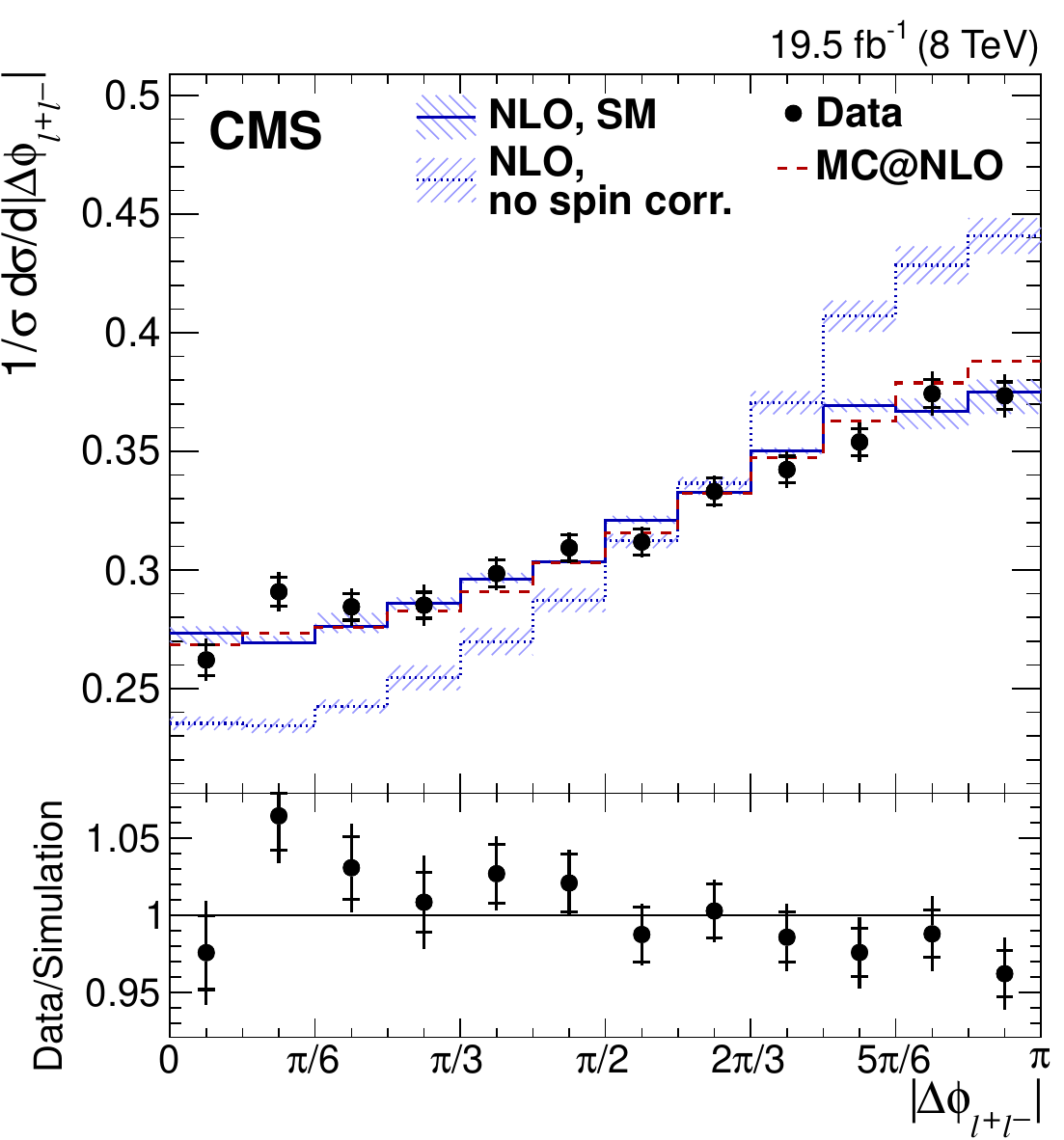}
\caption{
Normalized differential cross section as a function of the azimuthal opening angle between the two charged leptons in a \ttbar dilepton final state ($\abs{\Delta\phi_{\ell^+\ell^-}}$) from data (points); parton-level predictions from \textsc{MC@NLO} (dashed histograms); and theoretical predictions at NLO with (SM) and without (no spin corr.) spin correlations (solid and dotted histograms, respectively). The ratio of the data to the \textsc{MC@NLO} prediction is shown in the lower panel. The inner and outer vertical bars on the data points represent the statistical and total uncertainties, respectively. The hatched bands represent variations of $\mu_\text{R}$ and $\mu_\text{F}$ simultaneously up and down by a factor of 2. Figure from Ref.~\cite{CMS:2016piu}. }\label{fig:Figure_006c_top}
\end{figure*}

In Fig.~\ref{fig:Figure_006c_top}, the difference in azimuthal angle between the two charged leptons, $\Delta\phi(\ell,\ell')$ is presented as an illustration of how differential cross sections give access to the fundamental properties of the top quark. The SM predicts a correlation between the spins of the top quark and antiquark~\cite{Bernreuther:2015yna}. As the figure shows, the data are compatible with the standard model expectation, while a scenario without spin correlations is excluded. More recent measurements of spin correlations also show overall good agreement with the SM~\cite{CMS:2019nrx}.

\subsection{Top quark production in heavy ion collisions}\label{subsec:ttheavyion}

The set of $\sigma_\ttbar$ measurements performed by CMS is augmented with the first measurements of \ttbar production in $\Pp$Pb and PbPb collisions~\cite{CMS:2017hnw,CMS:2020aem}. These measurements bridge the SM and heavy ion physics programmes of the LHC with the potential to contribute to a better knowledge of the nuclear PDFs (nPDF) and the quark-gluon plasma (QGP)~\cite{dEnterria:2015mgr,Apolinario:2017sob}. Top quarks are a theoretically precise probe of the nuclear gluon density at high virtualities ($Q\sim m_\PQt$) and in a region of relatively unexplored Bjorken $x$ ($x>2m_\PQt/\sqrtsNN\approx 0.05$), where enhancement with respect to the free-proton PDF case (antishadowing) and ``EMC''~\cite{EuropeanMuon:1983wih} effects are expected~\cite{Eskola:2016oht}.
In both the pPb~\cite{CMS:2017hnw} and the PbPb~\cite{CMS:2020aem} data, the CMS analyses are limited by the small size of the data sets of $174\unit{nb}^{-1}$ and $1.7\unit{nb}^{-1}$, respectively. 
The \ttbar production has been observed with a significance above 5 standard deviations (s.d.) in pPb collisions and the cross section was measured with a relative uncertainty of 18\%, whereas in PbPb collisions the significance was 4~s.d.\,and the cross section was measured with a relative uncertainty of 33\%. Both results are somewhat lower than the corresponding SM expectations, albeit compatible within 1--2~s.d., and are still largely dominated by statistical effects. 
The measurement in $\Pp$Pb collisions is also in agreement with a more recent one made by the ATLAS Collaboration~\cite{ATLAS:2024qdu}.
The relevance of the top quark as a hard probe for nuclear PDFs (nPDFs) and the QGP is expected to gain relevance with larger data samples, as explored in Refs.~\cite{Dainese:2019rgk,CMS-PAS-FTR-18-027,Apolinario:2017sob}.

\subsection{Top quark production in association with vector bosons}\label{subsec:tvecbos}

Rare processes, such as the associated production of the top quark with vector bosons, have become accessible with the larger data samples of Run~2. Such processes offer the possibility to directly probe the EW couplings of the top quark and explore the sensitivity of the data to several BSM extensions. The production cross sections are typically small ($<$1\pb) owing to both the high mass of the state produced and the weaker couplings of the vector bosons with respect to QCD. The CMS Collaboration has either observed or found experimental evidence for all processes in which either \ttbar or single top quarks are produced in association with vector bosons (\PZ,\PW,\cPgg) or the Higgs boson (setting aside $\PQt\PH\PQq$). The measurements of associated production with the Higgs boson are later discussed in Section~\ref{sec:higgs}, whereas associated $\PQt\PW$ was already discussed in Section~\ref{subsec:singletop}.

\begin{figure*}[!htp]
\centering
\includegraphics[width=.32\textwidth]{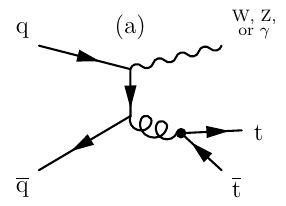}
\includegraphics[width=.32\textwidth]{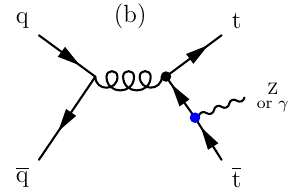}
\includegraphics[width=.32\textwidth]{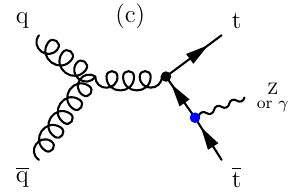}
\caption{
Example Feynman diagrams for the production of \ttbar with a vector boson through initial state radiation (a) or a direct coupling to the top quark (b and c).
The latter is only possible for neutral bosons $\Vzero=\cPgg,\PZ$.}\label{fig:Figure_009_top}
\end{figure*}

\begin{figure*}[!htp]
\centering
\includegraphics[width=.32\textwidth]{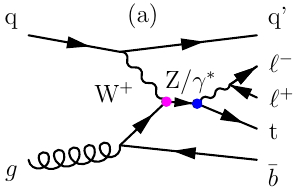}
\includegraphics[width=.32\textwidth]{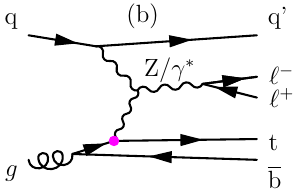}
\includegraphics[width=.32\textwidth]{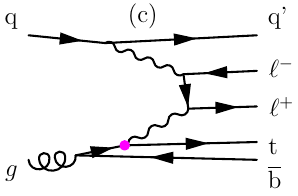}
\caption{
Example Feynman diagrams for the production of $\PQt\cPZ\PQq$.
}\label{fig:Figure_010_top}
\end{figure*}

Processes with neutral bosons \Vzero in the final state ($\Vzero=\cPgg,\PZ$) share similar diagrams that can be studied to examine the different EW dipole operators of the top quark~\cite{Schulze:2016qas}, or in background estimations~\cite{CMS:2017jrd}. Examples of these Feynman diagrams are shown in Fig.~\ref{fig:Figure_009_top} where the \Vzero is pictured as arising either from initial state radiation (ISR) or from a direct coupling to the top quark. Figure~\ref{fig:Figure_009_top}(a) depicts the possibility of a \PW~boson being produced by ISR only.
Some additional differences between $\cPgg$ and \PZ~bosons arise from an enhanced probability that the \cPgg may be radiated from a final-state charged particle, because it is massless. Conversely, the dilepton states typically explored in the \PZ~boson analysis, can be produced by additional off-shell and $\gamma^*\to\ell\ell$ contributions. 
In the data analyses, such additional contributions are typically suppressed by the requirement that the dilepton invariant mass $m(\ell\ell)$ is reconstructed in the vicinity of the \PZ~boson pole mass. 
These differences are also present in single top quark associated production with \Vzero, illustrated in Fig.~\ref{fig:Figure_010_top}, where contributions from $\PW\PW\PZ$ and $\PW\PW\cPgg$ TGCs may be present, as well as nonresonant dilepton contributions (Fig.~\ref{fig:Figure_010_top}(c)).   
Therefore, single top quark associated production has the potential of providing additional handles for EW fits of aTGCs.
Besides the obvious interest in the couplings of the top quark and the EW sector, the presence of the \Vzero  introduces an additional intrinsic asymmetry in the \ttbar system at LO level, which is a clean probe of BSM effects. The asymmetry arises from the increase of the relative contribution of $\PQq\cPaq$-initiated processes~\cite{Aguilar-Saavedra:2014vta}. The $\ttbar \Vzero$ processes receive background contributions from $\PQt\PW \Vzero$ processes, and at NLO, interference terms between $\ttbar \Vzero$ and $\PQt\PW \Vzero$ arise, in analogy to the inclusive case of $\PQt\PW$ and \ttbar described in Section~\ref{subsec:singletop} above. The cross section for $\PQt\PW\PZ$ is expected to be about 15\% of that for $\ttbar\PZ$~\cite{CMS:2023krq}. CMS obtained evidence for the $\PQt\PW\PZ$ process with an observed $3.4\sigma$ statistical significance~\cite{CMS:2023krq}. The result is in agreement with the SM expectation within one standard deviation.

\begin{figure*}[!htp]
\centering
\includegraphics[width=.99\textwidth]{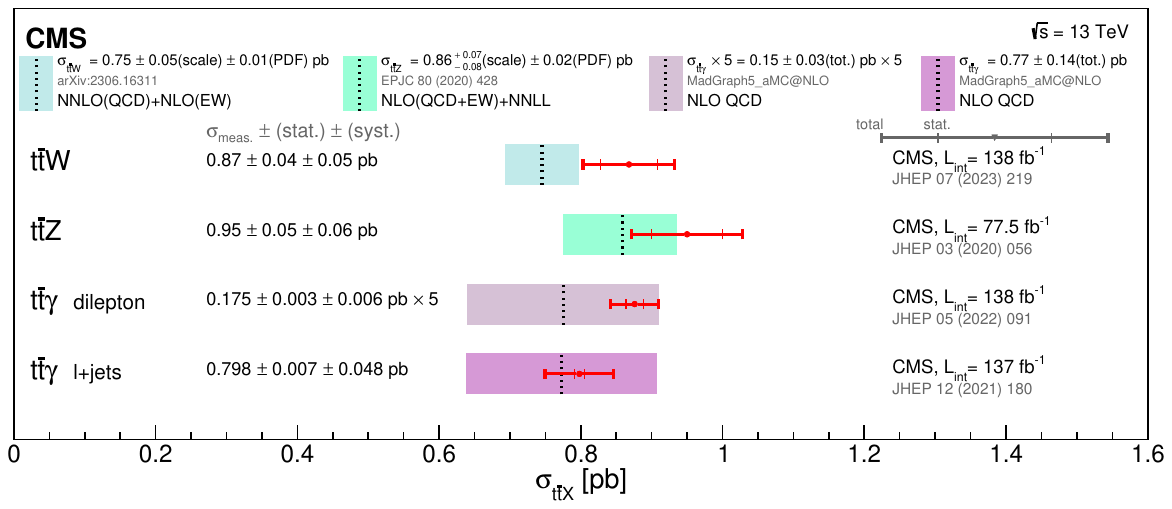}
\caption{
Summary of CMS $\ttbar\PW$ and $\ttbar\Vzero$ cross section measurements with respect to the SM prediction.
The horizontal bars display separately the statistical and the total uncertainties of the experimental measurements.
The uncertainty associated to the theory predictions is represented by shaded bands and includes the variations of the renormalization and factorization scales and parton density functions.
}\label{fig:Figure_011_top}
\end{figure*}

\begin{figure*}[!htp]
\centering
\includegraphics[width=.99\textwidth]{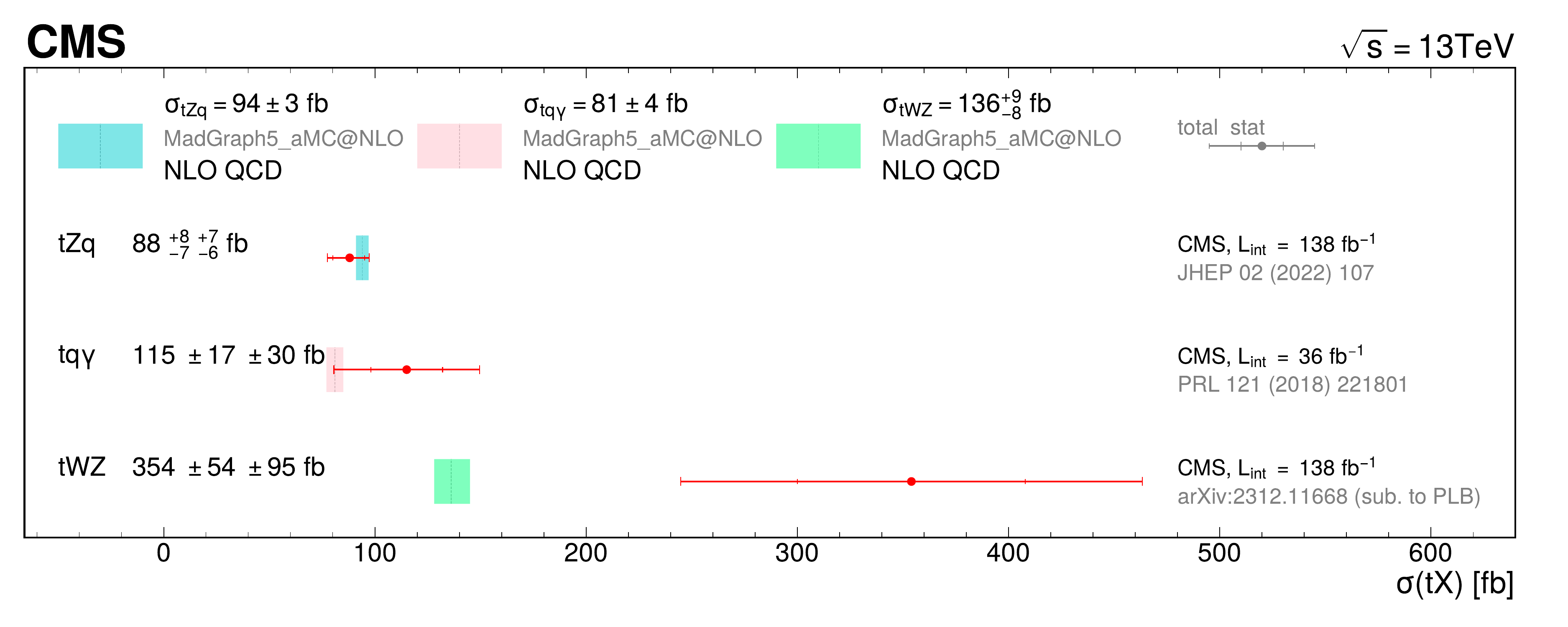}
\caption{
Summary of CMS measurements of $\PQt \Vzero \PQq $ ($\Vzero=\PZ,\cPgg$) cross sections at 13\TeV. 
The cross section measurements are compared with the NLO QCD theoretical calculation.  The horizontal bars display separately the statistical and the total uncertainties.
The uncertainty associated to the theory predictions is represented by shaded bands and includes the variations of the renormalization and factorization scales and parton density functions.
}\label{fig:Figure_012_top}
\end{figure*}

The CMS Collaboration has carried out several measurements of the $\ttbar \Vzero$ and $\PQt \Vzero \PQq$ processes; the results are summarized in Figs.~\ref{fig:Figure_011_top} and~\ref{fig:Figure_012_top}. 
Table~\ref{tab:tplusv0summary} summarizes the final states explored in these measurements, the corresponding references, and the NLO predictions.
Overall, good agreement between theory predictions and data is attained in these measurements.

The current uncertainties are about 8\% for the $\ttbar\PZ$ cross section, dominated by statistical and lepton-selection efficiency uncertainties~\cite{CMS:2019too}. In this analysis, the main background is from nonprompt leptons and $\PW\PZ$~boson production, modelled from dedicated control regions, and other associated top quark production $\PQt(\cPaqt)X$, modelled from simulation. The measurements of the $\PQt\PZ\PQq$ production cross section are mostly limited by the statistical uncertainty ($\approx$12\%) followed by systematic uncertainties related to backgrounds from $\PW\PZ$ and $\ttbar\PZ$ processes, from misidentified lepton candidates, jet energy scale, and lepton selection efficiencies~\cite{CMS:2021ugv}.

In the context of associated processes with photons, a total uncertainty of 3.5\% is achieved for the $\ttbar\cPgg$
process using all the available data at $\sqrts=13\TeV$, whereas the $\PQt\cPgg\PQq$ process has been measured
with 10\% total uncertainty (4.4~s.d.\,significance)~\cite{CMS:2018hfk} with an initial subset of the 13\TeV data.
Both are in agreement with the SM predictions at NLO.

\begin{table}[!htp]
\centering
\topcaption{Summary of final states covered experimentally in associated top quark and neutral boson production by CMS.
For each process listed in column (a), column (b) quotes the theoretical prediction at 13\TeV. 
Columns (c) and (d) summarize the different final states generated by the top quark (s) and boson decays with the corresponding branching fraction (B) listed in column (f). The combined results for the \PW~and \PZ~boson Bs include the propagation of $\PGt$-leptonic decays.
The nomenclature assigned to these channels is shown in column (e) with  SS (OS) used as a shorthand for same- (opposite-) charge lepton pairs.  
The CMS measurements of these channels are listed in column (g).
The theoretical uncertainties include the PDF+\alpS{} and scale choice. Symbols provide additional information:
($\dagger$) predicted at NLO accuracy using  \MGvATNLO v2.6.5, and corresponding to the fiducial region~\cite{CMS:2021klw};
($\bullet$) the quoted fiducial $\PQt{}\cPgg{}$ cross section is predicted at NLO QCD accuracy~\cite{Alwall:2014hca} corresponding to the selection of Ref.~\cite{CMS:2018hfk};
($\ast$) - computed at NLO including QCD+EW effects and NNLL QCD effects~\cite{Kulesza:2020nfh};
($\star$) - computed at NLO QCD and EW accuracy~\cite{LHCHiggsCrossSectionWorkingGroup:2016ypw,Frixione:2015zaa,Frederix:2018nkq};
($\diamond$) - computed at NLO QCD accuracy in the 5FS~\cite{Alwall:2014hca}, in the phase space of~\cite{CMS:2021aly}.
($\delta$) - computed at NLO QCD accuracy~\cite{Frixione:2015zaa,Frederix:2012ps}.
}\label{tab:tplusv0summary}
\renewcommand{\arraystretch}{1.2}
\maybeCmsTable{
\begin{tabular}{lclcccl}
\multicolumn{1}{c}{(a)} & \multicolumn{1}{c}{(b)} & \multicolumn{1}{c}{(c)} & \multicolumn{1}{c}{(d)} & \multicolumn{1}{c}{(e)} & \multicolumn{1}{c}{(f)} & \multicolumn{1}{c}{(g)} \\
Process & 
$\sigma$ or $\sigma_\text{fid}$ (fb) &
\ttbar{} decay &
Boson decay &
Channel &
B &
Measurements \\
\hline
\multirow{2}{*}{$\ttbar\cPgg$}
& $773 \pm 135$ $^\dagger$ & ($\ell^\pm\nu\PQb$)($\PQq\cPaq\PQb$) & - & 1$\ell$ &  34.4\% & {\cite{CMS:2017tzb,CMS:2021klw}} \\
& $63 \pm 9$ $^\dagger$ & ($\ell^\pm\nu\PQb$)($\ell^\mp\nu\PQb$) & - & 2$\ell$OS & 6.5\% & {\cite{CMS:2022lmh}} \\
[\cmsTabSkip] 

$\PQt\cPgg(\PQq)$ & $81 \pm 4\,^{\bullet}$ 
& $(\ell^\pm\nu\PQb)$ & - & 1$\ell$ & 25.6\% & {\cite{CMS:2018hfk}} \\
[\cmsTabSkip] 

\multirow{4}{*}{$\ttbar\PZ$} & \multirow{4}{*}{$840 \pm 100\,^\star$}
& $(\ell^\pm\nu\PQb)(\PQq\cPaq\PQb)$       & $\PQq\cPaq$             & 1$\ell$ & 24.1\% & {\cite{CMS:2022hjj}} \\
& & $(\ell^\pm\nu\PQb)(\ell^\mp\nu\PQb)$   & $\PQq\cPaq$             & 2$\ell$OS & 4.6\% & {\cite{CMS:2015uvn}} \\
& & $(\ell^\pm\nu\PQb)(\PQq\cPaq\PQb)$    & $\ell^\pm\ell^\mp$ & 3$\ell$ & 2.3\% & {\cite{CMS:2013vyn,CMS:2014fhw,CMS:2015uvn,CMS:2017ugv,CMS:2019too,CMS:2021aly}}\\
& & $(\ell^\pm\nu\PQb)(\ell^\mp\nu\PQb)$ & $\ell^\pm\ell^\mp$ & 4$\ell$ & 0.4\% & {\cite{CMS:2014fhw,CMS:2015uvn,CMS:2017ugv,CMS:2019too,CMS:2021aly}} \\
[\cmsTabSkip]

$\PQt\PZ(\PQq)$ & $94 \pm 3.1\,^{\diamond}$
& $(\ell^\pm\nu\PQb)$ & $\ell^\pm\ell^\mp$ & 3$\ell$ & 1.7\% & {\cite{CMS:2017wrv,CMS:2018sgc,CMS:2021aly,CMS:2021ugv}} \\
[\cmsTabSkip] 

\multirow{2}{*}{$\PQt\PW\PZ$} & \multirow{2}{*}{$136^{+9}_{-8}\,^{\delta}$} 
& $(\ell^\pm\nu\PQb)$ & $(\PQq\cPaq)(\ell^\pm\ell^\mp)$ & \multirow{2}{*}{3$\ell$} & \multirow{2}{*}{1.4\%} & \multirow{2}{*}{\cite{CMS:2023krq}} \\
& & $(\PQq\cPaq\PQb)$ & $(\ell^\pm\nu)(\ell^\pm\ell^\mp)$ & &  \\
\end{tabular}}
\end{table}

The $\ttbar\PW$ process, depicted in Fig.~\ref{fig:Figure_009_top}(a), is particularly interesting because the \ttbar pair is produced via gluon splitting from a $\PQq\cPaq$ initial state. Because of the proton PDFs, it is expected that  $\sigma(\ttbar\PWp)\approx 1.9 \sigma(\ttbar\PWm)$ at LO, \ie it is a charge-asymmetric process.
With the inclusion of higher orders in perturbation theory new production channels open up, and hence new colour-flow and flavour structures, and this results in a significant increase of the cross section. The PS predictions used to model this process have NLO accuracy in QCD for the production and are limited to on-shell decays, with the top quark decay modelled at LO~\cite{LHCHiggsCrossSectionWorkingGroup:2016ypw,Frixione:2015zaa,Frederix:2018nkq}.
More advanced fixed-order calculations, including off-shell effects, emission of extra partons, and NNLL contributions, are available but not employed yet.
Some effects, such as EW corrections, are larger in $\ttbar\PW$ than in $\ttbar\PZ$ production, making the  $\ttbar\PW$ process especially interesting. In Ref.~\cite{Bevilacqua:2021tzp}, it is estimated that NLO+PS cross sections, such as the one quoted in Table~\ref{tab:tplusv0summary}, fall short by 10--35\% with respect to a calculation at the same order, including the missing full off-shell effects. The experimental measurements of $\ttbar\PW$ production  are currently about 20\% higher than the SM prediction and thus provide important input in a phase-space region where theory is actively evolving.

In CMS, the measurements of the $\ttbar\PW$ process have mostly focused on multilepton final states, in particular those comprising either a same-sign dilepton pair or three leptons. 
A multitude of different competing processes constitute the background ranging from \ttbar, dibosons, nonprompt leptons, and rare \ttbar associated production processes, but also conversions of photons into electron pairs, and incorrect lepton charge measurements. These need to be estimated from data themselves. The events are analysed in different categories that enhance the different contributions, typically using jet or \PQb jet multiplicities, total lepton charge, \PZ~bosons reconstructed with same-sign lepton candidates, or leptons with loosened identification criteria. 
To reduce the uncertainties in lepton selection and background contamination, dedicated MVA methods have been employed.
The most precise measurement of the $\ttbar\PW$ cross section has a relative uncertainty of 7.5\%, dominated by the statistical component and the modelling of signal and backgrounds, specifically $\ttbar\PH$. The interplay between the $\ttbar\PW$ and $\ttbar\PH$ processes is discussed in Section~\ref{sec:higgs}. 
The measured charge asymmetry, $\sigma_{\ttbar\PWp}/\sigma_{\ttbar\PWm}=1.61^{+0.17}_{-0.16}$, is slightly below the SM prediction.
Table~\ref{tab:tplusw} summarizes the $\ttbar\PW$ measurements performed so far by the CMS Collaboration, and Fig.~\ref{fig:Figure_011_top} includes a comparison of the most precise $\ttbar\PW$ measurement with the theory prediction.

\begin{table}[!htp]
\centering
\topcaption{Summary of final states covered experimentally in associated $\ttbar\PW$ production. 
The structure of the table is similar to that of Table~\ref{tab:tplusv0summary}.
The cross section column cites the prediction at 13\TeV computed at NLO including QCD (up to two jets) and EW contributions~\cite{Frederix:2021agh}.
}\label{tab:tplusw}
\renewcommand{\arraystretch}{1.2}
\begin{tabular}{lcccccc}
Process & 
$\sigma$ (fb) &
\ttbar{} decay &
Boson decay &
Channel &
B &
Measurements \\
\hline
\multirow{2}{*}{$\ttbar\PW$} & \multirow{2}{*}{
$722^{+71}_{-78}$
}
& ($\ell^\pm\nu\PQb$)($\PQq\cPaq\PQb$) & $\ell^\pm\nu$ & 2$\ell$SS & 4.4\% & {\cite{CMS:2013vyn,CMS:2014fhw,CMS:2015uvn,CMS:2017ugv,CMS:2022tkv}} \\
& & ($\ell^\pm\nu\PQb$)($\ell^\mp\nu\PQb$) & $\ell^\pm\nu$ & 3$\ell$ & 1.7\% &  {\cite{CMS:2015uvn,CMS:2022tkv}} \\
\end{tabular}
\end{table}

\subsection{Associated production of \texorpdfstring{\ttbar}{ttbar} with jets}\label{subsec:topandjets}

Measurements of \ttbar with jets are typically performed as differential cross section measurements and interpreted as tests of perturbative QCD.
The CMS Collaboration has produced several such measurements at different $\sqrts$, using different final states and exploring the correlation with the kinematics of the top quark, the \ttbar system, and other event variables,
as outlined in Refs.~\cite{CMS:2014jya,CMS:2015ilk,CMS:2018htd,CMS:2018mdd,CMS:2018ypj,CMS:2020grm,CMS:2021vhb}.
The sensitivity of these distributions to the UE, PS modelling, and the ME-PS matching is explored in conjunction with ancillary measurements to improve the theoretical modelling and to validate new models. Recent examples are available in Ref.~\cite{CMS:2019csb}, where the best agreement with data is found for the \MGvATNLO matrix element generator and the \textsc{FxFx} matching scheme using \PYTHIA8, and in Ref.~\cite{CMS:2020dqt} where good agreement is found between data and the \POWHEG{}+\HERWIG7 setup.

\begin{figure*}[!htp]
\centering
\includegraphics[width=.32\textwidth]{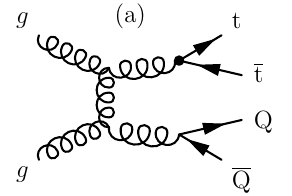}
\includegraphics[width=.32\textwidth]{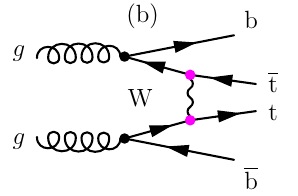}
\includegraphics[width=.32\textwidth]{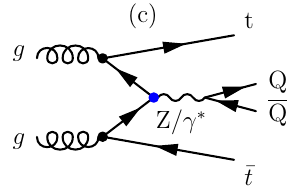}
\caption{
Feynman diagrams contributing to the associated production of top quarks with heavy-flavoured jets.
}\label{fig:Figure_012abc_top}
\end{figure*}

When the additional jets are heavy-flavoured, these processes are particularly important to understand, since they constitute backgrounds to the measurements of processes such as
$\ttbar\PH(\to\PQb\cPaqb)$ and \ttbar\ttbar.
The final states of $\ttbar\PQb\cPaqb$ and $\ttbar\PQqc\cPaqc$ are complex, as they comprise many jets. The additional heavy-flavour quark pair arises typically from  gluon splitting and the jets in the final state end up being soft in \pt and close in the $\eta$--$\phi$ plane. A gluon splitting Feynman diagram is shown in Fig.~\ref{fig:Figure_012abc_top}(a).
With the exception of the $\ttbar\PH$ measurements, described in Section~\ref{sec:higgs}, the analyses do not distinguish whether the origin of a jet is from gluon splitting, boson decay or another multiparton interaction. Two of these cases are represented in Figs.~\ref{fig:Figure_012abc_top}(b) and (c).

\begin{figure*}[!htp]
\centering
\includegraphics[width=.49\textwidth]{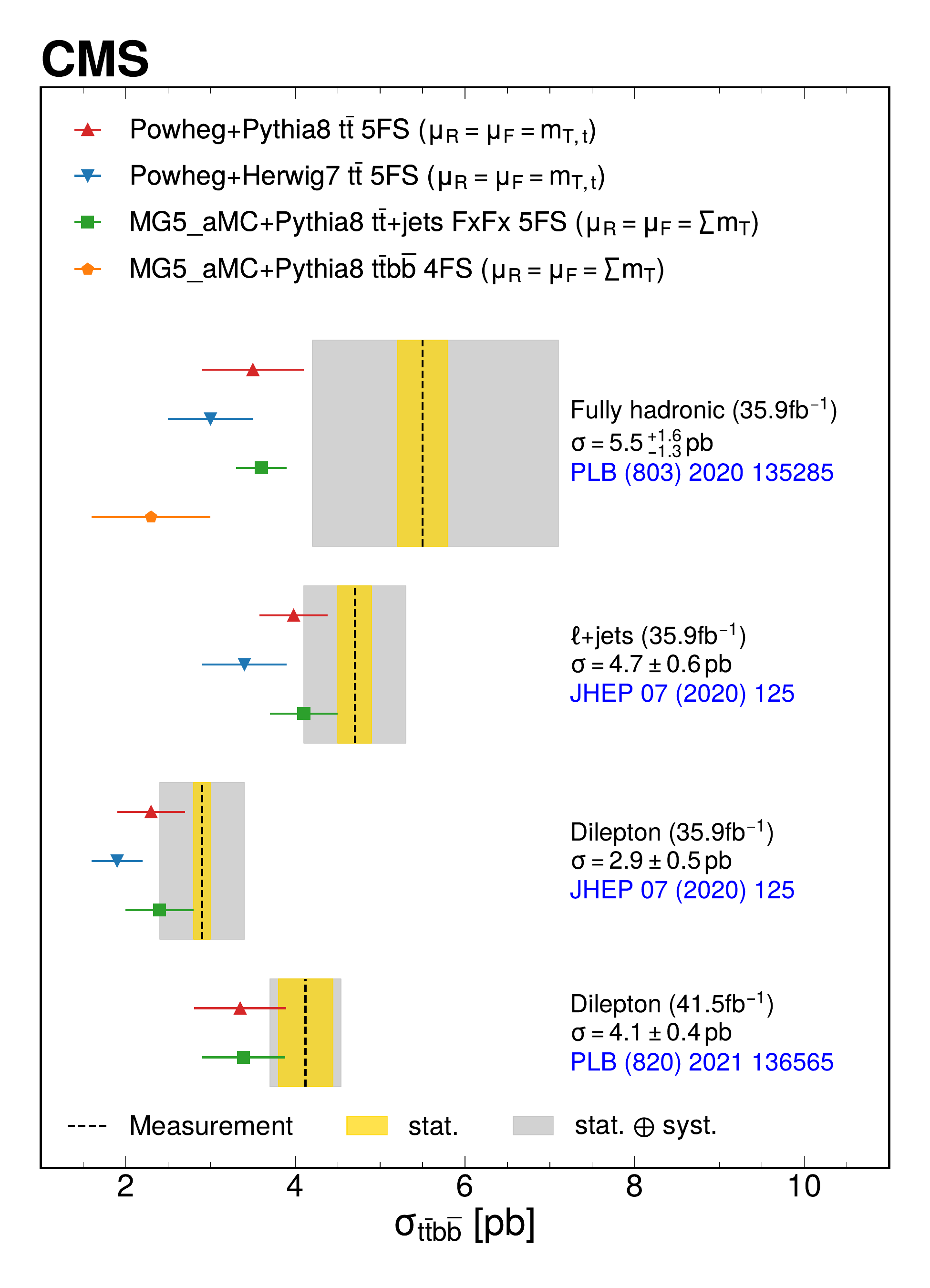}
\includegraphics[width=.49\textwidth]{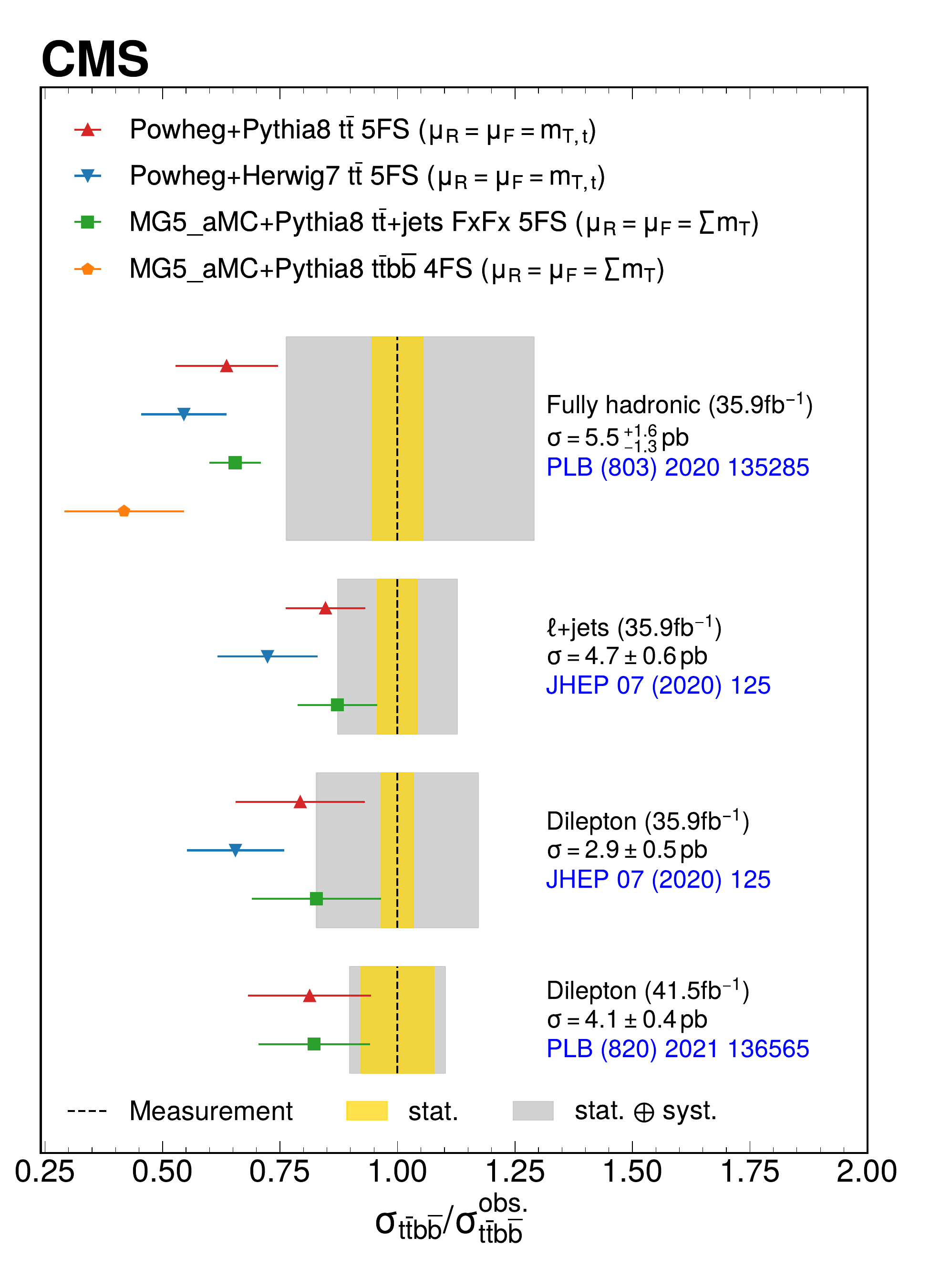}
\caption{
Summary of $\ttbar\PQb\cPaqb$ cross section measurements.
The left plot depicts the measurements performed in the full phase space using different final states and data sets, compared with different MC predictions.
The right plot shows the ratio between the theoretical and measured cross.
The statistical and total uncertainties on the measurements are represented by different shaded bands, while the uncertainty on the predictions are represented by error bars.
}\label{fig:Figure_013_top}
\end{figure*}

A summary of the $\ttbar\PQb\cPaqb$ measurements by CMS is given in Fig.~\ref{fig:Figure_013_top}.
The latest $\ttbar\PQb\cPaqb$~\cite{CMS:2019eih,CMS:2020grm,CMS:2023xjh} and $\ttbar\PQqc\cPaqc$~\cite{CMS:2020utv} measurements improve significantly over previous results because of higher statistics and better identification of heavy-flavoured jets. The achievement was made possible by the improved tracking capabilities of the upgraded pixel detector in the second part of Run~2 and the usage of more modern machine learning (ML) algorithms such as \textsc{DeepJet}~\cite{BTV-16-002,CMS:2021scf}. 

The measured cross sections are generally somewhat higher than the predictions.
Models that rely on parton showers for high jet multiplicities tend to underestimate the rate of events with three or more \PQb jets, indicating that either additional tuning or higher-order accuracy is needed. 
From the theoretical point of view, the calculations of these multiscale processes come with large NLO corrections, up to a factor of 2,
and a relatively large final uncertainty of typically 20\%~\cite{Denner:2020orv,Bevilacqua:2022twl}.
Because of the still large theoretical uncertainties, the difference of the experimental data with respect to theory has a reduced significance (1--2~s.d.).
Similar to previous discussions in Section~\ref{subsec:singletop}, the 5FS generally describes the observed rates better than the 4FS.  
The dominant experimental uncertainties are related to the efficiency of the flavour-tagging algorithms 
and to the modelling of the parton shower. 

Additional measurements, with larger data samples and exploring new jet algorithms which can probe the phase space typically vetoed by the hard jet selection constraints, will help to improve the description of these important processes.

\subsection{Four top quark production}\label{subsec:fourtops}

\begin{figure*}[!htp]
\centering
\includegraphics[width=.32\textwidth]{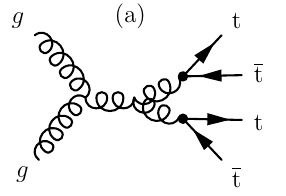}
\includegraphics[width=.32\textwidth]{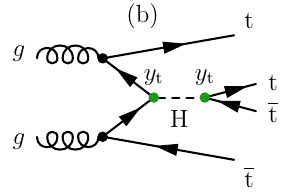}
\includegraphics[width=.32\textwidth]{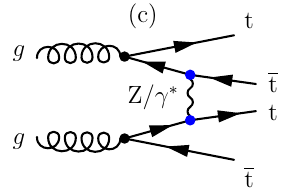}
\caption{
Representation of different Feynman diagrams contributing to \ttbar\ttbar production at the LHC. Diagrams that involve strong coupling vertices, shown in (a), are expected to dominate.
}\label{fig:Figure_014_top}
\end{figure*}

With a cross section that is five orders of magnitude lower than that of \ttbar production, four top quark production (\ttbar\ttbar) is among the rarest QCD processes established by the CMS experiment. At NLO plus next-to-leading logarithmic accuracy (NLO+NLL' QCD+EW), the expected cross section is
$\sigma_{\ttbar\ttbar}(13\TeV) = 13.4^{+1.0}_{-1.8}\unit{fb}$~\cite{vanBeekveld:2022hty}.
The large number of permutations of decay modes of the four \PW~bosons leads to a large number of different final states, all of which also contain four \PQb jets. 
Besides the dominant strong production mode, \ttbar\ttbar receives contributions from EW vertices, such as the ones involving the top quark Yukawa coupling as shown in Fig.~\ref{fig:Figure_014_top}(b). In addition, several BSM scenarios, such as supersymmetry, simplified dark matter models, and Type II Higgs doublet models, predict modifications to the SM \ttbar\ttbar production~\cite{Darme:2018dvz,Bauer:2017ota}.

The CMS Collaboration has analysed a large number of  decay channels, including the fully hadronic~\cite{CMS:2023zdh}, $1\ell$~\cite{CMS:2014ylv,CMS:2017nnq,CMS:2019jsc,CMS:2023zdh}, $2\ell$OS~\cite{CMS:2017nnq,CMS:2019jsc,CMS:2023zdh}, $2\ell$SS, and multilepton~\cite{CMS:2017ocm,CMS:2019rvj,CMS:2023ftu} final states. 
Various backgrounds contribute to each of these final states, some of them being common with the backgrounds of \ttbar{}+V associated production or \ttbar{}+jets. The correct modelling of \ttbar in association with vector bosons and with heavy flavours plays a crucial role, and control regions are established in data to validate the background estimations.

Among all these final states, the multilepton final states, specifically the $2\ell$SS and $3\ell$ channels, achieve the highest significance, owing to their purity.
In both Ref.~\cite{CMS:2019rvj} and Ref.~\cite{CMS:2023ftu} MVA discriminators are trained to separate the \ttbar\ttbar signal from the backgrounds. The cross section is measured from a combined fit using several categories. Although using the same data set, Ref.~\cite{CMS:2023ftu} improves over the results obtained in Ref.~\cite{CMS:2019rvj} because of the improved lepton and \PQb jet identification techniques. 
Observation-level significance above the background-only hypothesis is attained in Ref.~\cite{CMS:2023ftu}: 5.6~s.d.\,with 4.9~s.d.\,expected. The measured cross section $\sigma_{\ttbar\ttbar}(13\TeV)$ is $17.9\pm4.1\unit{fb}$, in agreement with the SM. The result is still statistically limited, and the main systematic uncertainties arise from the \PQb tagging efficiency (about 5\%) and the jet energy scale uncertainty (about 3\%).

The all-hadronic channel has also been explored by the CMS experiment for the first time~\cite{CMS:2023zdh}, making use of both resolved and boosted top quark reconstruction. A custom BDT and minimum $\eta$--$\phi$ separation is used in the resolved regime, whereas the boosted regime makes use of CMS's {DeepAK8} algorithm~\cite{CMS:2020poo}. 
The combination of the $\ell$+jets, $2\ell$OS, and all hadronic channels using full Run~2 data yield a significance of 3.9~s.d.\,with 1.5~s.d.\,expected; the excess is attributed to the full hadronic channel. After combination with the $2\ell$SS and multilepton analysis from Ref.~\cite{CMS:2019rvj} and the $2\ell$OS analysis from Ref.~\cite{CMS:2019jsc} the observed significance becomes 4.0~s.d.\,with 3.2~s.d.\,expected.

\begin{figure*}[!htp]
\centering
\includegraphics[width=.7\textwidth]{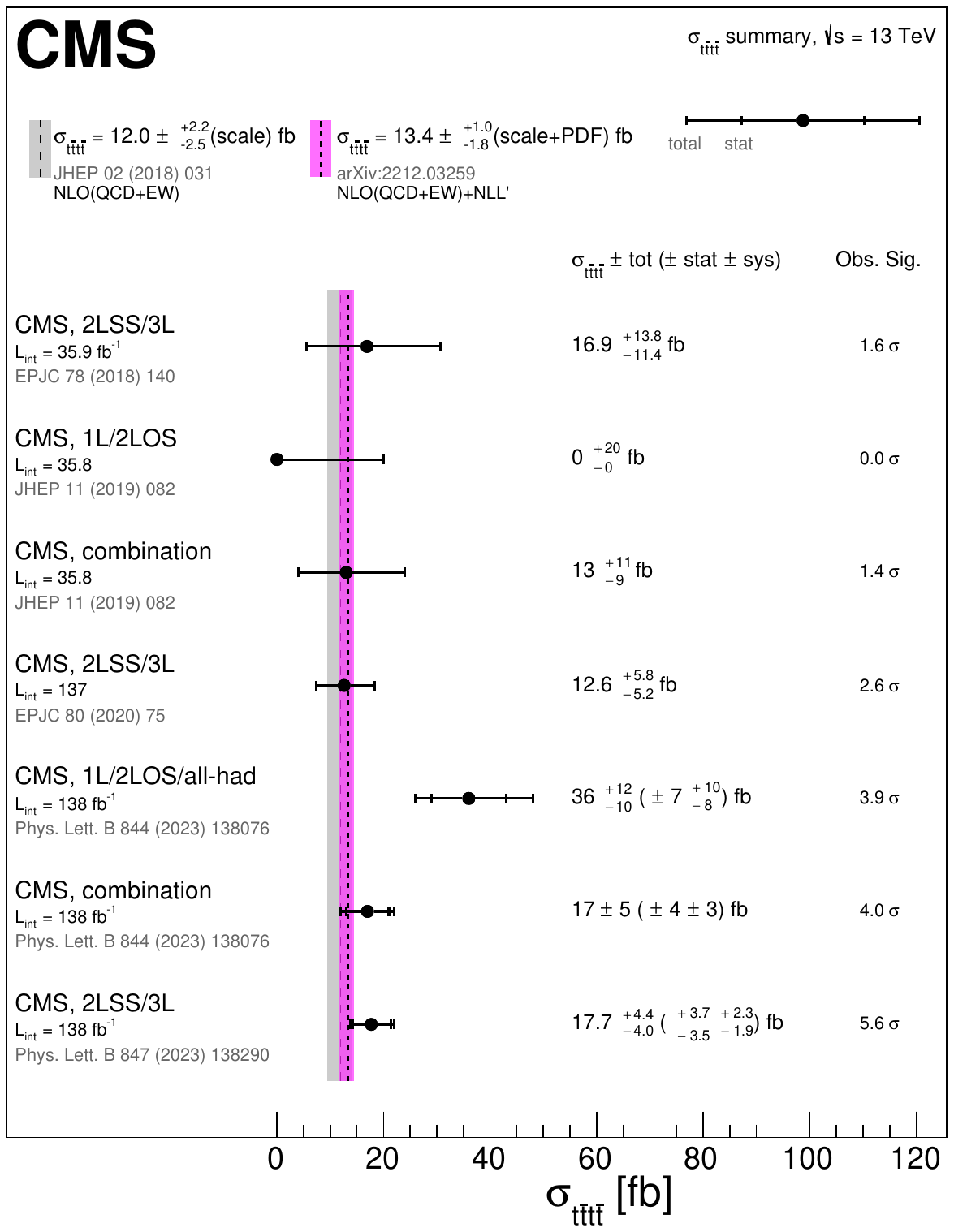}
\caption{
Summary of CMS measurements of the \ttbar\ttbar production cross section at 13\TeV in various channels. The total (statistical) uncertainty associated with the measurements is represented by the outer (inner) error bars.  The cross section measurements are compared with the NLO QCD and EW theoretical calculation. The theoretical band represents uncertainties due to renormalization and factorization scales. Complementary theory predictions are also available in Ref.~\cite{Frederix:2017wme}.
}\label{fig:Figure_015_top}
\end{figure*}

Figure~\ref{fig:Figure_015_top} summarizes all the $\ttbar\ttbar$ searches and measurements performed so far by CMS. They are consistent with the SM within the uncertainties. The most precise combination~\cite{CMS:2023ftu} shows a slightly larger measured cross section value and achieves observation of $\ttbar\ttbar$ production. 

Larger data sets will be used by CMS to further explore this process, to constrain fundamental parameters such as $y_\PQt$ and to look for BSM effects~\cite{Dainese:2019rgk}. Related analyses of the production of three top quarks in association with a jet or a \PW~boson will require data sets of higher integrated luminosity because of their small expected cross sections of about 0.47 and 0.73\unit{fb}, respectively~\cite{Barger:2010uw,Boos:2021yat}. This is analogous to the history of top quark cross section measurements in which the $\PQt\PW$ process was established long after that of $\ttbar$. The three-top quark processes share similar overlapping issues, albeit at a higher energy scale and top quark multiplicity. 

\subsection{Extraction of fundamental theory parameters from top quark cross sections}\label{subsec:topandsmconstants}

One of the main aims of inclusive cross section measurements is to extract information about fundamental SM parameters. Top quark production cross sections allow measurements of \alpS, $y_\PQt$ and $V_{\PQt\PQb}$. A short description of the precision achieved so far by CMS is given below. Direct measurements of $\ttbar\PH$ and the combined Higgs boson results to extract $y_\PQt$ are described later in Section~\ref{sec:higgs}. 

As noted in Section~\ref{subsec:ttpairs}, the $\sigma_\ttbar$ cross section is sensitive to both \alpS and $m^\text{pole}_\PQt$,
thus its measurement can be used to extract one of the two parameters while fixing the other. In addition, a choice has to be made related to the PDF set, and the corresponding fixed order and mass scheme.
In differential cross section measurements, \eg of the mass and rapidity of the \ttbar system, the three quantities (\alpS, $m^\text{pole}_\PQt$ and PDF) can be extracted simultaneously, as demonstrated in Ref.~\cite{CMS:2019esx}.

For the extraction of $\alpS$, the inclusive \ttbar cross section is used, and hence, residual uncertainties related to the extrapolation of the cross section from the fiducial phase space to the full phase space enter the measurement and cannot be constrained from data since they impact a region that is not accessible experimentally. The uncertainties include scale choices, PDF uncertainties, and the uncertainty in the LHC beam energy. Nonperturbative (NP) contributions related to the intrinsic $k_\text{T}$, but also to the modelling of the QCD colour charge carried by the top quark or antiquark (\ie colour reconnection~\cite{CMS:2022awf}) may contribute as well. Even though NP effects occur at a scale $\Lambda_\text{QCD}$ and in most cross section measurements $Q^2 \gg \Lambda_\text{QCD}$, NP effects may still be relevant if the selection is strict or involves a large extrapolation. The \pt distribution of the top quarks, discussed in the previous section, is also relevant. In most cases, cross section measurements using dilepton final states have been used in the determination of \alpS since they involve smaller extrapolations to the full phase space and have overall the best precision achieved so far. In the most precise measurements of \alpS from $\sigma_\ttbar$, summarized in the next paragraph, the dominant uncertainties turn out to be related to the QCD scale choice and the PDF.

\begin{figure*}[!htp]
\centering
\includegraphics[width=.8\textwidth]{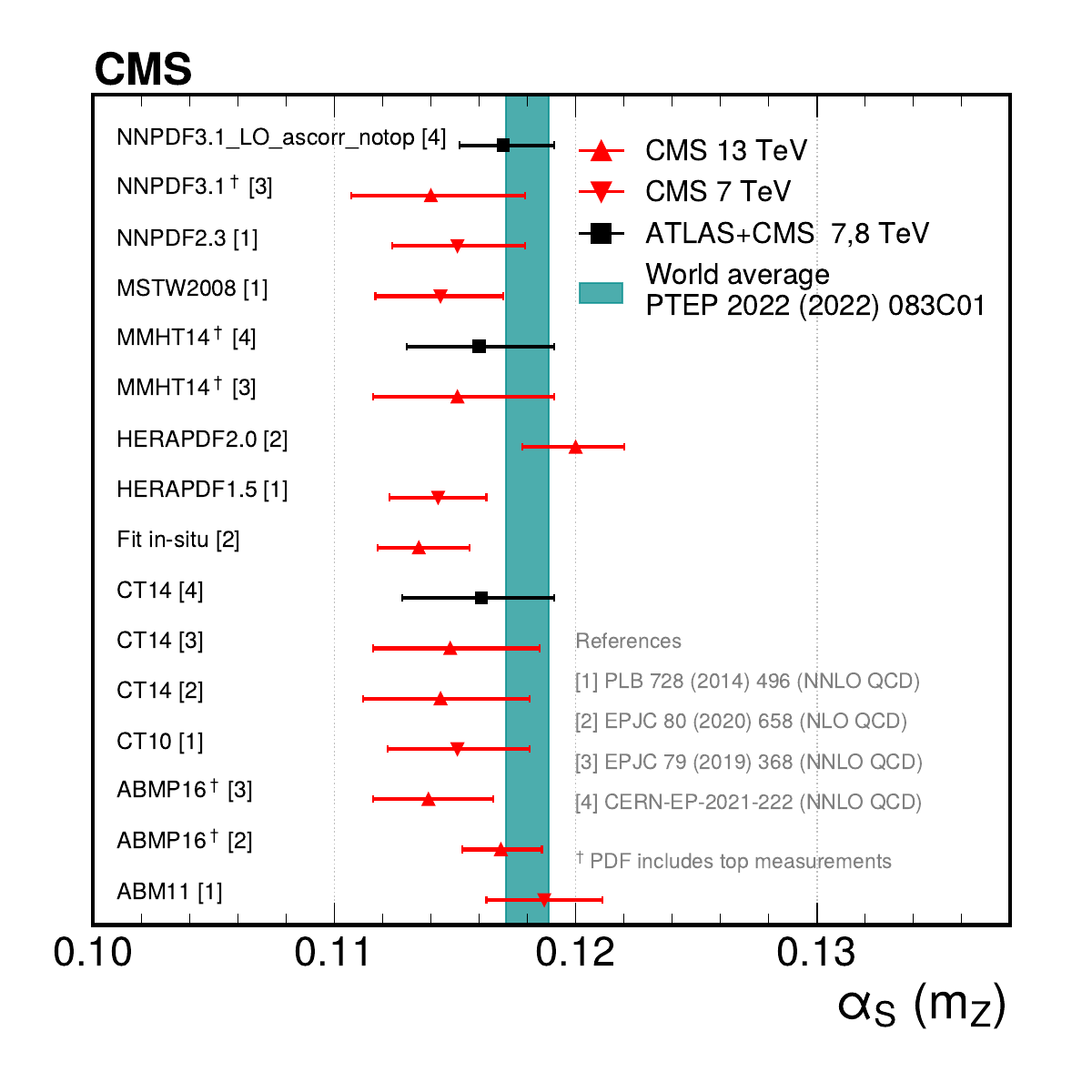}
\caption{
Summary of \alpS determinations from inclusive and differential top quark cross section measurements.
The error bars represent the total uncertainty of the measurements.
The results obtained with different PDF sets are compared with the world average~\cite{ParticleDataGroup:2022pth} and the reference \alpS in the corresponding PDF set. The 68\% confidence intervals are represented by the error bars and the coloured ranges.
The PDFs marked with a $\dagger$ include LHC top quark data in their fits.
}\label{fig:Figure_007_top}
\end{figure*}

The strong coupling \alpS is technically measured at the \ttbar scale, and one relies on the running of \alpS to translate the results to the $m_\cPZ$ scale. The measurement of $\alpS(m_Z)$ from $\sigma_\ttbar$ with the 7\TeV data has a total uncertainty of 2.4\%~\cite{CMS:2014rml} and with 13\TeV data a total uncertainty of 3.4\%~\cite{CMS:2016hbk}. 
Data sets at smaller centre-of-mass energy are more sensitive owing to the larger correlation between \alpS and $\sigma_\ttbar$. The most precise result to date comes from the combination of the CMS and ATLAS measurements at 7 and 8\TeV and achieves a total uncertainty of 1.8\%, as the main uncertainties in the individual measurements (\ttbar signal modelling and lepton identification and energy) are largely complementary between CMS and ATLAS~\cite{ATLAS:2022aof}.
The measurements are in agreement with the world average, as summarized in Fig.~\ref{fig:Figure_007_top}.

\begin{figure*}[!htp]
\centering
\includegraphics[width=.8\textwidth]{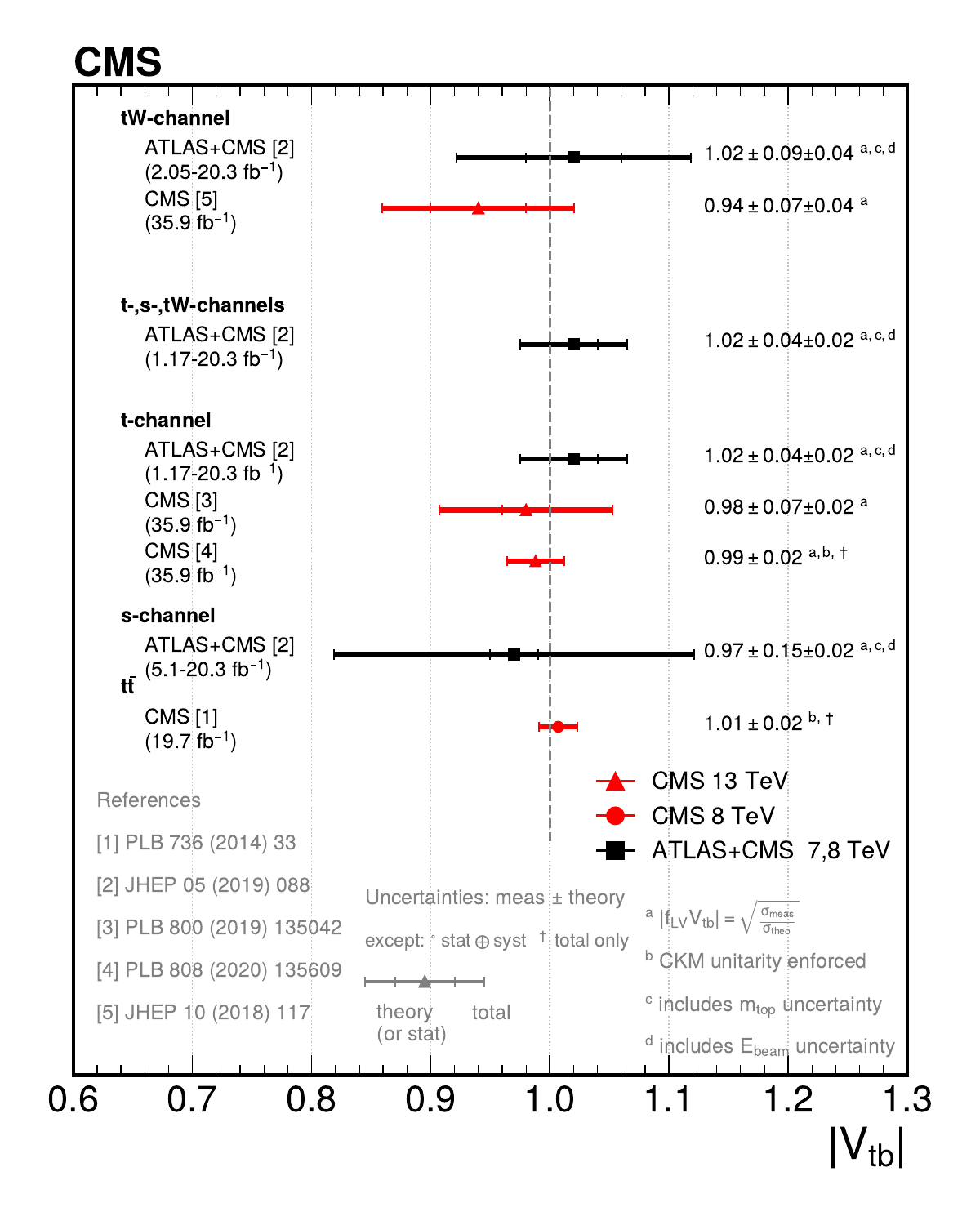}
\caption{
Summary of $\abs{V_{\PQt\PQb}}$ determinations from top quark events using different techniques.
The values measured and the corresponding references are given in the figures.
The error bars represent separately different uncertainties, as described in the legend.
In the LHC combinations, the reference theory cross section used in the $t$- and $s$-channel measurements is 
computed at NLO QCD accuracy~\cite{Kant:2014oha} with the PDF4LHC prescription for the 
PDF uncertainty using CT10nlo, MCSTW2008nlo, and NNPDF2.3nlo~\cite{Alekhin:2011sk},
whereas in the $\PQt\PW$ channel the theory reference is computed at 
NNLO+NNLL QCD accuracy~\cite{Kidonakis:2012rm} 
using the MSTW2008 NNLO PDF~\cite{Martin:2009iq}.
A line at $\abs{V_{\PQt\PQb}}=1$ is used as a common reference.
}\label{fig:Figure_008_top}
\end{figure*}

Another fundamental standard model parameter is $V_{\PQt\PQb}$. Since $V_{\PQt\PQb}$ is related to the EW coupling of the top quark, the measurement is carried out using the single top quark $t$-channel (Figs.~\ref{fig:Figure_000_top} (a) and (b)).
As noted in Section~\ref{subsec:singletop}, in $t$-channel processes, the $\PQt\PW\PQb$ vertices contribute in the production, giving rise to terms of the order $\abs{V_{\PQt\PQb}}^2$, and in the decay through $B(\PQt\to \PW\PQb)$ which, in the SM, equals $\abs{V_{\PQt\PQb}}^2/(\abs{V_{\PQt\PQb}}^2+\abs{V_{\PQt\PQs}}^2+\abs{V_{\PQt\PQd}}^2)$.
This results in an increased sensitivity with respect to the analysis of the top quark decays alone.
In practice, from the signal strength of the $t$-channel, \ie the ratio between observed and theoretical cross section, one extracts
$\abs{f_{LV}V_{\PQt\PQb}} =\sqrt{\frac{\sigma_\text{obs}}{\sigma_\text{theo}}}$, where in the SM the form factor $f_{LV} = 1$. For simplicity, we assume $f_{LV}=1$ in the following.
The CMS Collaboration has made several measurements at different \sqrts, the most precise result is achieved by combination of the results using this method on the 7\TeV and 8\TeV data: $\abs{V_{\PQt\PQb}} = 0.998 \pm 0.038\,(\text{exp})\pm 0.016\,(\text{theo})$~\cite{CMS:2014mgj}. The experimental uncertainty is dominated by the signal modelling and jet energy scale, as summarized in Section~\ref{subsec:singletop}. The combination with ATLAS results achieves a total uncertainty of 4.4\%~\cite{ATLAS:2019hhu}.
More recently, by performing a fit which includes the parameterized contributions of the different CKM matrix elements to the production and decay of single top quarks~\cite{CMS:2020vac}, a more precise measurement of $\abs{V_{\PQt\PQb}} = 0.988 \pm 0.024$ has been obtained. 
The uncertainty is limited by jet energy scale and PS scale uncertainties. The result is promising since it relaxes the SM-based assumptions used in the most precise measurement of $V_{\PQt\PQb}$ to date, based on the measurement $R_\PQb = B(\PQt \to \PW\PQb)/ B(\PQt \to \PW\PQq)$ in \ttbar events in which a limit of $V_{tb} > 0.975$ at 95\% confidence level was determined~\cite{CMS:2014mxl}. 
A direct measurement of $\abs{V_{\PQt\PQd}}^2+\abs{V_{\PQt\PQs}}^2=0.06 \pm 0.06$ is also made in~\cite{CMS:2020vac}. 
Figure~\ref{fig:Figure_008_top} summarizes the various measurements of $\abs{V_{\PQt\PQb}}$ performed by the CMS Collaboration. The combinations with ATLAS results are also included. All measurements are consistent with each other.

Finally, the top quark Yukawa coupling can be extracted from the $\ttbar\ttbar$ cross section as an almost independent measurement in which no other Higgs boson couplings intervene, given that at LO $\sigma_{\ttbar\ttbar} \propto \abs{y_\PQt / y_\PQt^\text{SM} }^4$, neglecting interference terms~\cite{Cao:2016wib}. There is, however, a contamination from the $\ttbar\PH$ background in the final sample. Its contribution (about 5\%) must also be taken into account for the final limit. The resulting upper limit is $\abs{y_\PQt / y_\PQt^\text{SM} } < 1.7$ at 95\% confidence level~\cite{CMS:2019rvj}. 
The value of $y_\PQt$ can also be extracted from the differential measurement of $m_{\ttbar}$ and $y_{\ttbar}$, attaining uncertainties of 20 to 40\% with 13\TeV data. This is possible owing to the contribution of diagrams where a virtual Higgs boson is exchanged between the \ttbar pair, giving sensitivity to $y_\PQt$ independently of other \PH{} couplings.
More details about the CMS measurements can be found in Refs.~\cite{CMS:2019art,CMS:2020djy}.

Additional constraints on the Higgs boson propagator can be obtained from \ttbar\ttbar production. The constraints are obtained after quantifying the modifications to $\sigma_{\ttbar\ttbar}$ with an effective field theory approach where additional contributions are added to the SM Lagrangian. These BSM contributions can be modelled with new operators proportional to $m_H/\Lambda^2$, where $\Lambda$ is the energy scale of new physics.
The so-called oblique $\hat{H}$-parameter falls in this category and modifies the Higgs boson propagator~\cite{Englert:2019zmt} inducing a parabolic variation of $\sigma_{\ttbar\ttbar}$ as a function of $\hat{H}$.
This dependency is used to obtain $\hat{H}<0.12$ at 95\% confidence level~\cite{CMS:2019jsc}. 
Even though it does not use the most precise $\sigma_{\ttbar\ttbar}$ measurement, this limit is better than that originally expected for the end of the HL-LHC~\cite{Englert:2019zmt}.

\subsection{Top quark summary}\label{subsec:topsummary}
The CMS experiment has observed or measured the majority of the expected production processes involving top quarks at the LHC.
The results are in good agreement with the SM predictions and, in some cases such as \ttbar\ttbar, are still dominated by statistical uncertainties. The inclusive cross section measurements have been used to extract or set independent constraints on fundamental parameters of the theory such as \alpS, $V_{\PQt\PQb}$, or $y_\PQt$. Furthermore, measurements of $\sigma_\ttbar$ and $t$-channel single top quark production provide important inputs for the determination of PDFs.

\begin{figure*}[!htp]
\centering
\includegraphics[width=.95\textwidth]{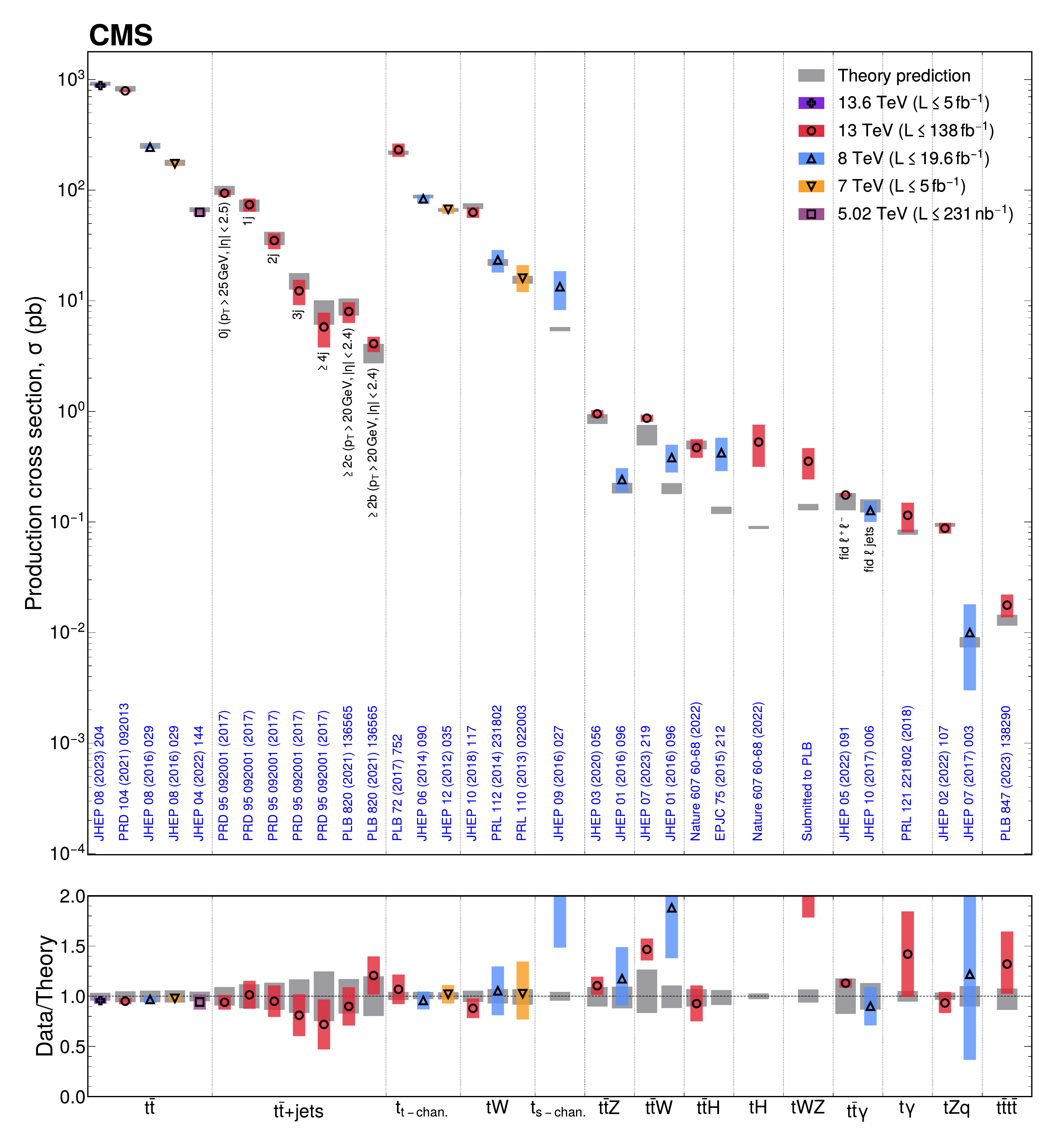}
\caption{
Summary of production cross section measurements involving top quarks.   Measurements performed at different LHC pp collision energies are marked by unique symbols and the coloured bands indicate the combined statistical and systematic uncertainty of the measurement.   Grey bands indicate the uncertainty of the corresponding SM theory predictions.   Shaded hashed bars indicate the excluded cross section region for a production process with the measured 95\% C.L. upper limit on the process indicated by the solid line of the same colour.
}\label{fig:Figure_016_top}
\end{figure*}

An overview of the main top quark cross section measurements at CMS is provided in Fig.~\ref{fig:Figure_016_top}.
Good overall agreement with the SM is observed.
Future measurements with increased statistics, improved experimental methodologies, and theoretical models are expected to contribute to finer tests of the SM along with the final goal to discover new physics.

\clearpage
\section{Measurements of Higgs boson production}\label{sec:higgs}
The discovery of the Higgs boson in 2012 by the CMS and ATLAS Collaborations~\cite{CMS:2012qbp,CMS:2013btf,ATLAS:2012yve} was a milestone in particle physics, leading to the experimental confirmation of the BEH EW symmetry-breaking mechanism and the first measurement of a fundamental parameter of the SM: the Higgs boson mass. The production of Higgs bosons at the LHC is dominated by gluon-gluon fusion (ggF) proceeding via a virtual top quark loop. Over the past decade, many studies have been performed in the form of precise measurements in order to characterize the nature of the Higgs boson. These started with the verification of the BEH mechanism through the observation of the direct Higgs boson decays to pairs of $\PW$ or $\PZ$ bosons~\cite{CMS:2012qbp,CMS:2013btf,CMS:2013zmy,CMS:2013fjq,CMS:2021ugl,CMS:2022uhn}, and the indirect decay to photon pairs through fermion and $\PW$ boson loops~\cite{CMS:2012qbp,CMS:2013btf,CMS:2014afl,CMS:2021kom}. An additional feature of this mechanism is that it grants masses to fermions through the Yukawa interaction, confirmed by the measurement of the Yukawa couplings of the Higgs boson to \PQb quarks and $\PGt$ leptons~\cite{CMS:2014suk,CMS:2017zyp,CMS:2018nsn,CMS:2022kdi} and tree-level $\PQt\PQt\PH$ production~\cite{CMS:2018uxb}. There is also evidence for other decay channels with smaller branching fractions, such as $\PH\to\PGm\PGm$~\cite{CMS:2020xwi} and $\PH\to \PZ\PGg$~\cite{CMS:2022ahq, ATLAS:2023yqk}. The Higgs boson mass is now known to the permille level ($125.38\pm0.14\GeV$~\cite{CMS:2020xrn}). The total Higgs boson width has been measured to be $\Gamma_{\PH}= 3.2^{+2.4}_{-1.7}\MeV$, in agreement with the SM expectation of $4.1\MeV$~\cite{CMS:2022ley}. The spin ($J$) and parity ($P$) were also found to be compatible with the SM prediction ($J^{P}=0^{+}$), already during Run 1~\cite{CMS:2012vby,CMS:2014nkk}. Further measurements have explored the Higgs boson spin and tensor structure~\cite{CMS:2022dbt,CMS:2022uox,CMS:2021sdq,CMS:2021nnc,CMS:2020cga} of its couplings to bosons and fermions~\cite{CMS:2022dwd}. Limits on the production cross section of pairs of Higgs bosons in a variety of final states and constraints on the Higgs boson
self-coupling have also been derived~\cite{CMS:2022hgz, CMS:2022cpr, CMS:2020tkr, CMS:2022omp, CMS:2022kdx, CMS:2022gjd, CMS:2022dwd}. A large number of direct and indirect searches for BSM physics connected to the Higgs sector have also probed the frontiers of the SM. With the current level of precision, the results are in agreement with the SM predictions.

The study of the cross section of the Higgs boson production at the LHC provides valuable insights into its
underlying production mechanisms and kinematics, a stringent test of the SM predictions. These cross section measurements are not only performed inclusively,
but also have been expanded to focus on obtaining a thorough description of the Higgs boson kinematics with the measurement of fiducial, differential, and 
double-differential cross sections.

 A detailed discussion of recent CMS measurements of Higgs boson production and decay is presented in Ref.~\cite{CMS:2022dwd}. In the next sections, the status of inclusive and differential cross sections of single Higgs boson production is reported, followed by a discussion of the current constraints on the production of pairs of Higgs bosons. These results are based on the \pp collision data collected by the CMS experiment during the Run~2 of the LHC, at a centre-of-mass energy of 13\TeV. When useful, comparisons to the corresponding 7 and 8\TeV results are made.  

\subsection{Inclusive cross sections for single Higgs boson production}\label{sec:HiggsProduction}
\begin{figure}[htbp]
    \centering{
      \includegraphics[width=1.0\textwidth]{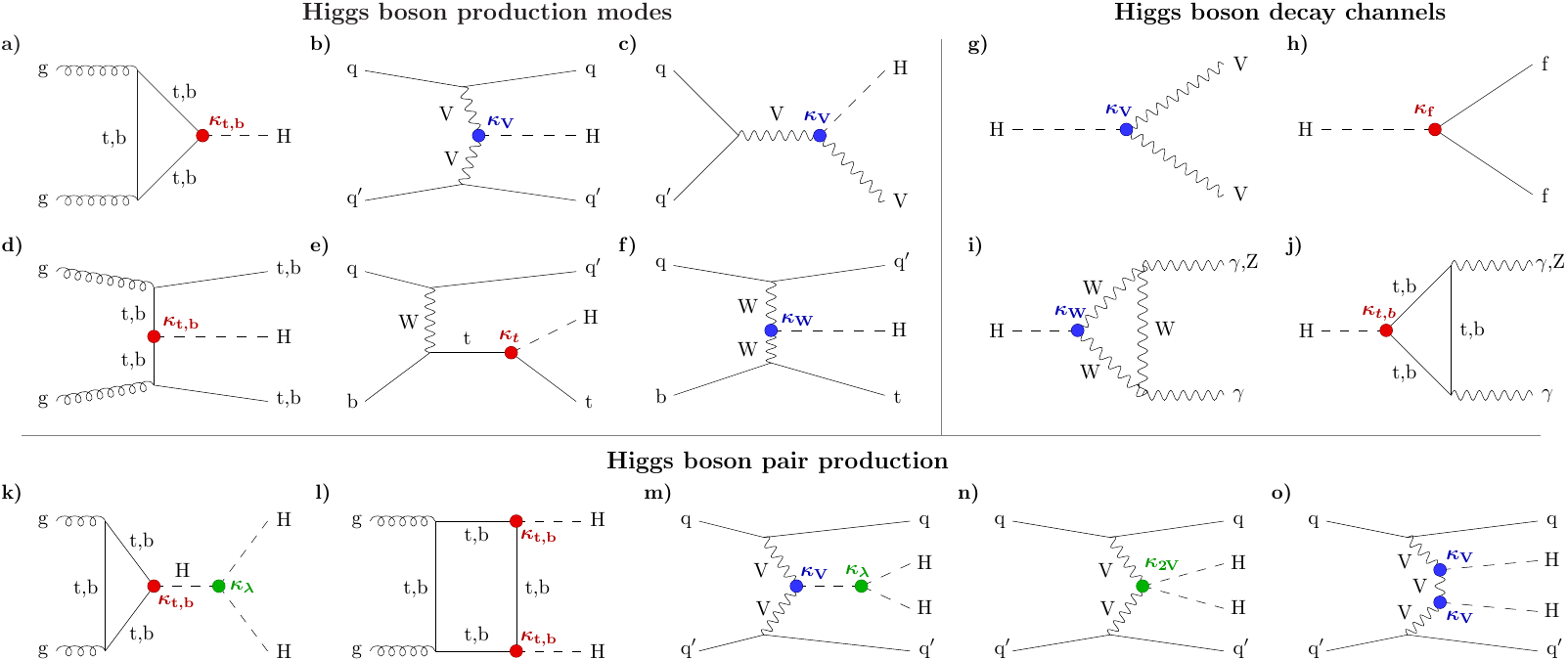}
    \caption{Higgs boson production in (a) gluon-gluon fusion (ggH), (b) vector boson fusion (VBF), (c) associated production with a $\PW$ or $\PZ$ ($\PV$) boson ($\PV\PH$), (d) associated production with a top or bottom quark pair ($\ttH$ or $\PQb\PQb\PH$), (e, f) associated production with a single top quark ($\tH$); with Higgs boson decays into (g) heavy vector boson pairs, (h) fermion-antifermion pairs, and (i, j) photon pairs or $\PZ\PGg$; Higgs boson pair production: (k, l) via gluon-gluon fusion, and (m, n, o) via vector boson fusion. The corresponding Higgs boson interactions are labelled with the coupling modifiers $\kappa$, and highlighted in different colours for Higgs-fermion interactions (red), Higgs-gauge-boson interactions (blue), and multiple Higgs boson interactions (green). The distinction between a particle and its antiparticle is dropped. Figure taken from Ref.~\cite{CMS:2022dwd}.}\label{fig:Feynman}}
\end{figure}

The main Feynman diagrams for the production and decay of the Higgs boson are shown in Fig.~\ref{fig:Feynman}. For a Higgs boson mass of 125.38\GeV, the total predicted
cross section for its production within the SM in \pp collisions at a centre-of-mass energy of 13\TeV is $55.4 \pm 2.6\unit{pb}$~\cite{LHCHiggsCrossSectionWorkingGroup:2016ypw}. In the dominant production mode, gluon-gluon fusion (ggF, Fig.~\ref{fig:Feynman} a), the Higgs boson is produced by the fusion of a pair of gluons, one from each of the colliding protons. With a cross section in the SM of $48.3 \pm 2.4\unit{pb}$, the ggF dominates over the other production modes. The next in relevance is vector boson fusion (VBF, Fig.~\ref{fig:Feynman} b), with a SM cross section of $3.77 \pm 0.80\unit{pb}$, where two quarks radiate virtual vector bosons ($\PW$ or $\PZ$), which then combine to produce a Higgs boson. As discussed in Section~\ref{subsec:EW}, a distinctive feature of VBF production is the presence of forward- and backward-scattered quarks that produce jets with large separation in rapidity. Other processes where the Higgs boson is produced in association with other SM particles have smaller cross sections. These include the associated production with vector bosons ($\PW\PH$ and $\PZ\PH$, Fig.~\ref{fig:Feynman} c, $1.359\pm0.028\unit{pb}$ and $0.877\pm0.036\unit{pb}$ in the SM, respectively), 
the associated production with pairs of top quarks ($\ttH$, Fig.~\ref{fig:Feynman} d, $0.503\pm0.028\unit{pb}$ in the SM) or single top quarks ($\tH$, Fig.~\ref{fig:Feynman} e and f, $0.092\pm0.008\unit{pb}$ in the SM), and the associated production with
bottom quarks ($\PQb\PQb\PH$, Fig.~\ref{fig:Feynman} d, $0.482\pm0.097\unit{pb}$ in the SM). The leading Higgs boson production modes (ggF, VBF, $\PV\PH$, $\tH$+$\ttH$) have been observed independently, with the measurements of the cross sections with precision at the 10--20\% level. 
The sensitivity of the LHC to the $\PQb\PQb\PH$ SM production is limited and this mode has not been extensively studied yet.

These production cross sections have been measured with dedicated analyses targeting the
decay to a pair of \PQb quarks (with the branching fraction in the SM~\cite{LHCHiggsCrossSectionWorkingGroup:2016ypw} of $\mathcal{B}(\PH\to \PQb\PAQb)=57.63\pm0.70\%$), $\PW$ bosons ($\mathcal{B}(\PH\to \WW)=22.00\pm0.33\%$), $\PGt$ leptons, ($\mathcal{B}(\PH\to\PGt\PGt)=6.21\pm0.09\%$), $\PZ$ bosons ($\mathcal{B} (\PH\to \ZZ)=2.71\pm0.04\%$),
and photons ($\mathcal{B}(\PH\to \PGg\PGg)=0.2\%$). 
These decay modes have all been measured~\cite{CMS:2018nsn,CMS:2022uhn,CMS:2022kdi,CMS:2021ugl,CMS:2021kom} and their branching fractions are in good agreement with the SM predictions in Ref.~\cite{LHCHiggsCrossSectionWorkingGroup:2016ypw}. Other decay modes, which are rarer or more challenging to observe experimentally, also have been studied. Examples include $\PH\to \PGm\PGm$~\cite{CMS:2020xwi}, $\PH\to \PQc\PAQc$~\cite{CMS:2022psv}, and $\PH\to \PZ\PGg$~\cite{CMS:2022ahq,ATLAS:2023yqk}.

Specific signatures associated with each decay mode and production mechanism are used to categorize the events. The reconstruction of Higgs boson candidates is based on the identification of pairs of photons, oppositely charged leptons ($\Pe$, $\PGm$, $\PGt$), or \PQb jets. Kinematic variables and their correlations are needed to discriminate against other SM processes with similar decay products that are produced more abundantly, such as the \PZ boson. Production modes other than ggF are distinguishable because of the additional objects in the event. The VBF events are characterized by the presence of two high-\pt jets with a large separation in rapidity, and $\PV\PH$ events by the identification of the $\PV$ decay through high-\pt charged leptons, jets, and/or \ptmiss. The $\ttH$ and $\tH$ signatures involve the decay of both the top quark and the Higgs boson, resulting in a rich variety of final states with the distinctive presence of multiple \PQb jets. Detailed descriptions of the event selection for each final state and production mode are presented in the references cited above. A brief summary was included in Ref.~\cite{CMS:2022dwd}.  

Measurements are compared with the predictions of the production and decay of the Higgs boson obtained using MC generators such as \POWHEG~2.0~\cite{Nason:2004rx,Frixione:2007vw,Alioli:2010xd},  
\MGvATNLO~\cite{Wiesemann:2014ioa,Alwall:2014hca}, 
\JHUGEN~\cite{Gao:2010qx, Bolognesi:2012mm, Anderson:2013afp, Gritsan:2016hjl,Gritsan:2020pib}, or \HJMINLO~\cite{Luisoni:2013kna,Hamilton:2012rf,Becker:2669113}. Events produced via the ggF mechanism are simulated at NLO with \POWHEG~2.0 and reweighted to match the predictions at NNLO in the strong coupling, including matching to the parton shower (\NNLOPS~\cite{NNLOPS1,NNLOPS2,NNLOPS3}) as a function of the $\ptH$ and of the number of jets in the event. 

The individual results featuring specific production and decay modes are combined for a global picture of Higgs boson production.  The overall statistical methodology used in this combination is described in Refs.~\cite{CMS:2018uag,ATLAS:2016neq}. 

As a first step towards quantifying the agreement of the observed Higgs boson signal with the expectation of the SM, the data from the various production modes and decay channels discussed are combined through a model that introduces signal strength parameters ($\mu$). These
parameters scale the observed signal yields relative to the SM predictions, while preserving the shape of the distributions. For specific initial and final states
$i\to f$, the corresponding signal strength is denoted as $\mu_{i}^{f}$. Signal strengths for individual production channels and decay modes are defined as functions of the cross section $\sigma_i$ and the branching fraction $\mathcal{B}_f$ as 
$\mu_{i} = \sigma_{i}/\sigma_{i}^\text{SM}$ and $\mu^{f} = \mathcal{B}_{f}/\mathcal{B}_{f}^\text{SM}$, respectively. A result in total agreement with the SM would be characterized by all signal strengths $\mu_{i}^{f}$ being equal to 1. 

\begin{figure}[htbp]
    \centering{
      \includegraphics[width=0.9\textwidth]{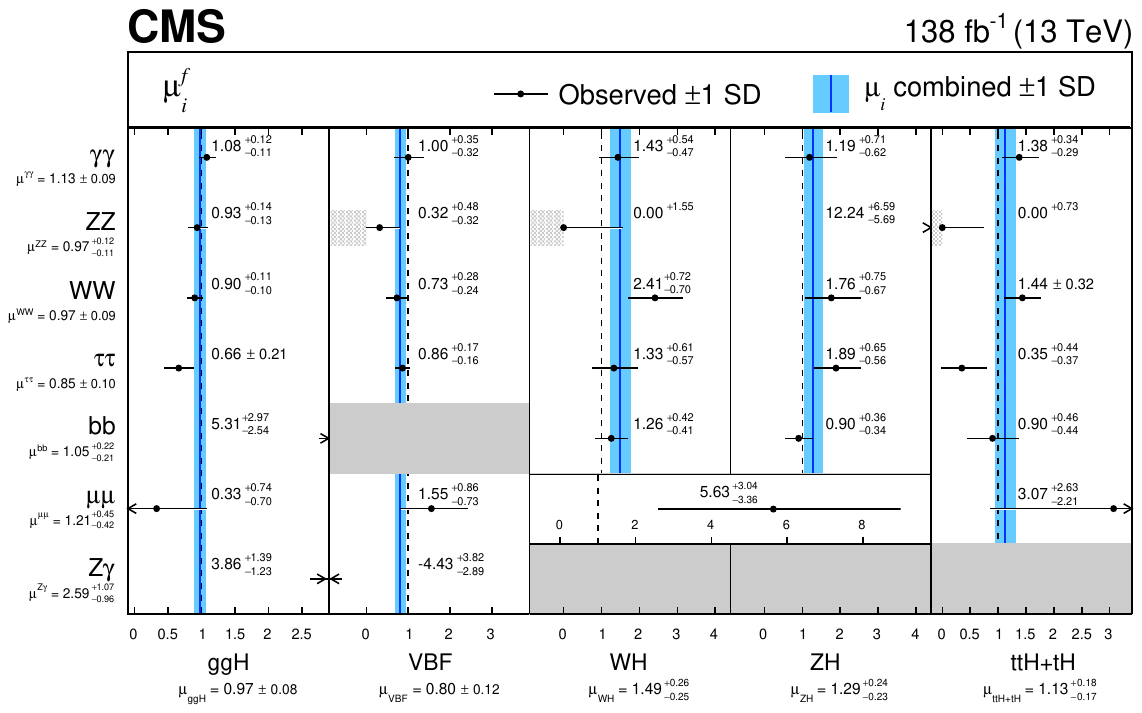}
    \caption{ 
Signal strength parameters per individual production mode and decay channel $\mu_{i}^{f}$, and combined per production mode $\mu_{i}$ and decay channel $\mu^{f}$. The SM expectation at 1~(dashed vertical lines) is shown as a reference. Light-grey shading indicates that $\mu$ is constrained to be positive. Dark-grey shading indicates the absence of a measurement. The measured value for each production cross section modifier obtained from the combination across the decay channels, $\mu_i$, is indicated by the blue vertical line. The corresponding 68\% \CL interval is indicated by the blue bands. The arrows indicate cases where the confidence intervals exceed the scale of the horizontal axis. Figure taken from Ref.~\cite{CMS:2022dwd}.\label{fig:SignalStrengths}}
    }
\end{figure}

We introduce different scenarios in which we incrementally increase the freedom allowed in the model, from considering a single signal strength parameter ($\mu$) that connects all the production and decay modes to allowing individual parameters ($\mu^f_i$) that modify individual channels independently. 
Figure~\ref{fig:SignalStrengths} summarizes the signal strength parameters per individual production mode and decay channel $\mu_i^{f}$, and combined per production mode $\mu_i$ and decay channel $\mu^{f}$. This result was obtained with the data collected at 13\TeV, corresponding to an integrated luminosity of 138\fbinv. Here the $\ttH$ and $\tH$ production modes are considered together. This global picture, including details of the production and decay of the Higgs boson, shows good agreement with the SM expectation. 

The measurements~\cite{CMS:2014fzn,CMS:2022dwd} of a common signal strength parameter are in excellent agreement with the SM:
\begin{linenomath}
\begin{displaymath}
\mu_\PH \textrm{(7 and 8\TeV)} = 1.00\pm0.008\thy \pm0.09\stat\pm0.07\syst,  
\end{displaymath}
\begin{displaymath}
\mu_\PH \textrm{(13\TeV)} =1.002\pm0.036\thy \pm0.029\stat\pm0.033\syst. 
\end{displaymath}
\end{linenomath}

For the 13\TeV measurement, the theoretical uncertainties in the signal prediction, as well as the experimental statistical and systematic uncertainties, are of comparable size. 

The theoretical uncertainties in the prediction of the production cross section impact the rate of events being produced and the kinematics of the Higgs boson and its decays. The signal strength parameters are relative measurements of the agreement with the SM, $\mu=\sigma/\sigma_\text{SM}$, and therefore fold in the total theoretical uncertainty in the prediction. In contrast, a cross section measurement is only subject to theoretical uncertainties in the acceptance, as discussed in Section~\ref{sec:intro}. As a result, production cross sections are less affected by theoretical uncertainties than the signal strength parameters. 

\begin{figure}[htbp]
    \centering{
      \includegraphics[width=0.9\textwidth]{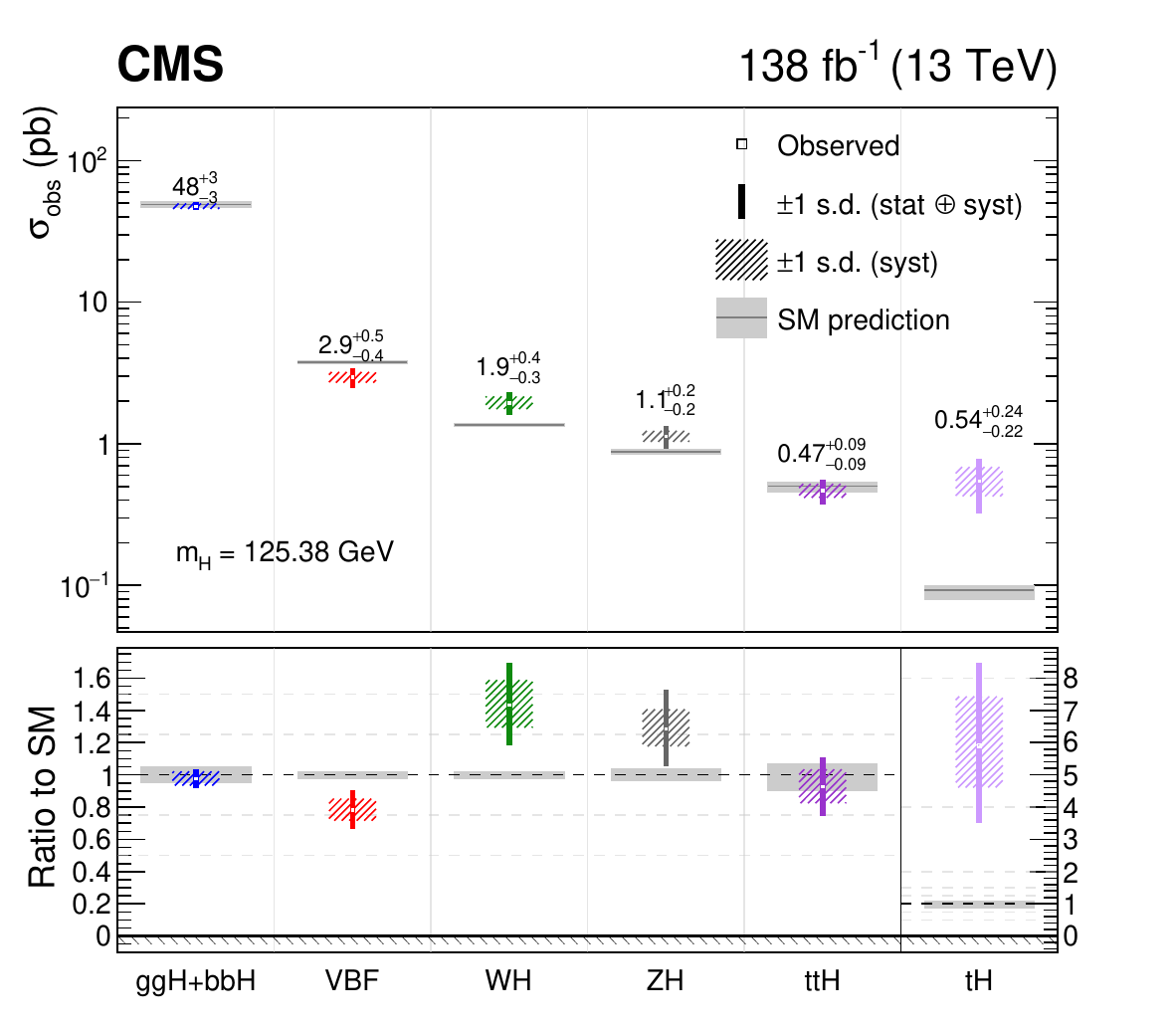}
    \caption{Measured cross sections for the main Higgs boson production modes. The best fit cross sections are plotted together with the respective 68\% confidence level intervals. The systematic components of the uncertainty in each parameter are shown by the coloured boxes. The grey boxes indicate the theoretical uncertainties in the SM predictions. The lower panel shows the ratio of the fitted values to the SM predictions. }\label{fig:HiggsXSecs}}
\end{figure}

The signal strength model with six $\mu_i$ parameters presented in Ref.~\cite{CMS:2022dwd} has been modified to obtain cross sections per production mode. The measurements of the inclusive cross sections at 13\TeV obtained deploying this method are represented graphically in Table~\ref{tab:HiggsXsecsInclusive} and Fig.~\ref{fig:HiggsXSecs}. Table~\ref{tab:HiggsXsecsInclusive} also lists the available measurements of inclusive cross section at 7 and 8\TeV. These have been derived by scaling the theoretical cross sections of Ref.~\cite{LHCHiggsCrossSectionWorkingGroup:2013rie} by the signal strengths published in Ref.~\cite{CMS:2014fzn}. The table also shows the corresponding SM prediction for the cross sections, taken from Ref.~\cite{LHCHiggsCrossSectionWorkingGroup:2013rie} and computed for $m_{\PH}=125\GeV$ for the 7 and 8 \TeV results, and from Ref.~\cite{CMS:2014fzn} and for $m_{\PH}=125.38\GeV$ for 13\TeV results, following the comparison done in the original publications. Overall, there is good agreement with the SM prediction in Ref.~\cite{LHCHiggsCrossSectionWorkingGroup:2016ypw}. 

\begin{table}[htbp]
  \centering
\topcaption{Measured inclusive cross sections for the main Higgs boson production modes. At 7 and 8\TeV, the measured cross sections are derived by scaling the theoretical cross sections of Ref.~\cite{LHCHiggsCrossSectionWorkingGroup:2013rie} by the signal strengths published in Ref.~\cite{CMS:2014fzn}. At $\sqrt{s} =13\TeV$, the cross sections are obtained from a global fit, as described in the text. The results are in good agreement with the predictions from Ref.~\cite{LHCHiggsCrossSectionWorkingGroup:2013rie} and Ref.~\cite{LHCHiggsCrossSectionWorkingGroup:2016ypw}, respectively.\label{tab:HiggsXsecsInclusive}}
\renewcommand{\arraystretch}{1.2}
  \begin{tabular}{ccll}
   $\sqrt{s}$ &  Production mode &  $\sigma(\PH)$ (pb) & $\sigma^\text{SM}(\PH)$ (pb) \\ 
  \hline
7\TeV & ggF  & $15.6^{+5.6}_{-5.0}$  & $15.13\pm 1.58$\\ 

       & VBF  & $2.2^{+1.2}_{-1.1}$  & $1.222\pm 0.038$\\ [\cmsTabSkip] 
8\TeV & ggF  & $15.2^{+3.7}_{-3.3}$  & $19.27\pm 2.01$ \\ 

       & VBF  & $1.61^{+0.62}_{-0.57}$  & $1.578\pm 0.035$\\

       & V$\PH$  & $1.08^{+0.46}_{-0.44}$ &  $1.120\pm 0.034$\\ 

       & $\ttH$ & $0.42^{+0.16}_{-0.13}$  &  $0.1293\pm 0.0078$ \\ [\cmsTabSkip] 
13\TeV  &  ggF+bbH  & $47.6^{+1.8}_{-1.8}\stat^{+2.3}_{-2.0}\syst$   & $48.80\pm2.46$ \\

  &  VBF & $2.94^{+0.37}_{-0.36}\stat^{+0.27}_{-0.25}\syst$ & $3.77\pm0.81$  \\
  
  &  $\PW\PH$  & $1.95^{+0.28}_{-0.28}\stat^{+0.21}_{-0.19}\syst$ & $1.359\pm 0.028$  \\
  
  &  $\PZ\PH$  & $1.13^{+0.18}_{-0.18}\stat ^{+0.11}_{-0.10}\syst$  & $0.877\pm 0.036$ \\
  
  &  $\ttH$ & $0.467^{+0.074}_{-0.072}\stat ^{0.054}_{-0.052}\syst$ & $0.503\pm 0.035$ \\

  &  $\tH$  & $0.54^{+0.19}_{-0.18}\stat ^{+0.14}_{-0.12}\syst$ & $0.092 \pm 0.008$ \\ 
  \end{tabular}
\end{table}

\begin{figure}[htbp]
    \centering{
      \includegraphics[width=0.9\textwidth]{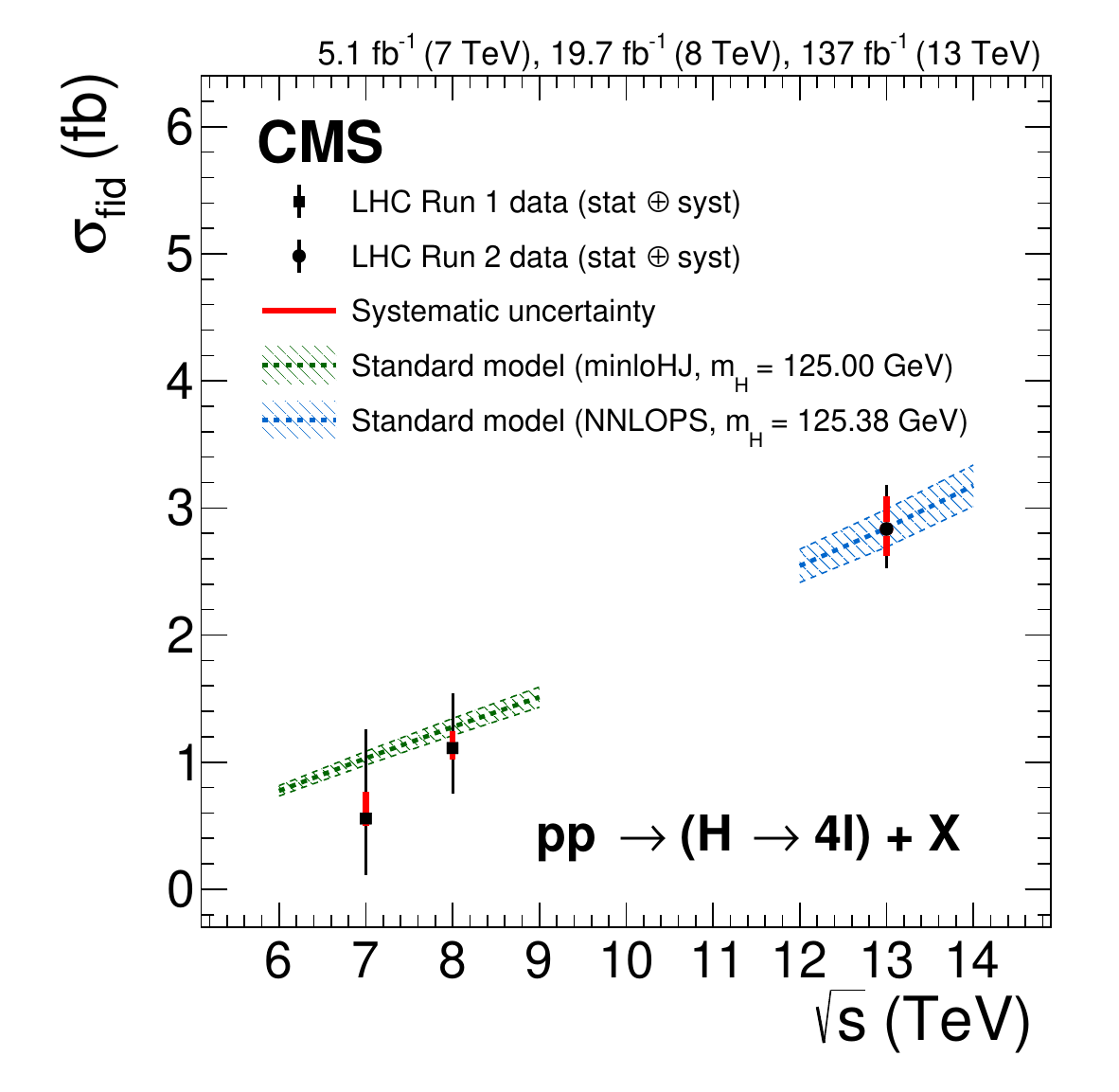}
    \caption{The measured inclusive fiducial cross section for $\PH\to \ZZ\to 4\ell$ as a function of $\sqrt{s}$. The acceptance is calculated using \textsc{HRes}~\cite{Grazzini:2013mca,deFlorian:2012mx} at 7 and 8\TeV, and \POWHEG at 13\TeV, and the total gluon fusion cross section and uncertainty are taken from Ref.~\cite{Anastasiou:2016cez}. The SM predictions and measurements are calculated at $m_{\PH}=125.0\GeV$ for $\sqrt{s}=6$--9\TeV, and at $m_{\PH}= 125.38\GeV$ for 12--14\TeV. Figure taken from Ref.~\cite{CMS:2021ugl}.}\label{fig:HZZEvolution}}
\end{figure}

In addition to this global view of Higgs boson production, fiducial production cross sections for specific decay modes have also been measured individually~\cite{CMS:2015qgt,CMS:2015zpx,CMS:2022wpo, CMS:2023gjz,CMS:2021gxc,CMS:2020dvg}. 
These fiducial cross sections correspond to well-defined regions of the phase space, and avoid the extrapolation to the full phase space necessary for the determination of total inclusive cross sections. Minimizing the differences in selection between the reconstructed- and particle-level objects facilitates a more model-independent comparison to theoretical calculations. Table~\ref{tab:tableFiducial} summarizes the available measurements at a centre-of-mass energy of 13\TeV, with an integrated luminosity of $138\invfb$. The table also lists the variables and the selection criteria that delineate the fiducial phase space.  The variables used to define it follow closely the event selection criteria of each analysis. These variables include the \pt and (pseudo)rapidities of the reconstructed Higgs boson and its visible decay products, the reconstructed invariant and transverse masses of the system, or the jet multiplicity. They are calculated at the MC generator level after parton showering and hadronization. The lepton momentum includes the momenta of photons radiated within a cone of $\Delta R<0.1 $ in the $\PW\PW$ and $\PGt\PGt$ analyses or $\Delta R <0.3$ in the $\ZZ$ case. Lepton or photon isolation ($\mathcal{I}^{\PGg}_\text{gen}$, $\mathcal{I}^{\ell}_\text{gen}$) is defined at the generator level as the sum of the energy of all stable hadrons produced in a cone of radius $\Delta R=0.3$ around the object. Additional details on the definition of the fiducial cross section are presented in the original references. Overall, there is remarkable agreement with the SM prediction. Figure~\ref{fig:HZZEvolution} shows the evolution of the fiducial cross section for $\PH\to \ZZ\to 4\ell$ from 7 and 8\TeV~\cite{CMS:2015zpx} to 13\TeV~\cite{CMS:2023gjz}.

\begin{table}[!htbp]
  \centering{
\topcaption{Measurements of the fiducial cross sections of  Higgs boson production in various decay modes  published by CMS using \pp data at a centre-of-mass energy of 13\TeV and an integrated luminosity of $138\invfb$. The reference Higgs boson mass is $125.38\GeV$. Isolation~($\mathcal{I}$) represents the sum of scalar $\pt$ of all stable particles within $\Delta R=0.3$ of the lepton or photon. Additional details on the fiducial phase space variables and on the calculation of the reference SM cross section are presented in the original references.\label{tab:tableFiducial}}
\renewcommand{\arraystretch}{1.1}
  \begin{tabular}{c c c c }
    Decay mode  &  Fiducial phase space  & $\sigma_\text{fid}(\PH)$ (fb) & $\sigma_\text{fid}^\text{SM}(\PH)$ (fb) \\
    \hline
     $\PH\to \PGg\PGg$ & $p_\text{T}^{\PGg_{1}}/m_{\PGg\PGg}>1/3$, & $73.4^{+5.4}_{-5.3}\stat^{+2.4}_{-2.2}\syst$ & $75.4\pm4.1$ \\
     {\cite{CMS:2022wpo}}  &  $p_\text{T}^{\PGg_{2}}/m_{\PGg\PGg}>1/4$, & & \\ 
     & $\mathcal{I}^\PGg_{\text{gen}}<10\GeV$, $\abs{\eta^{\PGg}}<2.5$ & & \\ [\cmsTabSkip] 
    
    $\PH\to \ZZ\to 4\ell$ & $p_\text{T}^{\text{lead}}>20\GeV$,  &  $2.73\pm0.22\stat\pm0.15\syst$ & $2.86\pm0.15$  \\
    {\cite{CMS:2023gjz}} & $p_\text{T}^{\text{sublead}}>10\GeV$, & & \\
    & $p_\text{T}^{\ell}>5(7)\GeV$ for $\PGm$~($\Pe$), & &  \\ 
    & $\abs{\eta^{\ell}}<2.4\, (2.5)$ for $\mu$~($\Pe$), & &  \\ 
    & $\mathcal{I}^{\ell}_{\text{gen}}<0.35 \pt$, & &  \\
    & $40<m_{\PZ1}<120\GeV$, & &  \\
    & $12<m_{\PZ2}<120\GeV$, & & \\
    & $\Delta R(\ell_i,\ell_j)>0.02$ for $i\neq j$,   & &  \\
    & $m_{\ell^{+}\ell'^{-}}>4\GeV$,& &  \\
    & $105<m_{4\ell}<160\GeV$   & & \\ [\cmsTabSkip] 
     
     $\PH\to \PGt\PGt$ & $\PGm\tauh$ ($\Pe\tauh$): $p_\text{T}^{\ell}>20\,(25)\GeV$, & $426\pm102$ &  $408\pm27$ \\
    {\cite{CMS:2021gxc}} & $p_\text{T,vis}^{\tauh}>30\GeV$, & &  \\
    & $\abs{\eta^{\ell}}<2.1$, $\abs{\eta^{\tauh}}<2.3$, & &  \\  
    & $m_\text{T}(\ell,\ptmiss)<50\GeV$, & &  \\
    & $\tauh\tauh$:  $p_{\text{T,vis}}^{\tauh}>40\GeV$, & &  \\
    & $\abs{\eta^{\tauh}}<2.1$, $n_{j\text{\,30\GeV}}\geq 1$   & &  \\
    & $\Pe\PGm$: $p_\text{T}^\text{lead}>24\GeV$,  & &  \\
    & $p_\text{T}^{\text{sublead}}>15\GeV$, $\abs{\eta^{\ell}}<2.4$, & &  \\
    &  $m_\text{T}(\Pe\PGm,\ptvecmiss)<60\GeV$ &  &  \\ [\cmsTabSkip] 
    
     $\PH\to \WW$~& $\Pe\PGm$, $p_\text{T}^\text{lead}>25\GeV$,  & $86.5\pm9.5$ & $82.5\pm4.2$ \\ 
     {\cite{CMS:2020dvg}}  &  $p_\text{T}^{\text{sublead}}>13\GeV$, & &  \\
     & $\abs{\eta_{\ell}}<2.5$, $m_{\ell\ell}>12\GeV$, & &  \\
     & $p_{\text{T}}^{\ell\ell}>30\GeV$, $m_\text{T}^{\ell2}>30\GeV$, & &  \\
     &  $m_{\text{T}}^{\PH}>60\GeV$ &  &  \\
  \end{tabular}
  }
\end{table}

\subsection{Differential cross sections for single Higgs boson production}\label{subsec:Higgsdiff}
The characterization of Higgs boson production cannot rely solely on measuring inclusive production cross sections. For a more complete picture of the nature of the boson,
a detailed mapping is needed of its production as a function of different observables, such as its transverse momentum, $\ptH$. The measurement of differential production cross sections with respect to key kinematic variables, compared with the corresponding theoretical expectations, provides a useful probe of the effects from higher-order corrections in perturbation theory or any deviation from the SM expectations. 

The CMS experiment has measured Higgs boson differential production cross sections in the principal decay modes: $\PH\to \PGg\PGg$~\cite{CMS:2015qgt,CMS:2018ctp,CMS:2022wpo}, $\PH\to \ZZ\to 4\ell$~\cite{CMS:2015zpx,CMS:2017dib,CMS:2023gjz},   $\PH\to \PGt\PGt$~\cite{CMS:2021gxc}, $\PH\to \WW$~\cite{CMS:2016ipg,CMS:2020dvg}, Lorentz-boosted $\PH\to \PQb\PAQb$~\cite{CMS:2017bcq,CMS:2020zge}. These measurements are complementary, as they probe different aspects of the Higgs boson production. As previously discussed, in the SM, the branching fraction for the Higgs boson decaying to a pair of photons or to four leptons is remarkably small. Nevertheless, because of the high precision of the invariant mass reconstruction and the fully reconstructed final state, the $\PH\to\PGg\PGg$ and $\PH\to \ZZ\to 4\ell$ decay channels provide the most comprehensive measurements of the Higgs boson differential production cross sections. These analyses probe a large number of observables, related to the measurement of the diphoton or four-lepton system, but also to the accompanying jets and event topology. These include the kinematics of the Higgs boson (\eg $\ptH$ or $\abs{y_{\PH}}$) and the accompanying jets (\eg $m_{jj}$ or the rapidity-weighted jet veto, $\mathcal{T}^\text{max}$, which provides a complementary way to divide the phase space into exclusive jet bins, allowing for an accurate comparison to theory predictions~\cite{Gangal:2014qda}). In the case of the four-lepton analysis, the measurements can also be performed as a function of matrix element discriminators targeting anomalous couplings ($D^{\text{dec}}$). Double-differential cross sections are also possible to measure for a selected number of variables.  

The larger branching fractions of the $\PH\to \PQb\PAQb$, $\PH\to \WW$, and $\PH\to \PGt\PGt$ decay modes allow studies in the areas of the phase space with smaller production cross sections. This is the case for high jet
multiplicities ($n_j$) and large Lorentz boosts of the Higgs boson. There is considerable interest in the measurement of Higgs bosons produced with very high \pt in the more dominant decay modes  (particularly in $\PH\to \PQb\PAQb$) since they yield significantly  better sensitivity than in $\PH\to \PGg\PGg$  and $\PH\to
\ZZ\to 4\ell$ final states. At the highest \pt, this measurement can resolve loop-induced contributions to the ggH process from BSM particles, which would be described by an effective ggH vertex at low \pt. Advances in the identification of large-radius jets~\cite{CMS:2020poo} resulting from massive colour-singlet particles with high
\pt and decaying to $\PQb\PAQb$ pairs have been fundamental for these measurements. 

These measurements of the differential cross sections in the different decay modes can be combined,  as shown in Ref.~\cite{CMS:2018gwt}, which incorporated the first measurements at 13\TeV, with $36\fbinv$ of $\PH\to \PGg\PGg$,  $\PH\to ZZ $  and $\PH\to \PQb\PAQb$ into a global measurement of the differential cross section as a function of observables, such as $\ptH$ or $n_j$. The $\PH\to \ZZ$, $\PH\to \PGg\PGg$, $\PH\to \WW$, and $\PH\to \PGt\PGt$ measurements have been updated using the full data sample collected during the second data-taking period of the LHC, 138\fbinv, and are summarized in  Table~\ref{tab:tableHiggsDiff}. Additional details of the observables targeted in each case are presented in the original references~\cite{CMS:2022wpo,CMS:2023gjz,CMS:2021gxc,CMS:2020dvg, CMS:2020zge}. Overall, they are in agreement with the SM predictions within uncertainties. 

\begin{figure}[!htbp]
    \centering{
      \includegraphics[width=0.62\textwidth]{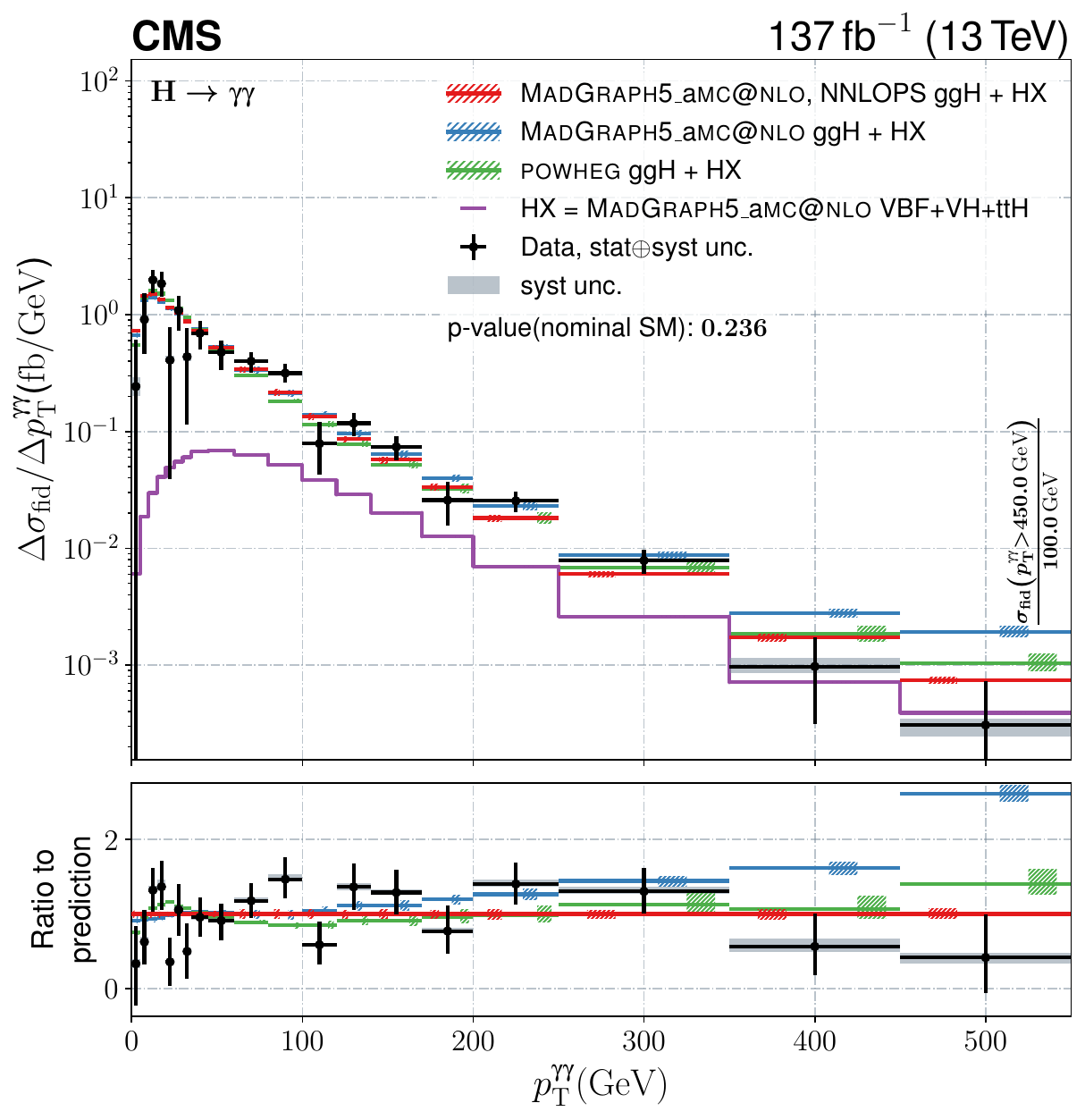}  \\
      \includegraphics[width=0.75\textwidth]{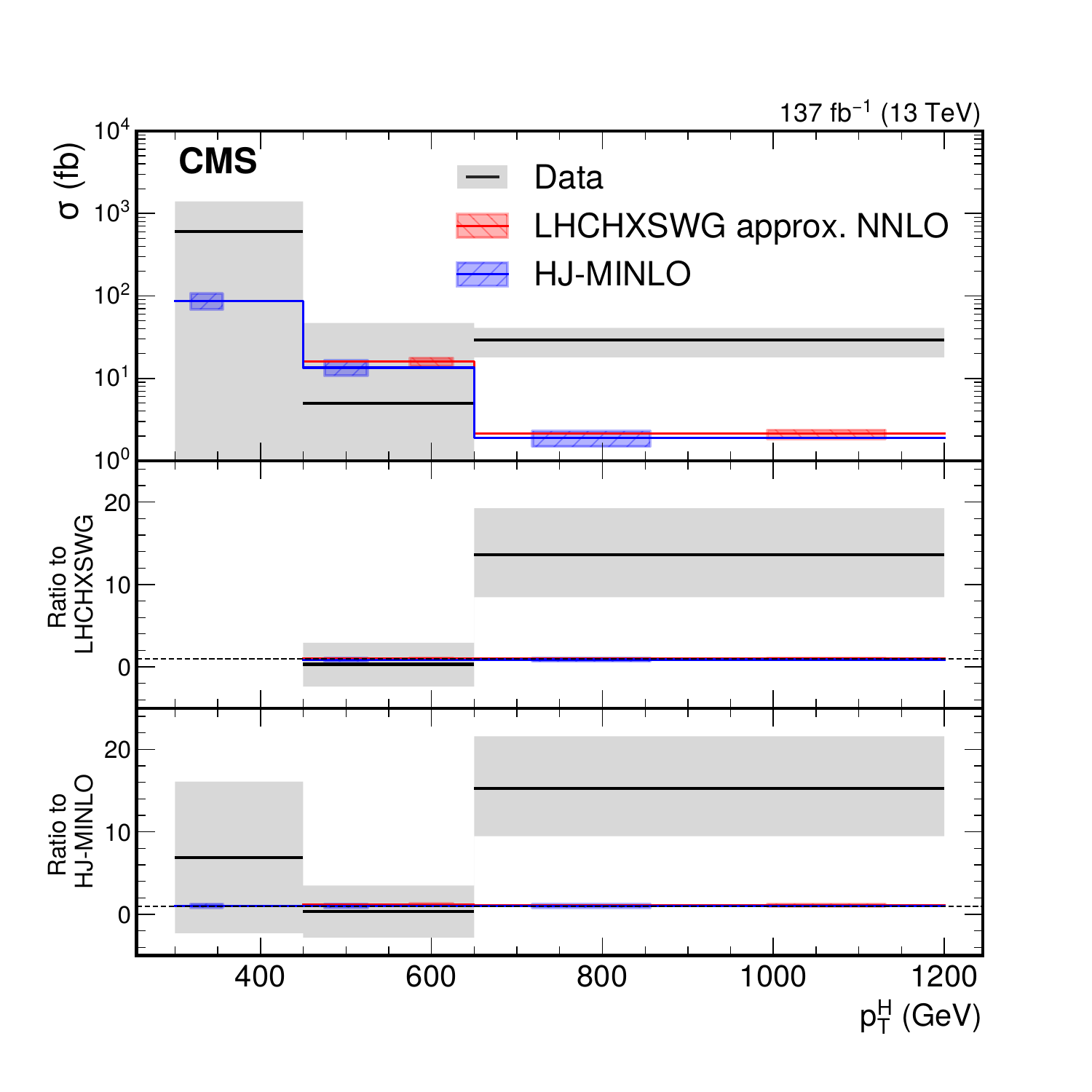}     
    \caption{ Differential fiducial cross sections for Higgs boson production in the $\PH\to\PGg\PGg$~\cite{CMS:2022wpo} (upper) and $\PH\to \PQb\PAQb$~\cite{CMS:2020zge} (lower) decay channels as functions of the transverse momentum of the Higgs boson \ptH. Figure compiled from Refs.~\cite{CMS:2022wpo,CMS:2020zge}.\label{fig:HiggsDiff}}
    }
\end{figure}

\begin{figure}[!htbp]
    \centering{
      \includegraphics[width=0.7\textwidth]{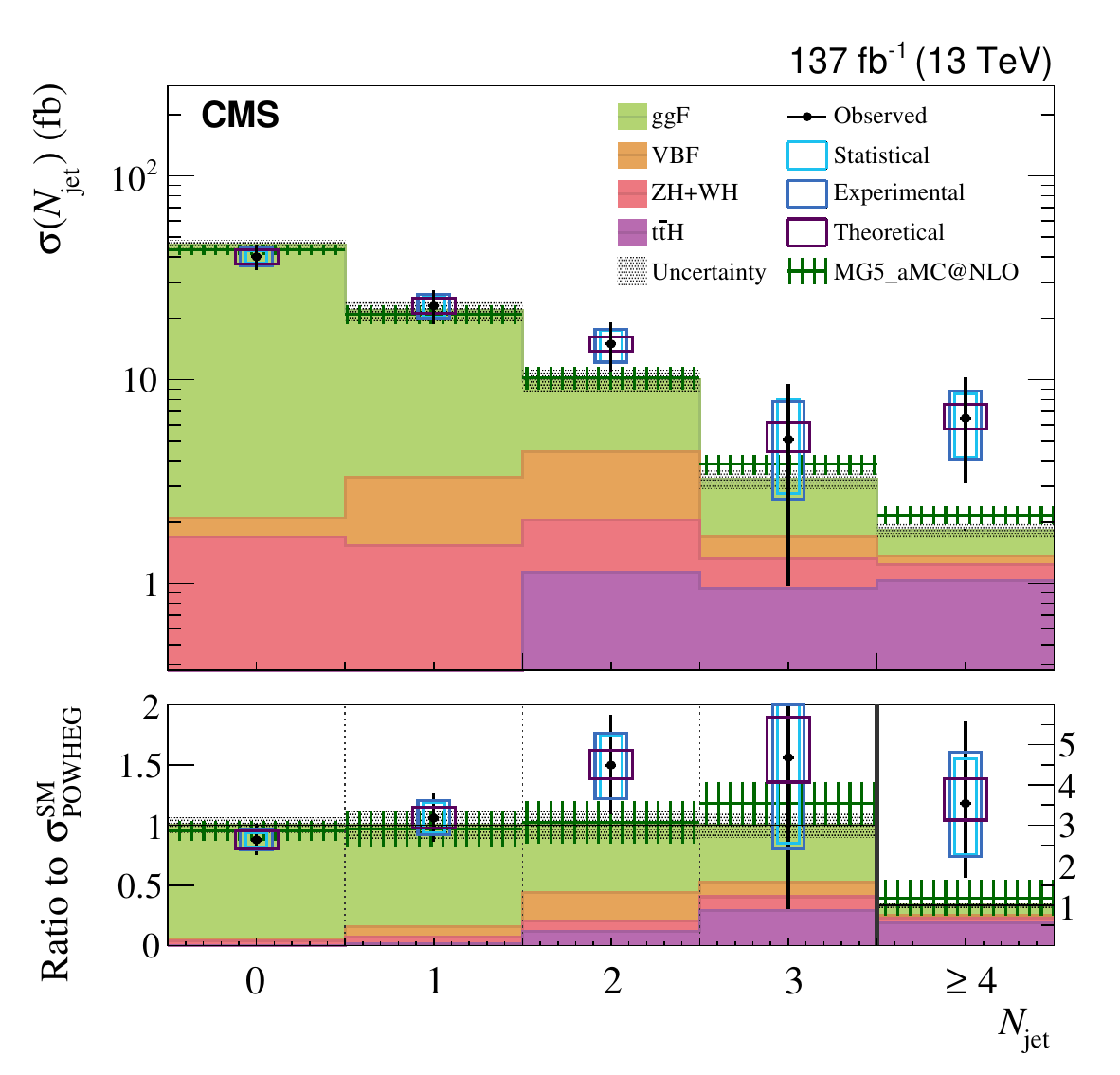}   \\
      \includegraphics[width=0.7\textwidth]{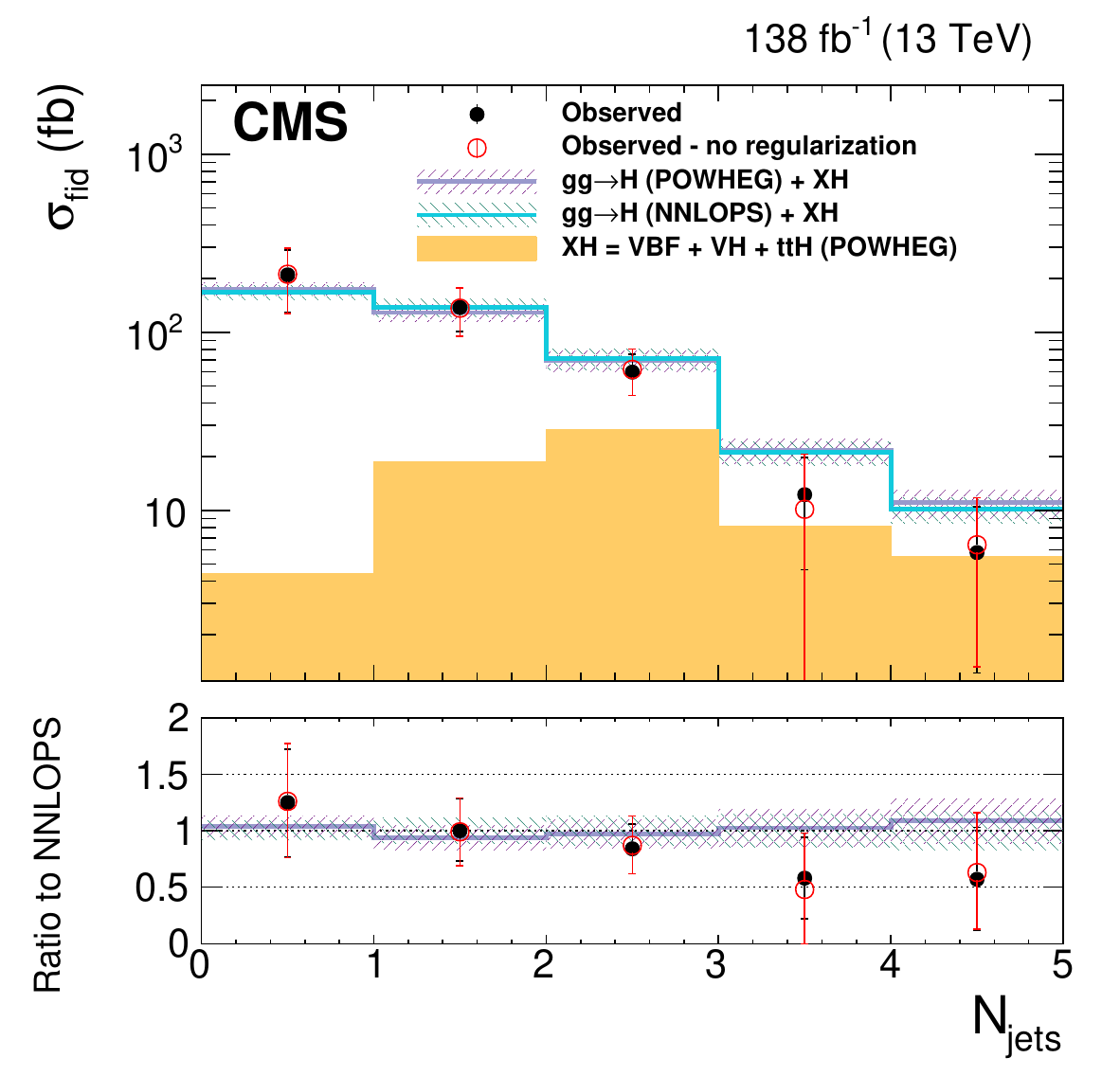}   
    \caption{ Differential fiducial cross sections for Higgs boson production  as functions of the number of jets in the event, for the  $\PH\to \WW$~\cite{CMS:2020dvg} (upper) and $\PH\to \PGt\PGt$~\cite{CMS:2021gxc} (lower) decay modes.
    Figure compiled from Refs.~\cite{CMS:2022wpo,CMS:2020zge}.\label{fig:HiggsDiffNJets}}
    }
\end{figure}

\begin{figure}[!htbp]
    \centering{
      \includegraphics[width=0.9\textwidth]{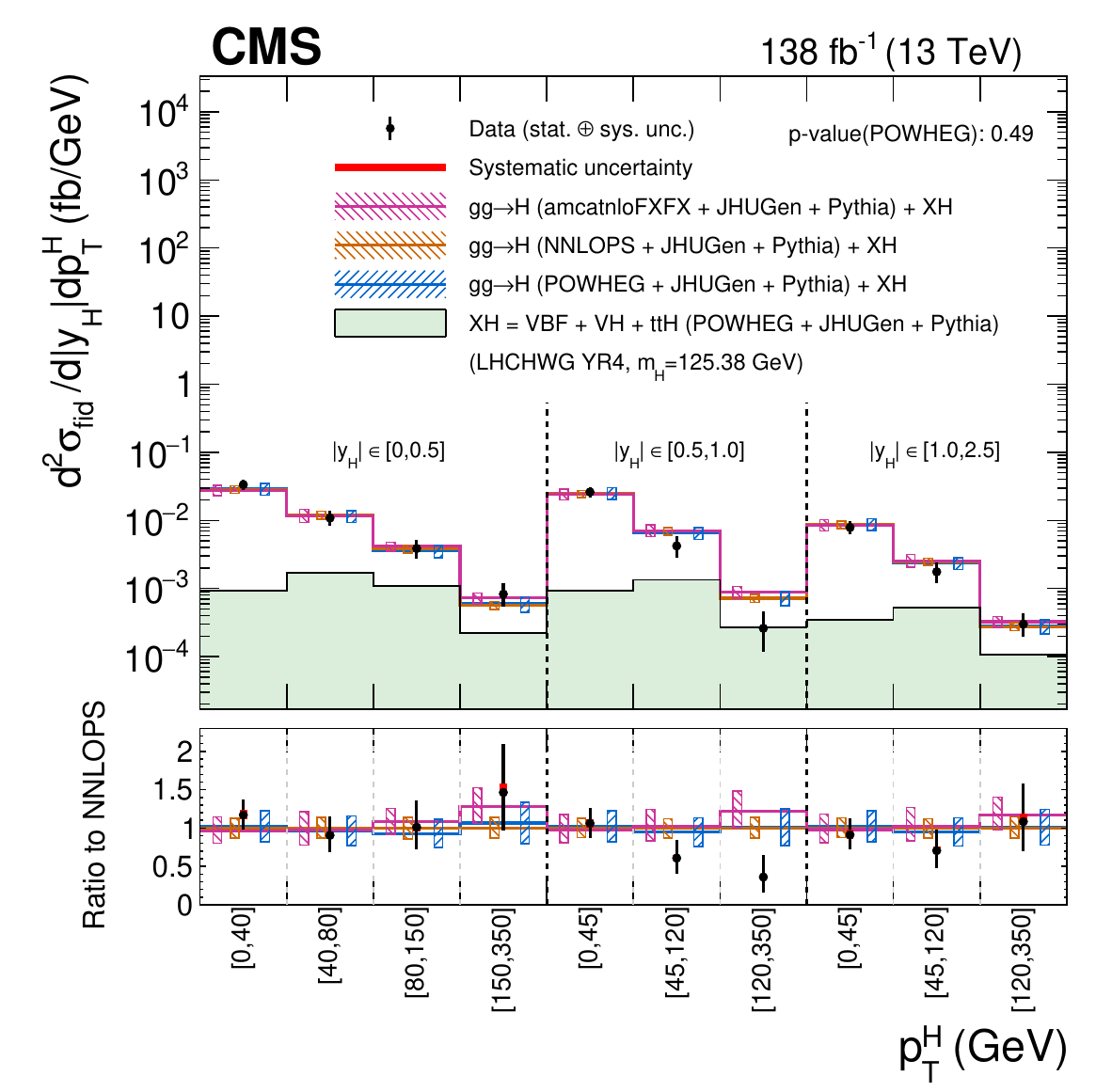}  
    \caption{Double-differential cross sections for Higgs boson production in the $\PH\to \ZZ\to 4\ell$ decay channel. The cross section is measured in bins of the rapidity of the Higgs boson $\abs{y_{\PH}}$, as a function of the Higgs boson transverse momentum $\ptH$. Figure taken from Ref.~\cite{CMS:2023gjz}.\label{fig:HiggsDiffDouble}}
    }
\end{figure}

Figures~\ref{fig:HiggsDiff} and~\ref{fig:HiggsDiffNJets} show the fiducial differential distributions as functions of the \pt of the Higgs boson and the number of jets in the event for the various decay modes, respectively.  Figure~\ref{fig:HiggsDiffDouble} is an example of a double-differential cross section; it shows the differential cross sections in bins of the absolute rapidity of the Higgs boson $\abs{y_{\PH}}$ as functions of the Higgs boson transverse momentum $\ptH$ in the $\PH\to \ZZ\to 4\ell$ decay channel. The measurements are compared with the predictions of the production and decay of the Higgs boson obtained using MC generators mentioned in the previous section. 

An alternative approach to characterize the production of the Higgs boson is the  ``simplified template cross sections'', STXS~\cite{Berger:2019wnu}. In this approach, fiducial cross sections are measured per production mode and in specific regions of phase space~(``bins''), 
defined in terms of specific kinematic variables ($\ptH$, $m_{jj}$, $p_\text{T}^{\PHjj}$, $p_\text{T}^\text{V}$). Their purpose is to reduce the theoretical uncertainties, that are directly folded into the measurements, as much as possible, while at the same time allowing for the combination of the measurements of different decay channels. The STXS approach offers convenient benchmarks for comparing theoretical predictions with experimental data to probe and understand the properties and interactions of the Higgs boson, while providing a well-defined platform to test for BSM deviations in kinematic distributions.

\begin{figure}[!htbp]
    \centering{
      \includegraphics[width=1.\textwidth, trim={1cm 0 2cm 0}]{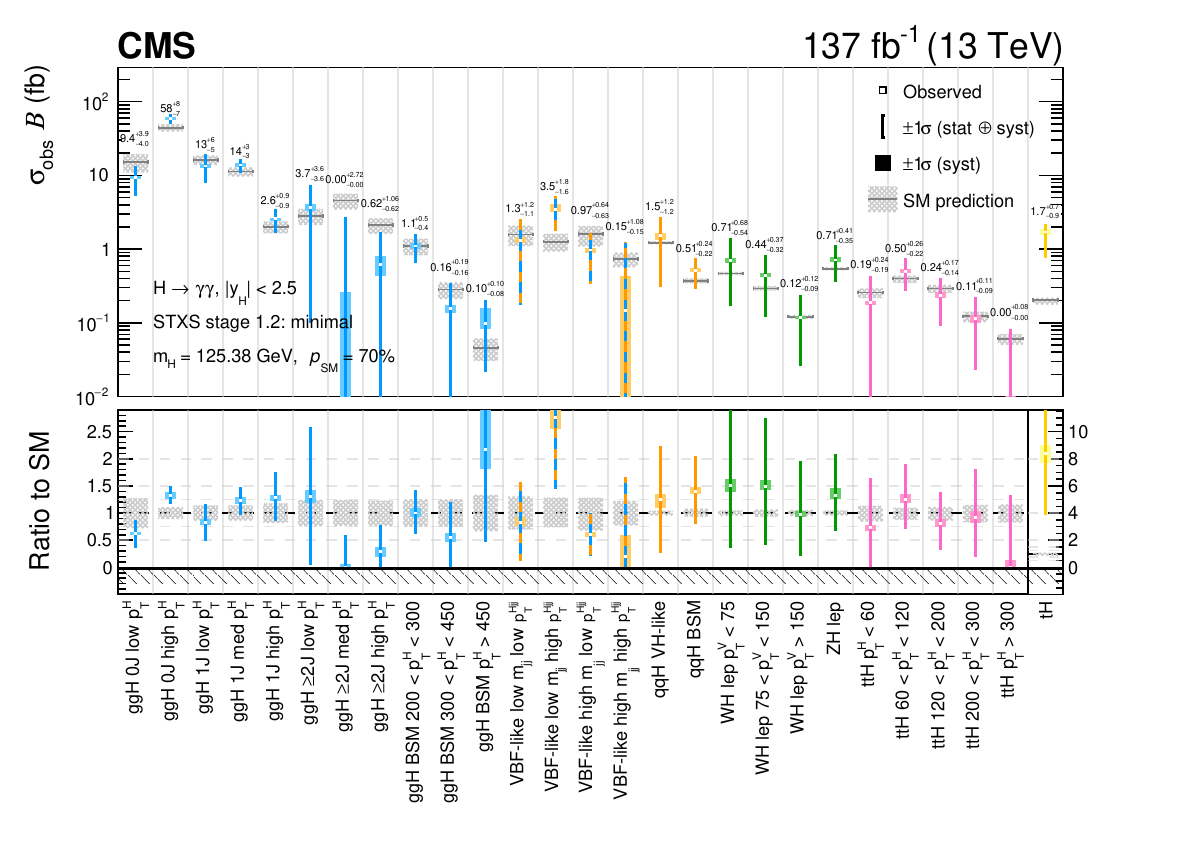}  
\caption{Observed results of the minimal merging scheme STXS fit for $\PH\to\PGg\PGg$ at 13\TeV. The best fit cross sections are plotted together with the respective 68\% confidence level intervals. Figure taken from Ref.~\cite{CMS:2021kom}.\label{fig:HiggsSTXS}}
     }
\end{figure}

The CMS experiment has measured STXS in the principal Higgs boson decay modes at 13\TeV: $\PH\to \PGg\PGg$~\cite{CMS:2021kom},  $\PH\to \ZZ\to 4\ell$~\cite{CMS:2021ugl}, $\PH\to \PGt\PGt$~\cite{CMS:2022kdi}, $\PH\to WW$~\cite{CMS:2022uhn}, and $\PH\to \PQb\PAQb$~\cite{CMS:2023vzh}. Figure~\ref{fig:HiggsSTXS} shows the STXS measurement for the $\PH\to \PGg\PGg$ process as an illustration. 

\begin{table}[htbp]
  \centering
\topcaption{Measurements of the  various fiducial cross sections of the Higgs boson for different decay modes  published by CMS using proton-proton data at a centre-of-mass energy of 13\TeV. Previous results at 7 and 8\TeV or with a partial data sample are not included in the table. 
The list of Higgs boson kinematic variables targeted in each case are listed.\label{tab:tableHiggsDiff}
}
\renewcommand{\arraystretch}{1.2}
  \begin{tabular}{c c c}
    Decay mode  &  Observables  & Data set \\
    \hline
     $\PH\to \PGg\PGg$~\cite{CMS:2022wpo} & $p_\text{T}^{\PGg\PGg}$,$n_j$,$\abs{y^{\PGg\PGg}}$,$\abs{\cos(\theta^*)}$, $\phi_{\eta}$, $n_{\PQb\text{ jet}}$, $n_\ell$, \ptmiss,
    & 137\fbinv \\
    &  $p_\text{T}^{j1}$, $\abs{y_{j1}}$, $\abs{\Delta \phi_{\PGg\PGg,j1}}$, $\abs{\Delta y_{\PGg\PGg,j1}}$, $\mathcal{T}^{j}_{C}$,  & \\
    & $p_\text{T}^{j2}$, $\abs{y_{j2}}$,$\abs{\Delta \Phi_{j1,j2}}$, $\abs{\Delta \Phi_{\PGg\PGg,j1j2}}$,    & \\
    & $\abs{\overline{\eta}_{j1 j2} -\eta_{\PGg\PGg}}$, $m_{jj}$, $\abs{\Delta\eta_{j1 j2}}$     & \\ [\cmsTabSkip] 
    $\PH\to \ZZ\to 4\ell$~\cite{CMS:2023gjz} & $\ptH$,$\abs{y_{\PH}}$, $n_j$, $p_\text{T}^{j1}$, $p_\text{T}^{j2}$, $m_{jj}$, &   138\fbinv \\
    & $\Delta \Phi_{jj}$, $\abs{\Delta \eta_{jj}}$, $m_{\PHj}$, $p_\text{T}^{\PHj}$,  $p_\text{T}^{\PHjj}$, $\mathcal{T}_{C}^\text{max}$,$\mathcal{T}_B^\text{max}$,  & \\
    & $m_{\PZ1}$,$m_{\PZ2}$,$\cos\theta^{*}$,$\cos\theta_{1}$, $\cos\theta_{2}$, $\Phi$, $\Phi_{1}$,  & \\ 
    & $D_{0-}^\text{dec}$, $D_\text{0h+}^\text{dec}$,  $D_{CP}^\text{dec}$, $D_\text{int}^\text{dec}$, $D_{\Lambda 1}^\text{dec}$,  $D_{\Lambda 1}^\text{$\PZ\PGg$,dec}$ & \\ [\cmsTabSkip] 
    $\PH\to \PGt\PGt$~\cite{CMS:2021gxc} & $\ptH$, $n_j$, $p_\text{T}^{j1}$ &   137\fbinv \\[\cmsTabSkip] 
    $\PH\to \WW$~\cite{CMS:2020dvg} & $\ptH$, $n_j$ &   137\fbinv \\ [\cmsTabSkip] 
    Boosted $\PH\to \PQb\PAQb$~\cite{CMS:2020zge}  & $\ptH$ $(\ptH>450\GeV)$  & 137\fbinv  \\ [\cmsTabSkip] 
    Combination $\PH\to \PGg\PGg$ & $\ptH$,$n_j$,$y_{\PH}$, $p_\text{T}^j$  & 36\fbinv \\
    $\PH\to \ZZ^{*}$, $\PH\to \PQb\PAQb$~\cite{CMS:2018gwt} & &  \\
  \end{tabular}
\end{table}

\subsection{Pair production of Higgs bosons}
The main mechanisms for Higgs boson pair production at the LHC were shown in Fig.~\ref{fig:Feynman}. This process has not been observed yet at the LHC because of its very small production cross section. In the SM, Higgs boson pairs are produced at the LHC  mainly via ggF, involving either couplings to a loop of virtual fermions, or the $\lambda_{\PH\PH\PH}$ coupling itself. The LO ggF Feynman diagrams shown in Fig.~\ref{fig:Feynman} have approximately the same amplitude but interfere destructively. This yields a very small SM cross section: $\sigma^{\PH\PH}_\text{ggF} =31.05^{+2.1}_{-7.2}~\fb$ at NNLO precision for a centre-of-mass energy of $\sqrt{s}=13\TeV$ and an $m_{\PH}$ of 125\GeV~\cite{Dawson:1998py,Borowka:2016ehy,Baglio:2018lrj,deFlorian:2013jea,Shao:2013bz, deFlorian:2015moa,Grazzini:2018bsd,Baglio:2020wgt}.
The CMS experiment has searched for this production in a variety of final states~\cite{CMS:2022hgz, CMS:2022cpr, CMS:2020tkr, CMS:2022omp, CMS:2022kdx, CMS:2022gjd, CMS:2022dwd} and placed limits at 95\% \CL on the production cross section and the self-coupling. The most sensitive final states are $\mathrm{HH}\to \PGg\PGg \PQb\PAQb$, $\PH\PH\to \PGt\PGt \PQb\PAQb$, $\PH\PH \to \PQb\PAQb \PQb\PAQb$, which benefit from the larger branching fraction of $\PQb\PAQb$ decays and the identification of the diphoton or ditau pair. 

\begin{figure}[htbp]
    \centering{
      \includegraphics[width=0.6\textwidth]{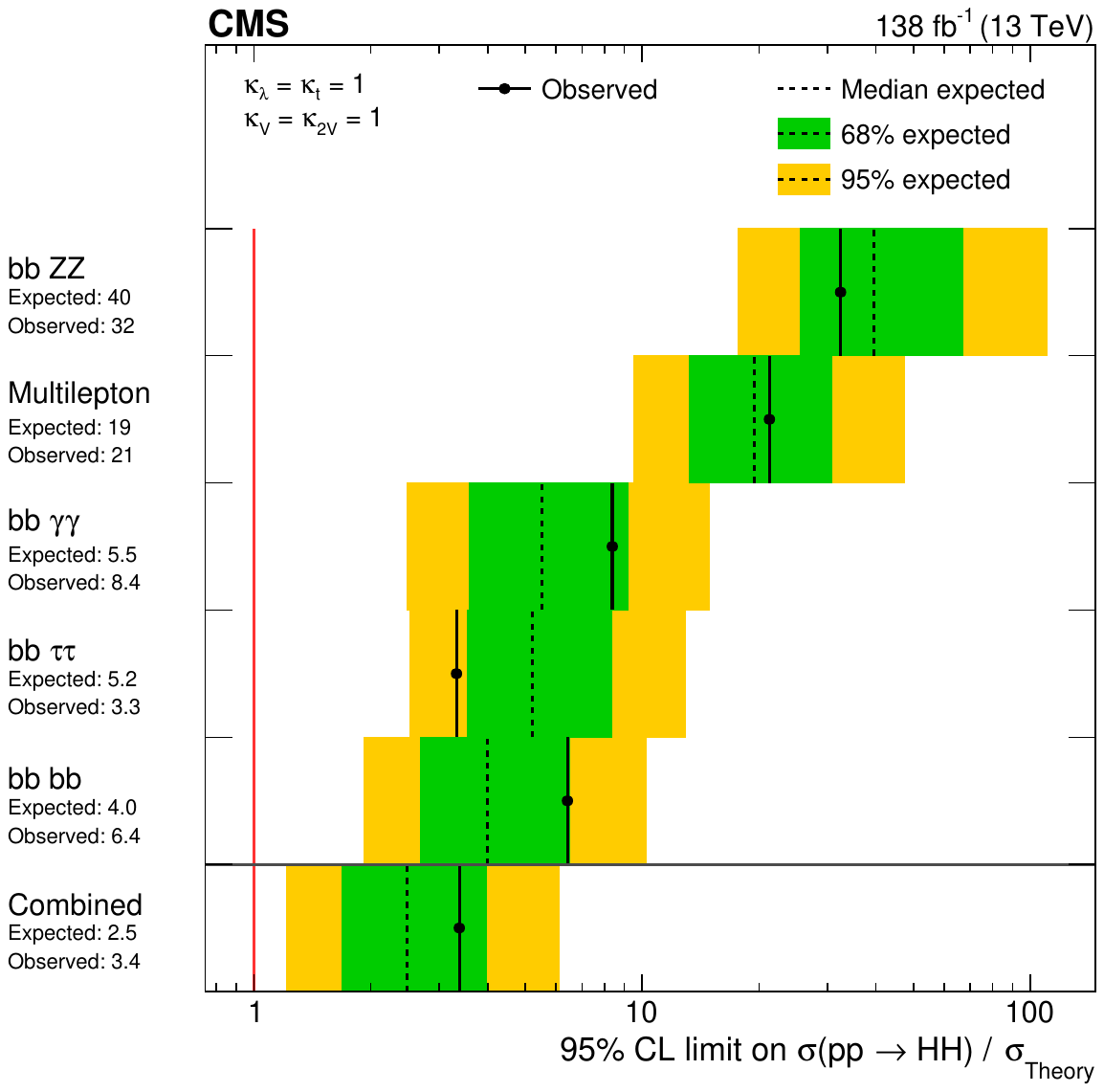}  
\caption{The expected and observed upper limits on the production of Higgs boson pairs. The results are expressed as a ratio to the SM prediction for the cross section ($\sigma(\Pp\Pp\to \PH\PH)/\sigma_\text{SM}$). A vertical red line at $\sigma(\Pp\Pp\to \PH\PH)/\sigma_\text{SM}=1$ is drawn to guide the eye. The search modes are ordered, from upper to lower, by their expected sensitivities from the least to the most sensitive. The overall combination of all searches is shown by the lowest entry. Figure taken from Ref.~\cite{CMS:2022dwd}.\label{fig:HHXS}}
}
\end{figure}

Figure~\ref{fig:HHXS} shows the expected and observed limits on Higgs boson pair production, expressed as ratios to the SM expectation, in searches using the different final states and their combination. With the current data set, and combining data from all currently studied modes and channels, the Higgs boson pair production cross
section is less than 3.4 times the SM expectation at 95\% \CL~\cite{CMS:2022dwd}.

\section{Prospects}\label{sec:prospects}

The upgraded High-Luminosity LHC machine (HL-LHC), scheduled to start running in 2029, is planned to deliver,
over its operational life, an integrated luminosity of 3000\fbinv at a collision energy of $\sqrt{s}=14\TeV$.
This will make available a data sample some 30 times larger than that used in this paper, making possible
measurements offering interesting and exciting prospects.
In addition, the CMS detector, with its trigger and readout, will be substantially upgraded for HL-LHC running, resulting in important improvements in performance. The larger data set will improve the cross section measurement of processes, where they are currently statistically limited. Constraints on PDFs at high values of $x$ will be improved, providing reduced PDF uncertainties in cross section measurements. The precision to which \alpS is known will also be improved. The larger data set will allow more detailed studies of backgrounds and allow tighter selection to reduce them, increasing the precision of the measurements of processes, where dealing with background contributes significantly to the uncertainty. It will enable a search of BSM particles some 200\GeV beyond their current mass limits in numerous suggested models. A discussion of the  physics potential of CMS during the HL-LHC can be found in Refs.~\cite{Dainese:2019rgk,ATL-PHYS-PUB-2022-018}. This section presents some of the highlights in terms of future measurements of cross sections and SM parameters. 

The remaining unobserved SM EW processes, such as production of $\ZZ\PZ$ and VBS $\ZZ$ are expected to be observed during LHC Run~3, but during the HL-LHC era
the cross section of some VBS final states will be measured with a precision similar to that of current measurements of diboson final states~\cite{ATL-PHYS-PUB-2022-018}. An interesting prospect for the full HL-LHC data set is the measurement of longitudinal VBS, a key process in establishing the mathematical consistency of the SM, because of the role played by the Higgs boson in taming the otherwise unphysical growth with energy of the calculated cross sections. Projection of the sensitivity for the full HL-LHC data set using simulation of the upgraded CMS Phase-2 detector indicates that a significance greater than $5\sigma$ can be expected for longitudinal VBS of $\WWSS$~\cite{CMS-PAS-FTR-21-001}. The uncertainty in the SM parameters, such as $\sin^2\theta^\text{eff}_{\text{lept}}$ will be reduced by a factor which may be as large as 4, due to improved statistical precision and improved constraints on PDFs. More details are reported in Section~6.1.1 of Ref.~\cite{ATL-PHYS-PUB-2022-018}.

The HL-LHC will enable better measurement of rare top quark processes, such as \ttbar\ttbar production, as discussed in Section~4.1.3 of Ref.~\cite{ATL-PHYS-PUB-2022-018}. With increased integrated luminosity for heavy ion collisions, the top quark is expected to produce significant results when used as a hard probe for nuclear PDFs, and for exploring the quark-gluon plasma~\cite{Dainese:2019rgk,CMS-PAS-FTR-18-027,Apolinario:2017sob}.

The HL-LHC will see the reduction of the uncertainties in the cross sections of all Higgs boson production modes, ranging from $<2\%$ for ggH to about 6\% for WH when both ATLAS and CMS results are combined~\cite{Dainese:2019rgk}.
A factor of 5 reduction is anticipated in the uncertainties in the measurements published so far of Higgs boson couplings to other SM particles. This will enable testing of BSM theories that predict only subtle differences in these couplings from the SM expectation.

The observation of Higgs boson pair production will be a landmark result. This process provides information on the exact shape of the BEH potential and is crucial for the understanding of the EW phase transition that occurred in the early universe, and its consequences~\cite{Isidori:2001bm}.
Projection of the $36\invfb$ analyses to $3000\invfb$ has shown that the combination of the CMS and ATLAS data sets could provide a signal significance in excess of 4 standard deviations for $\PH\PH$ production~\cite{Dainese:2019rgk}. The corresponding precision obtained on the Higgs boson self-coupling would be approximately 50\%.
The projections do not include all improvements expected from future detector upgrades. With the addition of future analysis developments, it can be hoped that the observation and first measurement of this process will take place during the HL-LHC era.
\clearpage

\section{Summary}
\begin{figure}
	\centering 
	\includegraphics[width=0.9\textwidth]{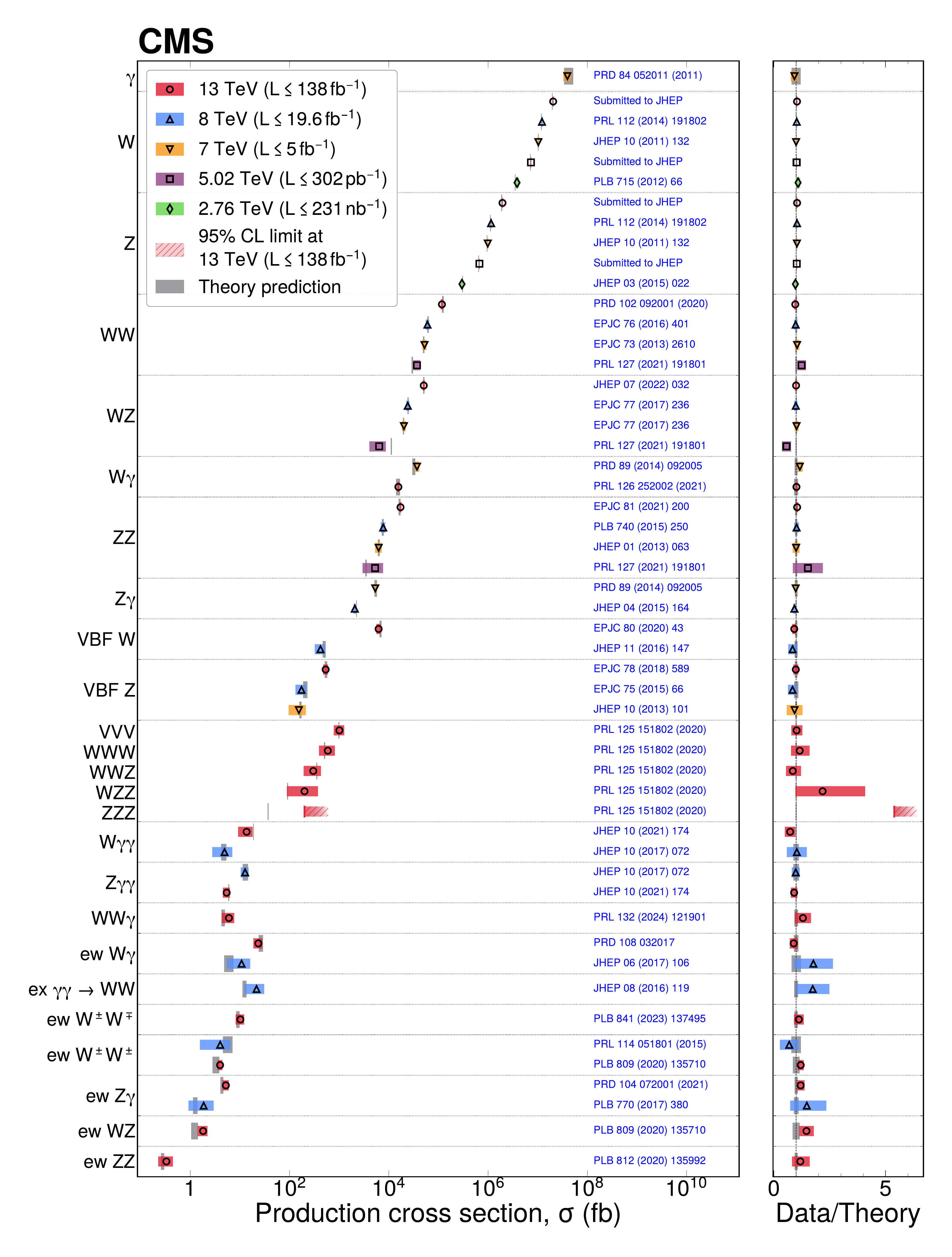}
	\caption{Summary of electroweak cross section measurements. Measurements performed at different LHC pp collision energies are marked by unique symbols and the coloured bands indicate the combined statistical and systematic uncertainty of the measurement.   Grey bands indicate the uncertainty of the corresponding SM theory predictions.   Shaded hashed bars indicate the excluded cross section region for a production process with the measured 95\% C.L. upper limit on the process indicated by the solid line of the same colour.\label{fig:XsEW}}
\end{figure}

\begin{figure}
	\centering 
	\includegraphics[width=0.9\textwidth]{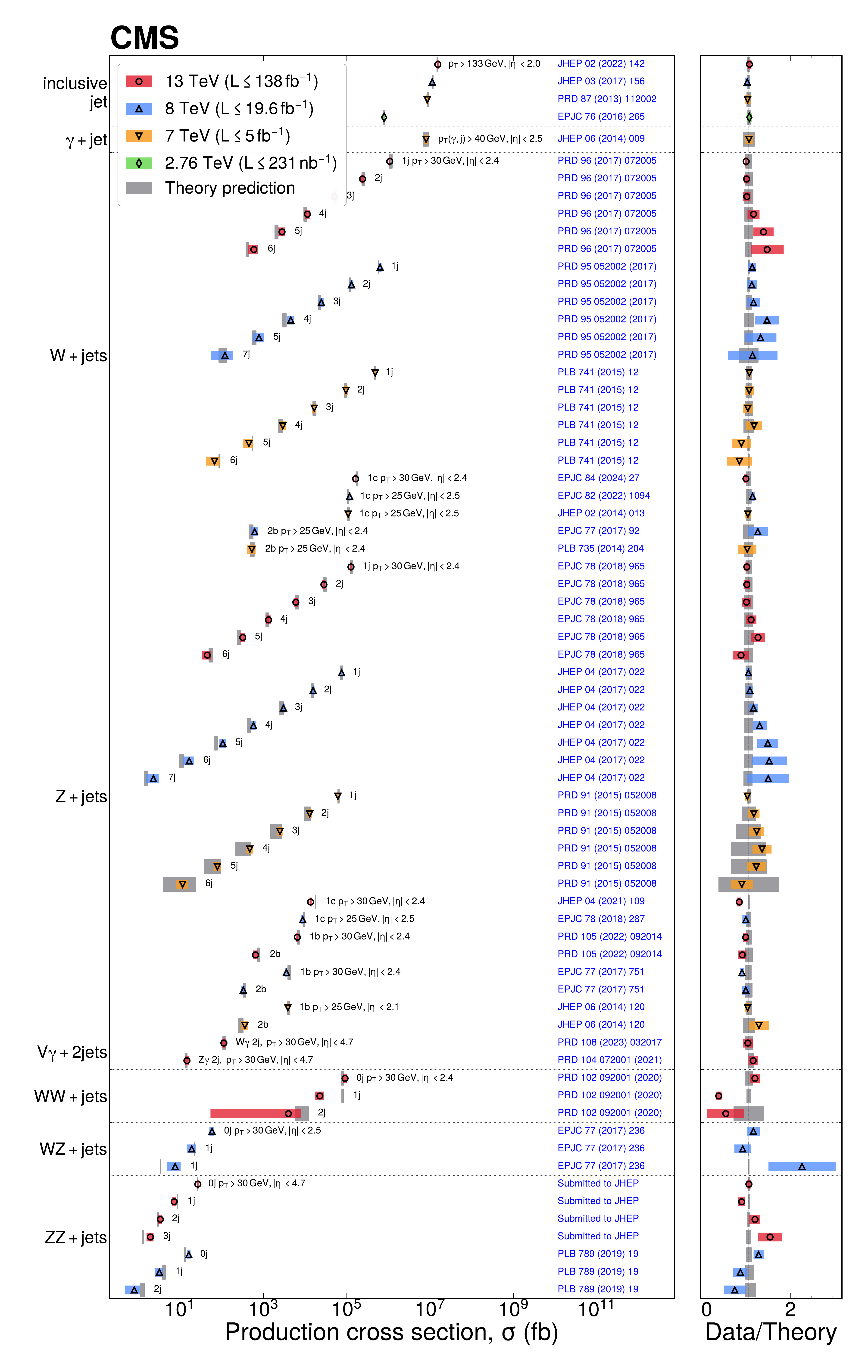}
	\caption{Summary of measurements of jet cross sections and electroweak processes in association with jets.}\label{fig:XsJetsEW}
\end{figure}

A wide selection of cross section measurements has been presented from the CMS programme of the quantum chromodynamics, electroweak, top quark, and Higgs physics. Summary plots of electroweak (Fig.~~\ref{fig:XsEW}), electroweak with jets (Fig.~\ref{fig:XsJetsEW}), top quark (Fig.~\ref{fig:XsTop}), and Higgs boson (Fig.~\ref{fig:XsHiggs}) production cross sections are shown below. No significant deviations from the standard-model (SM) predictions have been found in total or fiducial cross section measurements. Some deviations from the best predictions based on SM physics are found in differential measurements of difficult-to-model areas of phase space in events where multiple SM particles are produced including both light-flavour QCD jets and massive SM bosons or quarks. There is an expectation that improvements in the modelling of QCD and electroweak physics would result in better agreement in these measurements. These discrepancies present a challenge to improve our ability to model SM physics, rather than a sign of beyond-the-SM physics. Of particular note among the CMS cross section measurements are: the SM single $\PW$ boson production cross section determined with 1.9\% uncertainty; the ratios of $\PW$ to $\PZ$ production cross sections measured with 0.35\% accuracy; the measurement of the \WZ diboson cross section with 3.4\% precision; the measurement of the top quark pair production cross section with 3.2\% uncertainty; and the measurement of the inclusive Higgs boson production cross section with an uncertainty of 5.7\%. The achievement of sub-2\% level accuracy in production cross section measurements of massive SM particles is unprecedented at hadron colliders. The exploration of the Higgs boson through cross section measurements with high precision is one of the CMS physics programme's most exciting aspects, and the study of the Higgs boson, currently unique to the LHC, is one of our best prospects for finding signs of new physics. These CMS cross section measurements are an enduring legacy in particle physics.

\begin{figure}
	\centering 
	\includegraphics[width=0.9\textwidth]{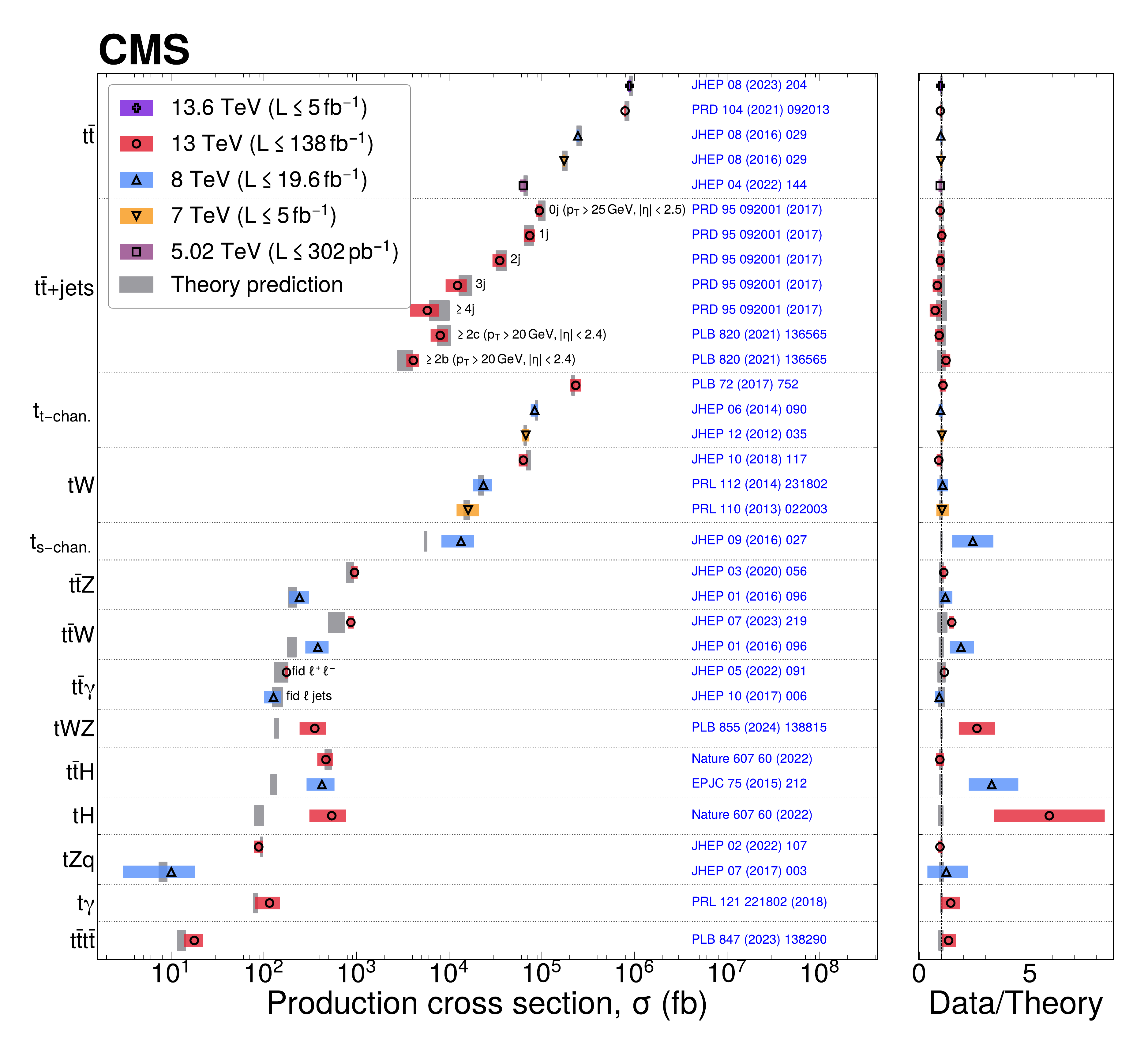}
	\caption{Summary of top quark production cross section measurements.}\label{fig:XsTop}
\end{figure}

\begin{figure}
	\centering 
	\includegraphics[width=0.9\textwidth]{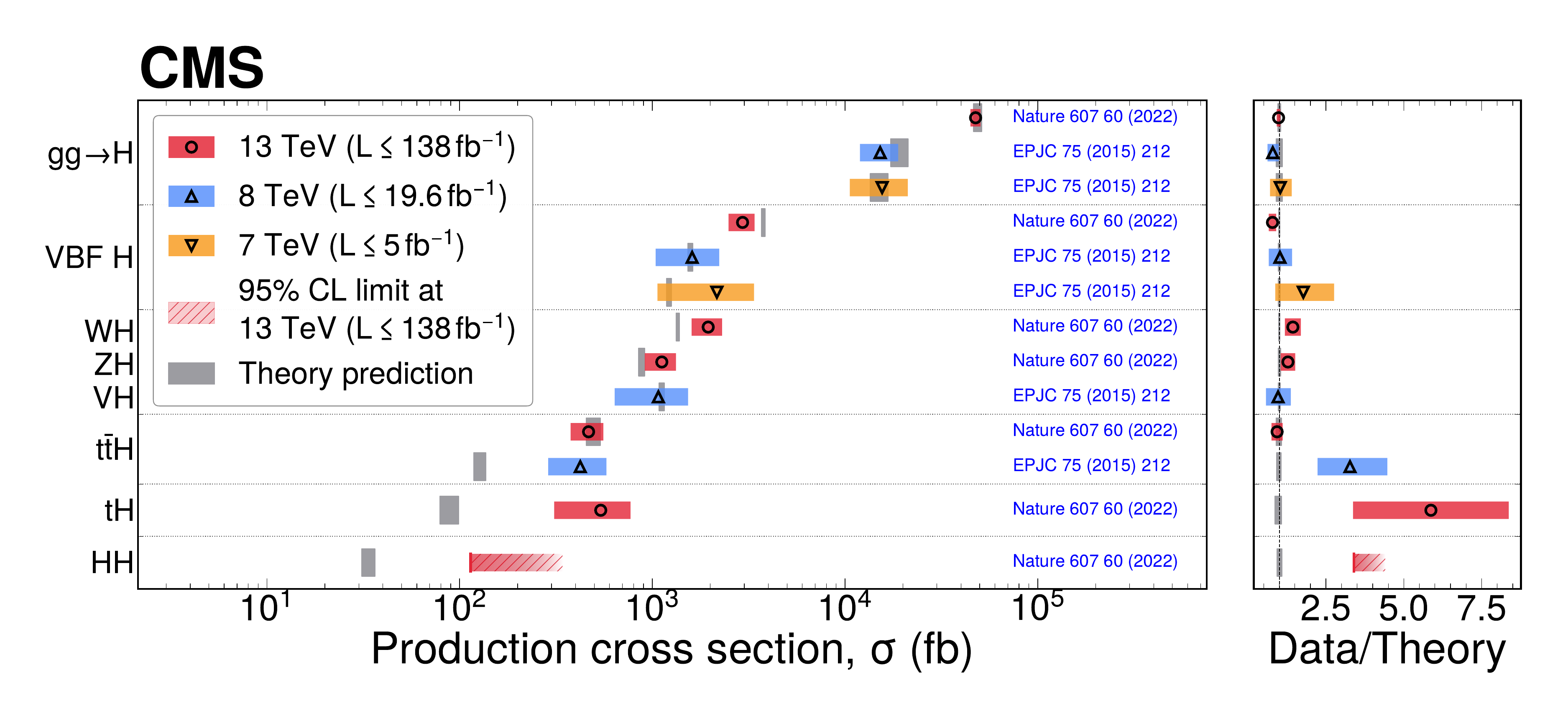}
	\caption{Summary of Higgs boson production cross section measurements.}\label{fig:XsHiggs}
\end{figure}

\begin{acknowledgments}
\hyphenation{Bundes-ministerium Forschungs-gemeinschaft Forschungs-zentren Rachada-pisek} We congratulate our colleagues in the CERN accelerator departments for the excellent performance of the LHC and thank the technical and administrative staffs at CERN and at other CMS institutes for their contributions to the success of the CMS effort. In addition, we gratefully acknowledge the computing centres and personnel of the Worldwide LHC Computing Grid and other centres for delivering so effectively the computing infrastructure essential to our analyses. Finally, we acknowledge the enduring support for the construction and operation of the LHC, the CMS detector, and the supporting computing infrastructure provided by the following funding agencies: the Armenian Science Committee, project no. 22rl-037; the Austrian Federal Ministry of Education, Science and Research and the Austrian Science Fund; the Belgian Fonds de la Recherche Scientifique, and Fonds voor Wetenschappelijk Onderzoek; the Brazilian Funding Agencies (CNPq, CAPES, FAPERJ, FAPERGS, and FAPESP); the Bulgarian Ministry of Education and Science, and the Bulgarian National Science Fund; CERN; the Chinese Academy of Sciences, Ministry of Science and Technology, the National Natural Science Foundation of China, and Fundamental Research Funds for the Central Universities; the Ministerio de Ciencia Tecnolog\'ia e Innovaci\'on (MINCIENCIAS), Colombia; the Croatian Ministry of Science, Education and Sport, and the Croatian Science Foundation; the Research and Innovation Foundation, Cyprus; the Secretariat for Higher Education, Science, Technology and Innovation, Ecuador; the Estonian Research Council via PRG780, PRG803, RVTT3 and the Ministry of Education and Research TK202; the Academy of Finland, Finnish Ministry of Education and Culture, and Helsinki Institute of Physics; the Institut National de Physique Nucl\'eaire et de Physique des Particules~/~CNRS, and Commissariat \`a l'\'Energie Atomique et aux \'Energies Alternatives~/~CEA, France; the Shota Rustaveli National Science Foundation, Georgia; the Bundesministerium f\"ur Bildung und Forschung, the Deutsche Forschungsgemeinschaft (DFG), under Germany's Excellence Strategy -- EXC 2121 ``Quantum Universe" -- 390833306, and under project number 400140256 - GRK2497, and Helmholtz-Gemeinschaft Deutscher Forschungszentren, Germany; the General Secretariat for Research and Innovation and the Hellenic Foundation for Research and Innovation (HFRI), Project Number 2288, Greece; the National Research, Development and Innovation Office (NKFIH), Hungary; the Department of Atomic Energy and the Department of Science and Technology, India; the Institute for Studies in Theoretical Physics and Mathematics, Iran; the Science Foundation, Ireland; the Istituto Nazionale di Fisica Nucleare, Italy; the Ministry of Science, ICT and Future Planning, and National Research Foundation (NRF), Republic of Korea; the Ministry of Education and Science of the Republic of Latvia; the Research Council of Lithuania, agreement No.\ VS-19 (LMTLT); the Ministry of Education, and University of Malaya (Malaysia); the Ministry of Science of Montenegro; the Mexican Funding Agencies (BUAP, CINVESTAV, CONACYT, LNS, SEP, and UASLP-FAI); the Ministry of Business, Innovation and Employment, New Zealand; the Pakistan Atomic Energy Commission; the Ministry of Education and Science and the National Science Centre, Poland; the Funda\c{c}\~ao para a Ci\^encia e a Tecnologia, grants CERN/FIS-PAR/0025/2019 and CERN/FIS-INS/0032/2019, Portugal; the Ministry of Education, Science and Technological Development of Serbia; MCIN/AEI/10.13039/501100011033, ERDF ``a way of making Europe", Programa Estatal de Fomento de la Investigaci{\'o}n Cient{\'i}fica y T{\'e}cnica de Excelencia Mar\'{\i}a de Maeztu, grant MDM-2017-0765, projects PID2020-113705RB, PID2020-113304RB, PID2020-116262RB and PID2020-113341RB-I00, and Plan de Ciencia, Tecnolog{\'i}a e Innovaci{\'o}n de Asturias, Spain; the Ministry of Science, Technology and Research, Sri Lanka; the Swiss Funding Agencies (ETH Board, ETH Zurich, PSI, SNF, UniZH, Canton Zurich, and SER); the Ministry of Science and Technology, Taipei; the Ministry of Higher Education, Science, Research and Innovation, and the National Science and Technology Development Agency of Thailand; the Scientific and Technical Research Council of Turkey, and Turkish Energy, Nuclear and Mineral Research Agency; the National Academy of Sciences of Ukraine; the Science and Technology Facilities Council, UK; the US Department of Energy, and the US National Science Foundation.

Individuals have received support from the Marie-Curie programme and the European Research Council and Horizon 2020 Grant, contract Nos.\ 675440, 724704, 752730, 758316, 765710, 824093, 101115353, 101002207, and COST Action CA16108 (European Union) the Leventis Foundation; the Alfred P.\ Sloan Foundation; the Alexander von Humboldt Foundation; the Belgian Federal Science Policy Office; the Fonds pour la Formation \`a la Recherche dans l'Industrie et dans l'Agriculture (FRIA-Belgium); the Agentschap voor Innovatie door Wetenschap en Technologie (IWT-Belgium); the F.R.S.-FNRS and FWO (Belgium) under the ``Excellence of Science -- EOS" -- be.h project n.\ 30820817; the Beijing Municipal Science \& Technology Commission, No. Z191100007219010; the Ministry of Education, Youth and Sports (MEYS) of the Czech Republic; the Shota Rustaveli National Science Foundation, grant FR-22-985 (Georgia); the Hungarian Academy of Sciences, the New National Excellence Program - \'UNKP, the NKFIH research grants K 131991, K 133046, K 138136, K 143460, K 143477, K 146913, K 146914, K 147048, 2020-2.2.1-ED-2021-00181, and TKP2021-NKTA-64 (Hungary); the Council of Scientific and Industrial Research, India; ICSC -- National Research Centre for High Performance Computing, Big Data and Quantum Computing and FAIR -- Future Artificial Intelligence Research, funded by the NextGenerationEU program (Italy); the Latvian Council of Science; the Ministry of Education and Science, project no. 2022/WK/14, and the National Science Center, contracts Opus 2021/41/B/ST2/01369 and 2021/43/B/ST2/01552 (Poland); the Funda\c{c}\~ao para a Ci\^encia e a Tecnologia, grant FCT CEECIND/01334/2018; the National Priorities Research Program by Qatar National Research Fund; the Programa Estatal de Fomento de la Investigaci{\'o}n Cient{\'i}fica y T{\'e}cnica de Excelencia Mar\'{\i}a de Maeztu, grant MDM-2017-0765 and projects PID2020-113705RB, PID2020-113304RB, PID2020-116262RB and PID2020-113341RB-I00, and Programa Severo Ochoa del Principado de Asturias (Spain); the Chulalongkorn Academic into Its 2nd Century Project Advancement Project, and the National Science, Research and Innovation Fund via the Program Management Unit for Human Resources \& Institutional Development, Research and Innovation, grant B37G660013 (Thailand); the Kavli Foundation; the Nvidia Corporation; the SuperMicro Corporation; the Welch Foundation, contract C-1845; and the Weston Havens Foundation (USA).
\end{acknowledgments}

\newpage
\appendix

\section{Glossary of terms}\label{sec:glossary}

\textit{Abbreviations:}
\begin{itemize}
  \item 4FS: four-flavour scheme ($\PQu\PQd\PQc\PQs$)
  \item 5FS: five-flavour scheme ($\PQu\PQd\PQc\PQs\PQb$)
  \item \alpS: the strong coupling
  \item aNNLL: approximate next-to-next-to-leading logarithmic (approximation)
  \item aQGC: anomalous quartic gauge boson couplings
  \item aTGC: anomalous triple gauge boson coupling
  \item BDT: boosted decision tree
  \item BSM: beyond the standard model
  \item CA: Cambridge--Aachen jet clustering algorithm
  \item CERN: Conseil Europ\'{e}en pour la Recherche Nucl\'{e}aire (English: European Council for Nuclear Research)
  \item CKM: Cabibbo--Kobayashi--Maskawa
  \item CMS: Compact Muon Solenoid
  \item CSV: Combined secondary vertex, a secondary vertex tagger used in CMS analyses 
  \item DEEPCSV: Deep learning based secondary vertex tagger used in CMS analyses 
  \item DIS: deep inelastic scattering
  \item DPS: double-parton scattering 
  \item DY: Drell--Yan quark-antiquark annihilation forming a virtual photon or \PZ boson which decays to a charged lepton-antilepton pair. Sometimes also used to refer to the similar process forming a \PW boson decaying to a lepton-antineutrino pair
  \item ECAL: electromagnetic calorimeter
  \item EW: electroweak
  \item FS: flavour schemes
  \item FSR: final-state radiation
  \item ggF: gluon-gluon fusion
  \item ggH: gluon-gluon fusion Higgs production
  \item ISR: initial-state radiation 
  \item IVF: inclusive vertex finder, secondary vertex tagger used in CMS analyses
  \item IP: interaction point
  \item IP5: interaction point 5, where the CMS experiment is located
  \item HCAL: hadron calorimeter
  \item HF: hadron forward calorimeter
  \item HL LHC: High-Luminosity LHC upgrade
  \item j: jet, also jj for two jets and jjj for three jets
  \item JES: jet energy scale
  \item JER: jet energy resolution
  \item $\ell$: charged lepton, typically an electron or a muon
  \item LHC: Large Hadron Collider
  \item LO: leading order, as in calculation in QCD or EW theory
  \item MC: Monte Carlo
  \item ME: matrix element
  \item ML: Machine learning
  \item MPI: Multiparton interactions
  \item MVA: Multivariate analysis
  \item NLL: next-to-leading logarithmic all-order resummation calculations in QCD theory. Typically used with an NLO calculation after matching the calculations to remove any overlaps.
  \item NLO: next-to-leading order, as in calculation in QCD or EW theory
  \item NNLL: next-to-next-to-leading logarithmic all order resummation calculations in QCD theory. In principle for use with an NNLO calculation but more often used as an addition to a NLO+NLL calculation.
  \item NNLO: next-to-next-to-leading order, as in calculation in QCD theory
  \item  nNNLO: NNLO QCD calculations matched to PS showers using the MiNNLO method
  \item \ncube: next-to-next-to-next-to-leading order, as in calculation in QCD theory
  \item NP: Nonperturbative, including underlying event, hadronization, and multiparton interactions
  \item nPDF: nuclear parton distribution functions
  \item os or OS: opposite-sign
  \item PB: Parton branching, as used in parton branching method transverse momentum dependent parton distribution functions PB-TMD PDFs
  \item PDF: parton (typically quark and gluon) distribution functions
  \item PF: particle flow, CMS global event reconstruction
  \item \pp: proton-proton
  \item $\Pp\PAp$: proton-antiproton
  \item pQDC: perturbative quantum chromodynamics
  \item PS: parton shower
  \item PU: pileup
  \item PUPPI: pileup-per-particle identification algorithm
  \item PV: primary vertex
  \item $Q$: momentum or energy transfer between partons in a collision
  \item QGC: Quartic gauge boson coupling
  \item QCD: quantum chromodynamics
  \item QED: quantum electrodynamics
  \item QGP: quark-gluon plasma
  \item RGE: renormalization group equation
  \item RP: Roman pot particle detectors
  \item sd: standard deviations 
  \item SM: standard model
  \item SPS: single-parton scattering 
  \item SSV: Simple secondary vertex, a secondary vertex tagger used in CMS analyses
  \item SV: Secondary vertex where a \PQb or \PQc hadron decays
  \item ss or SS: same-sign
  \item $SU$: special unitary, as in the special unitary groups $SU(2)$ and $SU(3)$
  \item TGC: triple gauge boson coupling 
  \item TMD: transverse momentum dependent, as used in parton branching method transverse momentum dependent parton distribution functions PB-TMD PDFs
  \item TPS: triple-parton scattering
  \item $U$; unitary, as in the unitary group $U(1)$
  \item UE: underlying event
  \item VBF: vector boson fusion
  \item VBS: vector boson scattering
  \item $x$: Bjorken $x$, momentum fraction of the proton carried by a parton
\end{itemize}

\textit{Units:}
\begin{itemize}
    \item b: barn $ = 1\times10^{-24}\cm^2$
    \item mb: millibarn $ = 1\times10^{-3}$\unit{b}
    \item $\mu$b: microbarn $ = 1\times10^{-6}$\unit{b}
    \item nb: nanobarn $ = 1\times10^{-9}$\unit{b}
    \item pb: picobarn $ = 1\times10^{-12}$\unit{b}
    \item fb: femtobarn $ = 1\times10^{-15}$\unit{b}
    \item eV: electronvolt $ = 1.60218\times10^{-19}$\unit{J}; energy gained by an electron traversing a potential difference of 1\unit{V}
    \item keV: kiloelectronvolt $ = 1\times10^{3}$\unit{eV}
    \item MeV: megaelectronvolt $ = 1\times10^{6}$\unit{eV}
    \item GeV: gigaelectronvolt $ = 1\times10^{9}$\unit{eV}
    \item TeV: teraelectronvolt $ = 1\times10^{12}$\unit{eV}
    \item Energy: typically given in \GeV
    \item Momentum: typically given in $\GeV$, which should be understood as $\GeV/c$
    \item Mass: typically given in $\GeV$, which should be understood as $\GeV/c^2$
\end{itemize}

\textit{Types of uncertainties in cross sections and other measurements:}
\begin{itemize}
  \item $(\alpS)$: uncertainties associated with the uncertainty in the strong coupling $(\alpS)$ (in this Report types of uncertainties are listed with parenthesis around the type)
  \item \experr: uncertainties associated with experimental sources
  \item \fit: fit uncertainty
  \item \lum: integrated luminosity uncertainty
  \item \model uncertainties associated with a model or comparisons between different models
  \item \num numerical uncertainties
  \item \param: parameter uncertainty
  \item \pdf: parton distribution function uncertainties
  \item \scale: factorization and renormalization scale uncertainties
  \item \stat: statistical uncertainty
  \item \syst: systematic uncertainty
  \item \thy: theoretical uncertainty
  \item \tot: total uncertainty
\end{itemize}

\textit{Monte Carlo simulation programs and production cross section and related process calculators. More details on the use of simulations for generating physics samples, on detector simulation, and the use of PDFs are given in Section~\ref{sec:simulation}.}
\begin{itemize}
    \item 2$\PGg$\textsc{NNLO}~\cite{Catani:2011qz}: NNLO diphoton production calculation
    \item BFG~\cite{Bourhis:1997yu}: Bourhis, Fontannazand, Guillet fragmentation functions for quarks and gluons into photons
    \item \BLACKHAT~\cite{Berger:2008ag}: Monte Carlo programs for automatic calculation of one-loop amplitudes for QCD cross sections
    \item CA3: \textsc{cascade}~\cite{CASCADE:2021bxe}: Monte Carlo event generator based on transverse momentum dependent (TMD) parton distribution functions
    \item \COMIX~\cite{Gleisberg:2008fv}: matrix element generator typically used with \SHERPA
    \item \COMPHEP~\cite{Boos:1992ap}: automatic calculation in high-energy physics from Lagrangians to collision events or particle decays
    \item \CSSHOWER~\cite{Schumann:2007mg}: parton shower program based on the Catani--Seymour dipole factorization, typically used with \SHERPA
    \item DGLAP: Dokshitzer--Gribov--Lipatov--Altarelli--Parisi~\cite{Gribov:1972ri,Lipatov:1974qm,Dokshitzer:1977sg,Altarelli:1977zs,Curci:1980uw,Furmanski:1980cm,Moch:2004pa,Vogt:2004mw} QCD evolution equations that describe the variation of PDFs with the energy scale
    \item \DYTURBO~\cite{Camarda:2019zyx}: fast predictions for Drell--Yan processes at NNLO and \ncube
    \item \FEWZ~\cite{Melnikov:2006di,Melnikov:2006kv,Gavin:2010az}: Fully Exclusive \PW and \PZ production generator
    \item \GAMJET~\cite{Baer:1989xj,Baer:1990ra}: NLL calculation of photon plus jet cross sections
    \item \GEANTfour~\cite{GEANT4:2002zbu}: toolkit for simulation of the passage of particles through matter  used for full detector simulations
    \item \GENEVA~\cite{Alioli:2015toa,Alioli:2012fc}: Monte Carlo program that combines NNLO matrix element calculations with NNLL-accuracy resummation
    \item  \textsc{HATHOR}~\cite{Aliev:2010zk,Kant:2014oha,Aliev:2010zk}: HAdronic Top and Heavy quarks crOss section calculatoR Monte Carlo program
    \item \HELACOnia~\cite{Shao:2012iz,Shao:2015vga}: onia production Monte Carlo generator
    \item \HERWIG and \HERWIGpp~\cite{Bahr:2008pv,Bellm:2015jjp}: general-purpose Monte Carlo generator
    \item \HJMINLO~\cite{Luisoni:2013kna,Hamilton:2012rf,Becker:2669113}: program for precise predictions for Lorentz-boosted Higgs boson production
    \item \JETPHOX~\cite{Catani:2002ny}: NLO photon production program
    \item \textsc{jhugen}~\cite{Gao:2010qx, Bolognesi:2012mm, Anderson:2013afp, Gritsan:2016hjl,Gritsan:2020pib}: program for simulating Higgs boson decays with full angular correlations
    \item \MADGRAPH 5 or MG5 and \MGvATNLO or MG5\_aMC~\cite{Alwall:2014hca}: automated computation of tree-level and NLO differential cross sections, matched to parton shower simulations
    \item  \textsc{matrix}~\cite{Grazzini:2017mhc}: Munich Automates qT-subtraction and Resummation to Integrate X-sections, fully automated NNLO QCD and NLO EW calculator
    \item \MCFM~\cite{Campbell:2010ff}: parton-level Monte Carlo program at NLO, NNLO, and \ncube in QCD
    \item \textsc{MiNNLO}~\cite{Buonocore:2021fnj}: nNNLO Monte Carlo simulation with NNLO QCD calculations matched to parton showers using the MiNNLO method
    \item \textsc{NLOJet++}~\cite{Nagy:2001fj,Nagy:2003tz} and \textsc{fastNLO}~\cite{Kluge:2006xs,Britzger:2012bs}: 3-jet NLO QCD calculator
    \item \NLLJET~\cite{Liu:2018ktv}: next-to-leading logarithmic cross section calculator for jet production
    \item \NNLOJET~\cite{Currie:2016bfm,Currie:2018xkj,Gehrmann:2018szu}: NNLO QCD calculator for single jet inclusive production
    \item \NNLOPS~\cite{NNLOPS1,NNLOPS2,NNLOPS3}: NNLO matched to parton shower simulation of Higgs boson production
    \item \OPENLOOPS~\cite{Buccioni:2019sur,Denner:2016kdg,Ossola:2007ax,vanHameren:2010cp}: matrix element calculator, typically used with \SHERPA for NLO+EW accuracy simulations
    \item \PHOJET~\cite{Bopp:1998rc}: Monte Carlo program for generating processes with large rapidity gaps
    \item \PHOTOS~\cite{Golonka:2005pn}: Monte Carlo program for precision simulation of QED radiation in decays. Used for description of final-state radiation
    \item \POWHEG and \POWHEGBOX~\cite{Nason:2004rx,Frixione:2007vw,Alioli:2010xd}: matching NLO QCD computations with parton shower simulations 
    \item \PYTHIA 6.4~\cite{Sjostrand:2006za}, 8.1~\cite{Sjostrand:2007gs}, 8.2~\cite{Sjostrand:2014zea}, Py: general-purpose LO Monte Carlo generator with simulation of parton showers, underlying event, and hadronization~\cite{Sjostrand:2014zea} 
   \item \SHERPA versions 1 and 2~\cite{Gleisberg:2008ta}: general-purpose Monte Carlo generator
    \item PB-TMD PDFs: transverse momentum dependent parton distribution functions~\cite{BermudezMartinez:2018fsv} based on the parton branching method~\cite{Hautmann:2017xtx,Hautmann:2017fcj}
    \item \VBFNLO \VBFNLO 2.7~\cite{Arnold:2008rz, Baglio:2014uba,Baglio:2011juf}: NLO vector boson fusion and vector boson scattering cross section Monte Carlo calculator
 \end{itemize}
\bibliography{auto_generated}
\cleardoublepage \section{The CMS Collaboration \label{app:collab}}\begin{sloppypar}\hyphenpenalty=5000\widowpenalty=500\clubpenalty=5000
\cmsinstitute{Yerevan Physics Institute, Yerevan, Armenia}
{\tolerance=6000
A.~Hayrapetyan, A.~Tumasyan\cmsAuthorMark{1}\cmsorcid{0009-0000-0684-6742}
\par}
\cmsinstitute{Institut f\"{u}r Hochenergiephysik, Vienna, Austria}
{\tolerance=6000
W.~Adam\cmsorcid{0000-0001-9099-4341}, J.W.~Andrejkovic, T.~Bergauer\cmsorcid{0000-0002-5786-0293}, S.~Chatterjee\cmsorcid{0000-0003-2660-0349}, K.~Damanakis\cmsorcid{0000-0001-5389-2872}, M.~Dragicevic\cmsorcid{0000-0003-1967-6783}, P.S.~Hussain\cmsorcid{0000-0002-4825-5278}, M.~Jeitler\cmsAuthorMark{2}\cmsorcid{0000-0002-5141-9560}, N.~Krammer\cmsorcid{0000-0002-0548-0985}, A.~Li\cmsorcid{0000-0002-4547-116X}, D.~Liko\cmsorcid{0000-0002-3380-473X}, I.~Mikulec\cmsorcid{0000-0003-0385-2746}, J.~Schieck\cmsAuthorMark{2}\cmsorcid{0000-0002-1058-8093}, R.~Sch\"{o}fbeck\cmsorcid{0000-0002-2332-8784}, D.~Schwarz\cmsorcid{0000-0002-3821-7331}, M.~Sonawane\cmsorcid{0000-0003-0510-7010}, S.~Templ\cmsorcid{0000-0003-3137-5692}, W.~Waltenberger\cmsorcid{0000-0002-6215-7228}, C.-E.~Wulz\cmsAuthorMark{2}\cmsorcid{0000-0001-9226-5812}
% ...
\par}
\vskip\cmsinstskip
\dag:~Deceased\\
$^{1}$Also at Yerevan State University, Yerevan, Armenia\\
$^{2}$Also at TU Wien, Vienna, Austria\\
% ...
\end{sloppypar}
\end{document}